\documentclass[twocolumn]{aastex63}
\usepackage{verbatim}
\usepackage{comment}
\usepackage{amsmath}
\usepackage{datetime2}
\newcommand{\July}[1]{{\color{black}  #1}}

\newcommand{\Eilen}{{\color{black}   January 29th, 2026}}
\newcommand{\Tanaan}{{\color{black} July 4th, 2026}}
\newcommand{\BigWave}{{\color{black} ``Big Wave''}}
\newcommand{\SolLC}{{\color{black} ``Solar dynamo light curve''}}
\newcommand{\Double}{``double-anomalies''}
\newcommand{\Ha}{HadISST1.1}
\newcommand{\Er}{ERSST~v6}
\newcommand{\sst}{SST}

\newcommand{\Ki}{} 
\newcommand{\Axel}{El Ni\~no}
\newcommand{\Data}{Ni\~no 4 index}
\newcommand{\WDE}{WD-effect}
\newcommand{\WDEtext}{''For any  sample window $\Delta T$,
  DCM inevitably detects the correct  $p(t)$  trend 
  and $h(t)$ \mbox{signal(-s)} when
  sample size $n$
 and/or data accuracy $\sigma$ increase.''}
\newcommand{\ConOne}{{\color{black}   \bf $\mathrm{C}_1$}}
\newcommand{\ConTwo}{{\color{black}   \bf $\mathrm{C}_2$}}
\newcommand{\ConThree}{{\color{black} \bf $\mathrm{C}_3$}}
\newcommand{\AL}[1]{{{\color{black}AL{#1}}}}
\newcommand{\UM}{{\color{black} {\bf \mathrm{UM}}}}         
\newcommand{\IF}{{\color{black} {\bf \mathrm{IF}}}}            
\newcommand{\AD}{{\color{black} {\bf \mathrm{AD}}}}         
\newcommand{\LP}{{\color{black} {\bf \mathrm{LP}}}}         
\newcommand{\SN}{{\mathrm{SN}}}
\newcommand{\PR}[1]{{\color{black}  \fontsize{8}{12}{\ttfamily #1}}}
\newcommand{\PRtext}[1]{{\color{black} {\ttfamily #1}}}
\newcommand{\Webcolour}{\color{blue}}
\newcommand{\Link}[2]{\href{#1}{\Webcolour #2}}
\newcommand{\M}{$\mathcal{M}$}
\newcommand{\REJ}{$Q_{\mathrm{F}}<\!10^{-16}$}
\newcommand{\QF}{Q_{\mathrm{F}}}
\newcommand{\SampleText}{The Seasonal Mean Nino 4 HadISST1.1 (NOAA PSL)}
\newcommand{\Cc}{$^{\mathrm{o}}\mathrm{C}$}
\newcommand{\TM}{\times}
\newcommand{\SignalOne}{S$_{11^{\mathrm{y}}}$} 
\newcommand{\SignalTwo}{S$_{10^{\mathrm{y}}}$} 
\newcommand{\SignalThree}{S$_{11.^{\mathrm{y}}86}$}

\newcommand{\Trigger}{\xrightarrow{\smash{\mathrm{Cli}}}}
\newcommand{\PM}{\pm}
\newcommand{\CModel}[1]{{\M$_{#1,\chi^2}$}}
\newcommand{\RModel}[1]{{\M$_{#1, R}$}}
\newcommand{\UU}{$\uparrow$}
\newcommand{\DD}{$\downarrow$}

\newcommand{\Bb}[1]{{\color{blue} \bf #1}}

\newcommand{\ArgOne}{{Argument 1}}
\newcommand{\ArgTwo}{{{Argument 2}}}
\newcommand{\ArgThree}{{{Argument 3}}}
\newcommand{\ArgFour}{{{Argument 4}}}
\newcommand{\CArgOne}{{{Coun\-ter\-argument 1}}}
\newcommand{\CArgTwo}{{{Coun\-ter\-argument 2}}}
\newcommand{\CArgThree}{{{Coun\-ter\-argument 3}}}
\newcommand{\Yy}{{\mathrm{~y}}}
\newcommand{\SignalNino}{S$_{12.^{\mathrm{y}}78}$}
\received{\DTMnow}
\shorttitle{Discrete Chi-square Method}
\shortauthors{Jetsu}
\begin{document}
\title{Discrete Chi-Square Method discovers solar forcing in \Axel ~time series: The Pacific Ocean as a
bolometer measuring the solar dynamo light curve in real-time
 }
\author[0000-0001-6418-5137]{L. Jetsu}
\email{lauri.jetsu@helsinki.fi}
\affil{Department of Physics, P.O. Box 64,
FI-00014 University of Helsinki,  Finland}
\begin{abstract}
\renewcommand{\Ki}{} 
     \Ki
     Discrete Chi-square Method (DCM) can
     detect multiple signals superimposed
     on an arbitrary trend.
\Ki
  DCM's backbone is Gau{\ss}-Markov theorem
  that Least Squares (LS) is
  the best unbiased estimator for linear regression models.
  \Ki
  DCM is robust because
  it computes 
  numerous linear model LS fits.
   \Ki
   Discrete Fourier Transform 
   and
     other frequency-domain methods
     have many application limitations.
  \Ki
  None of those limitations constrains DCM.
  \Ki
   Fisher-test provides signal significances
  and identifies the best  DCM model, which is validated by Forecast-test.
\Ki
Simulations verify the 
  Window Dimension Effect (\WDE):
     {\it \WDEtext}
     \Ki
     \WDE ~"sees through time".
     \Ki
     DCM's model analytical solution is ill-posed.
     \Ki
     We present a computational well-posed solution. 
\Ki
Mainstream considers \Axel ~phenomenon chaotic.
\Ki
Usual forecasts are probabilistic, not deterministic.
\Ki
We use \Axel ~time series to stress-test DCM. 
\Ki
It detects the multi-periodic \BigWave~superimposed
on     
global warming trend.
\Ki
This gives accurate \Axel ~forecasts.
\Ki
Our real-time forecast
  outperforms those of official agencies.
\Ki
Only solar forcing, not chaotic ocean-atmosphere coupling,
can cause the \BigWave ~cooling
  the Pacific Ocean at sunspot minima.
\Ki
The ocean acts like a giant bolometer
  measuring the deterministic \SolLC.
\Ki
DCM detects
    multi-periodicity
in the ocean and sunspot record.
\Ki
The mainstream stochastic dynamo
cannot cause this.
\Ki
Planetary
tidal forces may drive solar dynamo.
\Ki
Future \Axel ~models must integrate
 astrophysical cycles with chaotic climatological fluid dynamics.
\Ki
Validating our analysis now  can save trillions (USD)
in \Axel ~damages.
\end{abstract}

\section{Introduction} \label{SectInt}

\renewcommand{\Ki}{}
\citet{Jet20} formulated DCM.
  \Ki
  This current follow-up paper aims to demonstrate 
  that DCM outperforms all other parametric
  frequency-domain time series analysis methods.
  \Ki
    We validate this claim in our stress test,
  where DCM is applied to the notorious \Axel ~time series.
  \Ki
  The mainstream \Axel ~dynamical models rely
  on  supercomputers to simulate
  the chaotic and non-linear
  coupling of the ocean and atmosphere
  \citep{Tim18,Hu24}.
  \Ki
  The current short-term \Axel ~forecasts are usually probabilistic
  and typically fail beyond 18 months 
  \citep{Che04,Len24}.
  \Ki
  In contrast, our stress test will show that DCM can provide
  reliable, deterministic, long-term \Axel ~forecasts.

\citet{Che15} stated,
``Forecasting the evolution of complex systems is
noted as one of the 10 grand challenges of modern science.''
\Ki
We will show that DCM can detect the signal(-s) and
the trend in complex time series, 
and can also forecast the time series evolution.
\Ki
Our DCM is based on
the  LS method, which was originally formulated by  \cite{Leg05}.
\Ki
\cite{Gau09} connected this method to the principles
of probability and the normal distribution.
A few years later, he showed that the LS
method gives the best unbiased estimates for the free parameters
of linear models,
if the zero mean data errors are equal,
normally distributed and uncorrelated \citep{Gau21}.
The extended Gau{\ss}-Markov theorem states
that the LS method
gives the best estimates for the free parameters of linear models,
even if the data errors do not pass all the above-mentioned criteria 
\citep{Woo10}.
The Gau{\ss}-Markov theorem is the backbone of our
  DCM.
  This time series analysis method performs
  a massive number of LS fits
  to find the best model for the data.

\cite{Had02} defined the three conditions  for a well-posed problem.
The solution for the problem determines these conditions.
\begin{itemize}
\item[] \ConOne   ~Existence: A solution exists.
\item[] \ConTwo   ~Uniqueness: The solution is unique.
\item[]\ConThree ~Stability: Small changes
  in the input data cause small changes in the solution. 
\end{itemize}
\noindent
Stability means that the solution behaves predictably
as the input varies.
In other words,  there exists a continuous mapping
from the input space to the solution space.
If any of these conditions are not satisfied,
the problem is classified as ill-posed.

Ill-posed problems arise  in many fields
  of science \citep{Rud76,Tik77,Lav86}.
  \Ki
  For example, \citet{Pis90}
solved the stellar surface imaging 
inverse problem using the \citet{Tik63}
regularisation technique.
    Other
  typical examples are
  partial differential equation solutions
  \citep{Had23},
  non-linear parameter estimation
  \citep{Bar74}
  and
  regularisation of inverse problems
  \citep{Eng96,Vog02}.
  \Ki
  Ill-posed problems are encountered
  in time series analysis
\citep{Tim95,Bai12,Ber13,Box15}
because the models for the data
are often non-linear
\citep{Ton90,Tsa18}.
  \Ki
   The unknown free parameters of the model,
  namely the frequencies,
  are in the arguments of trigonometric functions
  representing the signals.
  \Ki
  This causes non-linearity because 
    the model partial derivatives contain
    free parameters of the model.

\renewcommand{\Ki}{}
\Ki
Our DCM is a frequency-domain
parametric time series analysis method.
\Ki
Here, we discuss shortly such methods
in the order of increasing model complexity.
\Ki
Numerous methods can be applied
if the time series is stationary (constant mean and variance).
\Ki
The Fast Fourier Transform (FFT) for evenly spaced data \citep{Coo65}
and DFT for unevenly spaced data \citep{Lom76,Sca82}
are the most widely-used methods.
\Ki
Astronomers have published several extensions of these methods
\citep{Zec09,Van18}.
\Ki
FFT and DFT assume that the
correct model for the time series is a pure sine.
\Ki
The spectral estimating techniques fail if the data are non-stationary.
\Ki
Therefore, trends
changing the sample mean and/or variance
must be removed
before applying spectral estimation \citep{Ner64}.
\Ki
There are numerous methods for
detecting one signal
superimposed on a trend, especially for
evenly spaced data
\citep{Cle90,Shu06,Bro09}.
\Ki
\citet{Kay81} compared many
spectral estimating techniques that can detect
more than one signal from evenly spaced data.
\Ki
They discussed how the  sample window $(\Delta T)$
limits the frequency resolution for detecting many signals
and causes leakage that shifts power from one spectral peak
to another.
\Ki
  Exactly correct frequency values may not be detected
because the leakage can shift the periodogram peaks.
\Ki
\citet{Gha21} presented different
techniques for reducing  spectral leakage.
\Ki
They also discussed techniques for
detecting many signals in unevenly spaced data,
if these signals are pure sines.
\Ki   
Astronomers apply
   DFT pre-whitening technique to detect
  many signals from unevenly spaced data \citep{Rei13,Zhu18}.
\Ki
  The data
must be detrended before applying this technique,
which detects one frequency at a time, until it
detects no new frequency.
\Ki
There are more complex modelling problems.
\Ki
For example, 
the signals are not necessarily pure sines
or the sample window $\Delta T$ is shorter
than the signal period(-s).

 \renewcommand{\Ki}{}
\Ki
Reliable forecasting is
the ultimate test for any time series analysis method
\citep{Che15,Jet25}.
\Ki
Forecasting can work for linear and stationary processes,
but it is challenging for non-linear or non-stationary ones.
\Ki
Other forecasting challenges are overfitting
and forecast error estimation
\citep{Mak87,Lef89,Gru17,Pet22}.

 \renewcommand{\Ki}{}
 \Ki
 Our DCM is a frequency-domain parametric time series analysis method.
  \Ki
Such methods
have their  own particular limitations that constrain their
applications.
    \Ki
     Here is our list of those application limitations (\AL{}).

      \begin{enumerate}

       \item Data errors (level of noise) are unknown.
       \label{LimNoise} 

            \item Data error information is not utilised. 
              \label{LimWeights}

       \item Data must be evenly spaced. 
       \label{LimSpacing} 

     \item Model parameter errors are unknown.
       \label{LimErrors}

      \item Model and forecast errors are unknown.
        \label{LimConfidence} 

      \item   Sample window is shorter than signal \mbox{period(s).} 
       \label{LimSampleWindow} $\!$ 

       \item Presence and shape of trend are unknown.
       \label{LimTrend} 

       \item Sample window causes leakage. 
       \label{LimWindow}

       \item Leakage weakens frequency resolution.
       \label{LimResolution} 

       \item Signal shapes are not pure sines.
         \label{LimNotSine} 
     
       \item Number of signals is unknown.
       \label{LimSignals} 

       \item Correct model alternative is unknown.
       \label{LimBestModel} 
        
       \item Signal significances are unknown.
       \label{LimSignificance} 
       
       \item Model solution is ill-posed.
       \label{LimIllPosed} 

       \item Complex non-linear model forecasts fail.
       \label{LimForecast}
        
  \end{enumerate}

  \noindent
  We will show that these \AL{s}
  do not constrain the DCM.
  
%
 \renewcommand{\Ki}{}

 The modelling and forecasting of \Axel ~is considered extremely challenging
  \citep{Thi24,Lu25,Cai14,Hu24,Liu23,Lud13,Tim18,Lia21}.
   \Ki
   In the mainstream climatological models,
   \Axel ~emerges from the non-linear  dynamical coupling 
   between the Pacific Ocean and atmosphere
   \citep{Bje69,Zeb87}. 
   \Ki
   Supercomputers must be used to solve millions of
   complex fluid dynamic equations of
   this chaotic interaction \citep{Can86,Bar12}.
\Ki
  The ``spring predictability barrier'' reduces
  forecast accuracy from March to May
  due to high ocean-atmosphere instability
  \citep{Web92,Lau96}.
  \Ki
  Even the best \Axel ~forecasts fail in 1.5-years
  \citep{Che04,Zha24}.


  The annual cost of \Axel ~damage
is roughly one trillion dollars \citep{Cal23,Liu23A}.
\Ki
An accurate \Axel ~prediction for just
one year ahead would save the global economy trillions of dollars 
by mitigating long-term productivity losses
\citep{Hsi11,Xu26}.
\Ki
If decadal reliable El Niño forecasts were possible,
they would allow us to prepare for these climatological disasters
well in advance.
\Ki
Such forecasts would stabilise global agriculture and economy by replacing
reactive guesswork with  proactive planning
\citep{Sol22}.

\renewcommand{\Ki}{}

The total solar irradiance varies by about
0.1\% during the solar cycle 
\citep{Wil91,Fro98}.
\Ki
Some studies have proposed a 
connection between solar cycle and climate temperature
\citep{Fri91,Sch11A,Had13,Con21}.
\Ki
Recent research has confirmed that solar irradiance
fluctuations
have a negligible impact on global warming
\citep{Gra10,Kop11,Loc20}.
\Ki
In the mainstream solar dynamo models,
the stochastic and non-stationary solar cycle
cannot cause any strictly periodic solar forcing
\citep{Uso17,Cha20}.
\Ki
The alternative planetary tidal dynamo models could provide
a precise solar cycle “clocking” mechanism
\citep{Abr12,Kle23,Mou25,Ste25,Jet25,Ste26}.
\Ki
Our DCM has already detected 
extremely significant, strictly periodic
signals in the sunspot record \citep{Jet25}.
\Ki
Reliable decadal \Axel ~forecasts may succeed,
if the solar forcing is discovered to cause strictly periodic
changes in  the \Data ~data.
\Ki
This would challenge the mainstream 
climate and solar dynamo theories.

  This study proceeds through following stages.
  We present DCM formulation (Section \ref{SectDCM}).
  The \WDE ~is discussed (Section \ref{SectWDE}).
Complex {time series} are simulated using DCM model
(Sect. \ref{SectSimulations}).
DFT and its {\AL{}s} 
are presented  (Section \ref{SectDFT}).
\Ki
 We compare our DCM to
 the renowned DFT.
\Ki
The time series analysis
of simulated data sets shows that
the  {\AL{}s} of DFT do not constrain DCM
(Sections \ref{SectModelOne}-\ref{SectSummary}). 
Then, we present
the identification of the best DCM
model for the data (Section \ref{SectBestModel}),
DCM significance estimates (Section  \ref{SectSignificance}),
the well-posed computational 
DCM model solution (Section \ref{SectIllPosed}) 
and DCM forecasts (Section \ref{SectForecast}).
We summarise why DCM outperforms other frequency-domain
parametric time series analysis methods (Section \ref{SectDiscussion}).
In our stress test, we show that DCM can model
  and forecast the \Axel ~time series (Section \ref{SectStress}).
  Many tables and figures are
    published only in Supplementary material.
  
\section{Methods}

The observations are
$y_i=y(t_i) \pm \sigma (t_i)$,
where $t_i$ are the observing
times and $\sigma_i$ are the errors
$(i=1,2, ..., n)$. 
The  sample window
is $\Delta T=t_n-t_1$.
The mid point is $t_{\mathrm{mid}}=t_1+\Delta T/2$.
The mean and the standard
deviation of all $y_i$ values are denoted with $m$ and $s$.

\subsection{Discrete Chi-square method (DCM)}
\label{SectDCM}

\cite{Jet20} introduced DCM which
is a frequency-domain parametric time series analysis method.
The primary objective of this current paper is to show
that DCM outperforms all other similar methods.

DCM model  is
\begin{eqnarray}
  g(t) = g(t,K_1,K_2,K_3)
   =  h(t) + p(t),
\label{Eqmodel}
\end{eqnarray}
where the integer values $K_1$, $K_2$ and $K_3$ are called
the model orders.
The notation $g_{K_1,K_2,K_3}(t)$ is used to specify these orders.
DCM model is a sum of two functions. These functions
are the periodic function
\begin{eqnarray}
    h(t)  =  h(t,K_1,K_2)  & = &
               \begin{cases}
                 0 , \text{if } K_1=0 \\
  \sum_{i=1}^{K_1} h_i(t),   \text{if }  K_1\ge 1 \label{Eqharmonicone}
  \end{cases} \\
h_i(t)=h_i(t,f_i) & = & \sum_{j=1}^{K_2} 
                 B_{i,j} \cos{(2 \pi j f_i t)}  \label{Eqharmonictwo} \\
             & + & C_{i,j} \sin{(2 \pi j f_i t)} \nonumber
\end{eqnarray}
and the aperiodic function
\begin{eqnarray}
  p(t)=p(t,K_3)=
  \begin{cases}
    0,                      & \text{if } K_3=-1           \\
   \sum_{k=0}^{K_3} p_k(t),  & \text{if } K_3=0, 1, 2, ...
 \end{cases}
\label{EqPolyOne}                               
\end{eqnarray}
where
\begin{eqnarray}
p_k(t) & =  & M_k \left[
                    {{2(t-t_{\mathrm{mid}}}) \over {\Delta T}}
                    \right]^k.
\label{EqPolyTwo}
\end{eqnarray} 

\noindent
For $k \ge 1$, the $p_k(t)$ function full range is
\begin{eqnarray}
  \begin{cases}
    2|M_k|,                  & \text{if $k$ is odd}           \\
    |M_k|,                   & \text{if $k$ is even.}
  \end{cases}
\end{eqnarray}

The free parameters of $g(t)$ model  are
\begin{eqnarray}
\beta & = &[\beta_1, \beta_2, ...,  \beta_{\eta}] \\
                   & = & [B_{1,1},C_{1,1},f_1, ...,  B_{K_1,K_2},  C_{K_1,K_2}, f_{K_1}, \nonumber \\
                    & &  M_0, ..., M_{K_3}]. \nonumber
\end{eqnarray}
The number of free parameters is
\begin{eqnarray}
\eta = K_1 \times (2K_2+1) + K_3+1.
\label{EqEta}
\end{eqnarray}
 We divide the free parameters $\beta$ into
 two groups $\beta_{I}$ and $\beta_{II}$.
The first group are the frequencies
\begin{eqnarray}
\bm{\beta}_{I}  & = & [f_1, ..., f_{K_1}]. \label{EqBetaOne} 
\end{eqnarray}
Due to this group,
all  free parameters are not eliminated from 
all partial derivatives $\partial g / \partial \beta_i$.
This makes the $g(t)$ model {\it non-linear}.
If the  $\beta_{I}$  frequencies have constant values,
the multipliers $ 2 \pi  j f_i$  in Equation \ref{Eqharmonictwo}
become constants.
In this case,  the model becomes {\it linear} because
the partial derivatives
$\partial g / \partial \beta_i$ 
no longer contain any free parameters.
The LS solution for the second group of remaining free parameters
\begin{eqnarray}
\beta_{II} & = & [B_{1,1}C_{1,1}, ..., 
                      B_{K_1,K_2}, C_{K_1,K_2},   \label{EqBetaTwo}\\
                          &      & M_0, ..., M_{K_3}] \nonumber
\end{eqnarray}
becomes {\it unique}.
This solution passes the \ConOne, \ConTwo ~and
\ConThree ~conditions
of a well-posed problem.

Let us assume that we search for periods
between $f_{\mathrm{min}}$ and  $f_{\mathrm{max}}$.
The non-linear $g(t)$ model becomes linear,
if the tested $\beta_{I}$
frequencies are fixed to any constant values.
The sum $h(t)$ of signals
$h_1(t,f_1)$, $h_2(t,f_2)$ ... and $h_{K_1}(t, f_{K_1})$
does not depend on the order in which these signals are added.
For example, the two signal $K_1=2$ model symmetry is
\begin{eqnarray}
  h(t)& = & h_1(t,f_1)+h_2(t,f_2)  \label{EqSymmetry}  \\
       & = & h_1(t,f_2)+h_2(t,f_1) \nonumber
\end{eqnarray}
Both $(f_1,f_2)$ and   $(f_2,f_1)$ combinations
give the same value for the $g(t)$ model.
Since this  symmetry applies to any
$K_1$ number of signals, we compute the linear $g(t)$ models
only for all tested frequency combinations
\begin{eqnarray}
  f_{\mathrm{max}} \ge f_1 > f_2 > ... > f_{K_1} \ge f_{\mathrm{min}}.
\label{EqCombinations}
\end{eqnarray}

\renewcommand{\Ki}{}

\Ki
\noindent
This frequency space symmetry idea  reduces CPU
consumption dramatically.
\Ki
For example,
were this symmetry not exploited,
the four signal $K_1=4$ search would
give $4!=24$ exactly
the same solutions from four-dimensional
tested grid in frequency space.
\Ki
This would cause unnecessary
use of CPU in the preliminary
testing all possible frequency
combinations inside
a tesseract (a four-dimensional cube).
\Ki
The search for the best frequency combination
  from $4!=24$ different non-linear iterations
  would be a pointless exercise  (see Equation \ref{EqIteration}).

\noindent
DCM model residuals
\begin{eqnarray}
  \epsilon_i=y(t_i)-g(t_i)= y_i-g_i.
  \label{EqResiduals}
\end{eqnarray}
give the sum of squared residuals
\begin{eqnarray}
R=\sum_{i=1}^n \epsilon_i^2,
\label{EqR}
\end{eqnarray}
and the Chi-square
\begin{eqnarray}
  \chi^2=\sum_{i=1}^n {{\epsilon_i^2} \over {\sigma_i^2}}.
  \label{EqChi}
\end{eqnarray}
\noindent
For every tested 
 $\beta_{I}=[f_1, f_2, ..., f_{K_1}]$ frequency combination,
    the LS fit gives DCM test statistic
  \begin{eqnarray}
    z & = z(f_1, f_2, ..., f_{K_1})=\sqrt{R/n}, &
                                                  \text{unknown $\sigma_i$}~~~~ 
    \label{EqzR} \\
    z & = z(f_1, f_2, ..., f_{K_1})=\sqrt{\chi^2/n}, &
                                                 \text{known $\sigma_i$}
    \label{EqzChi}
  \end{eqnarray}
  \noindent
computed  for a linear model.
The value of $z$ is unique.
The errors can be unknown in Equation \ref{EqzR}
  (\AL{\ref{LimNoise}}).
  For known errors, DCM uses this information
  in Equation \ref{EqzChi}
  (\AL{\ref{LimWeights}}).
In the preliminary long search, we test an evenly spaced
grid of $n_{\mathrm{L}}$ frequencies between
$f_{\mathrm{min}}$ and  $f_{\mathrm{max}}$.
This search gives the best frequency candidates
$f_{\mathrm{1,mid}}, ... f_{\mathrm{K_1,mid}}$.

In the final short search, we test a denser grid
of $n_{\mathrm{S}}$ frequencies within an interval
\begin{eqnarray}
[f_{\mathrm{i,mid}}-a,f_{\mathrm{i,mid}}+a],
\label{Eqc}
\end{eqnarray}
where  $i=1,..,K_1$, $a = c (f_{\mathrm{min}}-f_{\mathrm{max}})/2$
and $c=0.05$.

The total number of all tested long and short search
frequency combinations is
    \begin{eqnarray}
      \binom{n_f}{K_1}={{n_f!}\over{K_1! (n_f-K_1)!}},
      \label{EqCPU}
    \end{eqnarray}
where $n_f=n_L$ and $n_f=n_S$, respectively.
The even or uneven data spacing
  is irrelevant because 
  the LS fit result
  for every tested frequency combination
  does not depend on this spacing
  (\AL{\ref{LimSpacing})}.

The global periodogram minimum 
\begin{eqnarray}
  z_{\mathrm{min}}=
  z(f_{\mathrm{1,best}},f_{\mathrm{2,best}},...,
  f_{\mathrm{K_1,best}})
  \label{Eqzmin}
\end{eqnarray}
is at the tested frequencies
$f_{\mathrm{1,best}},$ $f_{\mathrm{2,best}},...,$
$f_{\mathrm{K_1,best}}$. 
This tested frequency combination
gives the best linear model for the data.
The periodogam value $z$ is a scalar,
which is  computed from $K_1$ frequency values.
It is possible to plot the $K_1=2$ two signal periodogram $z(f_1,f_2)$
as a map,
where $f_1$ and $f_2$ are the coordinates,
and $z=z(f_1,f_2)$ is the height.
For more than two signals,
there is no direct graphical $z$ plot   
because that requires more than three dimensions.
Our solution for this dimensional problem is simple.
We plot only the
following one-dimensional
slices of the full periodogram
\begin{eqnarray}
  z_1(f_1) & = & z(f_1,f_{\mathrm{2,best}}, ...,f_{\mathrm{K_1,best}})  \label{EqPeriodograms} \\
  z_2(f_2) & = & z(f_{\mathrm{1,best}},f_2, f_{\mathrm{3,best}}, ...,f_{\mathrm{K_1,best}}) \nonumber \\
  z_3(f_3) & = & z(f_{\mathrm{1,best}},f_{\mathrm{2,best}},f_3, f_{\mathrm{4,best}},...,f_{\mathrm{K_1,best}}) ~\nonumber \\
  z_4(f_4) & = & z(f_{\mathrm{1,best}},f_{\mathrm{2,best}},f_{\mathrm{3,best}},f_4, f_{\mathrm{5,best}},f_{\mathrm{K_1,best}}) \nonumber \\
  z_5(f_5) &  = & z(f_{\mathrm{1,best}},f_{\mathrm{2,best}},f_{\mathrm{3,best}},f_{\mathrm{4,best}},f_5,f_{\mathrm{K_1,best}}) \nonumber \\
  z_6(f_6) & = & z(f_{\mathrm{1,best}},f_{\mathrm{2,best}},f_{\mathrm{3,best}},f_{\mathrm{4,best}},f_{\mathrm{5,best}},f_6). \nonumber 
\end{eqnarray}
In the above-mentioned $K_1=2$ map,
the slice $z_1(f_1)$ would represent the height $z$ at
the location $(f_1,f_{\mathrm{2,best}})$
when moving along the
straight constant line $f_2=f_{\mathrm{2,best}}$
that crosses the global
minimum $z_{\mathrm{min}}$ (Equation \ref{Eqzmin})
at the coordinate point $(f_{\mathrm{1,best}},f_{\mathrm{2,best}})$.
These one-dimensional periodogram slices (Equation \ref{EqPeriodograms})
  allow us "to see inside''  the multi-dimensional structure
  of  DCM test statistic $z$
 (Equations  \ref{EqzR} and  \ref{EqzChi}).
 This kind of visualisation is important in time series analysis
 \citep{Su24}.

The short search gives the best $f_{\mathrm{1,best}}, ..., $
$f_\mathrm{K_1,best}$
frequencies for the data.
These  frequencies
are the  {\it unique} initial values for the first group of free parameters
$\beta_{\mathrm{I,initial}}=$
$[f_{\mathrm{1,best}}, ..., $ $f_\mathrm{K_1,best}]$
(Equation \ref{EqBetaOne}).
The linear model for these constant
$[f_{\mathrm{1,best}}, ..., f_\mathrm{K_1,best}]$
frequencies gives the {\it unique}
initial values for
the second group 
$\beta_{\mathrm{II,initial}}$ of free parameters
(Equation \ref{EqBetaTwo}).
The non-linear iteration
\begin{eqnarray}
  \beta_{\mathrm{initial}}=[\beta_{\mathrm{I,initial}},
  \beta_{\mathrm{II,initial}}] \rightarrow
   \beta_{\mathrm{final}}
\label{EqIteration}
\end{eqnarray}
gives the final free parameter
values $\beta_{\mathrm{final}}$.

\Ki
\citet{Fur20} emphasise that
  the analytical error estimates for the
  non-linear model free parameters can be tricky.
  \Ki
  They compare different analytical free parameter error
  estimating methods by using the computational statistical
  bootstrap technique \citep{Efr86}.
  \Ki
  They conclude  that the analytical error estimates become less reliable
   if the number of free parameters increases.
   \Ki
  For our non-linear DCM model,
  we determine the $i$:th signal parameters 
\begin{eqnarray}
P_i  & = & 1/f_i =  \text{Period} \label{EqP} \\
A_i & =  & \text{Peak to peak amplitude}  \label{EqA} \\
t_{\mathrm{i,min,1}} & = & \text{Deeper primary minimum}  \label{EqMinOne} \\
t_{\mathrm{i,min,2}} & =  & \text{Secondary minimum  (if present)}  \label{EqMinTwo} \\
t_{\mathrm{i,max,1}}& = & \text{Higher primary maximum}    \label{EqMaxOne} \\
t_{\mathrm{i,max,2}} & = & \text{Secondary maximum  (if present)~~~~} 
  \label{EqMaxTwo} 
\end{eqnarray}
and the trend parameters
\begin{eqnarray}
M_k & = & \text{Polynomial coefficients.} \label{EqMk}
\end{eqnarray}

\renewcommand{\Ki}{}

\Ki
Of these model parameters,
all $P_i=1/f_i$ and all $M_k$ estimates are among the free parameters
$\beta$.
\Ki
However, the $A_i$,
$t_{\mathrm{i,min,1}}$,
$t_{\mathrm{i,min,2}}$,
$t_{\mathrm{i,max,1}}$
and
$t_{\mathrm{i,max,2}}$ model parameters
depend on $\beta$ values.
\Ki
The analytical solutions for these model parameters are simple
for pure sines $(K_2=1)$, but become quite complicated
for double waves $(K_2=2)$.
\Ki
Clearly, the analytical solution for the errors of
DCM model parameters would be tedious.
\Ki
The analytical solution for the model error
\begin{eqnarray}
  g(t) \pm \sigma_{ g(t) }
  \label{EqModelError}
\end{eqnarray}
would be practically impossible because the number of DCM model free
parameters is large, especially if the data contains many signals.

\Ki
We use
the computational statistical bootstrap technique \citep{Efr86}
to solve the above-mentioned analytical problems.
\Ki
In our bootstrap, the tested frequencies are
the same as in the short search.
  \Ki
We select
a random sample $\bm{\epsilon}^*$
from
the residuals $\bm{\epsilon}$
of  DCM model (Equation \ref{EqResiduals}).
\Ki
Any $\epsilon_i$ value
can be chosen to the $\bm{\epsilon}^*$ sample
as many times as
the random selection 
happens to favour it.
\Ki
  We create $J=1, 2, ..., n_{\mathrm{B}}$
  residual random samples $\bm{\epsilon}^*_{J}$.
  Every $\bm{\epsilon}^*_{J}$ sample gives one
 {\it artificial} bootstrap data set
\begin{eqnarray}
\bm{y}^*_{J}=\bm{g}+\bm{\epsilon}^*_J.
\label{EqBoot}
\end{eqnarray}
Each 
$\bm{y}^*_J$ sample gives
{\it one} free parameter estimate $\beta_J$.
This bootstrap procedure
gives  $n_{\mathrm{B}}$ free parameter estimates $\beta_J$.
The standard deviations for all $n_{\mathrm{B}}$ estimates 
for $P_i=1/f_i$ and all $M_k$ give the errors of these  model parameters.
We use each $\beta_J$
to compute the model for a dense grid of time points.
This gives us  $n_{\mathrm{B}}$ {\it numerical}
$A_i$
$t_{\mathrm{i,min,1}}$,
$t_{\mathrm{i,min,2}}$,
$t_{\mathrm{i,max,1}}$
and
$t_{\mathrm{i,max,2}}$ estimates.
 The standard deviation of those $n_{\mathrm{B}}$ 
 estimates is the error of these model parameters.
 The  $\beta_J$ values give $n_{\mathrm{B}}$ estimates
 for $g(t)$ at any time  $t$.
 The standard deviation of these $g(t)$ estimates gives
 the error $\sigma_{g(t)}$ of Equation \ref{EqModelError}.
 Note that these  $\beta_J$ values can also be used
to compute the errors for $h(t), h_i (t), p(t)$ and $p_k(t)$.
Our computational statistical bootstrap approach
  gives not only the errors estimates for DCM model
  parameters (\AL{\ref{LimErrors}}),
  but also the model error 
    inside $\Delta T$ and the forecast error 
  outside $\Delta T$ (\AL{\ref{LimConfidence}}: Equation \ref{EqModelError}).

\renewcommand{\Ki}{}
  Every bootstrap procedure gives slighly different numerical
  values for the standard
  error estimates of the model parameters.
\Ki
This effect could be nearly eliminated by testing $n_{\mathrm{B}}\ge 1000$
bootstrap samples (Equation \ref{EqBoot}).
\Ki
We perform the bootstrap procedure for over one hundred models,
and $n_{\mathrm{B}}\ge 1000$ values are
practically impossible for  more complex DCM models.
\Ki
  \citet{Efr86} noted that $50 \le n_{\mathrm{B}} \le 100$ already
 give reliable standard error estimates,
 and that even $n_{\mathrm{B}}=25$ can give ``reasonable'' estimates.
\Ki
 The number of tested DCM models increases
 radically for a larger $K_1$ number of signals (see Equation \ref{EqMassive}).
 \Ki
 Therefore, we use $n_{\mathrm{B}}=50$ for one and two signal models.
 \Ki
 For three and four signal models, we use $n_{\mathrm{B}}=25$ 
 because some of them require several months of CPU.
 \Ki
 With enough CPU and patience,
 any one can use our Python code to
 re-compute more accurate DCM model parameter standard error
 estimates  by using higher $n_{\mathrm{B}}$ values 
  (Section \ref{SectGeneral}: see footnote \ref{FootBoot}).

    There are, of course, totally wrong
    DCM models for the data.
    For example,
    DCM can be forced search for too few or too many $K_1$
    signals,
    or
    the selected $p(t)$
    trend order $K_3$ can be wrong,
    as shown in Figures 5-10 by 
    \cite{Jet20}.
    Such DCM models are unstable and
    we denote them with ``$\UM$'',
    like in \cite{Jet25}.
     These unstable models have three signatures
  \begin{itemize}
  \item[]  Intersecting frequencies $(\IF)$
  \item[]  Dispersing amplitudes $(\AD)$
  \item[] Leaking periods  $(\LP)$
  \end{itemize}
 Intersecting
  frequencies  occur when the signal frequencies in the
  data are very close to each other.
  We give the following example of how  this instability can arise in
  the two signal model.
   If the frequency $f_1$ approaches the frequency $f_2$,
   both $h_1(t)$ and $h_2(t)$ signals become essentially one and
   the same signal.
   The LS fit fails because it makes no sense
   to model the same signal twice.

   Dispersing amplitudes instability can occur,
   if the two signal
   frequencies are too close to each other.
   The LS  fit finds a model,
   where two high amplitude signals 
   nearly cancel out each other.
   The low amplitude signal,
   the sum of these two high amplitude signals,
   fits to the data.

    There are DCM models  where 
    the detected frequency $f$ is
    outside the tested frequency interval
    between $f_{\mathrm{min}}$ and  $f_{\mathrm{max}}$.
    This leaking periods instability may indicate
    that the chosen tested period range is wrong.

\begin{table*}
  \caption{Model 1: DCM analysis between
    $P_{\mathrm{min}}=0.63$
    and
    $P_{\mathrm{max}}=5.70$.
    (1) Simulated $P_1$, $A_1$, $t_{\mathrm{1,min,1}}$,  $t_{\mathrm{1,max,1}}$
    and $M_0$ values.
    (2-4) Detected values for different $n$ and $\SN$ combinations.
    Two lowest lines specify electronic
    data files and DCM analysis control files.
  }\label{TableModelOne}
\begin{scriptsize}
\begin{center}
\begin{tabular}{lccccccc}
  \hline
(1)                                        & (2)                                          & (3)                                            & (4)                               \\
                                            &$n=50$                                   &$n=50$                                     &$n=100$                       \\
Model 1                                &$\SN=10$                                &$\SN=50$                                  &$\SN=100$                    \\
\hline
 $P_1=1.9$                          &$1.58\pm0.32$                         &$1.822\pm0.039$                     &$1.863\pm0.020$    \\
$A_1=2.0$                           &$1.64\pm0.64$                        &$1.894\pm0.059$                     &$1.941\pm0.031$    \\
$t_{\mathrm{1,min,1}}=1.35$&$1.20\pm0.16$                        &$1.312\pm0.019$                      &$1.332\pm0.010$    \\
$t_{\mathrm{1,max,1}}=0.40$&$0.4098\pm0.0062$                &$0.4007\pm0.0013$                &$0.40056\pm0.00066$\\
$M_0=1.0$                           &$1.20\pm0.33$                        &$1.061\pm0.031$                    &$1.030\pm0.016$    \\
 Data file                               & \PR{Model1n50SN10.dat}       &\PR{Model1n50SN50.dat}        & \PR{Model1n100SN100.dat}\\
 Control file                           & \PR{dcmModel1n50SN10.dat} &\PR{dcmModel1n50SN50.dat} & \PR{dcmModel1n100SN100.dat} \\
\hline
\end{tabular}
\end{center}
\end{scriptsize}
\end{table*}

\begin{figure*}
\vspace{0.02\textwidth}
\centerline{\hspace*{0.005\textwidth}
 \includegraphics[width=0.46\textwidth,clip=]{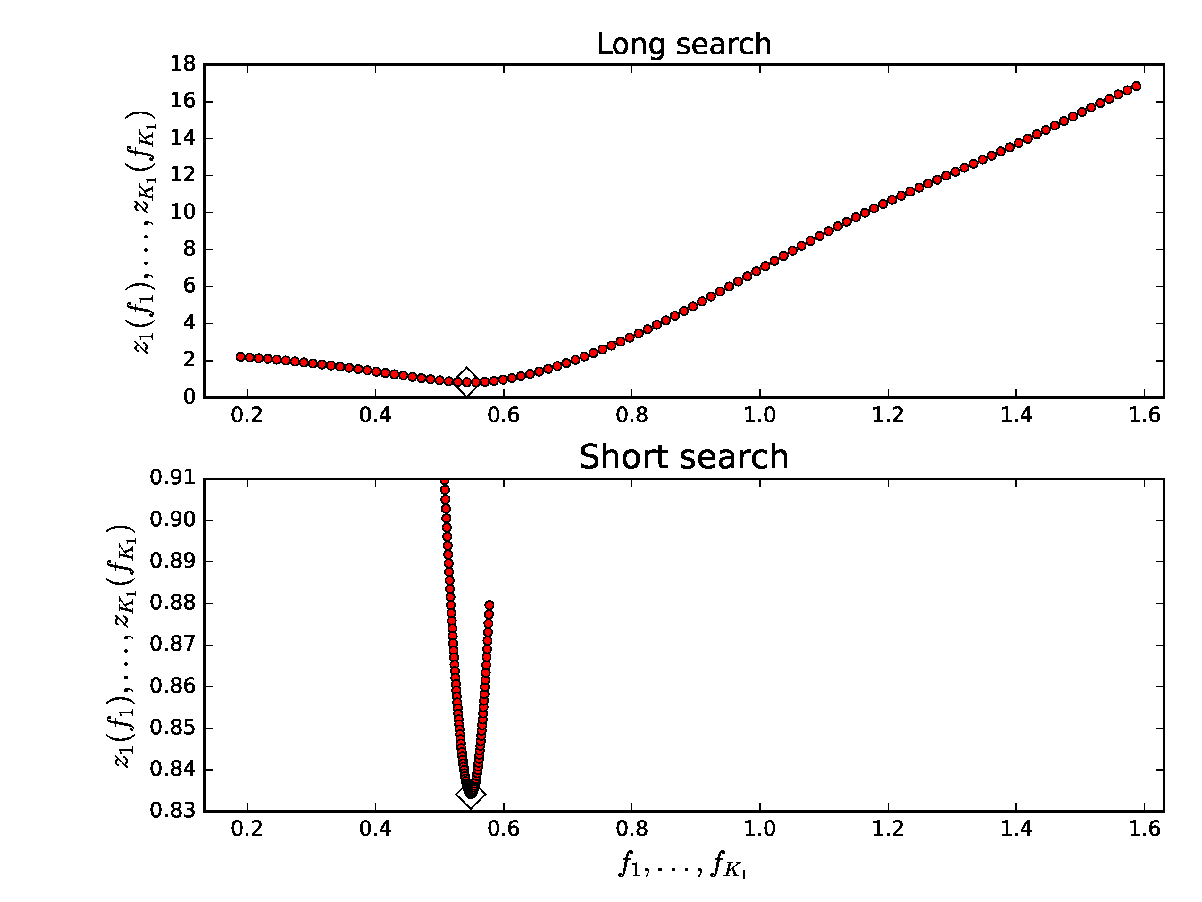} 
 \hspace*{-0.01\textwidth}
 \includegraphics[width=0.46\textwidth,clip=]{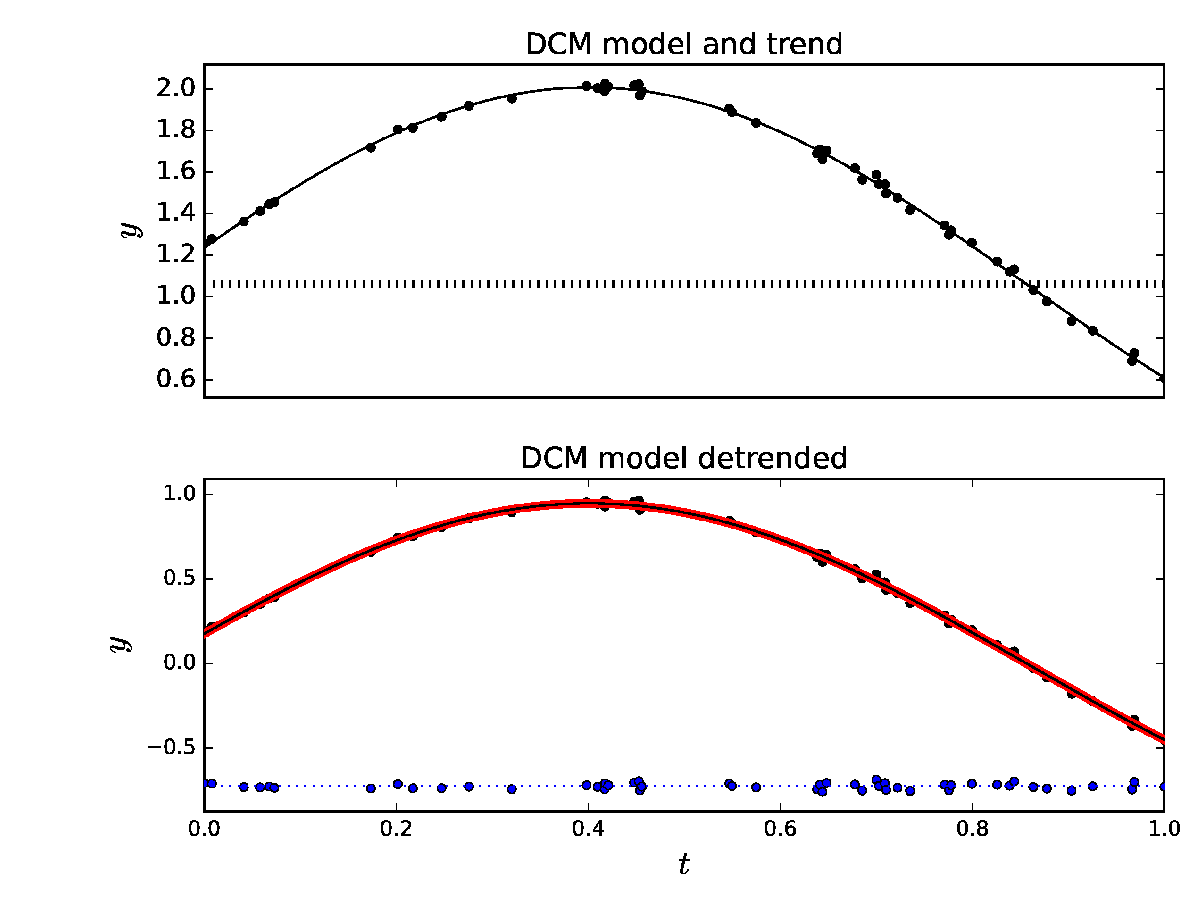} 
        }
\vspace{-0.32\textwidth}
\centerline{\Large \bf 
\hspace{0.49\textwidth}  \color{black}{(a)}
\hspace{0.39\textwidth}  \color{black}{(c)}
\hfill}
\vspace{0.118\textwidth}
\centerline{\Large \bf 
\hspace{0.50\textwidth}  \color{black}{(b)}
\hspace{0.38\textwidth}  \color{black}{(d)}
\hfill}
\vspace{0.12\textwidth}
\centerline{\hspace*{0.005\textwidth}
 \includegraphics[width=0.46\textwidth,clip=]{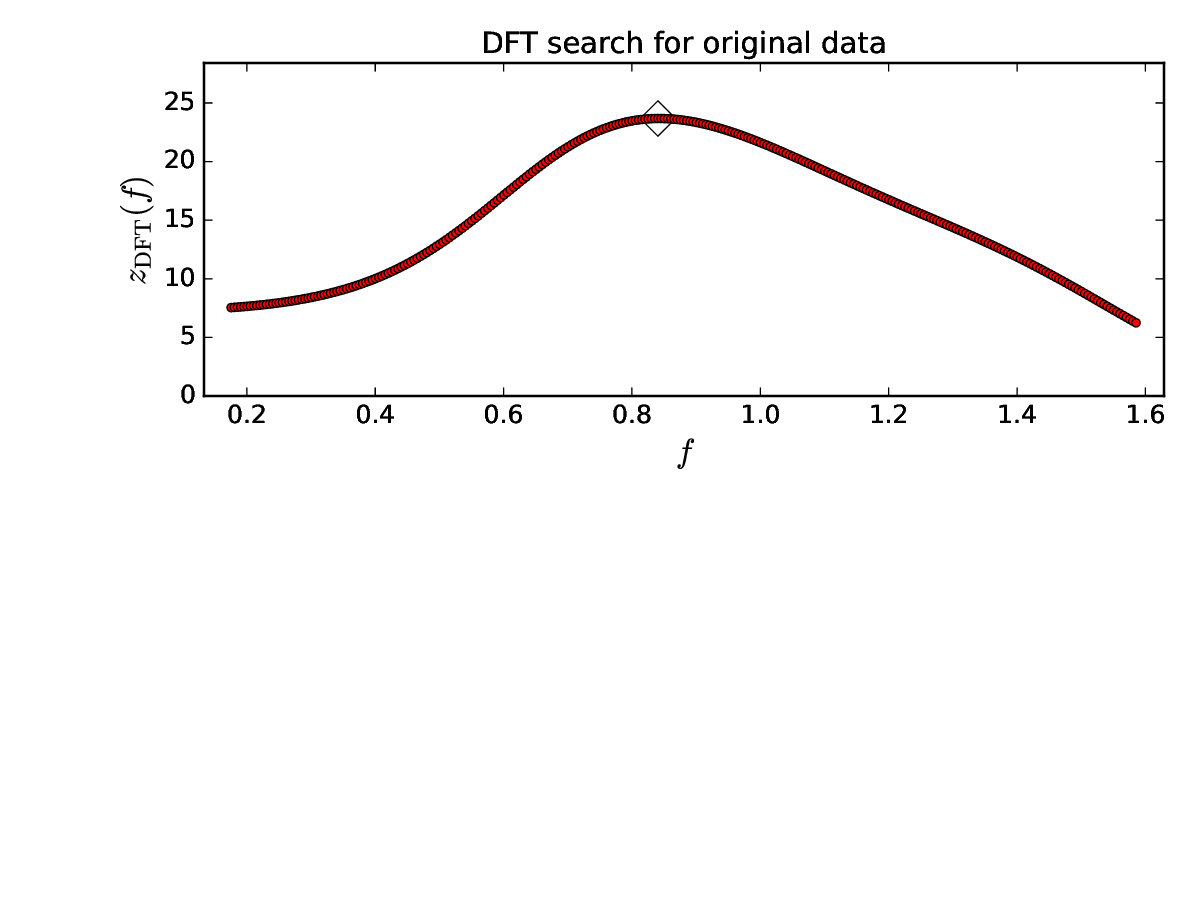}  
 \hspace*{-0.01\textwidth}
 \includegraphics[width=0.46\textwidth,clip=]{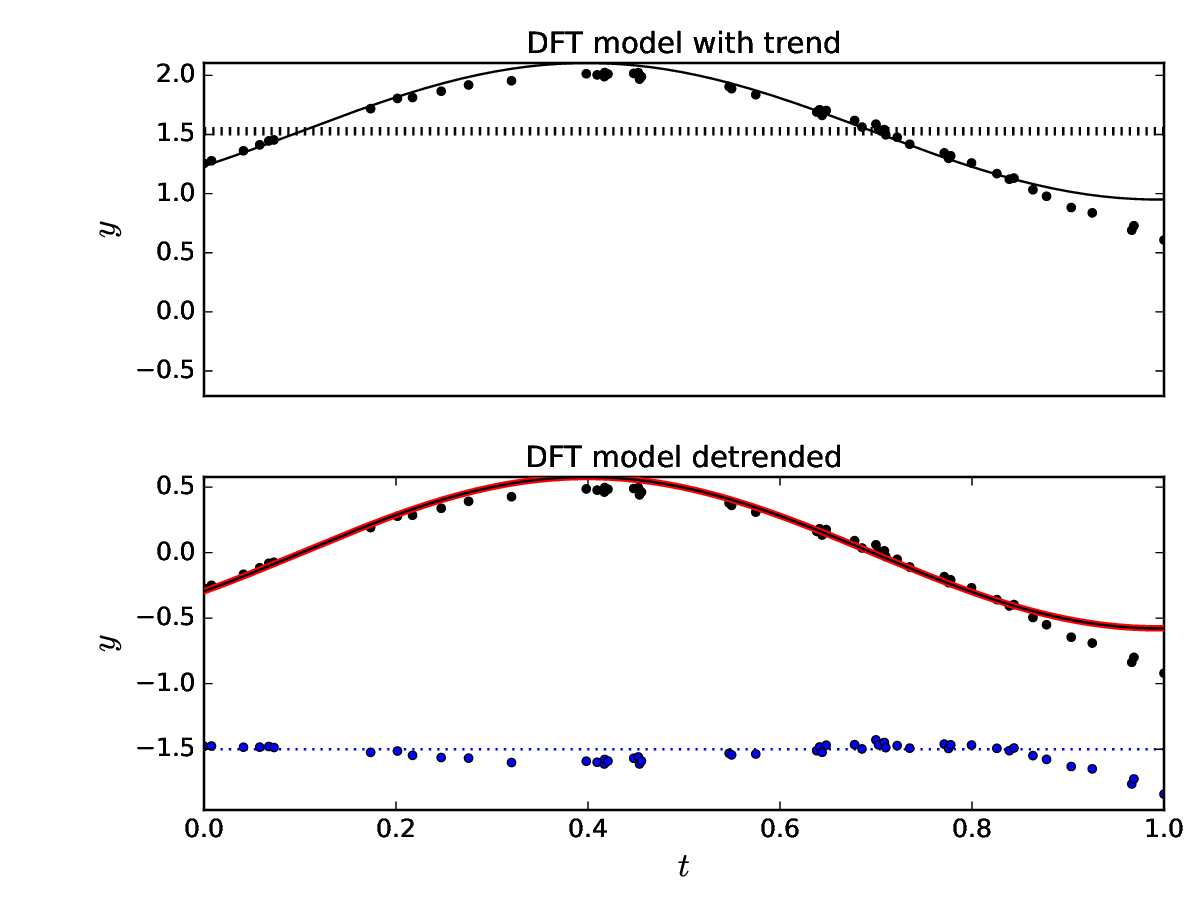}  
        }
\vspace{-0.32\textwidth}
\centerline{\Large \bf 
\hspace{0.49\textwidth}  \color{black}{(e)}
\hspace{0.39\textwidth}  \color{black}{(f)}
\hfill}
\vspace{0.12\textwidth}
\centerline{\Large \bf 
\hspace{0.96\textwidth}  \color{black}{(g)}
\hfill}
\vspace{0.15\textwidth}
\caption{Model 1 (Table \ref{TableModelOne}: $n=50$, $\SN=50$ simulation).
  (a) DCM long search periodogram $z_1(f_1)$ gives
  best period at 1.843 (diamond).
  (b) DCM short search periodogram $z_1(f_1)$ gives
  best period at 1.822 (diamond).
  (c) DCM model $g(t)$ (black continuous line),
  DCM trend $p(t)$ (black dashed line)
  and data $y_i$ (black dots).
  (d) DCM model detrended $g(t)-p(t)$ (black continuous line),
  DCM signal $h_1(t)$ (red thick continuous line),
  detrended data $y(t_i)-p(t_i)$
  (black dots) and DCM
  model residuals $y(t_i)-g(t_1)$ (blue dots)
  offset to -0.65 level (blue dotted line).
  (e) DFT periodogram $z_{\mathrm{DFT}}(f)$ gives
  best period at 1.190 (Diamond).
  (f) DFT model $g_{\mathrm{DFT}}(t)$
  (black continuous line), DFT trend $p_{\mathrm{DFT}}(t)$
  (black dashed line) and data $y_i$ (black dots).
  (g) DFT model detrended
  $g_{\mathrm{DFT}}(t)-p_{\mathrm{DFT}}(t)$
  (black continuous line),
  DFT pure sine $s_{\mathrm{DFT}}(t)$  (red thick continuous line),
  detrended data $y(t_i)-p_{\mathrm{DFT}}(t_i)$ (black dots)
  and DFT model
  residuals (blue dots) offset to -1.5 level (blue dotted line).}
\label{FigModelOne}
\end{figure*}

 DCM model (Equation \ref{Eqmodel})
  is more sophisticated than the models of our former
  time series analysis methods,
  the Three Stage Period Search \citep[][TSPA, Equation 1]{Jet99}
  and
  the Continuous Period Search \citep[][CPS, Equation 3]{Leh11}.
  TSPA and CPS can detect only one signal $(K_1=1)$
  from stationary {time series}  $(K_3=0)$.
  TSPA can detect pure sine and double wave signals
  ($K_2=1$ or 2).
  The extension of TSPA, the CPS method,
  tests three
  alternatives: $K_2=1$ or 2,
  or no signal at all.
  DCM model is more sophisticated because it can have
  any arbitrary $K_1$, $K_2$ and $K_3$ combination.
  DCM sum $g(t)$ $=h(t)+p(t)$ of arbitrary periodic and aperiodic
  functions represents a universal model  because the innumerable
$K_1$, $K_2$ and $K_3$ combinations
  allow unlimited complexity.
  \Ki
  Even more complex signal shape $(K_2)$ combinations could be used.
  \Ki
  For example,
  these shapes could be $K_{2,i=1}=3$ (3rd harmonic),
  $K_{2,i=2}=1$ (pure sine) and  $K_{2,i=3}=2$ (double wave),
  where   if $K_{2,i}$ denotes the shape of $i$:th signal.

\renewcommand{\Ki}{}

\subsection{\WDE: Spearhead of DCM}  \label{SectWDE}

In this section, we discuss what causes the \WDE ~defined
  in our abstract:
  \begin{itemize}
  \item[] {\it ``\WDEtext''}
  \end{itemize}
\Ki
The consequences of this effect are also discussed.

\Ki
The $t_i$, $y_i$ and $\sigma_i$ data 
     are inside the rectangle  $\Delta T \times \Delta y=(t_n-t_1)  \times
     [{\mathrm{max}}(y_i+\sigma_i)-{\mathrm{min}}(y_i-\sigma_i)]$.
\Ki
The LS fit results do
not depend on $\Delta T$.
\Ki
     If the measuring time intervals $\delta t_i$ for each observation $y_i$
     fulfil $\delta t_i \ll \Delta T$,
     the $R$ and $\chi^2$ values obtained from {\it all} LS
     (Equation  \ref{EqCPU})
     depend {\it only} on $y_i$ changes
      in $\Delta Y$ direction, but not on the $t_i$ changes in $\Delta T$ direction.
    \Ki
    For fixed $y_i$ and $\sigma_i$,
      the residuals $\epsilon_i$ determine the $R$ and $\chi^2$ values.
       \Ki
     These residuals measure {\it only} the $\Delta y$ direction.
      \Ki
      Hence, the periodogram $z$ values
(Equation \ref{EqzR} or   \ref{EqzChi})
      do not depend on $\Delta T$.
    \Ki
  If $\Delta T$ decreases, the $R$ and $\chi^2$ estimates
    can measure the $g_i$ model details
    inside the $\Delta T \times \Delta y$  rectangle
    only if $n$ increases and/or $\sigma$ decreases.
  \Ki
  Better data reveal these model details.

\renewcommand{\Ki}{}
\Ki
Plenty of concrete examples 
will confirm that
the \WDE ~is real
(Sections \ref{SectModelOne}-\ref{SectModelSeven}).
\Ki
We will show that DCM can detect signals(-s) when
the sample window
$\Delta T$ is shorter than the period $P$ \mbox{value(-s)}.
\Ki
DCM surpasses  \AL{\ref{LimSampleWindow}}.
\Ki
The data window $\Delta T$ length is irrelevant.
\Ki
This means that DCM can model an infinitesimal time series,
as well as forecast its future and past, 
if the sample size $(n)$ and/or
the data accuracy $(\sigma)$ are sufficient.
\Ki
This revolutionary achievement
  allows DCM to ``see through time''.
     
    \subsection{Simulated DCM model time series
      samples}\label{SectSimulations}

    We use seven different $g(t)$ models
    to simulate 21 artificial complex {time series}
(Sections \ref{SectModelOne}-\ref{SectModelSeven}).
The sample window
of all simulated {time series} is  $\Delta T=1.$
The $n^{\star}$ simulated time points $t_i^{\star}$ are
drawn from a random uniform distribution $U(0,\Delta T,n^{\star})$.
The first and last time point values are then modified to
$t_1^{\star}=0$ and $t_n^{\star}=\Delta T$.
Hence,
the distance between independent frequencies
\citep{Lou78,Kay81,Pre92}
is always
\begin{eqnarray}
  f_0=1/\Delta T=1.
  \label{Equationfzero}
\end{eqnarray}
\noindent
The $n^{\star}$ residuals $\epsilon^{\star}(t_i^{\star})$
of simulated model are drawn from a random normal distribution
$N(0,\sigma,n^{\star})$,
where $\sigma$ is the accuracy of simulated data.
The simulated data are
\begin{eqnarray}
  y^{\star}(t_i^{\star})& = &g(t_i^{\star})+\epsilon^{\star}(t_i^{\star})
                              \nonumber \\
                        & = & h(t_i^{\star})+p(t_i^{\star}) + \epsilon^{\star}(t_i^{\star}).
                              \nonumber
\end{eqnarray}

The peak to peak amplitudes of all simulated signals is $A=2.$
Our definition for the signal to noise ratio is
\begin{eqnarray}
\SN = (A/2)/\sigma.
\end{eqnarray}

\subsection{Discrete Fourier Transform (DFT)}
\label{SectDFT}

\renewcommand{\Ki}{}

  DFT is one of the most prevalent frequency-domain
  parametric time series
  analysis methods for unevenly spaced time series.
  \Ki
  It searches for the best pure sine model for the data.
   \Ki
   The equivalent DCM model has the orders $K_1=1, K_2=1$ and $K_3=0$.
   \Ki
   Any time series analysis method must be remarkable if it
  performs better than the distinguished DFT.
  \Ki
  Therefore,
  we will compare how DCM and DFT perform
  in the analyses of simulated {time series}
     (Sections \ref{SectModelOne}-\ref{SectModelSeven}). 
   \Ki
   We search for signals in these simulated {time series}
   by using the frequently
   applied DFT version
formulated by \cite{Hor86}\footnote{This paper had
  over 2600 citations in December 2025}.
\Ki
Our notation for DFT test statistic is  $z_{\mathrm{PDF}}(f)$
 \cite[][Equation 1]{Hor86}.
The notations for  DFT model are
   \begin{eqnarray}
     g_{\mathrm{DFT}}(t)=s_{\mathrm{DFT}}(t)+p_{\mathrm{DFT}}(t),
   \end{eqnarray}
   where $s_{\mathrm{DFT}}(t)$ is the sum of pure sine signals
   and $p_{\mathrm{DFT}}(t)$ is the trend.
   The  pure sine signals for the $y_i$ data and the $\epsilon_i$
   residuals are denoted with
   $s_{y,\mathrm{DFT}}(t)$
   and $s_{\epsilon,\mathrm{DFT}}(t)$, respectively.
Our DFT analyses of simulated {time series}
may fail due to  \AL{\ref{LimSampleWindow}}-\AL{\ref{LimNotSine}}.

The {time series} is ``too short''
\citep[][\AL{\ref{LimSampleWindow}}]{Kay81,Sca82,Sca89}
\begin{eqnarray}
  P > \Delta T.
\label{EqTooShort}
\end{eqnarray}

The {time series} is non-stationary
because it contains ``a trend'' \citep[][\AL{\ref{LimTrend}}]{Ner64,Chi05,Kim09}
   \begin{eqnarray}
     p_{\mathrm{DFT}}(t) \neq 0.
\label{EqTrend}
   \end{eqnarray}

   Due to the leakage caused by
     the sample window  (\AL{\ref{LimWindow}}),
     the signal frequencies are ``too close''
     \citep[][\AL{\ref{LimResolution}}]{Lou78,Kay81,Mar90}.
 \begin{eqnarray}
  |f_1-f_2|< f_0= {\Delta T}^{-1}.
\label{EqTooClose}
 \end{eqnarray}

 The signals are not ``pure sines''
 \citep[][\AL{\ref{LimNotSine}}]{Bre88,Koe95,Bal09}
 \begin{eqnarray}
   K_2 \neq 1.
 \label{EqSine}
 \end{eqnarray}

 \renewcommand{\Ki}{}

  \Ki
  Many parametric frequency-domain time series analysis methods,
  like DFT, can be applied only to stationary data.
\Ki
Trends changing the {time series} mean or variance
must be removed before applying these methods
   \citep{Ner64}.
   \Ki
   DFT can detect only one period at the time.
   \Ki
   Such frequency-domain parametric
   time series analysis methods are hereafter
   called ``one-dimensional''.
   \Ki
   Our DFT analysis of simulated data proceeds through two stages.
   \Ki
   First,
   the $p_i=p_{\mathrm{DFT}}(t_i)$
   trend is
   removed from the simulated data $y_i=y(t_i)$.
   \Ki
   Then, the iterative pre-whitening technique
   \citep{Rei13,Zhu18,Gha21}
   is applied to search for the pure sine signals.
   \Ki
   We search for the first signal
   by applying DFT
   to the detrended $y_i-p_i$ simulated data.
   \Ki
   In the second signal search,
   the DTF is applied to
   the original model residuals
   $\epsilon_i=(y_i-p_i)-g_{\mathrm{y,DFT}}(t_i)$.
   \Ki
   At any stage, this combination of detrending and iterative pre-whitening analysis
   may fail and corrupt the results obtained at the next stages.
   \Ki
   For example, all DFT analyses of simulated {time series}
   will fail already at the detrending stage
   (Sections \ref{SectModelOne}-\ref{SectModelSeven}). 
   \Ki
   Our DCM has no corrupting separate stages because
  the trend and the signal(-s) are detected (modelled) simultaneously.

  \section{Results of
    DCM analysis for simulated data} \label{SectResults}

The validity of the \WDE ~is critical to the credibility of our work.
  We confirm this validity thoroughly
  by applying DCM and DFT to seven different simulated time series
(Sections \ref{SectModelOne} - \ref{SectModelSeven}).
As we proceed, these
  {time series} become increasingly complex.
DCM analysis succeeds for all {time series}.
DFT analysis fails for every {time series}.

\subsection{Model 1 } \label{SectModelOne}

Our first {time series} simulation model is the one signal model
\begin{eqnarray}
  g(t)=
  (A_1/2)
  \cos{
  \left[
  {
  {2 \pi (t-t_{\mathrm{1,max,1}})}
  \over
  {P_1}
  }
  \right]
  }+M_0.
\end{eqnarray}
We give the $P_1$, $A_1$, $t_{\mathrm{1,max,1}}$ and $M_0$
values in Table \ref{TableModelOne} (Column 1).
This sample is ``too short''
(Equation \ref{EqTooShort})
because
$P_1=1.9 \Delta T$.
The constant mean level $M_0$ is unknown (Equation \ref{EqTrend}).
We perform DCM and DFT time series  analysis between
$P_{\mathrm{min}}=P_1/3=0.63$ and
$P_{\mathrm{max}}=3P_1=5.70$.

Model 1 is a DCM model, where $K_1=1$, $K_2=1$, $K_3=0$,
$B_{1,1}=$ $(A_1/2) \cos{(2\pi f_1 t_{\mathrm{1,max,1}})}$ and
$C_{1,1}=$ $(A_1/2) $ $\sin{(2\pi f_1 t_{\mathrm{1,max,1}})}.$
We give DCM analysis results for three samples
having different $n$ and $\SN$ combinations
(Table \ref{TableModelOne}: Columns 2-4).
For each sample, this table specifies
the electronic data file
and the electronic DCM
control file.\footnote{See declaration ``Code and data availability''.}
The  detected $P_1$, $A_1$, $t_{\mathrm{1,max,1}}$ and $M_0$
values are correct
and accurate even for the combination $n=50$ and $\SN=50$.
Regardless of $\Delta T < P_1$,
these model parameter values become more accurate
and converge to the correct simulated values
when $n$ and $\SN$ increase.
This confirms the \WDE.

A graphical presentation of DCM analysis results is shown for Model 1
simulated {time series}, where
$n=50$ and $\SN=50$  (Figures \ref{FigModelOne}a-d).
DCM long search $z_1(f_1)$ periodogram  minimum is at 
$P_1=1.843$
(Figure \ref{FigModelOne}a: diamond). 
DCM short search periodogram $z_1(f_1)$ gives
the best period at $P_1=1.822$ 
(Figure \ref{FigModelOne}b: diamond).
The continuous black line denoting DCM model $g(t)$ 
fits perfectly to the black dots denoting the data $y_i$ 
(Figure \ref{FigModelOne}c).
The mean $p(t)=M_0=1.061\pm0.036$
is correct
(Figure \ref{FigModelOne}c: dashed black line).
The detrended model $g(t)-p(t)$ (black continuous line),
the detrended data $y(t_i)-p(t_i)$ (black dots) and
the pure sine signal $h_1(t)$ (red thick continuous line)
are shown in Figure \ref{FigModelOne}d.
Note that the thick continuous red line stays
under the thin continuous black line
because $h_1(t)=g(t)-p(t)$.
DCM residuals (blue dots) are offset to
the level of -0.65 (blue dotted line).
These residuals are stable and display no trends.

DFT detects the wrong period $P_1=1.190$
(Figure \ref{FigModelOne}e: Diamond).
DFT mean level estimate  $M_0=1.527$ is also wrong
(Figure \ref{FigModelOne}f: Dashed black line).
The  black dots denoting the data $y_i$ deviate from
the continuous
black line denoting DFT model $g_{\mathrm{DFT}}(t)$.
The detrended model $g_{\mathrm{DFT}}(t)-p_{\mathrm{DFT}}(t)$
(continuous black line),
the detrended data $y(t_i) - p_{\mathrm{DFT}}(t_i)$
(black dots)
and
the signal $s_{\mathrm{y,DFT}}(t)$
(continuous thick red line)
are shown in Figure \ref{FigModelOne}g.
The thin black line covers the thick red line because
$s_{\mathrm{y,DFT}}(t)=g_{\mathrm{DFT}}(t)-p_{\mathrm{DFT}}(t)$.
DFT residuals (blue dots) offset to the level of -1.5 (blue dotted line)
display obvious trends, especially at the end of analysed sample.

For the simulated {time series} of Model 1,
DCM analysis succeeds,
but DFT analysis fails.

\begin{table*}
  \caption{Model 2. DCM analysis between
    $P_{\mathrm{min}}=0.63$
    and
    $P_{\mathrm{max}}=4.70$.
Notations as in Table \ref{TableModelOne}.  }\label{TableModelTwo}
\begin{scriptsize}
 \begin{center}
\begin{tabular}{lcccc}
  \hline
  (1)                                     & (2)                                               & (3)                                          & (4)                               \\
                                            &$n=1~000$                                 &$n=10~000$                          &$n=10~000$              \\
Model 2                                &$\SN=100$                                  &$\SN=100$                             &$\SN=200$                  \\
\hline
$P_1=1.9$                           &$1.98\pm0.28$                            &$1.852\pm0.078$                   &$1.933\pm0.033$    \\
$A_1=2.0$                           &$2.4\pm2.1$                                &$1.84\pm0.30$                       &$2.13\pm0.14$        \\
$t_{\mathrm{1,min,1}}=1.35$&$1.39\pm0.14$                            &$1.326\pm0.039$                   &$1.367\pm0.016$    \\
$t_{\mathrm{1,max,1}}=0.40$&$0.4015\pm0.0017$                   &$0.40007\pm0.00068$           &$0.40017\pm0.00064$\\
$M_0=1.0$                           &$0.8\pm1.0$                               &$1.08\pm0.14$                      &$0.937\pm0.066$       \\
$M_1=0.25$                         &$0.29\pm0.17$                           &$0.229\pm0.038$                  &$0.266\pm0.016$     \\
$M_2=0.50$                         &$0.62\pm0.42$                           &$0.451\pm0.094$                  &$0.540\pm0.041$      \\
Data file                                &\PR{Model2n1000SN100.dat}      &\PR{Model2n10000SN100.dat}&\PR{Model2n10000SN200.dat}\\
Control file                            &\PR{dcmModel2n1000SN100.dat}&\PR{dcmModel2n10000SN100.dat}&\PR{dcmModel2n10000SN200.dat}\\
  
\hline
\end{tabular}
\end{center}
\end{scriptsize}
\end{table*}

\begin{figure*}
\vspace{0.02\textwidth}
\centerline{\hspace*{0.005\textwidth}
 \includegraphics[width=0.46\textwidth,clip=]{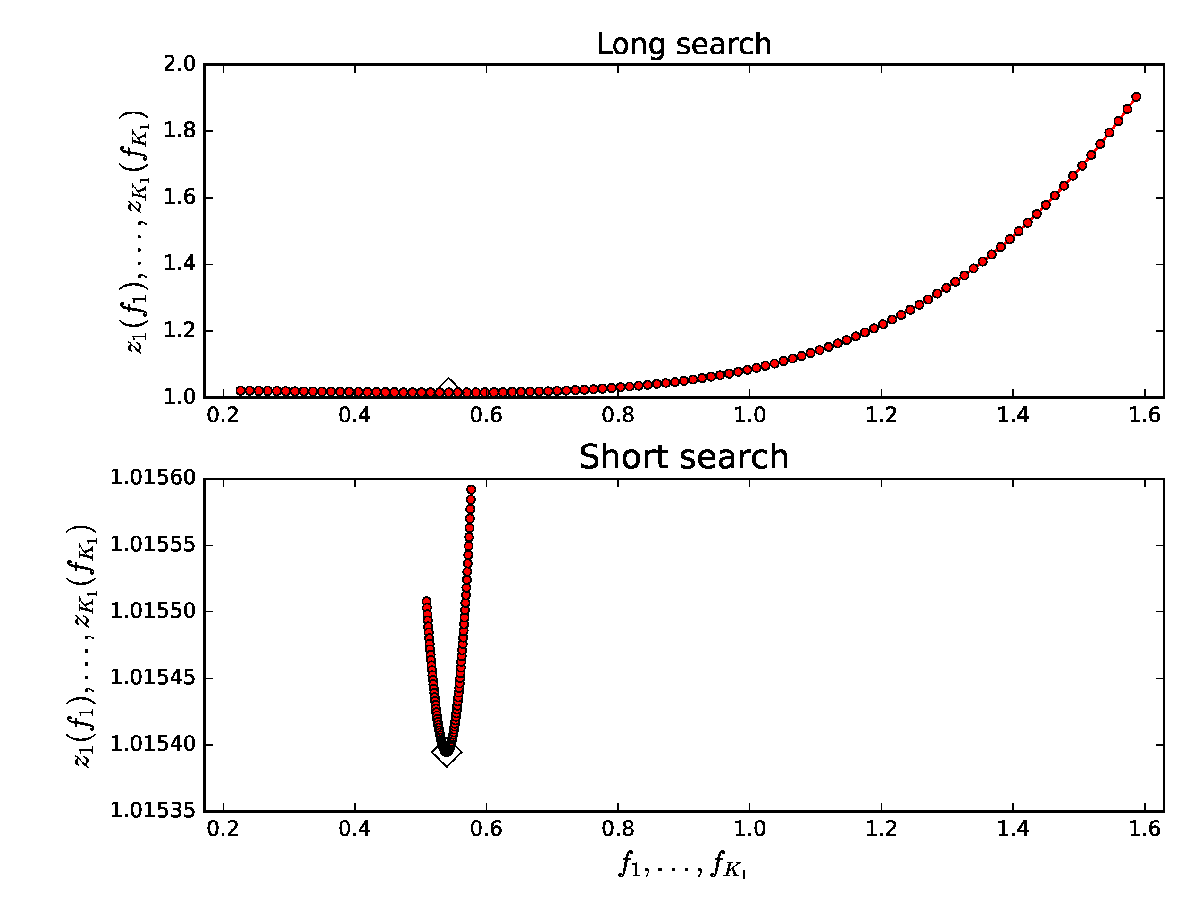}
 \hspace*{-0.01\textwidth}
 \includegraphics[width=0.46\textwidth,clip=]{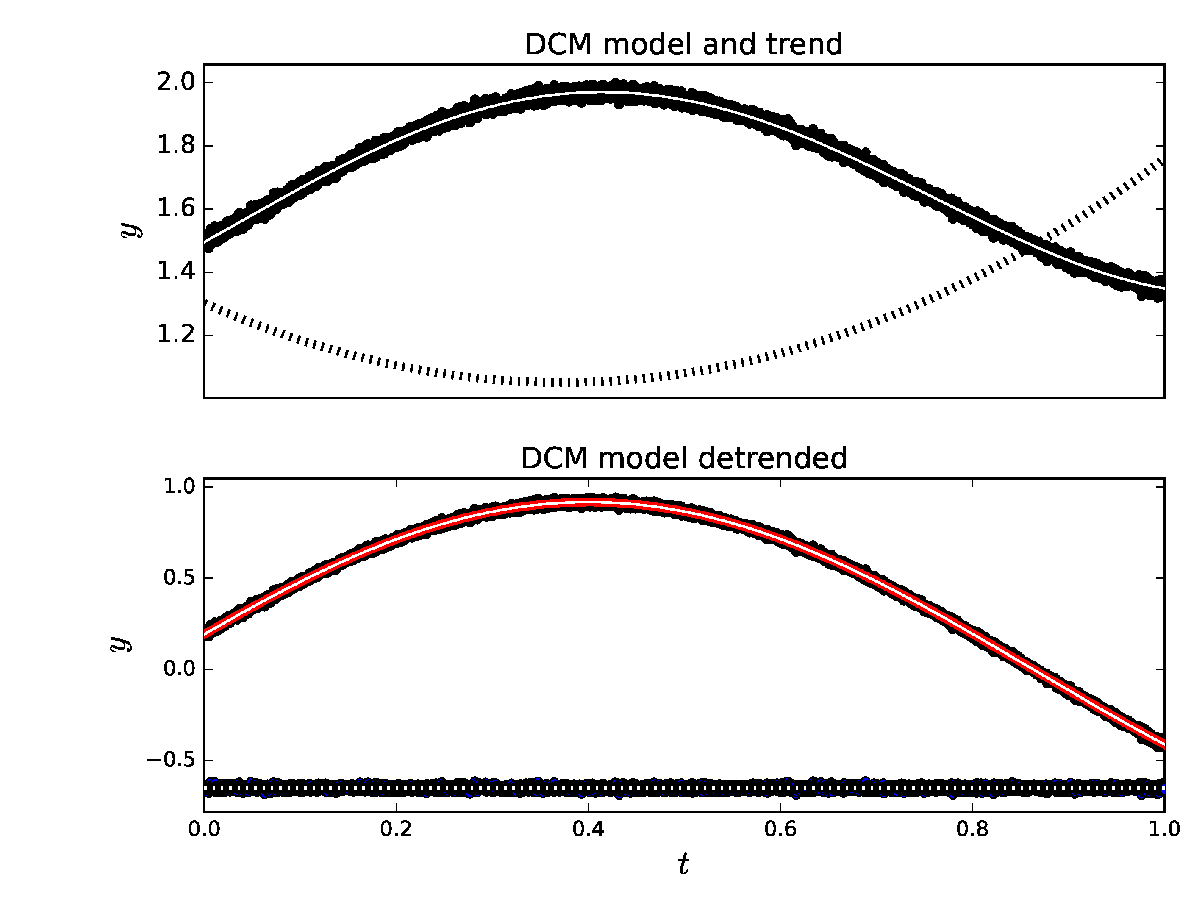}
        }
\vspace{-0.32\textwidth}
\centerline{\Large \bf 
\hspace{0.49\textwidth}  \color{black}{(a)}
\hspace{0.39\textwidth}  \color{black}{(c)}
\hfill}
\vspace{0.118\textwidth}
\centerline{\Large \bf 
\hspace{0.495\textwidth}  \color{black}{(b)}
\hspace{0.385\textwidth}  \color{black}{(d)}
\hfill}
\vspace{0.12\textwidth}
\centerline{\hspace*{0.005\textwidth}
 \includegraphics[width=0.46\textwidth,clip=]{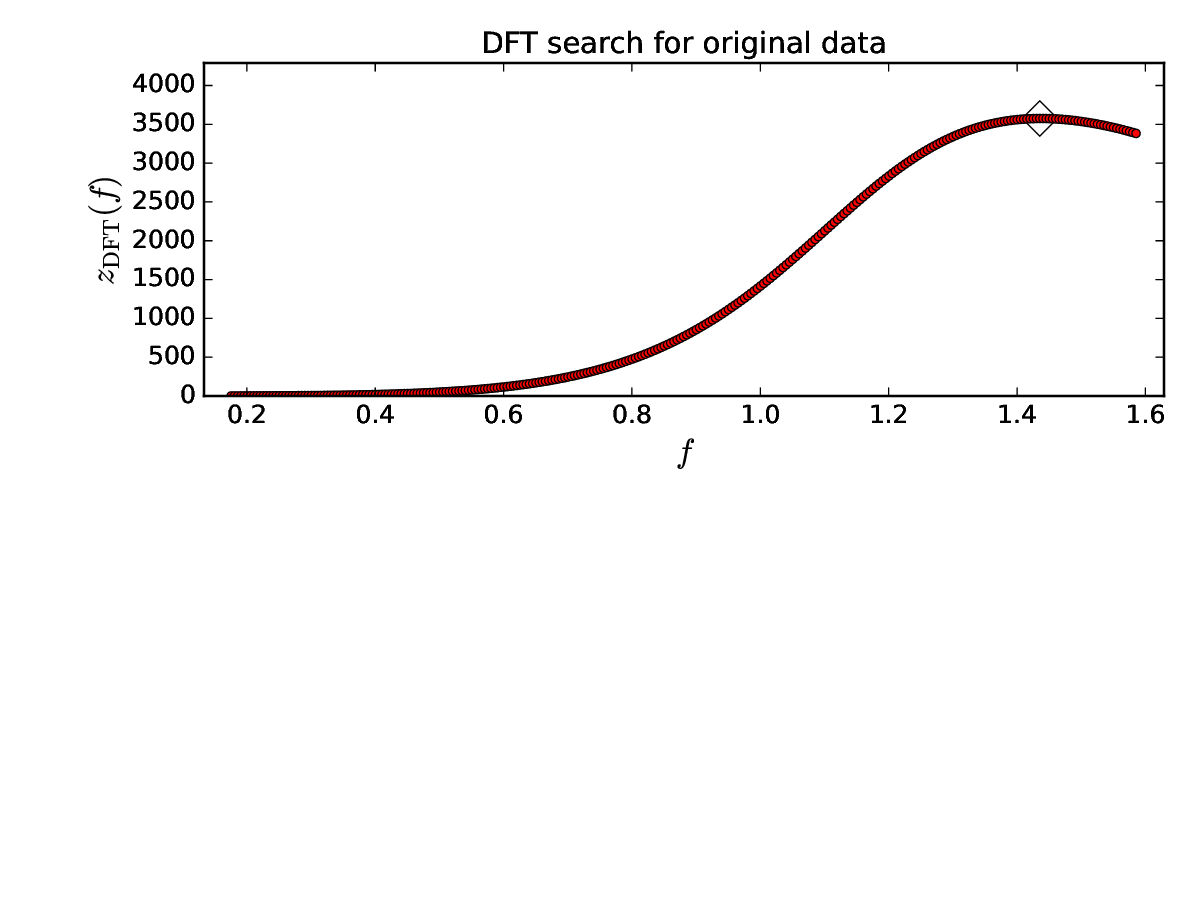}
 \hspace*{-0.01\textwidth}
 \includegraphics[width=0.46\textwidth,clip=]{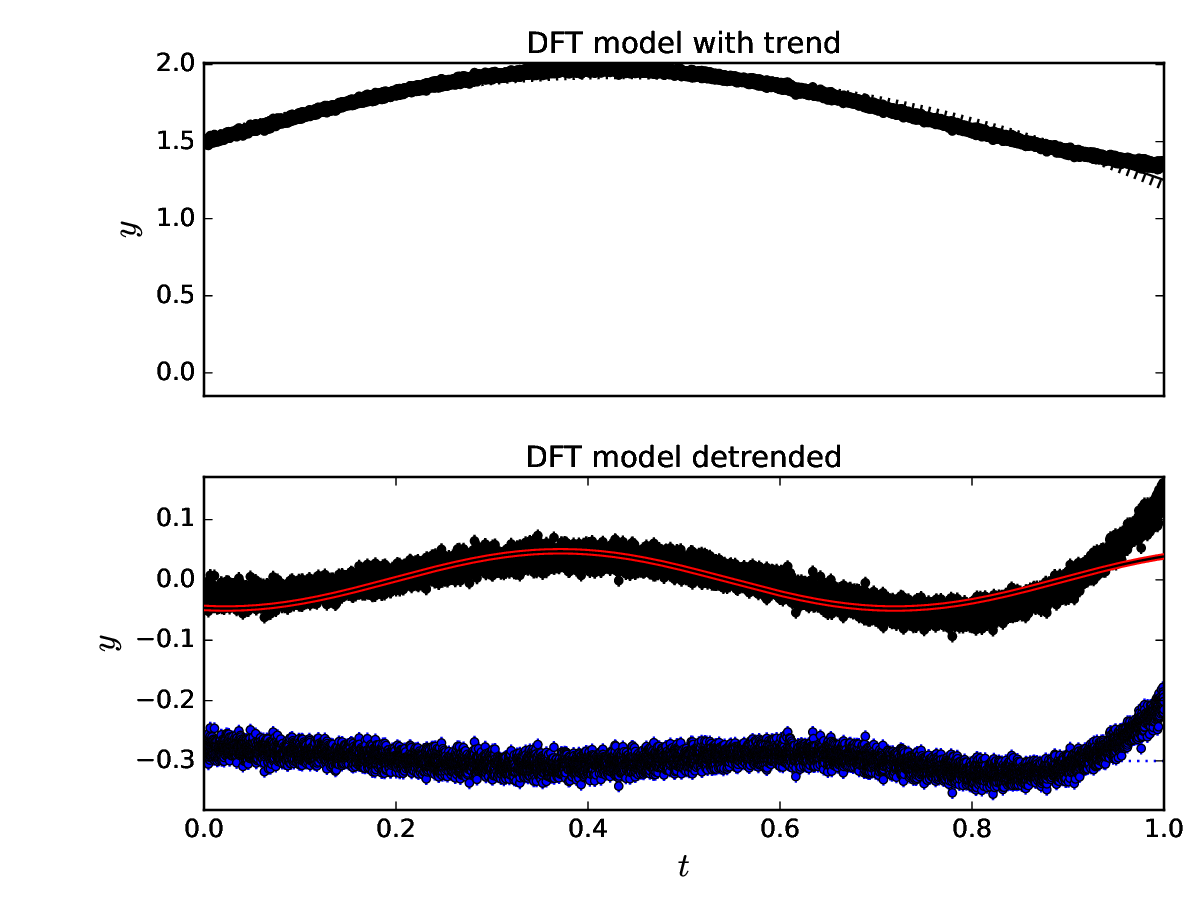}
        }
\vspace{-0.32\textwidth}
\centerline{\Large \bf 
\hspace{0.495\textwidth}  \color{black}{(e)}
\hspace{0.39\textwidth}  \color{black}{(f)}
\hfill}
\vspace{0.12\textwidth}
\centerline{\Large \bf 
\hspace{0.96\textwidth}  \color{black}{(g)}
\hfill}
\vspace{0.15\textwidth}
\caption{Model 2 (Table \ref{TableModelTwo}: $n=10~000$, $\SN=100$ simulation).
  (c) Colour of  $g(t)$ line has been changed from black to white.
  (d) Colour of  $g(t)$ line has been changed from black to white.
  Colour of offset level -0.3 dotted line has been changed from blue to white.
  Locations of best periods (diamonds) are explained
  in text (Section \ref{SectModelTwo}).
  Otherwise, notations are as in Figure \ref{FigModelOne}.}
\label{FigModelTwo}
\end{figure*}

\subsection{Model 2}  \label{SectModelTwo}

Our next one signal {time series} simulation model is
\begin{eqnarray}
  g(t) & = &
  (A_1/2)
  \cos{
  \left[
  {
  {2 \pi (t-t_{\mathrm{1,max,1}})}
  \over
  {P_1}
  }
  \right]
  } \\
 & + & M_0+M_1T +M_2T^2 \nonumber
\end{eqnarray}
where $T=[2(t-t_{\mathrm{mid}})] /\Delta T$.
We give the
$P_1$, $A_1$, $t_{\mathrm{1,max,1}}$, $M_0$, $M_1$ and $M_2$
values in Table \ref{TableModelTwo}.
 As a DCM model,
 the orders of  Model 2 are $K_1=1$, $K_2=1$ and $K_3=2$.
The simulated {time series} is  ``too short'' because
the period $P_1$ is $1.9 \times \Delta T$
(Equation \ref{EqTooShort}).
The parabolic trend $p(t)$ is unknown
(Equation \ref{EqTrend}).
Again, we use 
DCM and DFT time series analysis methods
to search for periods between
$P_{\mathrm{min}}=P_1/3=0.63$
and
$P_{\mathrm{max}}=3P_1=4.70$. 

DCM analysis results are given
in Table \ref{TableModelTwo}.
These results
are not very accurate for the $n=100$ and $\SN=100$ combination,
but they definitely improve for larger $n$ and $\SN$ values.
The \WDE ~ensures that the detected values converge to
  the correct simulated model parameter values.
  The short sample window, $\Delta T < P_1$, does not
mislead DCM analysis.

For Model 2, DCM analysis results  are illustrated for the
$n=10~000$ and $\SN=100$ combination
(Figures \ref{FigModelTwo}a-d).
DCM long search $z_1(f_1)$ periodogram  minimum is at 
$P_1=1.843$ (Figure \ref{FigModelTwo}a: diamond).
DCM short search gives the value $P_1=1.852$
(Figure \ref{FigModelTwo}b: diamond).
DCM model $g(t)$ is so good that it's continuous
black line is totally covered by the
black dots representing the $y_i$ data (Figure \ref{FigModelTwo}c).
Therefore, the colour of this $g(t)$ line has been
changed from black to white.
The results for the parabolic trend
coefficients
$M_0=1.079\pm0.096$,
$M_1=0.229\pm0.026$
and
$M_2=0.451\pm0,064$
of the dashed black $p(t)$ line are correct.
In Figure \ref{FigModelOne}d,
the white continuous line shows
the detrended model $g(t)-p(t)$.
The black dots show the detrended data $y(t_i)-p(t_i)$
and
the red thick continuous line shows
the pure sine signal $h_1(t)$.
Note that the red thick line is under the white thin line
because $h_1(t)=g(t)-p(t)$.
DCM residuals (blue dots) are offset
to the level of -0.65.
The colour of dotted line,
which denotes
this offset level,
has been changed from blue to white.
The distribution of these DCM model residuals is stable,
as expected for a random normal distribution.

The wrong period
$P_1=0.697$
is detected by DFT
(Figure \ref{FigModelTwo}e: Diamond).
DFT estimates for the trend $p(t)$ coefficients,
$M_0=1.92$, $M_1=-0.15$ and $M_2=-0.56$
are also wrong
(Figure \ref{FigModelTwo}f: dashed black line).
The data $y_i$ (black dots) deviate from
DFT model $g_{\mathrm{DFT}}(t)$ (black continuous line),
especially in the end of the sample (Figure \ref{FigModelTwo}g).
For the detrended DFT model,
the thin black line covers the thick red line
because
$s_{\mathrm{y,DFT}}(t)=g_{\mathrm{DFT}}(t)-p_{\mathrm{DFT}}(t)$
(Figure \ref{FigModelTwo}g).
The  $s_{\mathrm{y,DFT}}(t)$ sine curve peak to peak
amplitude is far below the correct $A_1=2.0$ value.
DFT residuals (blue dots) are offset to the level
of -1.5 (blue dotted line).
The trends of these residuals confirm that DFT analysis fails.

 Only DCM (not DFT)  succeeds in
 the analysis of Model 2 simulated {time series}.

\begin{figure*}
\vspace{0.02\textwidth}
\centerline{\hspace*{0.005\textwidth}
 \includegraphics[width=0.46\textwidth,clip=]{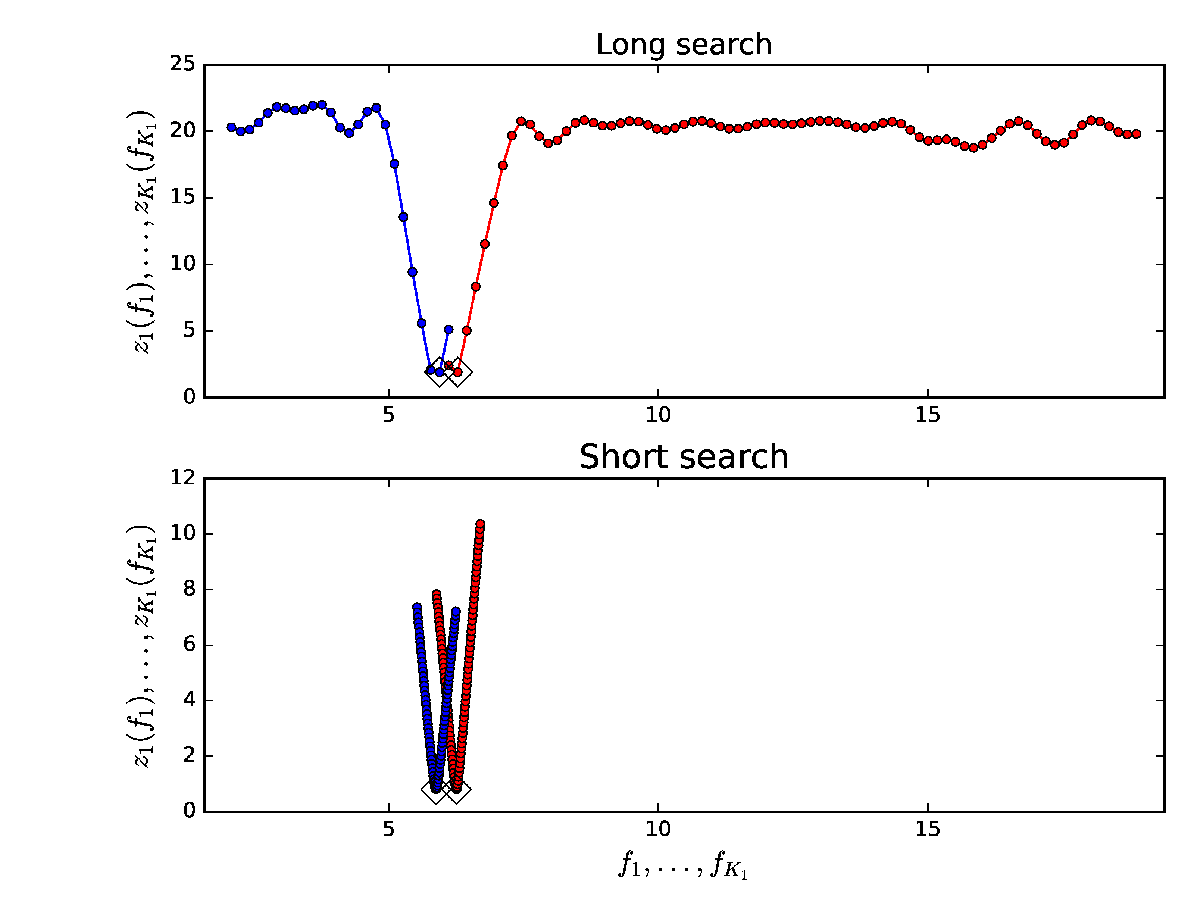}
 \hspace*{-0.01\textwidth}
 \includegraphics[width=0.46\textwidth,clip=]{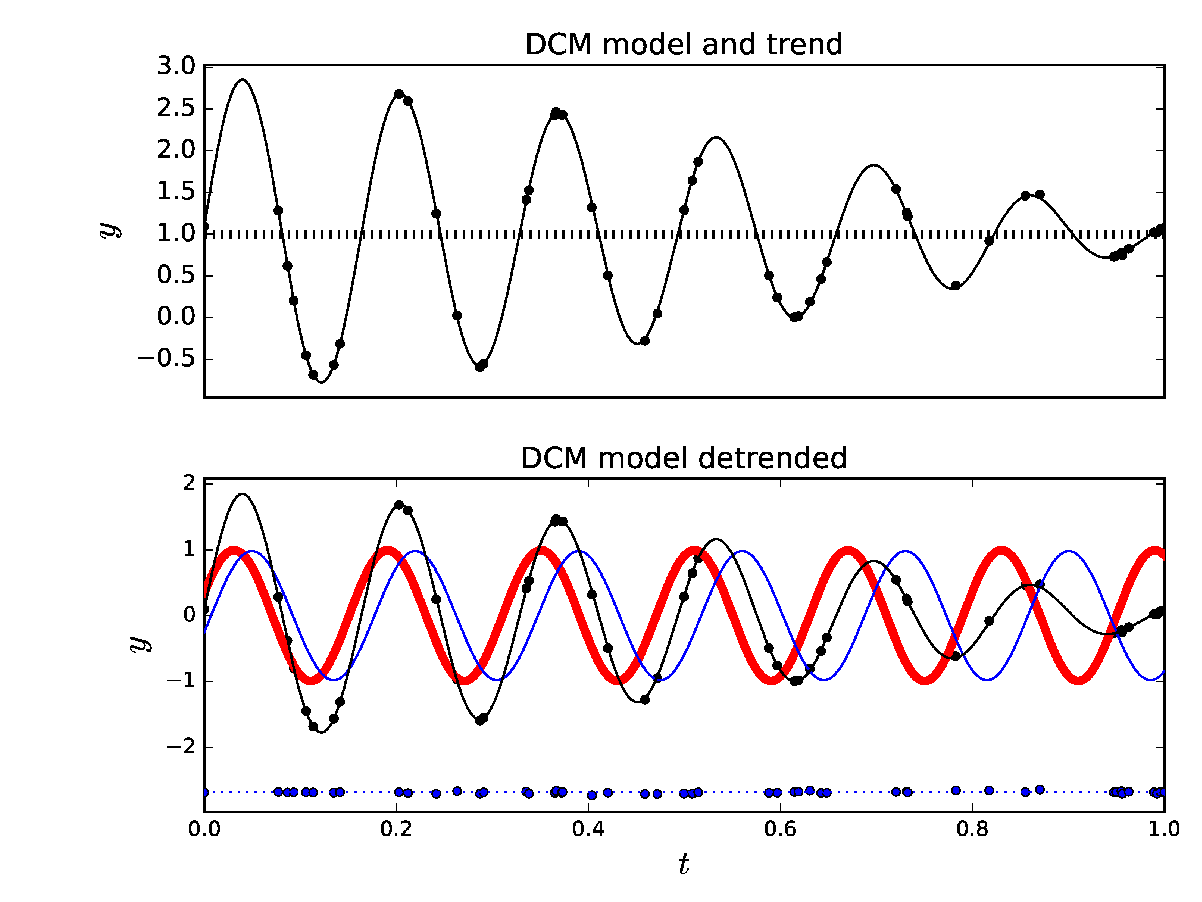}
        }
\vspace{-0.32\textwidth}
\centerline{\Large \bf 
\hspace{0.50\textwidth}  \color{black}{(a)}
\hspace{0.39\textwidth}  \color{black}{(c)}
\hfill}
\vspace{0.118\textwidth}
\centerline{\Large \bf 
\hspace{0.50\textwidth}  \color{black}{(b)}
\hspace{0.39\textwidth}  \color{black}{(d)}
\hfill}
\vspace{0.12\textwidth}
\centerline{\hspace*{0.005\textwidth}
 \includegraphics[width=0.46\textwidth,clip=]{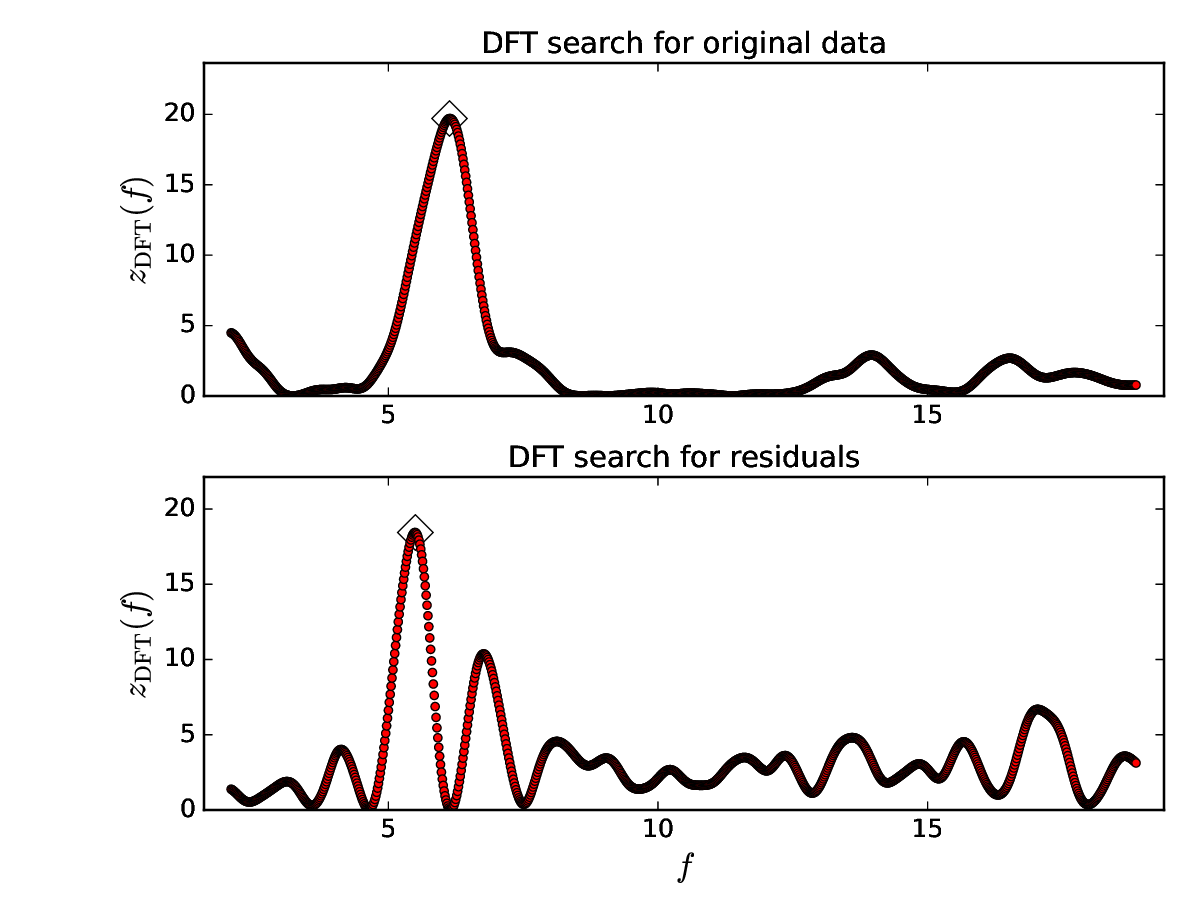}
 \hspace*{-0.01\textwidth}
 \includegraphics[width=0.46\textwidth,clip=]{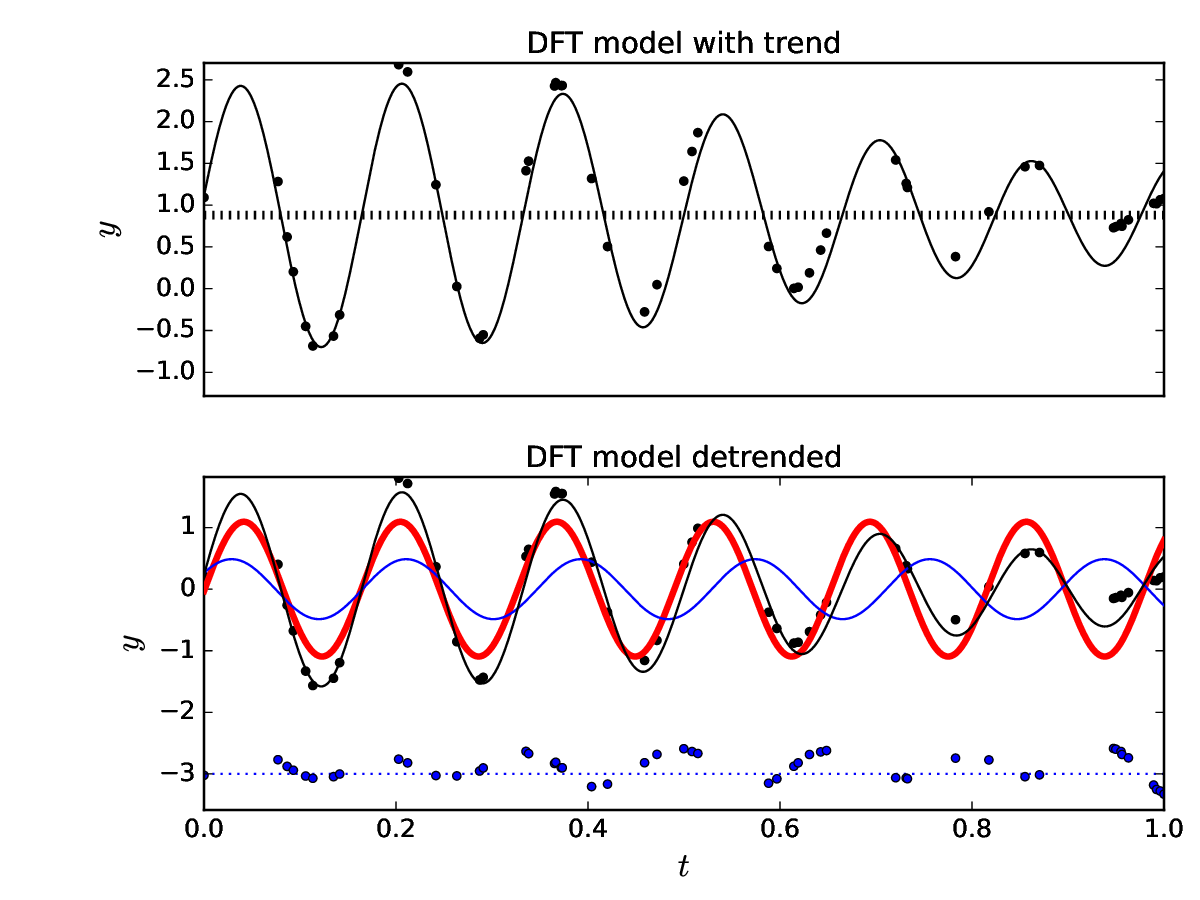}
        }
\vspace{-0.32\textwidth}
\centerline{\Large \bf 
\hspace{0.50\textwidth}  \color{black}{(e)}
\hspace{0.385\textwidth}  \color{black}{(g)}
\hfill}
\vspace{0.12\textwidth}
\centerline{\Large \bf 
\hspace{0.495\textwidth}  \color{black}{(f)}
\hspace{0.395\textwidth}  \color{black}{(h)}
\hfill}
\vspace{0.15\textwidth}
\caption{Model 3 (Table \ref{TableModelThree}: $n=50$, $\SN=50$ simulation).
  (a) DCM long search periodograms $z_1(f_1)$ (red) and $z_2(f)$ (blue)
  give best periods  at  0.160 and  0.168 (diamonds).
 (b) DCM short search periodograms $z_1(f_1)$ (red)
  and $z_2(f)$ (blue) give
  best periods at 0.160 and 0.170 (diamonds). 
(c) DCM model $g(t)$ (black continuous line),
  DCM trend $p(t)$ (black dashed line)
  and data $y_i$ (black dots).  
  (d) DCM model detrended $g(t)-p(t)$ (black continuous line),
  DCM signal $h_1(t)$ (red thick continuous line),
  DCM signal $h_2(t)$ (blue continuous thin line),
  detrended data $y(t_i)-p(t_i)$
  (black dots) and DCM
  model residuals $y(t_i)-g(t_1)$ (blue dots)
  offset to -3.0 level (blue dotted line)
  (e) DFT periodogram $z_{\mathrm{DFT}}(f)$ for the original data gives
  best period at 0.163 (diamond).
  (f) DFT periodogram $z_{\mathrm{DFT}}(f)$ for the
  sine model residuals gives
  best period at 0.182 (diamond).
  (g) DFT model $g_{\mathrm{DFT}}(t)$
  (black continuous line), DFT trend $p_{\mathrm{DFT}}(t)$
  (black dashed line) and data $y_i$ (black dots). 
  (h) DFT model detrended
  $g_{\mathrm{DFT}}(t)-p_{\mathrm{DFT}}(t)$
  (black continuous line),
  DFT pure sine model for original data $s_{y,\mathrm{DFT}}(t)$
  (red thick continuous line),
  DFT pure sine model for first residuals $s_{\epsilon,\mathrm{DFT}}(t)$ 
  (blue continuous thin line),
  detrended data $y(t_i)-p_{\mathrm{DFT}}(t_i)$ (black dots)
  and DFT model
  residuals (blue dots) offset to -3.0 level (blue dotted line). }      
\label{FigModelThree}
\end{figure*}

\begin{table*}
  \caption{Model 3. DCM analysis between
    $P_{\mathrm{min}}=0.053$
    and
    $P_{\mathrm{max}}=0.480$.
    Notations as in Table \ref{TableModelOne}.}\label{TableModelThree}
  \begin{scriptsize}
  \begin{center}
\begin{tabular}{lccc}
  \hline
  (1)                                      & (2)                                        &  (3)                                            & (4)                   \\
                                            &$n=50                                   $&$n=50$                                    &$n=100$                        \\
Model 3                                &$\SN=10                               $&$\SN=50$                                  &$\SN=100$                    \\
\hline
$P_1=0.16$                         &$0.15898\pm0.00078           $&$0.15989\pm0.00015              $&$0.159950\pm0.000084$\\
$A_1=2.0$                           &$1.85\pm0.12                       $&$1.987\pm0.035                      $&$1.994\pm0.016           $\\
$t_{\mathrm{1,min,1}}=0.11$&$0.1122\pm0.0023              $&$0.11080\pm0.00083              $&$0.11004\pm0.00034   $\\
$t_{\mathrm{1,max,1}}=0.03$&$0.0328\pm0.0026              $&$0.03086\pm0.00089              $&$0.03007\pm0.00037   $\\
$P_2=0.17$                         &$0.17025\pm0.00063           $&$0.17018\pm0.00021              $&$0.16996\pm0.00010$\\
$A_2=2.0$                           &$2.001\pm0.099                    $&$1.962\pm0.043                      $&$2.002\pm0.017         $\\
$t_{\mathrm{2,min,1}}=0.135$&$0.1312\pm0.0026             $&$0.13461\pm0.00085             $&$0.13495\pm0.00035   $\\
$t_{\mathrm{2,max,1}}=0.05$ &$0.0461\pm0.0027             $&$0.04952\pm0.00094              $&$0.04997\pm0.000039   $\\
 $M_0=1.0$                           &$0.996\pm0.012                  $&$0.9985\pm0.0030                 $&$0.9995\pm0.0013$  \\
 Data file                                & \PR{Model3n50SN10.dat}     &\PR{Model3n550.dat}         &\PR{Model3n100SN100.dat} \\
 Control file                            & \PR{dcmModel3n50SN10.dat}&\PR{dcmModel3n50SN50.dat} & \PR{dcmModel3n100SN100.dat} \\  
\hline
\end{tabular}
\end{center}
\end{scriptsize}
\end{table*}

\subsection{Model 3}  \label{SectModelThree}

Our third {time series} simulation model is the
two signal model 
\begin{eqnarray}
  g(t) & = &
  (A_1/2)
  \cos{
  \left[
  {
  {2 \pi (t-t_{\mathrm{1,max,1}})}
  \over
  {P_1}
  }
  \right]
  } \\
  & + &
  (A_2/2)
  \cos{
  \left[
  {
  {2 \pi (t-t_{\mathrm{2,max,1}})}
  \over
  {P_2}
  }
  \right]
  }
  +M_0. \nonumber
\end{eqnarray}
We give the $P_1$, $P_2$, $A_1$,$A_2$,
$t_{\mathrm{1,max,1}}$,
$t_{\mathrm{2,max,1}}$ and
$M_0$ values in Table \ref{TableModelThree} (Column 1).
DCM orders of Model 3 are $K_1=2$, $K_2=1$ and $K_3=0$.
The simulated {time series}
is not ``too short'' (Equation \Ref{EqTooShort})
because $P_1<P_2<\Delta T$.
The $p(t)$ trend mean level $M_0$ is unknown
(Equation \ref{EqTrend}).
The two frequencies
are ``too close''  because $\Delta f=f_1-f_2=f_0/2.72$
(Equation \ref{EqTooClose}).
We apply DCM and DFT time series analyses
to search for periods between
$P_{\mathrm{min}}=P_1/3=0.053$
and
$P_{\mathrm{max}}=3P_1=0.480$. 

DCM analysis results for different $n$ and $\SN$ combinations
are given in Table \ref{TableModelThree} (Columns 2-4).
The results are surprisingly accurate even for the small
$n=50$ sample having a low $\SN=10$.
Due to the \WDE, the detected model parameter values
converge to the correct simulated values when
$n$ and $\SN$ increase.

DCM results for Model 3 are illustrated in
Figures \ref{FigModelOne}a-d
($n=50$ and $\SN=50$ combination).
DCM long search $z_1(f_1)$ periodogram (red)
and $z_2(f)$ periodogram (blue)
minima
are  at the best periods $P_1=0.159$ and  $P_2=0.168$
(Figure \ref{FigModelThree}a: diamonds).
DCM short search gives
$P_1=0.160$ and  $P_2=0.170$
(Figure \ref{FigModelThree}b: diamonds).
The black continuous DCM model $g(t)$ curve
crosses trough the black dots of $y_i$ data
(Figure \ref{FigModelOne}c).
The result $M_0=0.9985 \pm 0.0032$ for the
dashed black $p(t)$ trend line
is correct.
Our Figure \ref{FigModelThree}d shows
the detrended DCM model $g(t)-p(t)$ (black continuous line),
the detrended data $y(t_i)-p(t_i)$ (black dots),
the pure sine signal $h_1(t)$ (red thick continuous line)
and
the pure sine signal $h_2(t) $ (blue thin continuous line).
DCM residuals (blue dots) are offset to
the level of -3.0 (blue dotted line).
These residuals are small and their level is stable.

DFT detects the wrong period $P_1=1.163$
for the original data (Figure \ref{FigModelThree}e: diamond).
This is an expected result because the detected
period should be close to $(P_1+P_2)/2$ when
the peak to peak amplitudes of the simulated data,
$A_1=A_2$, are equal
\citep{Jet25}.
The two DFT periodogram $z_{\mathrm{DFT}}(f)$ peaks
at frequencies $1/P_1$ and $1/P_2$ overlap and merge into one peak.
The period $P_2=0.182$ detected for the residuals is
also wrong (Figure \ref{FigModelThree}f: diamond).
DFT trend $p_{\mathrm{PDF}}(t)$ estimate $M_0=0.880$ fails
(Figure \ref{FigModelThree}f: dashed black line).
The black dots $y_i$ show minor deviations from
the continuous black  DFT model $g_{\mathrm{DFT}}(t)$ line
(Figure \ref{FigModelThree}g).
The detrended model $g_{\mathrm{DFT}}(t)-p_{\mathrm{DFT}}(t)$
(continuous black line),
the detrended data $y(t_i) - p_{\mathrm{DFT}}(t_i)$
(black dots),
the pure sine signal $s_{\mathrm{y,DFT}}(t)$ for the data
(continuous thick red line)
and
the pure sine signal $s_{\mathrm{\epsilon,DFT}}(t)$ for the residuals
(continuous blue thin line)
are shown in Figure \ref{FigModelThree}h.
Note that the $s_{\mathrm{y,DFT}}(t)$ and $s_{\mathrm{\epsilon,DFT}}(t)$
signal amplitudes are far from equal.
DFT residuals (blue dots) offset to the level
of -3.0 (blue dotted line)
are not stable.

DCM analysis of Model 3 simulated {time series}
succeeds, but DFT analysis does not.

\begin{table*}
  \caption{Model 4. DCM analysis between
    $P_{\mathrm{min}}=0.053$
    and
    $P_{\mathrm{max}}=0.480$.
  Notations as in Table \ref{TableModelOne}.}\label{TableModelFour}
\begin{scriptsize}
 \begin{center}
\begin{tabular}{lccc}
  \hline
  (1)                                             & (2)                                      & (3)                                           & (4)                               \\
                                                    &$n=50                                $&$n=50$                                  &$n=100$                      \\
 Model 4                                       &$\SN=10                             $&$\SN=50$                               &$\SN=100$                   \\
\hline
$P_1=0.16$                                 &$0.1606\pm0.0089             $&$0.15982\pm0.00018           $&$0.16001\pm0.000051$\\
$A_1=2.0$                                   &$1.97\pm0.25                      $&$1.958\pm0.040                      $&$2.012\pm0.013           $\\
$t_{\mathrm{1,min,1}}=0.11$   &$0.1134\pm0.0038              $&$0.11059\pm0.00077          $&$0.10977\pm0.00028   $\\
$t_{\mathrm{1,max,1}}=0.03$  &$0.0331\pm0.0042              $&$0.03068\pm0.00083          $&$0.02976\pm0.00030  $\\
$P_2=0.17$                                 &$0.1722\pm0.0012              $&$0.17009\pm0.00029           $&$0.169958\pm0.000073$\\
$A_2=2.0$                                   &$1.74\pm0.25                       $&$1.982\pm0.046                   $&$2.018\pm0.015          $\\
$t_{\mathrm{2,min,1}}=0.135 $&$0.1321\pm0.0045               $&$0.13420\pm0.00089              $&$0.13528\pm0.00027   $\\
$t_{\mathrm{2,max,1}}=0.05 $ &$0.0460\pm0.0050               $&$0.0492\pm0.0010                $&$0.05031\pm0.00031   $\\
  $M_0=1.0$                                &$1.015\pm0.022                   $&$0.9997\pm0.0042                $&$1.0012\pm0.0011$  \\
  $M_1=0.25$                              &$0.219\pm0.023                    $&$0.2565\pm0.0048               $&$0.2495\pm0.0014$\\
  $M_2=0.5$                                &$0.485\pm0.047                    $&$0.511\pm0.011                   $&$0.4983\pm0.0024$\\
 Data file                                      &\PR{Model4n50SN10.dat}        &\PR{Model4n50SN50.dat}       &\PR{Model4n100SN100.dat}\\
 Control file                                  &\PR{dcmModel4n50SN10.dat}  &\PR{dcmModel4n50SN50.dat}& \PR{dcmModel4n100SN100.dat} \\  
\hline
\end{tabular}
\end{center}
\end{scriptsize}
\end{table*}

\begin{figure*}
\vspace{0.02\textwidth}
\centerline{\hspace*{0.005\textwidth}
 \includegraphics[width=0.46\textwidth,clip=]{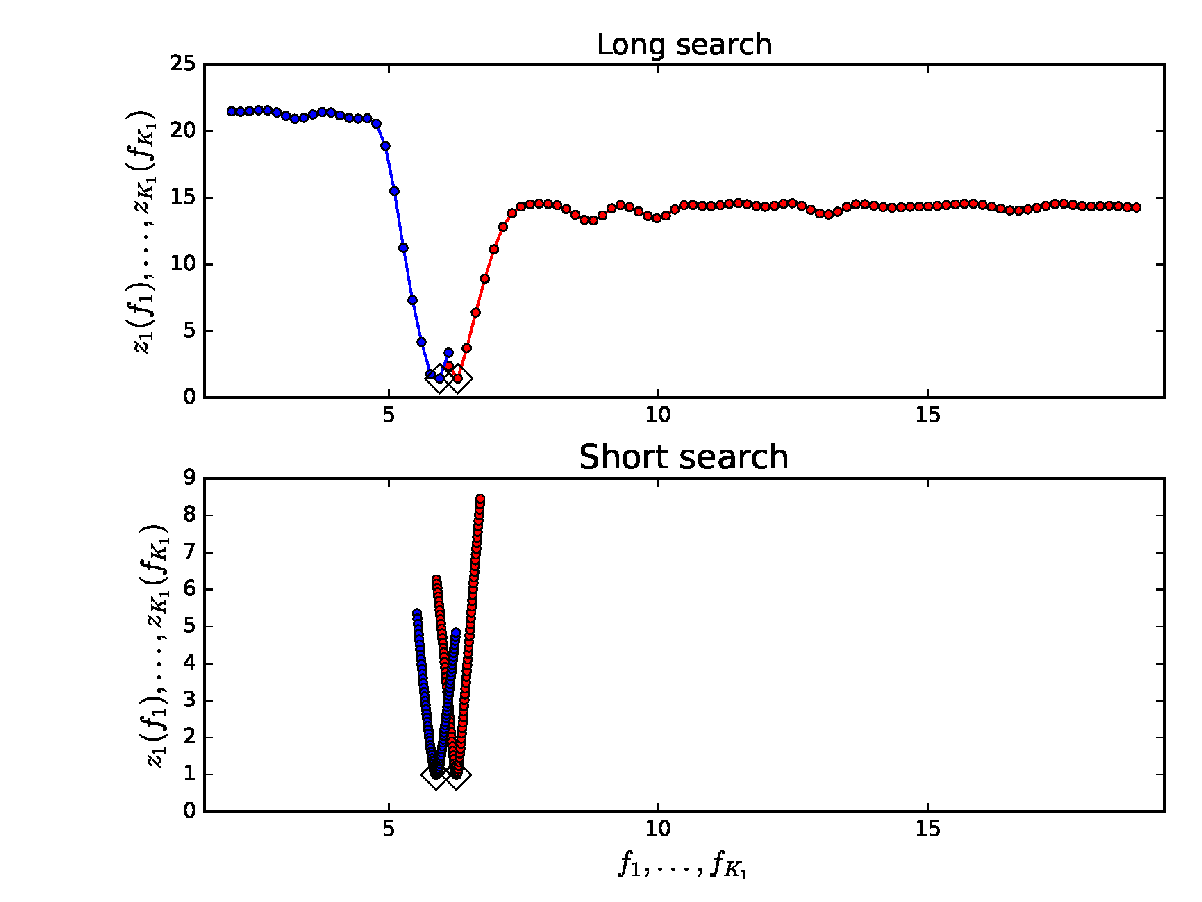}
 \hspace*{-0.01\textwidth}
 \includegraphics[width=0.46\textwidth,clip=]{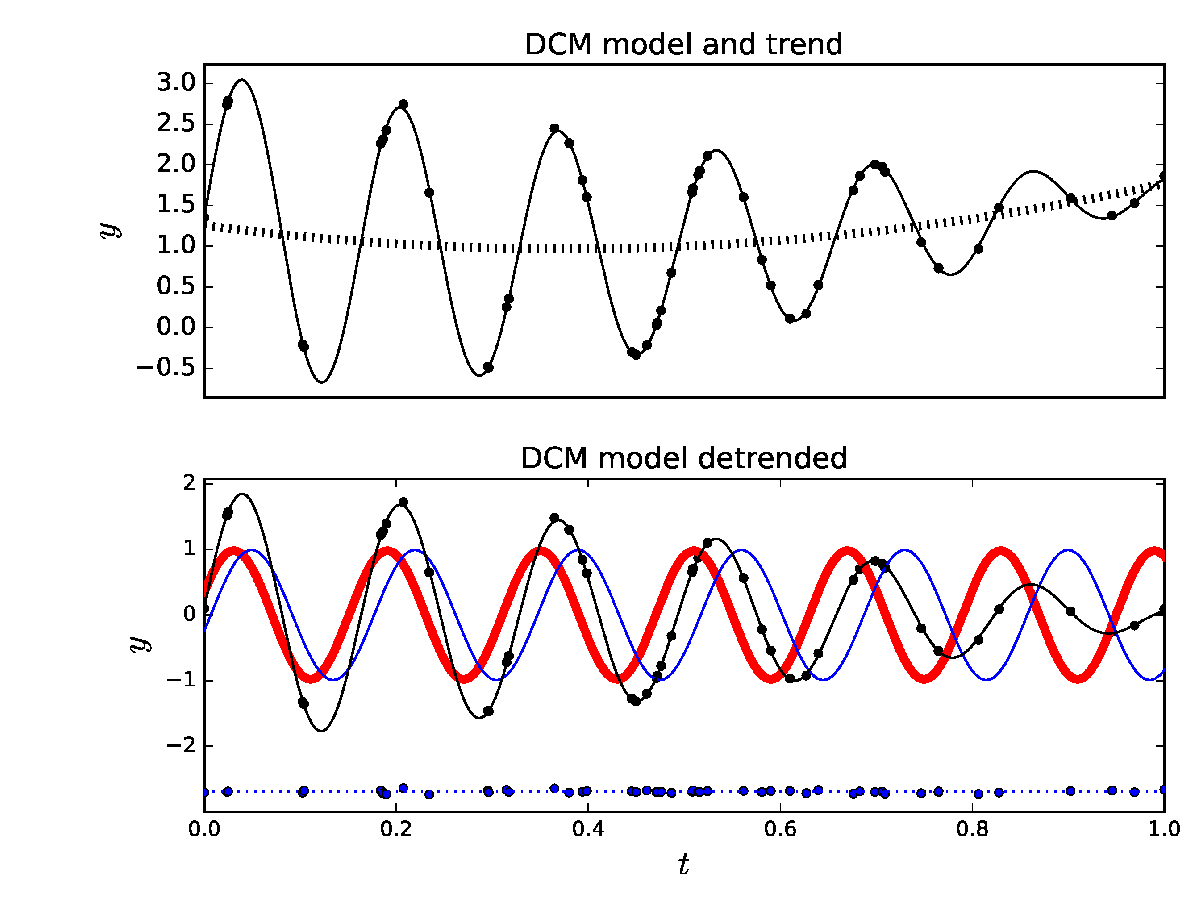}
        }
\vspace{-0.32\textwidth}
\centerline{\Large \bf 
\hspace{0.49\textwidth}  \color{black}{(a)}
\hspace{0.39\textwidth}  \color{black}{(c)}
\hfill}
\vspace{0.118\textwidth}
\centerline{\Large \bf 
\hspace{0.495\textwidth}  \color{black}{(b)}
\hspace{0.39
  \textwidth}  \color{black}{(d)}
\hfill}
\vspace{0.12\textwidth}
\centerline{\hspace*{0.005\textwidth}
 \includegraphics[width=0.46\textwidth,clip=]{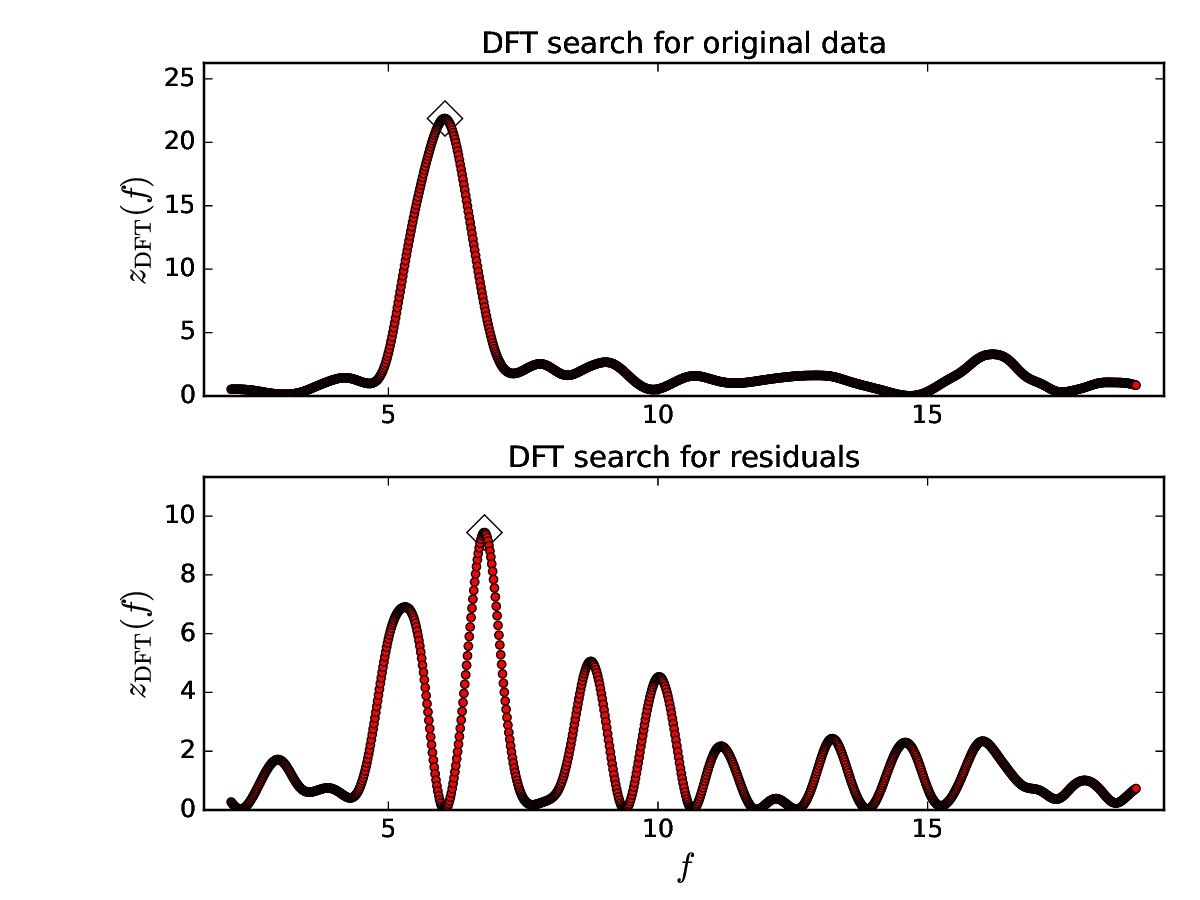}
 \hspace*{-0.01\textwidth}
 \includegraphics[width=0.46\textwidth,clip=]{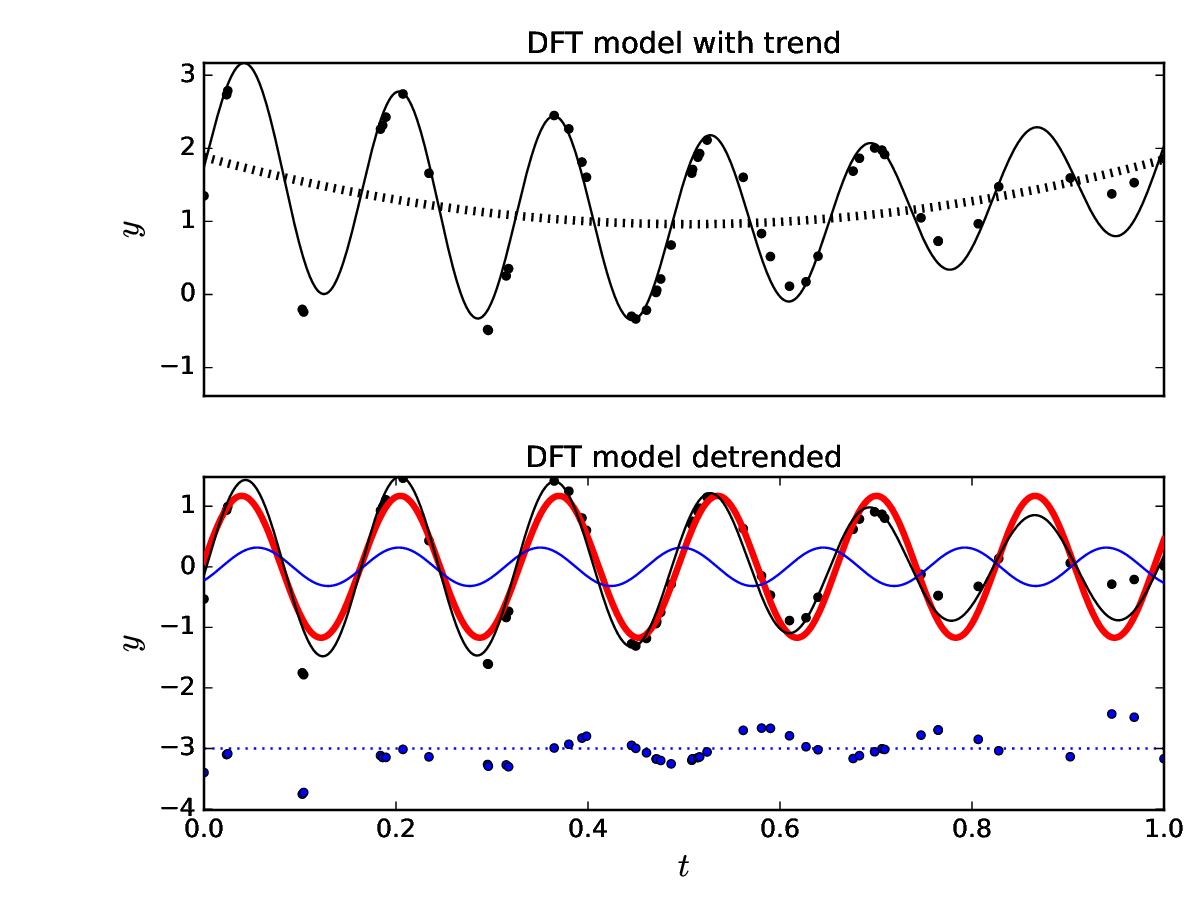}
        }
\vspace{-0.32\textwidth}
\centerline{\Large \bf 
\hspace{0.495\textwidth}  \color{black}{(e)}
\hspace{0.395\textwidth}  \color{black}{(g)}
\hfill}
\vspace{0.12\textwidth}
\centerline{\Large \bf 
\hspace{0.495\textwidth}  \color{black}{(f)}
\hspace{0.395\textwidth}  \color{black}{(h)}
\hfill}
\vspace{0.15\textwidth}
\caption{Model 4 (Table \ref{TableModelFour}: $n=50$, $\SN=50$ simulation).
  Notations as in Figure \ref{FigModelThree}.
Best periods (diamonds) are explained in Section \ref{SectModelFour}. }
\label{FigModelFour}
\end{figure*}

\subsection{Model 4}  \label{SectModelFour}

The next {time series} simulation model is 
\begin{eqnarray}
  g(t) & = &
  (A_1/2)
  \cos{
  \left[
  {
  {2 \pi (t-t_{\mathrm{1,max,1}})}
  \over
  {P_1}
  }
  \right]
  } \label{EqModelFour} \\
  &  + & 
  (A_2/2)
  \cos{
  \left[
  {
  {2 \pi (t-t_{\mathrm{2,max,1}})}
  \over
  {P_2}
  }
  \right]
  } \nonumber \\
 & + & M_0+M_1T + M_2T^2, \nonumber
\end{eqnarray}
where $T=[2(t-t_{\mathrm{mid}})] /\Delta T$.
In this model,
two signals are superimposed on an unknown parabolic trend.
The $P_1$, $P_2$, $A_1$,$A_2$,
$t_{\mathrm{1,max,1}}$,
$t_{\mathrm{2,max,1}}$,
$M_0$, $M_1$ and $M_2$ values are given
in Table \ref{TableModelFour}.
This simulated sample is not ``too short'' (Equation \ref{EqTooShort}).
The polynomial trend $p(t)$ is unknown
(Equation \ref{EqTrend}).
The  $\Delta f=f_1-f_2=f_0/2.72$ difference means that
the frequencies are ``too close'' (Equation \ref{EqTooClose}).
DCM and DFT time series analysis methods
are used to search for periods between
$P_{\mathrm{min}}=P_1/3=0.053$
and
$P_{\mathrm{max}}=3P_1=0.480$.

Model 4 is a DCM model having orders
$K_1=2$, $K_2=1$ and $K_3=2$.
Our DCM analysis results for different $n$ and $\SN$ combinations
are given in Table \ref{TableModelFour} (Columns 2-4).
DCM  detects the correct
$P_1$, $P_2$, $A_1$, $A_2$,
$t_{\mathrm{max,1}}$, $t_{\mathrm{max,2}}$,
$M_0$, $M_1$ and $M_2$ values
even for the lowest $n=50$ and $\SN=10$ combination.
As the simulated data $n$ and $\SN$ values increase,
the \WDE ~ensures that all detected values
converge to the correct simulated model parameter values.

Our Figures \ref{FigModelFour}a-d illustrate
DCM analysis results for one sample of simulated {time series}
(Model 4: $n=50$ and $\SN=50$).
DCM long search best periods 
are  at $P_1=0.159$ and  $P_2=0.168$
(Figure \ref{FigModelFour}a: diamonds).
DCM short search values are
$P_1=0.160$ and  $P_2=0.170$
(Figure \ref{FigModelFour}b: diamonds).
The continuous  DCM model $g(t)$  black line crosses
through all black dots representing the $y_i$ data
(Figure \ref{FigModelFour}c).
DCM detects the correct  polynomial trend $p(t)$ coefficients
$M_0=0.9997\pm0.0030$,
$M_1=0.2565\pm0.0066$
and
$M_2=0.511\pm0.011$
(Figure \ref{FigModelFour}c: dashed black line).
The detrended DCM model $g(t)-p(t)$ (black continuous line),
the detrended data $y(t_i)-p(t_i)$ (black dots),
the pure sine signal $h_1(t)$ (red thick continuous line)
and
the pure sine signal $h_2(t) $ (blue thin continuous line) are
shown in Figure \ref{FigModelFour}d.
DCM residuals (blue dots) offset to the level of -3.0
show no trends and are extremely stable.

Since the peak to peak amplitudes of the simulated data signals
are equal,
$A_1=A_2$,
the expected result for  DFT analysis of original
data is  $(P_1+P_2)/2$, where $P_1$ and $P_2$ are the
simulated signal periods \citep{Jet25}.
DFT detects this expected wrong period $P_1=1.165$
for the original data (Figure \ref{FigModelFour}e: diamond).
A wrong period $P_2=0.147$ is also detected for the residuals
(Figure \ref{FigModelFour}f: diamond).
DFT estimate $M_0=0.962$ for the $p(t)$ trend
is close to the correct value $M_0=1$,
but the $M_1=-0.017$ and $M_2=0.903$
estimates are wrong
(Figure \ref{FigModelFour}g: dashed black line).
The continuous black  DFT model $g_{\mathrm{DFT}}(t)$ line
deviates from the black dots of data $y_i$,
especially in the beginning and end
of the sample.
In our Figure \ref{FigModelFour}h,
the black dots are 
the detrended data $y(t_i) - p_{\mathrm{DFT}}(t_i)$ and
the continuous black line is 
the detrended DFT model $g_{\mathrm{DFT}}(t)-p_{\mathrm{DFT}}(t)$.
The continuous thick red line is
the pure sine signal $s_{\mathrm{y,DFT}}(t)$ for the original data
and 
the continuous thinner blue line is the 
pure sine signal $s_{\mathrm{\epsilon,DFT}}(t)$ for the residuals.
The $s_{\mathrm{\epsilon,DFT}}(t)$ signal amplitude
is far below the simulated correct value $A_2=2.0$.
The blue dots representing DFT model
residuals are offset to the level of -3.0 and show clear trends.

Our  Model 4 simulated {time series} analysis succeeds
for DCM and fails for DFT.

\begin{table*}
  \caption{Model 5. DCM analysis between $P_{\mathrm{min}}=0.47$
    and $P_{\mathrm{max}}=4.20$. 
    Notations as in Table \ref{TableModelOne}.}\label{TableModelFive}
  \begin{scriptsize}
    \begin{center}
  \begin{tabular}{lccc}
  \hline
     (1)                                  & (2)                                                   & (3)                                                    & (4) \\
                                           &$n=10~000                                     $&$n=10~000                                     $&$n=100~000$  \\
Model 5                               &$\SN=1~000                                    $&$\SN=10~000                                  $&$\SN=10~000$ \\
\hline
$P_1=1.4$                          &$1.404\pm0.045                              $&$1.3926\pm0.0092                              $&$1.4019\pm0.0035$  \\
$A_1=2.0$                          &$2.27\pm0.44                                  $&$1.975\pm0.070                                   $&$1.994\pm0.029$ \\
$t_{\mathrm{1,min,1}}=1.1$  &$1.129\pm0.022                             $&$1.1030\pm0.0029                            $&$1.09766\pm0.00082$ \\
$t_{\mathrm{1,max,1}}=0.4$ &$0.4267\pm0.0013                         $&$0.4066\pm0.0052                            $&$0.3967\pm0.0020$ \\
$P_2=1.9$                          &$5.24\pm0.79                                  $&$2.023\pm0.091                                $&$1.854\pm0.023$ \\
$A_2=2.0$                          &$5\pm25                                          $&$2.35\pm0.31                                    $&$1.918\pm0.074 $\\
$t_{\mathrm{2,min,1}}=1.55$&$3.23\pm0.56                                  $&$1.599\pm0.034                                $&$1.5292\pm0.0066$\\
$t_{\mathrm{2,max,1}}=0.6$ &$0.61\pm0.40                                 $&$0.587\pm0.016                                 $&$0.6024\pm0.0064$ \\
  $M_0=1.0$                        &$-2\pm12                                        $&$0.81\pm0.15                                     $&$1.055\pm0.035$      \\
  $M_1=0.25$                      &$-1.27\pm0.60                                $&$0.230\pm0.011                                 $&$0.2590\pm0.0029$ \\
  $M_2=0.5$                        &$3.8\pm1.9                                     $&$0.596\pm0.078                                 $&$0.470\pm0.020$\\
 Data file                              &\PR{Model5n10000SN1000.dat}      & \PR{Model5n10000SN10000.dat}         & \PR{Model5n100000SN10000.dat}\\
 Control file                          &\PR{dcmModel5n10000SN1000.dat}& \PR{dcmModel5n10000SN10000.dat} & \PR{dcmModel5n100000SN10000.dat} \\
\hline
  \end{tabular}
  \end{center}
  \end{scriptsize}
\end{table*}

\begin{figure*}
\vspace{0.02\textwidth}
\centerline{\hspace*{0.005\textwidth}
 \includegraphics[width=0.46\textwidth,clip=]{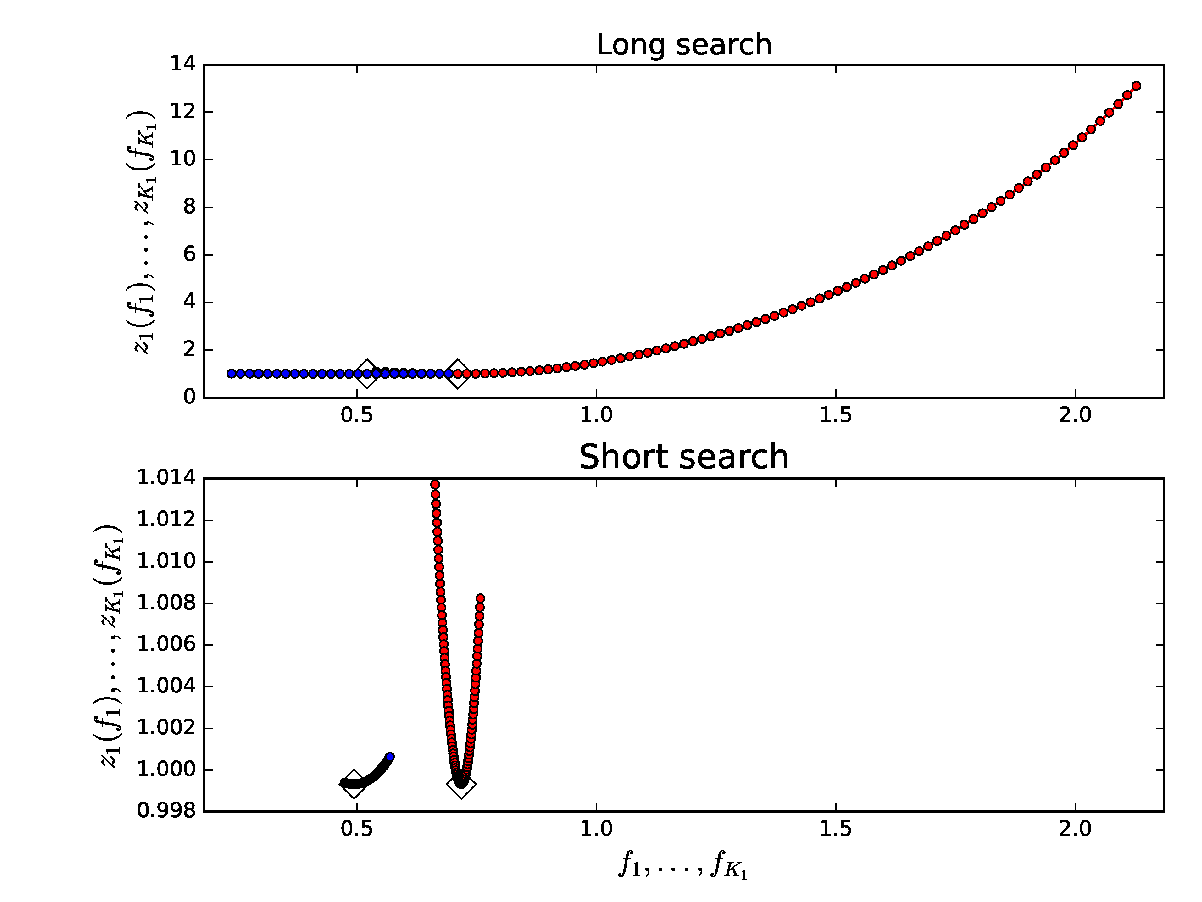}
 \hspace*{-0.01\textwidth}
 \includegraphics[width=0.46\textwidth,clip=]{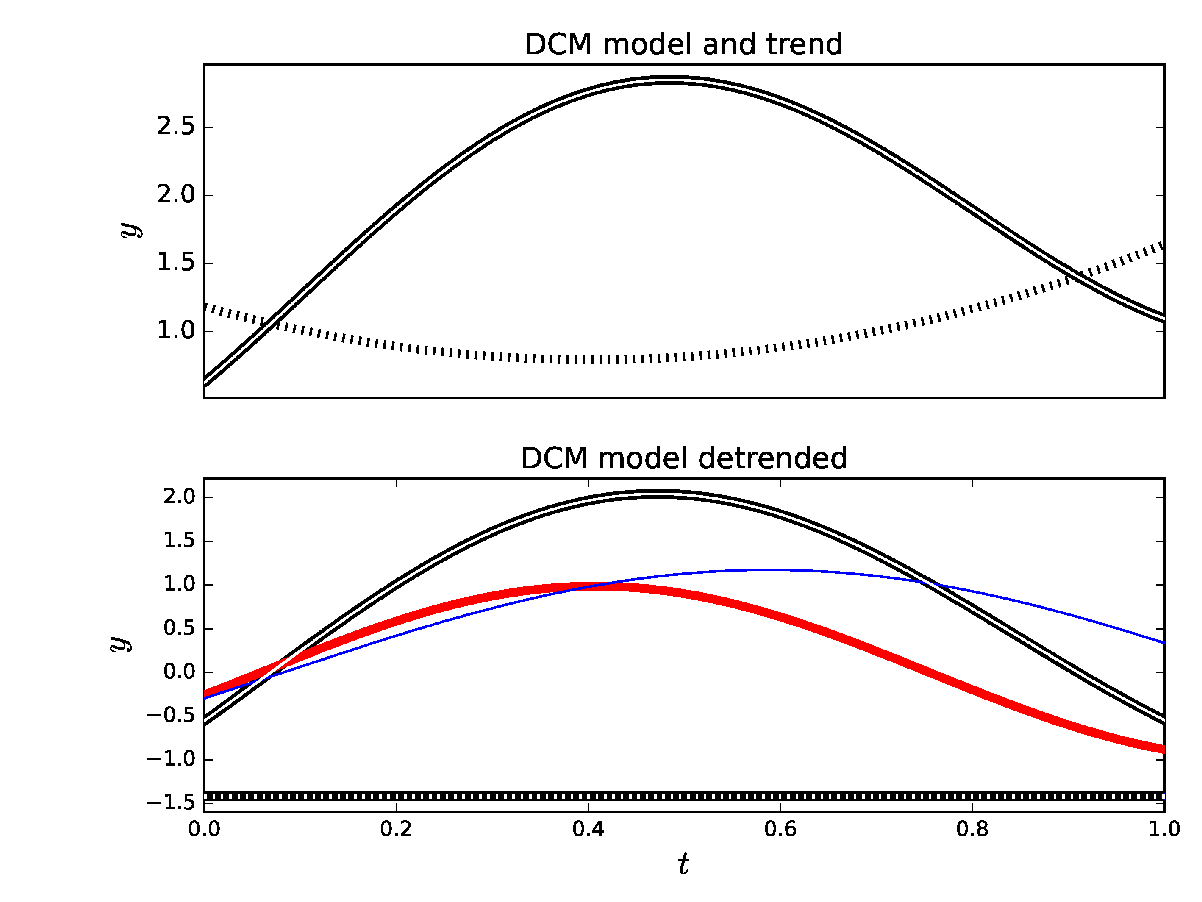}
        }
\vspace{-0.32\textwidth}
\centerline{\Large \bf 
\hspace{0.49\textwidth}  \color{black}{(a)}
\hspace{0.39\textwidth}  \color{black}{(c)}
\hfill}
\vspace{0.118\textwidth}
\centerline{\Large \bf 
\hspace{0.495\textwidth}  \color{black}{(b)}
\hspace{0.385\textwidth}  \color{black}{(d)}
\hfill}
\vspace{0.12\textwidth}
\centerline{\hspace*{0.005\textwidth}
 \includegraphics[width=0.46\textwidth,clip=]{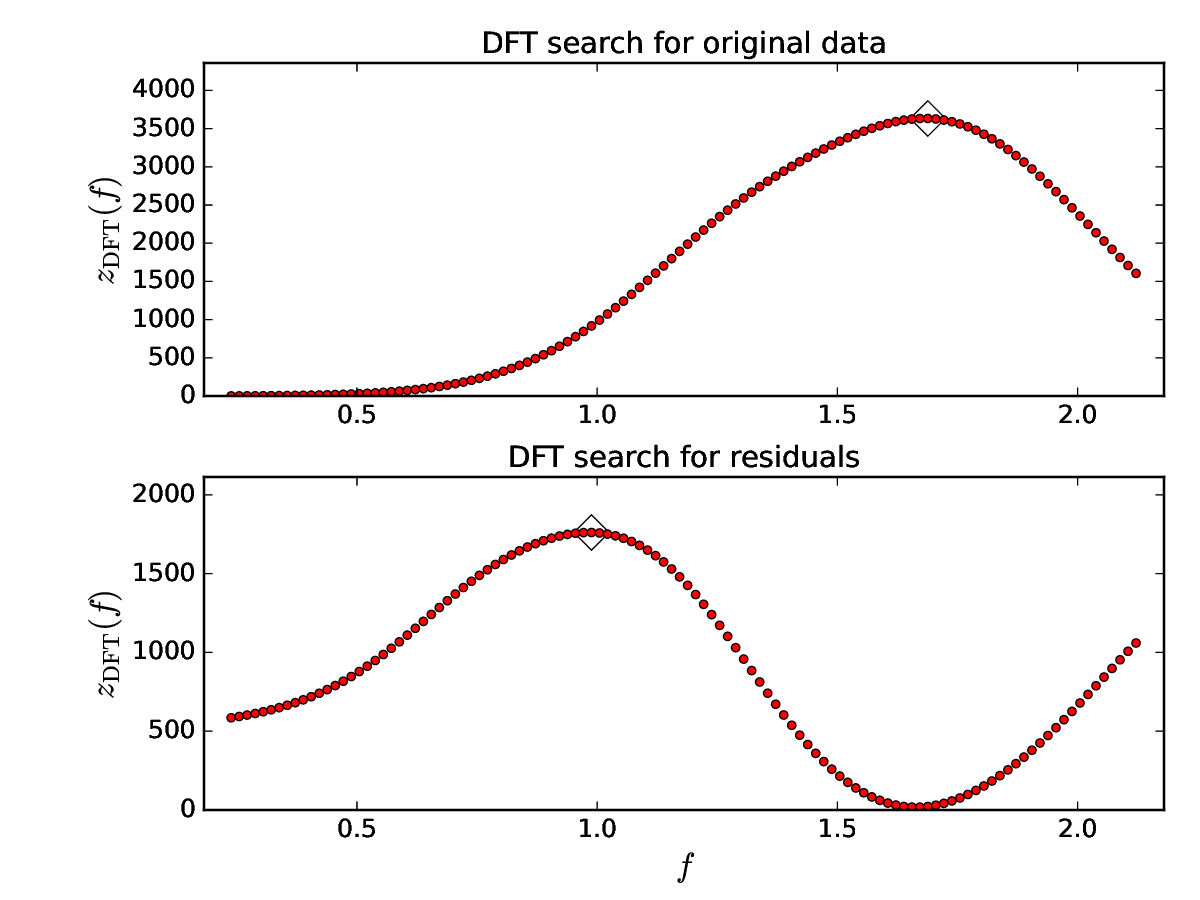}
 \hspace*{-0.01\textwidth}
 \includegraphics[width=0.46\textwidth,clip=]{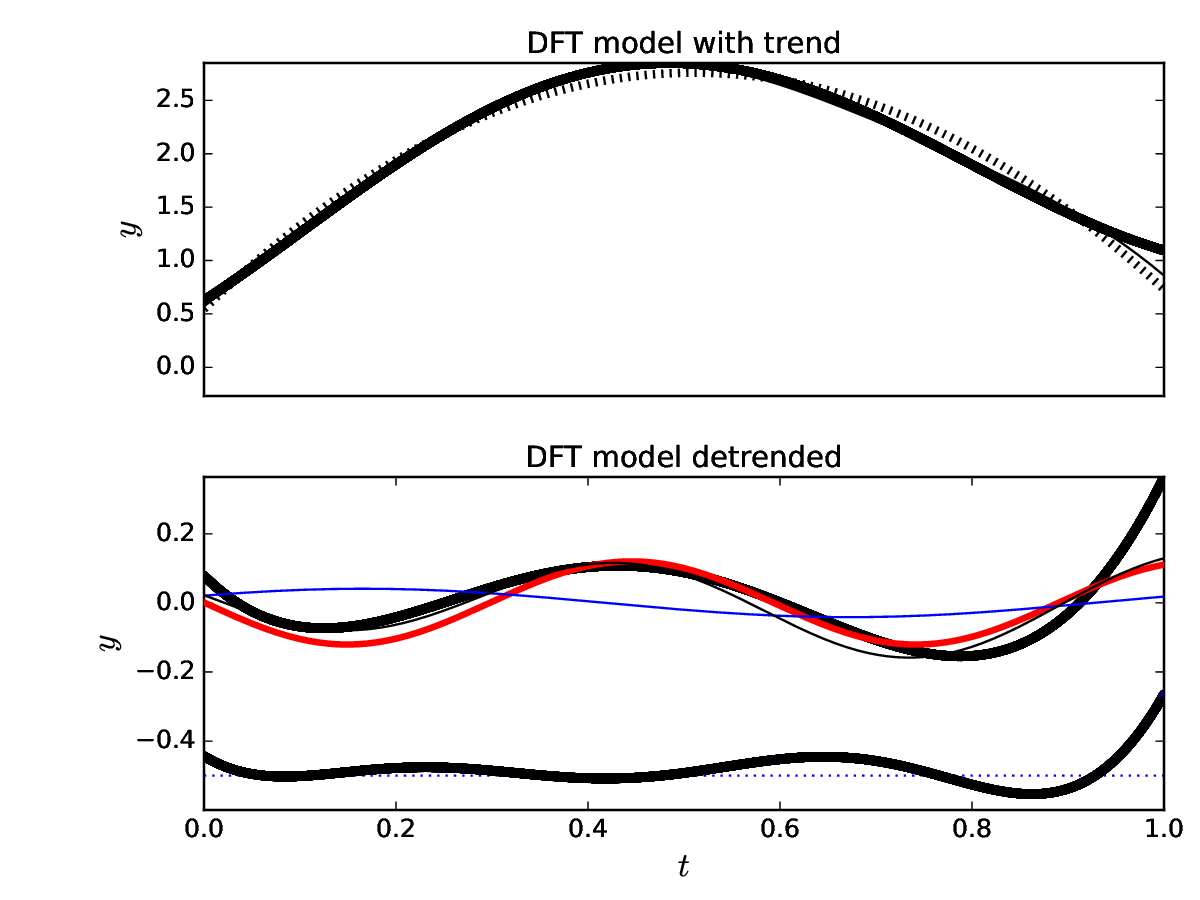}
        }
\vspace{-0.32\textwidth}
\centerline{\Large \bf 
\hspace{0.495\textwidth}  \color{black}{(e)}
\hspace{0.39\textwidth}  \color{black}{(g)}
\hfill}
\vspace{0.12\textwidth}
\centerline{\Large \bf 
\hspace{0.495\textwidth}  \color{black}{(f)}
\hspace{0.39\textwidth}  \color{black}{(h)}
\hfill}
\vspace{0.15\textwidth}
\caption{Model 5 (Table \ref{TableModelFive}:
  $n=10~000$, $\SN=10~000$ simulation).
  Notations as in Figure \ref{FigModelThree}.
  Best periods (diamonds) are explained
  in Section \ref{SectModelFive}. }
\label{FigModelFive}
\end{figure*}

\begin{table*}
  \caption{Model 6. DCM analysis between $P_{\mathrm{min}}=0.053$
    and $P_{\mathrm{max}}=0.480$. 
    Notations as in Table \ref{TableModelOne}.}\label{TableModelSix}
\begin{scriptsize}
  \begin{center}
  \begin{tabular}{lccc}
  \hline
     (1)                                      & (2)                                         & (3)                                        & (4) \\
                                               &$n=50                                    $&$n=50$                               &$n=100$\\
Model 6                                   &$\SN=10                                 $&$\SN=50$                            &$\SN=100$\\
\hline
$P_1=0.16$                             &$0.15966\pm0.00020            $&$0.160022\pm0.000021      $&$0.160024\pm0.000019$\\
$A_1=2.0$                               &$1.983\pm\pm0.041             $&$2.0031\pm0.0086               $&$2.0013\pm0.0032$\\
$t_{\mathrm{1,min,1}}=0.0979$&$0.09868\pm0.00057           $&$0.097933\pm0.000096       $&$0.097934\pm0.000079$\\
$t_{\mathrm{1,min,2}}=0.0225$&$0.02283\pm0.00070         $&$0.022563\pm0.000077          $&$0.022403\pm0.000071$\\
$t_{\mathrm{1,max,1}}=0.1421$&$0.14241\pm0.00064          $&$0.142099\pm0.000079        $&$0.142100\pm0.000077$\\
$t_{\mathrm{1,max,2}}=0.0575$&$0.05827\pm0.00057         $&$0.05761\pm0.00011             $&$0.057448\pm0.000074$\\
$M_0=0                                  $&$0.003\pm0.012                 $&$-0.0055\pm0.0026              $&$0.00176\pm0.00088$\\
 Data file                                   &\PR{Model6n50SN10.dat}      &\PR{Model6n50SN50.dat}      &\PR{Model6n100SN100.dat}\\
 Control file                               &\PR{dcmModel6n50SN10.dat}&\PR{dcmModel6n50SN50.dat}&\PR{dcmModel6n100SN100.dat}\\
\hline
  \end{tabular}
  \end{center}
  \end{scriptsize}
\end{table*}

\begin{figure*}
\vspace{0.02\textwidth}
\centerline{\hspace*{0.005\textwidth}
 \includegraphics[width=0.46\textwidth,clip=]{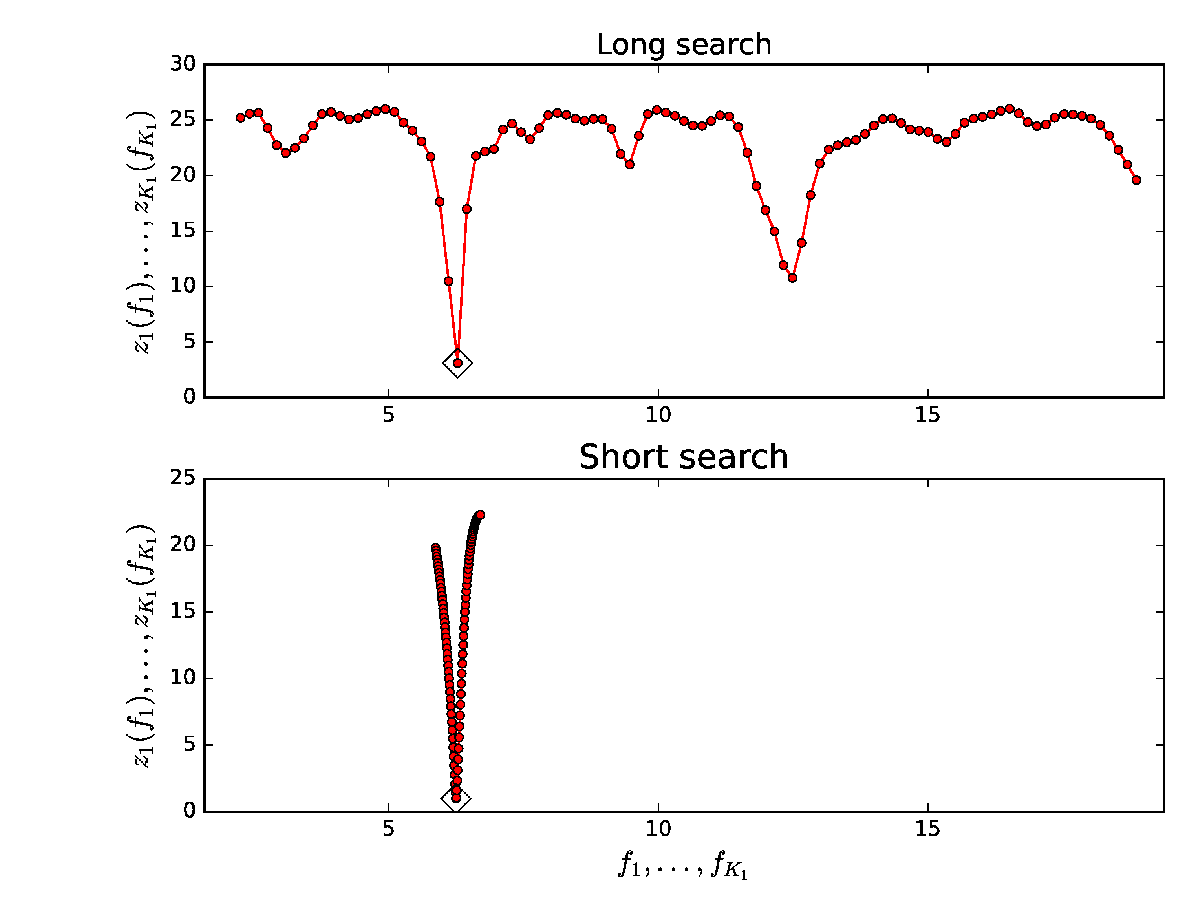}
 \hspace*{-0.01\textwidth}
 \includegraphics[width=0.46\textwidth,clip=]{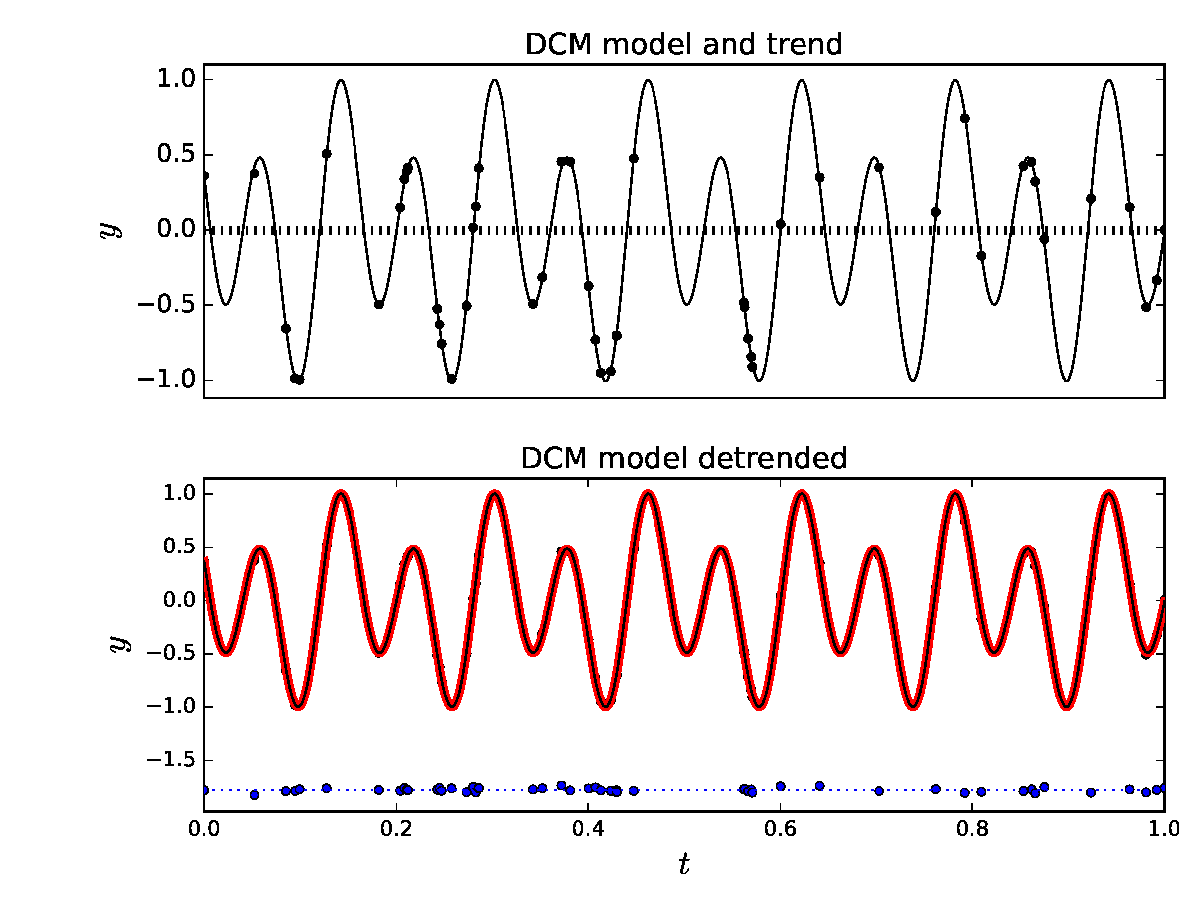}
        }
\vspace{-0.32\textwidth}
\centerline{\Large \bf 
\hspace{0.49\textwidth}  \color{black}{(a)}
\hspace{0.39\textwidth}  \color{black}{(c)}
\hfill}
\vspace{0.118\textwidth}
\centerline{\Large \bf 
\hspace{0.495\textwidth}  \color{black}{(b)}
\hspace{0.385\textwidth}  \color{black}{(d)}
\hfill}
\vspace{0.12\textwidth}
\centerline{\hspace*{0.005\textwidth}
 \includegraphics[width=0.46\textwidth,clip=]{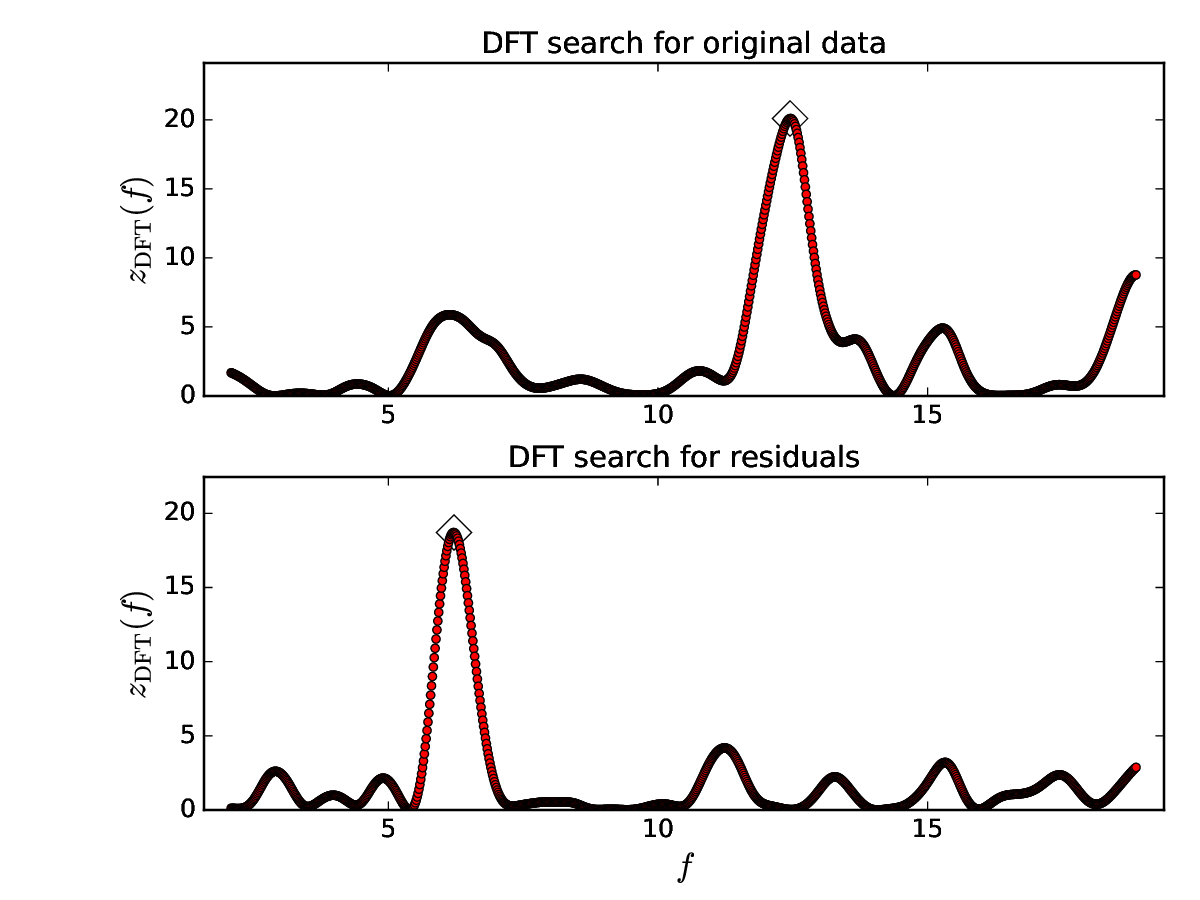}
 \hspace*{-0.01\textwidth}
 \includegraphics[width=0.46\textwidth,clip=]{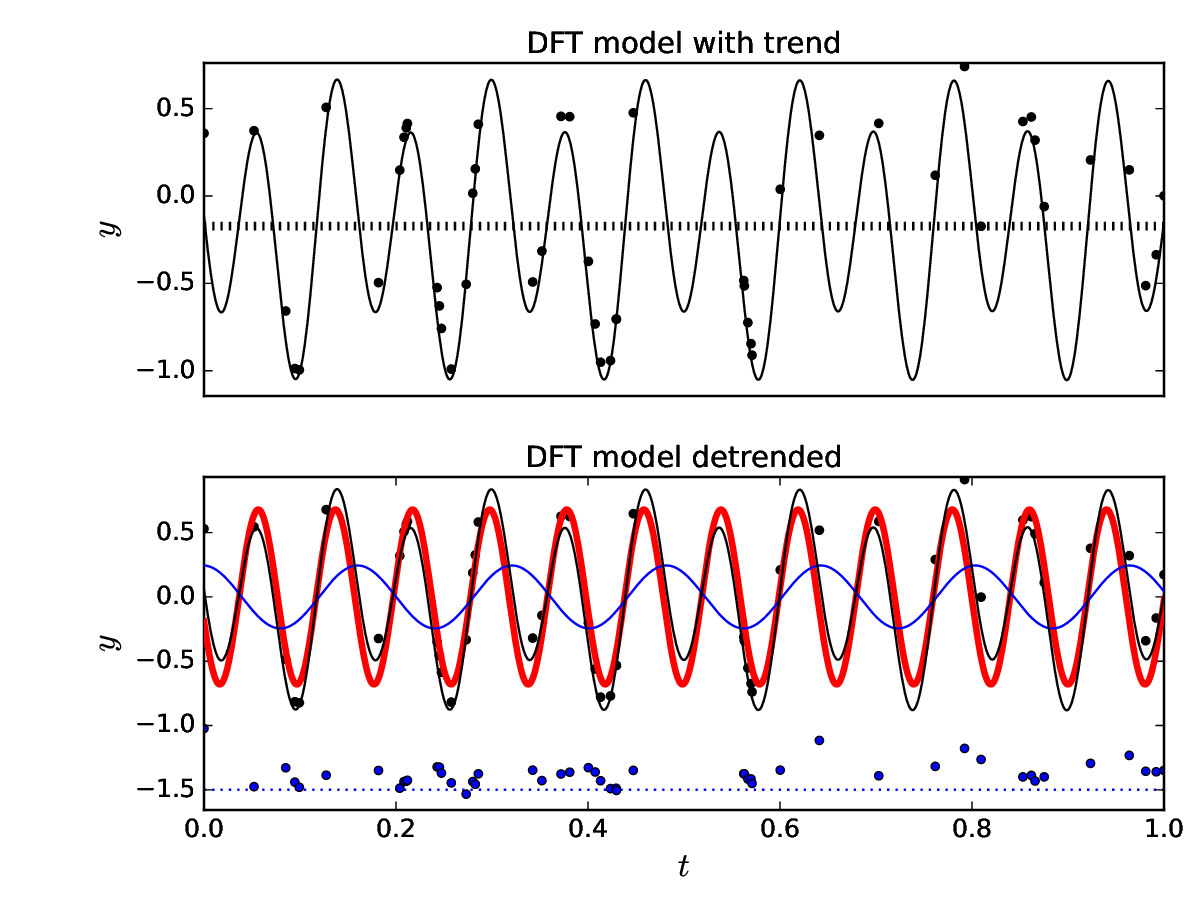}
        }
\vspace{-0.32\textwidth}
\centerline{\Large \bf 
\hspace{0.495\textwidth}  \color{black}{(e)}
\hspace{0.390\textwidth}  \color{black}{(g)}
\hfill}
\vspace{0.12\textwidth}
\centerline{\Large \bf 
\hspace{0.495\textwidth}  \color{black}{(f)}
\hspace{0.390\textwidth}  \color{black}{(h)}
\hfill}
\vspace{0.15\textwidth}
\caption{Model  6 (Table \ref{TableModelSix}:
  $n=50$, $\SN=50$ simulation).
  Notations as in Figure \ref{FigModelThree}.
  Best periods (diamonds) are explained
  in Section \ref{SectModelSix}. }
\label{FigModelSix}
\end{figure*}

\begin{table*}
  \caption{Model  7. DCM analysis between $P_{\mathrm{min}}=0.4$
    and $P_{\mathrm{max}}=3.6$. 
    Notations as in Table \ref{TableModelOne}. }\label{TableModelSeven}
  \begin{scriptsize}
    \begin{center}
\begin{tabular}{lccc}
  \hline
     (1)                                      & (2)                                          & (3)                                                          & (4) \\
                                               & $n=100$                                &$n=1~000                                               $&$n=10~000$  \\
Model 7                                   & $\SN=5~000~000$                &$\SN=1~000~000                                $&$\SN=1~000~000$  \\
\hline
$P_1=1.2$                               &$1.19917\pm0.00070            $&$1.20000\pm0.00063                           $&$1.20000\pm0.00024$  \\
$A_1=2.0$                               &$2.20\pm0.20                        $&$1.94\pm0.14                                       $&$1.990\pm0.057       $  \\
$t_{\mathrm{1,min,1}}=0.6892$&$0.6835\pm0.0050                $&$0.6900\pm0.0043                               $&$0.6888\pm0.0019    $  \\
$t_{\mathrm{1,min,2}}=0.1134$&$0.1235\pm0.0098                $&$0.1160\pm0.0067                               $&$0.1128\pm0.0026     $  \\
$t_{\mathrm{1,max,1}}=1.0195$&$1.0253\pm0.0052               $&$1.0176\pm0.0039                              $&$1.0188\pm0.0018      $  \\
$t_{\mathrm{1,max,2}}=0.3779$&$0.3705\pm0.0082               $&$0.3804\pm0.0050                              $&$0.3780\pm0.0020     $  \\
$P_2=1.4$                               &$1.4050\pm0.0048               $&$1.3986\pm0.00034                             $&$1.3997\pm0.0014     $  \\
$A_2=2.0$                               &$1.871\pm0.098                   $&$2.049\pm0.095                                   $&$2.009\pm0.040        $  \\
$t_{\mathrm{2,min,1}}=0.9262$&$0.938\pm0.012                   $&$0.9231\pm0.0090                               $&$0.9252\pm0.0036    $  \\
$t_{\mathrm{2,min,2}}=0.2766$&$0.260\pm0.016                   $&$0.281\pm0.012                                  $&$0.2771\pm0.0046    $  \\
$t_{\mathrm{2,max,1}}=1.3109$&$1.3165\pm0.0049              $&$1.3105\pm0.0039                             $&$1.3101\pm0.0014    $\\
$t_{\mathrm{2,max,2}}=0.5864$&$0.5817\pm0.0037              $&$0.5860\pm0.0042                             $&$0.5865\pm0.0015    $\\
$M_0=1.0                                $&$0.950\pm0.050                 $&$1.011\pm0.035                                $&$1.002\pm0.013         $\\
$M_1=0.25                               $&$0.2447\pm0.0067            $&$0.2548\pm0.0039                           $&$0.2509\pm0.0018    $\\
$M_2=0.5                                 $&$0.551\pm0.050                $&$0.488\pm0.035                               $&$0.498 \pm0.014      $\\
 Data file                            & \PR{Model7n100SN5000000.dat}&\PR{Model7n1000SN1000000.dat}      & \PR{Model7n10000SN1000000.dat}\\
 Control file                  & \PR{dcmModel7n100SN5000000.dat}&\PR{dcmModel7n1000SN1000000.dat}& \PR{dcmModel7n10000SN1000000.dat}\\
\hline
\end{tabular}
\end{center}
\end{scriptsize}
\end{table*}

\begin{figure*}
\vspace{0.02\textwidth}
\centerline{\hspace*{0.005\textwidth}
 \includegraphics[width=0.46\textwidth,clip=]{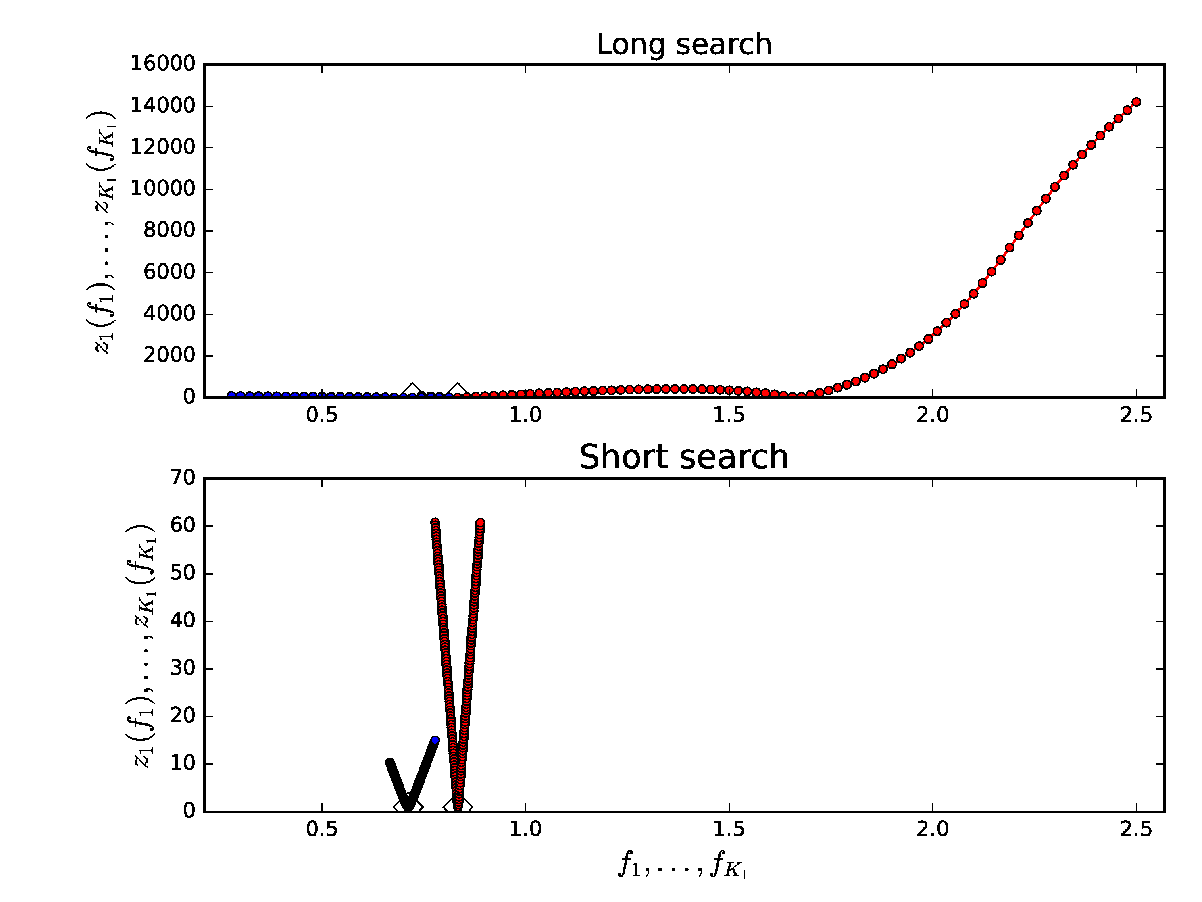}
 \hspace*{-0.01\textwidth}
 \includegraphics[width=0.46\textwidth,clip=]{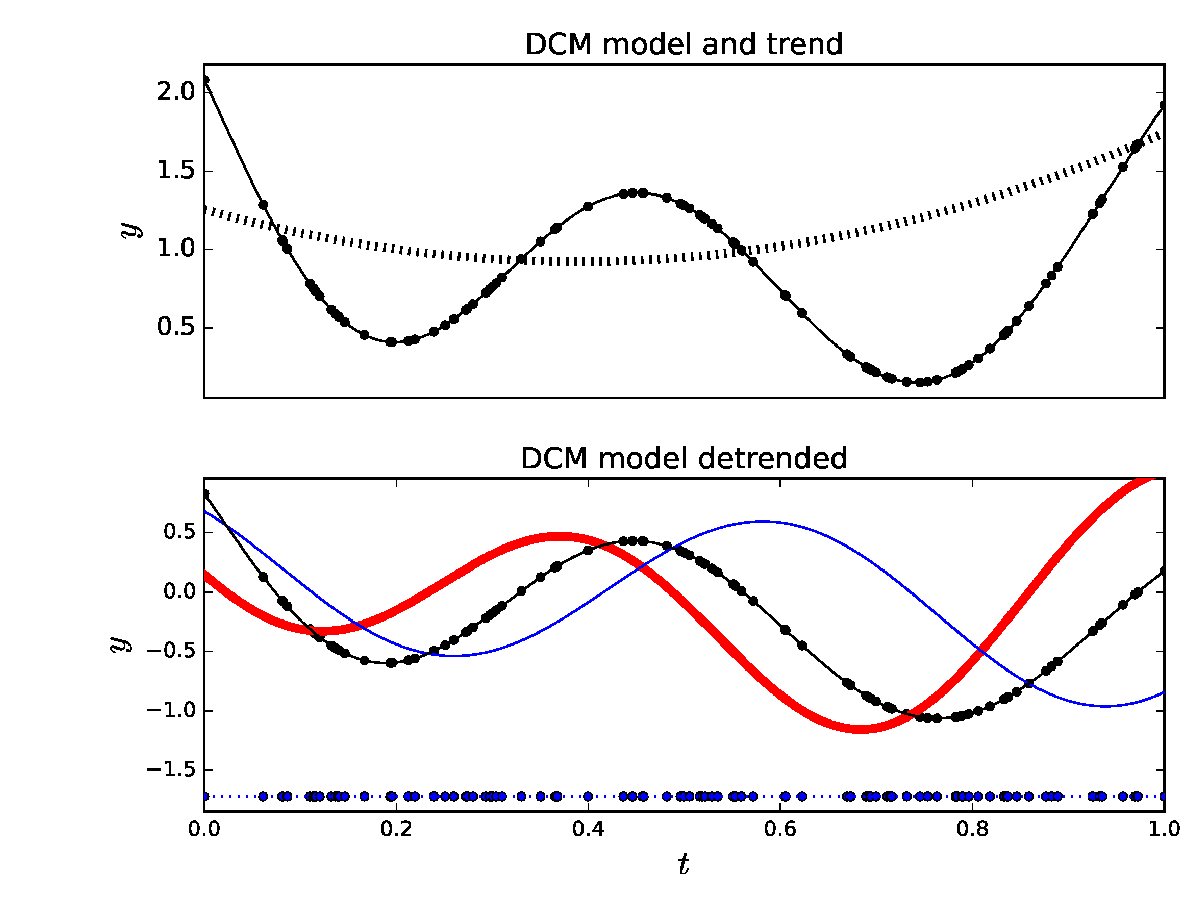}
        }
\vspace{-0.32\textwidth}
\centerline{\Large \bf 
\hspace{0.49\textwidth}  \color{black}{(a)}
\hspace{0.39\textwidth}  \color{black}{(c)}
\hfill}
\vspace{0.118\textwidth}
\centerline{\Large \bf 
\hspace{0.495\textwidth}  \color{black}{(b)}
\hspace{0.385\textwidth}  \color{black}{(d)}
\hfill}
\vspace{0.12\textwidth}
\centerline{\hspace*{0.005\textwidth}
 \includegraphics[width=0.46\textwidth,clip=]{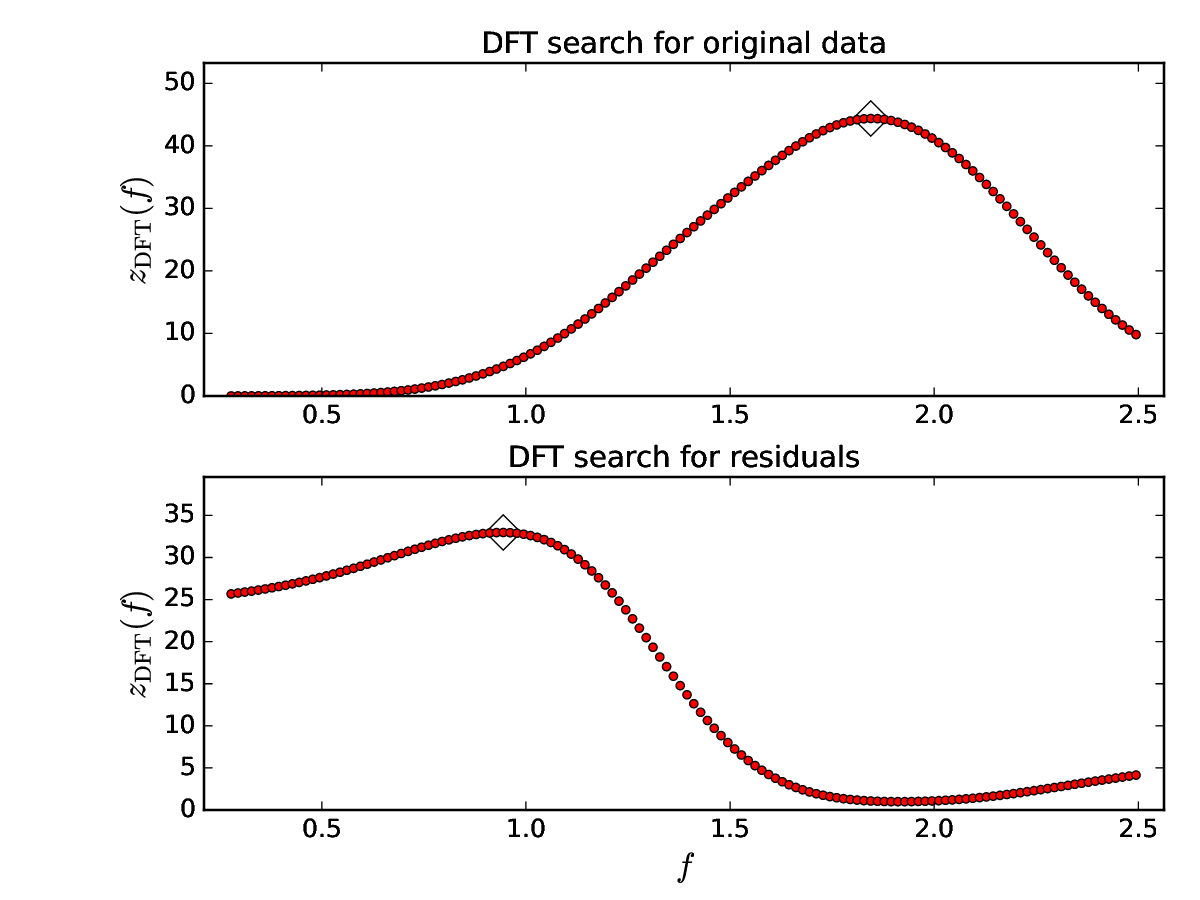}
 \hspace*{-0.01\textwidth}
 \includegraphics[width=0.46\textwidth,clip=]{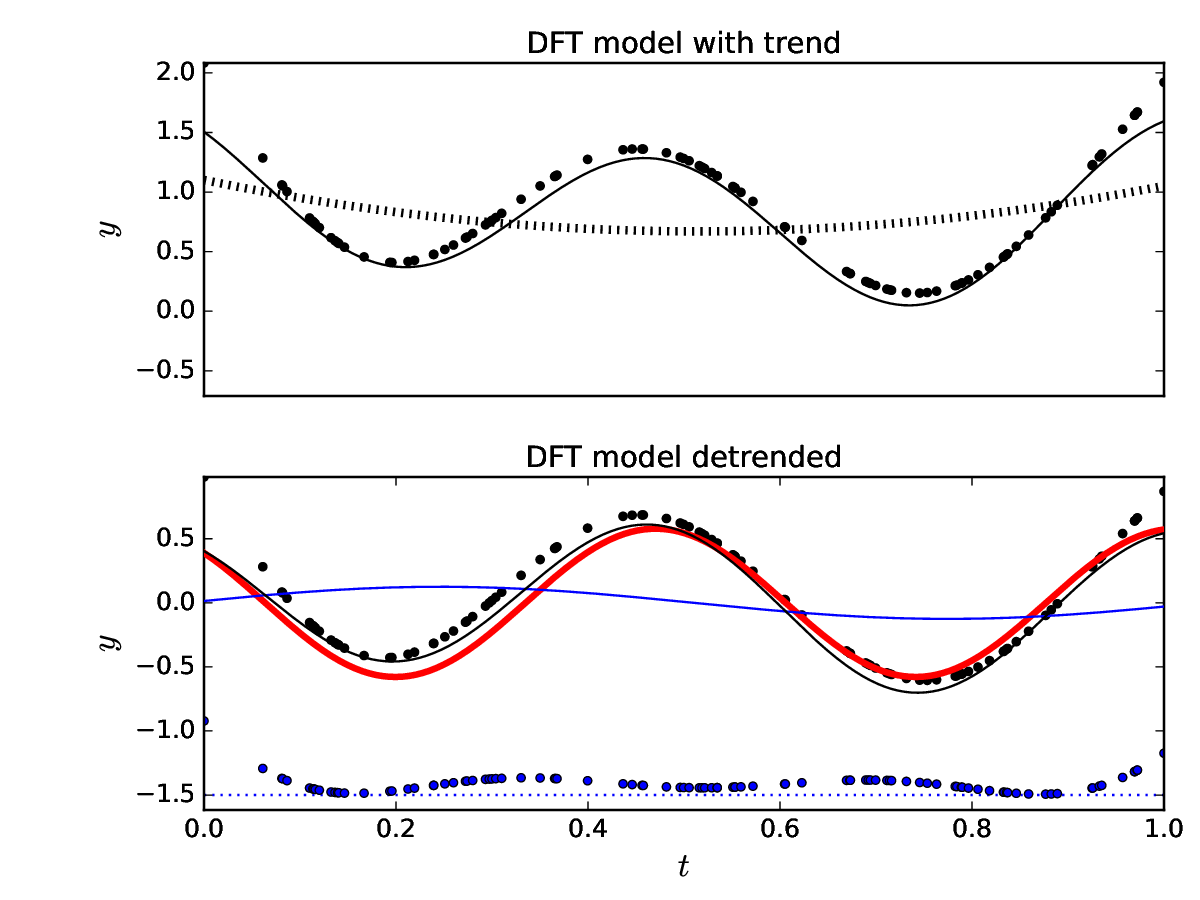}
        }
\vspace{-0.32\textwidth}
\centerline{\Large \bf 
\hspace{0.495\textwidth}  \color{black}{(e)}
\hspace{0.39\textwidth}  \color{black}{(g)}
\hfill}
\vspace{0.12\textwidth}
\centerline{\Large \bf 
\hspace{0.495\textwidth}  \color{black}{(f)}
\hspace{0.39\textwidth}  \color{black}{(h)}
\hfill}
\vspace{0.15\textwidth}
\caption{Model 7 (Table \ref{TableModelSeven}:
  $n=100$, $\SN=5~000~000$ simulation).
  Notations as in Figure \ref{FigModelThree}.
  Best periods (diamonds) are explained
  in Section \ref{SectModelSeven}. }
\label{FigModelSeven}
\end{figure*}

\subsection{Model 5}  \label{SectModelFive}

The mathematical {time series} equation
\begin{eqnarray}
  g(t) & = &
  (A_1/2)
  \cos{
  \left[
  {
  {2 \pi (t-t_{\mathrm{1,max,1}})}
  \over
  {P_1}
  }
  \right]
             } \label{EqModelFive} \\
             &  + & 
  (A_2/2)
  \cos{
  \left[
  {
  {2 \pi (t-t_{\mathrm{2,max,1}})}
  \over
  {P_2}
  }
  \right]
  } \nonumber \\
 & + & M_0+M_1T + M_2T^2 \nonumber
\end{eqnarray}
for our fifth model  is the same as for
Model 4 (Equation \ref{EqModelFour}).
However, this new Model 5 differs from the earlier Model 4
because we use totally different $P_1$, $P_2$, $A_1$,$A_2$,
$t_{\mathrm{1,max,1}}$,
$t_{\mathrm{2,max,1}}$,
$M_0$, $M_1$ and $M_2$ values
(Table \ref{TableModelFive}, Column 1).
The simulated {time series} is ``too short'' for both
$P_1$ and $P_2$ periods (Equation \ref{EqTooShort}).
There is ``a trend''  $p(t)$ (Equation \ref{EqTrend}).
The signal frequencies $f_1$ and $f_2$ are ``too close'' (Equation \ref{EqTooClose}).
The three main reasons that can cause the failure
of DFT analysis are present.
We perform DCM and DFT time series analysis between
$P_{\mathrm{min}}=P_1/3=0.47$
and
$P_{\mathrm{max}}=3P_1=4.20$.

Model 5 has DCM orders $K_1=2$, $K_2=1$ and $K_3=2$.
We give DCM analysis results
for different $n$ and $\SN$ combinations
in Table \ref{TableModelFive} (Columns 2-4).
DCM  fails to detect the correct
$P_1$, $P_2$,  ..., $M_1$ and $M_2$ values for the
the lowest $n=10~000$ and $\SN=1~000$ combination
(Table \ref{TableModelFive}, Column 2).
This shows that DCM can fail,
just like any other time series analysis method,
if the quality of data is too low.
Due to the \WDE,
DCM results for higher $n$ and $\SN$ combinations
are correct (Table \ref{TableModelFive}, Columns 3-4).

We show DCM analysis results
for one sample of Model 5 simulated {time series}
(Figures \ref{FigModelFive}a-d: $n=10~000$ and $\SN=10~000$).
The long and short DCM searches give
$P_1=1.407$ and  $P_2=1.917$
(Figure \ref{FigModelFive}a: diamonds),
and $P_1=1.393$ and  $P_2=2.024$
(Figure \ref{FigModelFive}b: Diamonds).
The continuous line denoting the model $g(t)$ is
totally covered by 
the black dots of $y_i$ data
(Figure \ref{FigModelFive}c).
Therefore, we use white colour to highlight 
this  DCM model $g(t)$  line.
The detected  polynomial trend $p(t)$ coefficients
$M_0=0.81\pm0.16$,
$M_1=0.230\pm0.013$
and
$M_2=0.596\pm0.084$ are correct
(Figure \ref{FigModelFive}c: dashed black line).
We show 
the detrended DCM model $g(t)-p(t)$ (white continuous line),
the detrended data $y(t_i)-p(t_i)$ (black dots),
the signal $h_1(t)$ (red thick continuous line)
and
the signal $h_2(t) $ (blue thin continuous line)
in Figure \ref{FigModelFive}d.
DCM residuals (blue dots) are offset to the level of -1.5.
These blue dots appear black because there are 10~000 of them.
For obvious reasons, the -1.5 level  of these residuals
is highlighted by a white dotted line.
DCM model residuals are stable and show no trends.

DFT detects the wrong periods 
for the original data
(Figure \ref{FigModelFive}e: diamond at $P_1=0.592$)
and for the residuals
(Figure \ref{FigModelFive}f: diamond at $P_2=1.012$).
DFT estimates for the $p(t)$ trend,
$M_0=2.758$, $M_1=0.091$
and
$M_2=-2.120$,
are also wrong (Figure \ref{FigModelFive}g: dashed black line).
DFT model $g_{\mathrm{DFT}}(t)$ black continuous line
deviates from the black dots of data $y_i$,
especially in the end
of the simulated data sample.
The black dots denoting
the detrended data $y(t_i) - p_{\mathrm{DFT}}(t_i)$ and
the continuous black line denoting 
the detrended DFT model $g_{\mathrm{DFT}}(t)-p_{\mathrm{DFT}}(t)$
are shown in Figure \ref{FigModelFive}h.
DFT model $g_{\mathrm{DFT}}(t)$  gives very low amplitudes for
the pure sine signal $s_{\mathrm{y,DFT}}(t)$
(continuous thick red line)
and
pure sine signal $s_{\mathrm{\epsilon,DFT}}(t)$
(continuous thinner blue line).
These amplitudes are far below the correct simulated values
$A_1=A_2=2.$
The offset level for  DFT model
residuals (blue dots) is -0.5.
These residuals show strong trends.

DCM analysis succeeds for Model 5 simulated {time series}, but
DFT analysis fails.

\subsection{Model 6}   \label{SectModelSix}

Our sixth {time series} simulation model is
\begin{eqnarray}
  g(t)& = &
  c_1
  \cos
  {
  \left[
  {
  {2 \pi t }
  \over
  {P_1}
  }
  \right]
  } \label{EqModelSix} \\
  & + &
  c_2
  \cos
  {
  \left[
  {
  {4 \pi (t-c_3) }
  \over
  {P_1}
  }
  \right]
  }
+ M_0, \nonumber
\end{eqnarray}
where $P=0.16$ and $M_0=0$. The coefficients
$c_1=0.3655$, $c_2=0.7310$ and $c_3=0.3000$
determine the $A_1$,
$t_{\mathrm{1,min,1}}$,
$t_{\mathrm{1,min,2}}$,
$t_{\mathrm{1,max,1}}$ and
$t_{\mathrm{1,max,2}}$
values given in Table \ref{TableModelSix} (Column 1). 
This simulated {time series} is not ``too short''
because $P_1< \Delta T$
(Equation \ref{EqTooShort}).
There is no trend because $p(t)=M_0=0$ 
(Equation \ref{EqTrend}).
This simulated {time series} contains only one signal
(Equation \ref{EqTooClose}).
However, this simulation Model 6 is not a $K_2=1€ $ pure sine model
(Equation \ref{EqSine}).
This double wave simulation model
has two unequal minima and maxima.
Its DCM orders
are $K_1=1$, $K_2=2$ and $K_3=0$.
We perform DCM and DFT time series  analysis between
$P_{\mathrm{min}}=P_1/3=0.053$ and
$P_{\mathrm{max}}=3P_1=0.480$. 

DCM time series analysis results for
different $n$ and $\SN$ combinations are given
in Table \ref{TableModelSix} (Columns 2-4).
DCM detects the correct  $P_1$, $A_1$,
$t_{\mathrm{1,min,1}}$
$t_{\mathrm{1,min,2}}$
$t_{\mathrm{1,max,1}}$
$t_{\mathrm{1,max,2}}$
and $M_0$
values 
even for the lowest
$n=50$ and $\SN=10$ combination.
For increasing $n$ and $\SN$,
  the results for the model parameters converge to correct values
(\WDE).

We demonstrate 
DCM analysis results for
simulated {time series} having $n=50$ and $\SN=50$
(Figures \ref{FigModelSix}a-d).
The long and short searches give
$P_1=0.159$ (Figure \ref{FigModelSix}a: diamond)
and  $P_1=0.160$ 
(Figure \ref{FigModelSix}b: diamond).
DCM model $g(t)$ black line 
covers the black dots denoting the data $y_i$ 
(Figure \ref{FigModelSix}c).
The "trend" at $p(t)=M_0=-0.0055\pm0.0026$ is correct.
The detrended model $g(t)-p(t)$ (black continuous line),
the pure sine signal $h_1(t)$ (red thick continuous line)
and
the detrended data $y(t_i)-p(t_i)$ (black dots) 
are displayed in Figure \ref{FigModelSix}d.
The thick continuous red line stays
under the thin continuous black line
because  $h_1(t)=g(t)-p(t)$. 
DCM residuals (blue dots) are offset to the level
of -1.8 (blue dotted line).
These residuals show no trends and their level is stable.

DFT detects the wrong period $P_1=0.080$
(Figure \ref{FigModelSix}e: Diamond).
This results is exactly half of the correct simulated value $P_1=0.160$.
The reason for this ``detection'' is that the double wave dominates
because $c_2=2c_1$ in Model 6 (Equation \ref{EqModelSix}).
DFT mean level estimate $M_0=-0.172$ fails
(Figure \ref{FigModelSix}f: Dashed black line).
DFT analysis of the residuals
$\epsilon_i=$ $y(t_i)-[s_{\mathrm{y,DFT}}(t_i)+p_{\mathrm{DFT}}(t_i)]$
gives $P_2=0.161$,
which is nearly equal to the correct simulated $P_1=0.160$ value.
The  black dots denoting the data $y_i$ show minor 
deviations from
continuous black line denoting DFT model $g_{\mathrm{DFT}}(t)$
(Figure \ref{FigModelSix}g).
The detrended model $g_{\mathrm{DFT}}(t)-p_{\mathrm{DFT}}(t)$
(continuous black line),
the detrended data $y(t_i) - p_{\mathrm{DFT}}(t_i)$
(black dots)
and
the signal $s_{\mathrm{y,DFT}}(t)$
(continuous thick red line)
are shown in Figure \ref{FigModelSix}h.
DFT analysis residuals (blue dots) are offset to
the level of -1.5 (blue dotted line).
These residuals  display trends.

We conclude that DFT ``detects'' the $P_1/2$ and $P_1$ periods,
where $P_1$ is the correct simulated period value.
However,
DFT two signal model is not the correct model for these
Model 6 simulated {time series},
which contains only one signal.
If the correct period is $P$
and the correct model is a double wave $(K_2=2)$,
DCM pure sine model $(K_1=1)$ analysis
can also give the values $P/2$  and $P$. 

DCM analysis succeeds
for Model 6 simulated {time series}.
DFT analysis fails.

\subsection{Model 7}   \label{SectModelSeven}

In this section, we analyse our most complex {time series}.
The seventh simulation model is
\begin{eqnarray}
  g(t) & = &
  c_1
  \cos
  {
  \left[
  {
  {2 \pi t }
  \over
  {P_1}
  }
  \right]
  }
  +
  c_2
  \cos
  {
  \left[
  {
  {4 \pi (t-c_3) }
  \over
  {P_1}
  }
  \right]
             }
             \label{EqModelSeven} \\
       & + &
  c_4
  \cos
  {
  \left[
  {
  {2 \pi t }
  \over
  {P_2}
  }
  \right]
  }
  +
  c_5
  \cos
  {
  \left[
  {
  {4 \pi (t-c_6) }
  \over
  {P_2}
  }
  \right]
             }        \nonumber \\
       &  + & M_0+M_1T +M_2 T^2 \nonumber 
\end{eqnarray}
where
$T=[2(t-t_{\mathrm{mid}})] /\Delta T$,
$P_1=1.2$, $P_2=1.4$,  $M_0=1$ $M_1=0.25$ and $M_2=0.5$.
The coefficients
$c_1=0.3687$, $c_2=0.7374$, $c_3=0.4000$,
$c_4=0.3708$, $c_5=0.7416$ and $c_6=0.6000$
determine the
$A_1$,
$t_{\mathrm{1,min,1}}$,
$t_{\mathrm{1,min,2}}$,
$t_{\mathrm{1,max,1}}$,
$t_{\mathrm{1,max,2}}$,
$A_2$,
$t_{\mathrm{2,min,1}}$,
$t_{\mathrm{2,min,2}}$,
$t_{\mathrm{2,max,1}}$ and
$t_{\mathrm{2,max,2}}$
values given in Table \ref{TableModelSeven} (Column 1). 
This simulated {time series} is ``too short'' because
$\Delta T < P_1 < P_2$  (Equation \ref{EqTooShort}).
The parabolic $p(t)$ represents 
``a trend'' (Equation \ref{EqTrend}).
The signal frequencies $f_1$ and $f_2$
are ``too close''  because
$\Delta f=f_1-f_2=0.12< f_0=1$ (Equation \ref{EqTooClose}).
The two $h_1(t)$ and $h_2(t)$ signals are not ``pure sines''
(Equation \ref{EqSine}).
These signals are double waves having two unequal minima
and maxima.
All four main reasons that can cause the failure
of DFT analysis are present
(Equations  \ref{EqTooShort}-\ref{EqSine}).
Therefore, this simulated {time series} is the most complex one of all
analysed seven {time series}.
Our DCM and DFT time series analysis of Model 7 simulated
{time series} is performed between
$P_{\mathrm{min}}=P_1/3=0.4$
and
$P_{\mathrm{max}}=3P_1=3.6$.

DCM orders of Model 7 are $K_1=2$, $K_2=2$ and $K_3=2$.
This model has $\eta=13$ free parameters (Equation \ref{EqEta}).
We give DCM analysis results
for different $n$ and $\SN$ combinations
in Table \ref{TableModelSeven} (Columns 2-4).
DCM analysis results are displayed
for one sample of Model 7 simulated {time series}  
(Figures \ref{FigModelSeven}a-d: $n=100$ and $\SN=5~000~000$).
Since there are $\eta=13$ free model parameters,
this $n=100$ {time series} is quite small for time series analysis.
The long and short DCM searches give
$P_1=1.200$ and  $P_2=1.385$
(Figure \ref{FigModelSeven}a: diamonds),
and
$P_1=1.199$ and  $P_2=1.405$
(Figure \ref{FigModelSeven}b: Diamonds).
The black $g(t)$ model line goes
through all black $y_i$ data dots
(Figure \ref{FigModelSeven}c).
DCM detects the correct  polynomial trend $p(t)$ coefficients
$M_0=0.950\pm0.050$,
$M_1=0.2447\pm0.0067$
and
$M_2=0.551\pm0.050$
(Figure \ref{FigModelSeven}c: dashed black line).
The detrended DCM model $g(t)-p(t)$ (white continuous line),
the detrended data $y(t_i)-p(t_i)$ (black dots),
the signal $h_1(t)$ (red thick continuous line)
and
the signal $h_2(t) $ (blue thin continuous line)
are displayed in Figure \ref{FigModelSeven}d.
The $n=100$ residuals (blue dots) are offset to
the level of -1.8
(dotted blue line).
These DCM model residuals are small and their level is stable.
These results confirm that  DCM time series analysis method
can detect complex non-linear models $(\eta=13)$ 
from very small {time series}
$(n=100)$,
if the data are extremely accurate $(\SN=5~000~000).$ 
These results demonstrate the power of \WDE ~because both
  periods $P_1$ and $P_2$ are shorter than the sample window $\Delta T$,
the trend is a parabola and the signals are not pure sines.

DFT time series analysis gives the wrong periods 
for the original data
(Figure \ref{FigModelSeven}e: diamond at $P_1=0.542$
and for the residuals
(Figure \ref{FigModelSeven}f: diamond at $P_2=1.059$.
DFT also gives wrong $p(t)$ trend coefficients
$M_0=0.670$, $M_1=-0.025$
and
$M_2=0.406$
(Figure \ref{FigModelSeven}g: dashed black line).
DFT model $g_{\mathrm{DFT}}(t)$ (black continuous line)
deviates from the data $y_i$ (black dots).
This deviation is largest at the beginning and the end
of the simulated {time series}.
 In our Figure \ref{FigModelSeven}h,
the black dots denote
the detrended data $y(t_i) - p_{\mathrm{DFT}}(t_i)$,
the continuous black line denotes
the detrended DFT model $g_{\mathrm{DFT}}(t)-p_{\mathrm{DFT}}(t)$
and the continuous thick red line denotes pure sine signal
$s_{\mathrm{y,DFT}}(t)$ detected from the original data.
DFT model $g_{\mathrm{DFT}}(t)$  gives very low amplitude for
the pure sine signal $s_{\mathrm{\epsilon,DFT}}(t)$ detected from the residuals
(continuous thinner blue line).
The correct simulated peak to peak amplitude
for this second signal is much larger, $A_2=2.$
DFT model
residuals (blue dots) are offset to -1.5 level and show strong trends.

DCM analysis succeeds for Model 7 simulated {time series}.
DFT analysis fails,
as predicted by Equations  \ref{EqTooShort}-\ref{EqSine}.

\subsection{Summary of simulations} \label{SectSummary}

For all seven simulated time series,
all DCM analyses succeed and all DFT analyses fail
(Sections \ref{SectModelOne} - \ref{SectModelSeven}).

\renewcommand{\Ki}{}
\Ki
 DCM detects exactly the correct signal(-s) and trend.
 \Ki
  \Ki
  Clearly, the \AL{\ref{LimSampleWindow}}-\AL{\ref{LimNotSine}}
  do not constrain our DCM.
  \Ki
   We do admit that the simulated $n$ and $\SN$ values of Models 2, 5 and 7
   are extreme and unrealistic for most cases of real data.
   \Ki
   However,
   those model simulations are necessary for demonstrating
   the \WDE.
\Ki
  We could keep on adding complexity by simulating
a larger number of signals $(K_1)$,
more complex signal shapes $(K_2)$
and/or
higher order polynomial trends $(K_3)$.
\Ki
The \WDE ~ensures that DCM would detect
those complex models when the sample size
$(n)$ and/or the signal to noise $(\SN)$ increase.
\Ki
The other way round,
the correct model can be found
{\it if} the correct DCM model combination
$K_1$, $K_2$ and $K_3$ is tested.
\Ki
Furthermore, totally wrong $K_1$, $K_2$ and $K_3$ combinations
give unstable DCM models $("\UM")$.

\renewcommand{\Ki}{}
The analyses of seven simulated time series
  expose DFT's weaknesses.
  \Ki
  Every analysis fails!
  \Ki
  The \AL{\ref{LimSampleWindow}}-\AL{\ref{LimNotSine}}
  are not the only causes for these failures.
  \Ki
  In general, the solutions for one-dimensional
  time series analysis models are ill-posed,
  if the data contain an unknown trend and/or more than one signal.
  \Ki
  For example,
  the detrending of simulated data misleads DFT even when
  the correct $K_3$ value is known
  (Sections \ref{SectModelOne}-\ref{SectModelSeven}).
  \Ki
  This correct $K_3$ value can be unknown for real data.
  \Ki
  The combination of detrending and iterative pre-whitening 
  can fail at any stage.
  \Ki
  Even if the detrending succeeds, the leakage can cause
  the detection of wrong first signal frequency.
  \Ki
  This corrupts the model residuals and the whole analysis.
  \Ki
  For any one-dimensional time series analysis method,
  like DFT,
  the search for many signals superimposed
  on an unknown trend can fail when
  the combination of detrending and pre-whitening is applied.

\renewcommand{\Ki}{}
Our DCM outperforms the esteemed DTF.
  \Ki
  DCM detects the correct signal(s) and trend,
  but DFT does not.
  \Ki
  For the sunspot data,
  DCM detects
  signals superimposed on a constant trend  \citep{Jet25}.
  \Ki
  The simulations presented here support the conclusion
  that those solar signals and the trend are correct.
   \Ki
   This would explain why  DCM is the first
   time series analysis method that detects Jupiter's exact
   $11.^{\mathrm{y}}86$ period in the sunspot record.
  \Ki
  The one-dimensional time series analysis methods,
  like DFF, have failed to find exactly correct periods.

   \begin{table*}
    \centering
  \caption{Fisher tests. DCM analysis of Model 1
    simulated {time series} ($n=100, \SN=100$,
    data file \PR{Model1n100SN100.dat}). (1) Simple model:
    Model number \M, model $g_{K_1,K_2,K_3}$,
    number of free parameters $\eta$, chi-square $\chi^2$
    and control file name.
    (2-8) Complex model:
    Hypothesis H$_0$ rejected $\uparrow$ (Complex model better),
    Hypothesis H$_0$ not rejected $\leftarrow$ (Simple model better),
    test statistic $F$ and critical level $\QF$.
    Note that no Fisher-test is done between models \M=4 and
      \M=5 because $\eta_1=\eta_2=6$.
  } \label{TableFisher}
 \begin{center}
\begin{scriptsize}
\begin{tabular}{lccccccc}
  \hline
 (1)                                             & (2)                                  & (3)              & (4)          & (5)                      & (6)      & (7)      & (8) \\
                                                  & \multicolumn{7}{@{}c@{}}{Fisher-test $(\gamma=0.001)$} \\ 
                                                  & \multicolumn{7}{@{}c@{}}{Complex model} \\ 

  Simple model                           & \M=2                               & \M=3          &\M=4       & \M=5                  & \M=6 & \M=7 & \M=8  \\
  \hline
  \M=1                                        &$\uparrow$&$\uparrow$ &$\uparrow$&$\uparrow$         &$\uparrow$  &$\uparrow$      &$\uparrow$\\  
  $g_{1,,1, -1}$                             &$F=733$  &$F=367$     &$F=242  $&$F=242$            &$F=180  $    &$F=142$          &$F=121$\\
  $\eta=3$,$\chi^2=738$          &\REJ           &\REJ              &\REJ           &\REJ                     & \REJ             &\REJ                   &\REJ      \\
\PR{dcmModel1K11-1.dat}        &                 &                    &                &                           &                   &                         &            \\
  \hline
  \M=2                                        &  -              &$\leftarrow$ &$\leftarrow$  &$\leftarrow      $&$\leftarrow$ &$\leftarrow$      &$\leftarrow$\\
  $g_{1, 1, 0}$                              &  -              &$F=1.0451$&$F=0.5404$ &$F=0.5159    $ &$F=0.3553$&$F=0.2784$     &$F=0.7686$\\
  $\eta=4$,$\chi^2=84.7422$   & -               &$\QF=0.309$&$\QF=0.584$&$\QF=0.599 $&$\QF=0.785$&$\QF=0.891$  &$\QF=0.575$\\
 \PR{dcmModel1K110.dat}        &                 &                    &                    &                        &                    &                        &                     \\
  \hline
  \M=3                                        &  -              & -                 &$\leftarrow$   &$\leftarrow$      &$\leftarrow$   &$\leftarrow$   &$\leftarrow$\\
  $g_{1,1,,1}$                              &  -               & -                &$F=0.0464$  &$F=-0.0021$    &$F=0.0213$  &$F=0.0336 $ &$F=0.7027$\\
  $\eta=5$,$\chi^2=83.8104$   &  -              & -                  &$\QF=0.830$&$\QF=1$          &$\QF=0.979$&$\QF=0.992$  &$\QF=0.592$\\
 \PR{dcmModel1K111.dat}        &                 &                    &                     &                        &                     &                      &                       \\
  \hline
  \M=4                                        &  -              & -                 & -                   &$\leftarrow$      &$\leftarrow$  &$\leftarrow$    &$\leftarrow$\\
  $g_{1, 1, 2}$                             &  -               & -                & -                   &  No Test           &$F=-0.0033$&$F=0.0277$  &$F=0.9215$     \\
  $\eta=6$,$\chi^2=83.7686$    &  -              & -                & -                    &                       &$\QF=1      $&$\QF=0.973$&$\QF=0.434$\\
  \PR{dcmModel1K112.dat}        &                 &                  &                      &                       &                   &                       &                      \\
  \hline
  \M=5  $\UM: \AD$                    &  -              & -                & -                     & -                     &$\leftarrow$  &$\leftarrow$    &$\leftarrow$\\
  $g_{2, 1, -1}$                          &  -              & -               & -                     & -                     &$F=0.0447$ &$F=0.0515$   &$F=0.9377$\\
  $\eta=6$,$\chi^2=83.8123$    &  -              & -                & -                    & -                     &$\QF=0.833$&$\QF=0.950$&$\QF=0.426$ \\
\PR{dcmModel1K21-1.dat}         &                 &                  &                      &                       &                    &                       &                        \\
  \hline
  \M=6 $\UM:\IF,\AD$                   &  -              & -                    & -                   & -                       & -            &$\leftarrow$    &$\leftarrow$\\
  $g_{2, 1, 0}$                              &  -              & -                   & -                   & -                       & -                  &$F=0.0587$ &$F=1.3840$\\
  $\eta=7$,$\chi^2=83.7716$    &  -              & -                     & -                   & -                       &-                   &$\QF=0.809$&$\QF=0.256$\\
 \PR{dcmModel1K210.dat}         &                 &                      &                      &                         &                    &                     &                      \\
  \hline
M=7 $\UM:\IF,\AD$                    &  -              & -                    & -                   & -                       & -                  & -                       &$\leftarrow$\\
  $g_{2, 1, 1}$                               &  -               & -                   & -                   & -                       & -                  &-                        &$F=2.7081$\\
  $\eta=8$,$\chi^2=83.7176$    &  -              & -                     & -                   & -                       &-                   &-                        &$\QF=0.103$\\
\PR{dcmModel1K211.dat}          &                 &                      &                      &                         &                    &-                        &                         \\
 \hline
 \M=8 $\UM:\IF,\AD$                    &  -              & -                    & -                   & -                       & -                  & -                       & - \\
  $g_{2, 1, 2}$                             &  -               & -                   & -                   & -                       & -                  &-                        & - \\
  $\eta=9$,$\chi^2=81.2721$    &  -              & -                     & -                   & -                       &-                   &-                        &- \\
  \PR{dcmModel1K212.dat}        &                 &                      &                      &                         &                    &-                        &\\
  \hline
\end{tabular}
\end{scriptsize}
\end{center}
\end{table*}

\subsection{Best model} \label{SectBestModel}

The $K_1$, $K_2$ and $K_3$
orders of the best model are not necessarily known when some
time series analysis method
is applied to  the real {time series}. 
We know a priori the best DCM and DFT model orders
for the simulated {time series} of Models 1-7
(Sections \ref{SectModelOne}-\ref{SectModelSeven}).
It could be argued that our DCM analysis
succeeds only for this reason.

All alternative $K_1$, $K_2$ and $K_3$ order models are nested.
For example,
the {\it simple}  one signal $K_1=1$ model $g_1(t)$ is a special case
of the {\it complex}
two signal $K_1=2$ model $g_2(t)$ having $A_2=0$.
We use the Fisher-test to compare
any pair of simple $g_1(t)$ and complex $g_2(t)$ models.
The model parameters
 (Equation \ref{EqR}: $R_1, R_2$ ),
(Equation \ref{EqChi}: $\chi_1$,$\chi_2$)
and
(Equation \ref{EqEta}: $\eta_1 < \eta_2$)
give the test statistic of Fisher-test  
\begin{eqnarray}
F_R & = &
\left(
{
{R_1}
\over
{R_2}
}
-1
\right)
\left(
{
{n-\eta_2-1}
\over
{\eta_2-\eta_1}
}
\right)
          \label{EqFR} \\
 F_{\chi} & = &
\left(
{
{\chi_1^2}
\over
{\chi_2^2}
}
-1
\right)
\left(
{
{n-\eta_2-1}
\over
{\eta_2-\eta_1}
}
\right).
          \label{EqFChi}          
\end{eqnarray}
The Fisher-test is based on the null hypothesis
\begin{itemize}

\item[] $H_{\mathrm{0}}$: {\it ``The complex model $g_2(t)$ 
does not provide
a significantly better fit to the data
than the simple
model $g_{\mathrm{1}}(t)$.'' }

\end{itemize}
\noindent
Under this hypothesis,
the $F_R$ and $F_{\chi}$
test statistic parameters 
have an $F$ distribution with 
$\nu_1=\eta_2-\eta_1$
and $\nu_2=n-\eta_2$
degrees of freedom
\citep{Dra98}. 
The critical level
$Q_{F} = P(F_R \ge F)$ or $Q_{F} = P(F_{\chi} \ge F)$ is
the probability that
$F_R$ or $F_{\chi}$ exceeds the numerical value $F$. If
\begin{eqnarray}
  Q_F < \gamma_F=0.001,
\label{EqFisher}
\end{eqnarray}
we reject the $H_0$ hypothesis, which
means that the complex $g_2(t)$ model
is better than the simple $g_1(t)$ model.
The pre-assigned significance level
$\gamma_F=0.001$  represents the probability that
we falsely reject the
$H_0$ hypothesis when it is in fact true.

 Larger $F_R$ or $F_{\chi}$  values
 have smaller $Q_F$ critical levels.
Hence, the probability for the $H_0$ hypothesis
rejection increases when $F_R$ or $F_{\chi}$  increases.
If the  number of complex model free parameters $\eta_2$ increases,
the $R_2$ or $\chi^2_2$ values decrease.
This increases  the $F_R$ or $F_{\chi}$ values because
the  terms $(R_1/R_2-1)$ or
$(\chi^2_1/\chi^2_2-1)$ increase
(Equations \ref{EqFR} and \ref{EqFChi}: first terms).
At the same time, the $(n-\eta_2-1)/(\eta_2-\eta_1)$ penalty 
term decreases
(Equations \ref{EqFR} and \ref{EqFChi}: second terms),
and this decreases  the $F_R$ or $F_{\chi}$ values.
This penalty term prevents the use of
too high $\eta_2$ values (too complex models).

Here, we illustrate how the Fisher-test
finds the best model
from a group of numerous alternative nested DCM models.
The Fisher-test is used
to find the best model for the simulated {time series} of Model 1 combination
$n=100$ and $\SN=100$ (Table \ref{TableModelOne}: column 4).
In other words, we assume that the correct DCM model
orders $K_1$, $K_2$ and $K_3$ are unknown,
which can be the case for real {time series}.
 The eight tested models contain one or two signals,
 and no trend or a constant trend or a linear trend or a parabolic trend. 
 We compare all these eight models
 \M=1-8 against each other (Table \ref{TableFisher}).
 The special model number notation``\M''
 is used because the notations
 ``$M_0$, ..., $M_{K_3}$'' have already been reserved
 for  the $p(t)$ trend.

Model \M=2 has 
the known correct Model 1 orders 
$K_1=1$, $K_2=1$ and $K_3=0$.
For example, the Fisher-test between the simple model
\M=1 $(\eta=3$, $\chi^2=738)$
and
the complex model \M=2  $(\eta=4$, $\chi^2=84.7422)$
gives
$F=733$
(Equation \ref{EqFChi}).
The critical level\footnote{The highest achievable
    accuracy for the computational
    f.cdf subroutine in scipy.optimize
   python library is $10^{-16}$.}
 for this very large $F$ value is extremely significant, \REJ.
 This means that the $H_0$ hypothesis must be rejected,
 and the complex \M=2 model is certainly better
 than the simple \M=1 model
 (Equation \ref{EqFisher}).
 The upward arrow ``$\uparrow$''
 in Table \ref{TableFisher} indicates
 that  \M=2 model is a better model than  \M=1 model.
 A closer look at Table \ref{TableFisher}
 reveals that  \M=2 model
 is better than all other models because the 
 ``$\uparrow$''  and ``$\leftarrow$'' 
 arrows of all other models point toward \M=2 model.
 There is no need to test models having more than two signals
 because all two signal \M=5-8 models are already unstable
 (Table \ref{TableFisher}:''$\UM$'').
 The Fisher-test finds the correct DCM model for Model 1
 simulated {time series}.

 In the above example,
 the Fisher-test
 finds the correct number of signals $(K_1)$
 and the correct trend $(K_3)$
 for the pure sine signal alternative $(K_2=1)$.
 We do not test the double
 wave signal alternative $(K_2=2)$
 against the pure sine signal alternative $(K_1=1)$
 because the number of tested models would increase from 8 to 16, 
and Table \ref{TableFisher} would become excessively large.
One example of testing
the $K_2=2$ signal models against
each other can be found in \citet[][Table S7]{Jet25}.

 We conclude that
the best model for the real {time series} can be found
by applying the Fisher-test to any arbitrary number of
different nested DCM or DFT models.

\subsection{Significance estimates} \label{SectSignificance}

\renewcommand{\Ki}{}
\Ki
 \citet{Jet20} or \citet{Jet25} gave
 no signal significance estimate for the first detected period.
 \Ki
 Here, we use the Fisher-test for this purpose.
 \Ki
 The one signal model is the complex model.
 \Ki
 The logical simple model alternative is  $g(t)=p(t)=m \equiv$ white noise
 having standard deviation $s$.
 \Ki
 However, white noise is not the only possible alternative simple model.
 \Ki
 The other possible nested no signal DCM polynomial models are
 \begin{eqnarray}
   g(t)=p(t,K_3)
   \label{EqNoSignal}
 \end{eqnarray}
where $h(t)=0$, $K_3= 0, 1, 2..$.
The $K_3=0$ polynomial
$g(t)=p(t)=M_0$ represents white noise.
\Ki
The Fisher-test gives
the critical level $\QF$ for rejecting the $H_0$ hypothesis
when this hypothesis is in fact true
(Equation \ref{EqFisher}).
\Ki
Therefore,  this $Q_{\mathrm{F}}$ value represents the probability of false
signal detection.
\Ki
In cases \REJ, the signal detection is absolutely certain.

 \renewcommand{\Ki}{}
We use
the Model 1 simulated {time series} combination $n=100$ and $\SN=100$
to demonstrate the significance estimation for the first detected period.
\Ki
The correct model for the first detected signal is $g_{1,1,0}(t)$.
\Ki
The Fisher-test is used to compare 
this correct $g_{1,1,0}(t)$ model
to different  polynomial models $g(t)=p(t,K_3)$ having $K_3=0,1,..,7$
(Table \ref{TablePolynomials}: Equation \ref{EqNoSignal}).
\Ki
The \M=1, 2 and 3 models have less free parameters than
the correct  $g_{1,1,0}(t)$  model, but the $\chi^2$ values
of these three polynomial models
are so large
that the Fisher-test is merely a formality.
\Ki
The \M=4 model has the same number of free parameters
as the  correct $g_{1, 1,0}(t)$ model, but the
comparison of $\chi^2$ values reveals that this third order
$p(t)$ polynomial is not the correct model for the {time series}.
\Ki
The next model \M=5 has more free parameters
than the correct $g_{1, 1,0}(t)$ model.
This fourth order $p(t)$ polynomial \M=5 model
must be rejected
because it has a larger $\chi^2$ value 
than the correct $g_{1, 1,0}(t)$ model.
\Ki
The critical levels  $\QF \gg\gamma=0.001$
  for the remaining \M=6, 7 and 8 polynomial models
  are so large that these models must also be rejected.
  \Ki
The results in Table \ref{TablePolynomials} confirm
that the one signal model is better than any polynomial
model in Equation \ref{EqNoSignal}.
\Ki
Hence, the analysed Model 1 simulated {time series}
must contain at least the $h_1(t)$ pure sine signal.
\Ki
For the constant, linear or parabolic $p(t)$ trend alternatives,
the significance for this $h_1(t)$ signal is \REJ.

\renewcommand{\Ki}{}
\Ki
After the detection of the first strongest signal,
the Fisher-test critical level
$\QF$ values increase for the next detected weaker signals.
\Ki
In other words, the signal significances decrease.
\Ki 
Typical examples can be found  in \citet[][Tables S5-S16]{Jet25}.
\Ki
No new signals are detected when $\QF>\gamma=0.001$ because
the critical level exceeds the pre-assigned significance level and
the $H_0$ hypothesis is no longer rejected.
For the Model 1 simulated $n=100$ and $\SN=100$ combination,
Table \ref{TableFisher} shows that this time series contains only one signal.

For this Model 1 time series,
there is no need to discuss 
DFT significance estimates \citep[][their Equation 22]{Hor86}
because this method fails to detect the correct period.

We conclude that the Fisher-test identifies the correct DCM model
  (Section \ref{SectBestModel}), as well as
  gives the signal significance $\QF$ estimates
  (Section \ref{SectSignificance}).

\begin{table}
  \caption{Fisher tests for Model 1
    $n=100, \SN=100$ simulated {time series} (electronic
    data file \PR{Model1n100SN100.dat}). 
    (1) Polynomial model (Equation \ref{EqNoSignal}).
    (2) DCM model $g_{1,1,0}$.
    Note that \M=1-3 ~polynomials represent simple models,
    and \M=5-8 ~polynomials represent complex models.
    Otherwise as in Table \ref{TableFisher}.}\label{TablePolynomials}
  \begin{center}
\begin{tabular}{lc}
  \hline
(1)                                                          & (2)                                    \\
                                                              & Fisher-test $(\gamma=0.001)$ \\         
                                                             & $g_{1,1,0}$ \\
  Polynomial                                            & $\eta=4, \chi^2=84.7422$   \\
  \hline
  \M=1                                                     &   $\uparrow$                 \\
  $g(t)=p(t,K_3=0)$                                   &    $F=63218$                 \\
  $\eta=1$, $\chi^2=169260$                    & \REJ                             \\
  \hline
  \M=2                                                     &   $\uparrow$                \\
  $g(t)=p(t,K_3=1)$                                   &    $F=54366$           \\
  $\eta=2$, $\chi^2=97076$                      &          \REJ                   \\
  \hline
  \M=3                                                     &    $\uparrow$                \\
  $g(t)=p(t,K_3=2)$                                   &     $F=1945$               \\
  $\eta=3$, $\chi^2=1820$                       &            \REJ                 \\
  \hline
  \M=4                                                     &     $\uparrow$               \\
  $g(t)=p(t,K_3=3)$                                   &      No test                  \\
  $\eta=4$, $\chi^2=469$                         &                                   \\
  \hline
  \M=5                                                     &    $\uparrow$              \\
  $g(t)=p(t,K_3=4)$                                   &     $F=-0.3718$           \\
  $\eta=5$, $\chi^2=85.0787$                   &     $\QF=1$                 \\
  \hline
  \M=6                                                     &     $\uparrow$             \\
  $g(t)=p(t,K_3=5)$                                   &     $F=0.3403$              \\
  $\eta=6$, $\chi^2=84.1265$                   &     $\QF=0.712$          \\
  \hline
  \M=7                                                     &      $\uparrow$            \\
  $g(t)=p(t,K_3=6)$                                   &       $F=0.3524$            \\
  $\eta=7$, $\chi^2=83.7794$                    &    $\QF=0.787$          \\
  \hline
  \M=8                                                     &     $\uparrow$              \\
  $g(t)=p(t,K_3=7)$                                   &      $F=0.9394$          \\
  $\eta=8$, $\chi^2=81.3819$                      &   $\QF=0.445$                 \\
\hline
\end{tabular}
\end{center}
\end{table}

    \subsection{Ill-posed problem} \label{SectIllPosed}

The solution for the non-linear DCM model is an
ill-posed problem (Equations \ref{Eqmodel}-\ref{EqPolyTwo}).
We present a computational statistical solution that fulfils
the \ConOne, \ConTwo ~and
\ConThree ~conditions of a well-posed problem.
DCM model is just one special case of a non-linear model.
Our technique can be applied to solve
other non-linear models:
Use the free parameters
that make the model non-linear
(Equation \ref{EqBetaOne}: $\bm{\beta}_{I}$)
for solving the remaining other free parameters
(Equation \ref{EqBetaTwo}: $\bm{\beta}_{II}$).

\subsubsection{Existence (\ConOne)} \label{SectExistence}

\cite{Fou22} transformed the original function
into the frequency domain.
The modern DFT time series
analysis method transforms the original {time series}
into the frequency domain and gives the best
frequency for a pure sine model.
\cite{Gau21} presented the LS method,
which minimises the differences
between the data and the linear model.
Our DCM does the same by testing
a large number of linear models.
The data spacing,  even or uneven,
  is irrelevant for these models.
For every chosen DCM model,
the total
number of tested linear models is
\begin{equation}
  n_{\mathrm{Lin}}=\binom{n_{\mathrm{L}}}{K_1}+(1+n_{\mathrm{B}})
  \times  \binom{n_{\mathrm{S}}}{K_1},
 \label{EqMassive}
\end{equation}
where the number of tested
long and short search
frequencies is $n_{\mathrm{L}}$ and $n_{\mathrm{S}}$,
respectively.
The number of bootstrap samples is $n_{\mathrm{B}}$.
In other words, we solve this ill-posed problem by using
brute computational  force.
If the {time series} contains only zero mean white noise,
the Gau{\ss}-Markov theorem ensures that
a LS fit solution {\it exists} for
every tested frequency combination.

\subsubsection{Uniqueness (\ConTwo)}

 We make the most of the Gau{\ss}-Markov theorem. 
  When the numerical values of the
  tested frequencies $\bm{\beta}_{I}$ (Equation \ref{EqBetaOne})
  are fixed,
  the model becomes linear and the solution
  for the other free parameters 
  $\bm{\beta}_{II}$ is {\it unique} (Equation \ref{EqBetaTwo}).
  All possible  $\bm{\beta}_{I}$
  frequency combinations are tested
  (Equation \ref{EqCombinations}).
  For
  every tested frequency combination $\bm{\beta}_{I}$,
  the linear model
  gives a {\it unique} value for
  the test statistic $z$ (Equations~ \ref{EqzR} and \ref{EqzChi}).
  From all tested frequency combinations,
  we select the best frequency
  combination  $\bm{\beta}_{\mathrm{I,best}}$
  which minimises $z$.
  The linear model for this
  best frequency combination $\bm{\beta}_{\mathrm{I,best}}$
  gives {\it unique} values for the remaining other
  free parameters $\bm{\beta}_{\mathrm{II,best}}$.
  The only goal for the massive DCM search
(Equation \ref{EqMassive})
  is to
  find these {\it unique} initial free parameter values
  $\bm{\beta}_{\mathrm{initial}}=
  [\bm{\beta}_{\mathrm{I,best}},\bm{\beta}_{\mathrm{II,best}}]$
  for the non-linear iteration
  that gives the {\it unique} final free parameter
  values  $\bm{\beta}_{\mathrm{final}}$  (Equation \ref{EqIteration}).

 We  use the Fisher-test
  to compare many different non-linear
  DCM models against each other.
  The selection criterion
  for the best model is {\it unique} (Equation \ref{EqFisher}).
  The best DCM model is not necessarily the correct model,
  if this correct model is not among the compared models.
  The correct model must be able to forecast the future 
  and past data.
  We formulate the Forecast-test for alternative DCM models
  (Equation \ref{EqForez}).
  The order of Fisher- and Forecast-tests can be reversed.
  However, the former uses all data, while the latter 
  uses a subset of all data, the forecasting data.
  In ideal cases, both tests identify the same best
  and correct model,
  like in Section \ref{SectForecast}.
  
\subsubsection{Stability (\ConThree)} \label{SectStability}

   The artificial bootstrap data sets (Equation \ref{EqBoot})
  represent ``small changes in the input data'',
  while the bootstrap
  results for the model parameters represent
  ``small changes in the solution''
  (Section \ref{SectInt}: \ConThree ~condition formulation).
  We routinely
  check the {\it stability}
  of these solutions \citep[][Figure S5]{Jet25}.
  The unstable models, where the model parameter changes
  are large, are rejected  (Section \ref{SectDCM}: ``$\UM$'' models).
  
  There are additional signatures of {\it stability}.
  The $z$  periodogram solution is unique
  for every tested
   $\bm{\beta}_{I}$ 
  frequency combination.
  If these periodograms are continuous
  and their changes are not irregular
  (e.g., like in Figs. \Ref{FigModelOne}a-b),
  DCM model solution is {\it stable}
  because it  does not change by increasing
  the number of tested frequencies
  $n_{\mathrm{L}}$ and  $n_{\mathrm{S}}$.
  Furthermore, the solutions for all seven
  simulated {time series} converge when the $n$ and $\SN$
  values increase
  (Tables \ref{TableModelOne} - \ref{TableModelSeven}).
  The different $n$ and $\SN$ combinations
  give the same {\it stable} DCM model solution.

  We conclude that our computational statistical
  DCM model solution
fulfils the \ConOne,
\ConTwo ~and \ConThree
~conditions of the solution for a well-posed problem
(Sections \ref{SectExistence} - \ref{SectStability}).

\begin{figure*}  
\vspace{0.02\textwidth}
\centerline{\hspace*{0.005\textwidth}
 \includegraphics[width=0.90\textwidth,clip=]{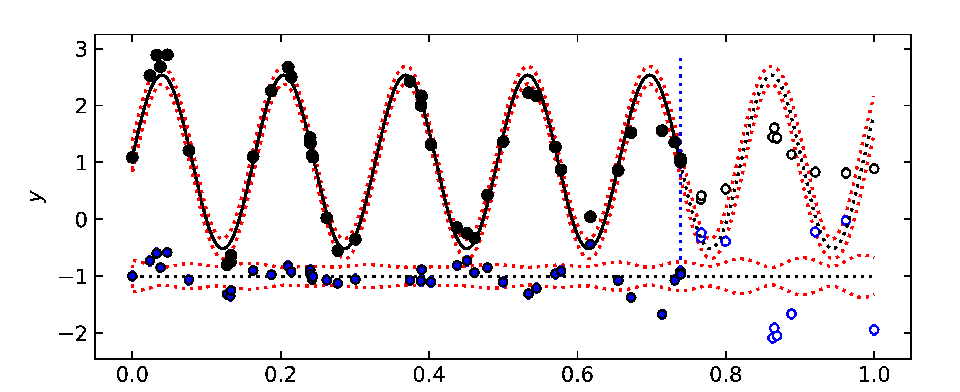}
        }
\vspace{-0.37\textwidth}
\centerline{\Large \bf 
\hspace{0.85\textwidth}  \color{black}{(a)}
\hfill}
\vspace{0.35\textwidth}
\centerline{\hspace*{0.005\textwidth}
 \includegraphics[width=0.90\textwidth,clip=]{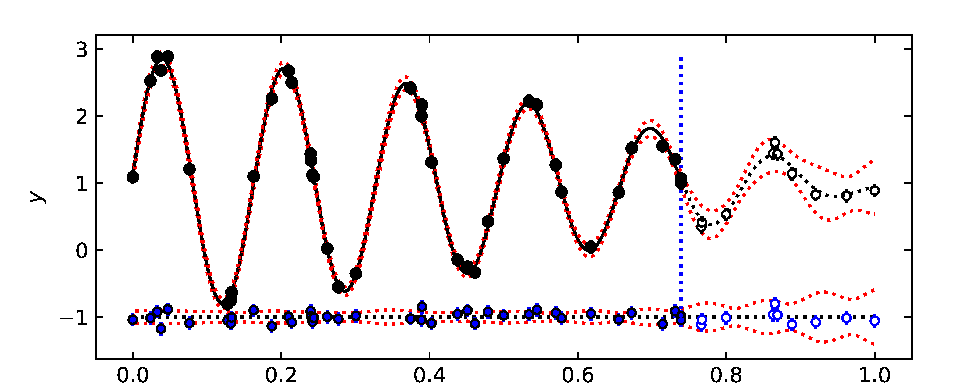}
        }
\vspace{-0.365\textwidth}
\centerline{\Large \bf 
  \hspace{0.85\textwidth}  \color{black}{(b)}
\hfill}
\vspace{0.35\textwidth}
\centerline{\hspace*{0.005\textwidth}
 \includegraphics[width=0.90\textwidth,clip=]{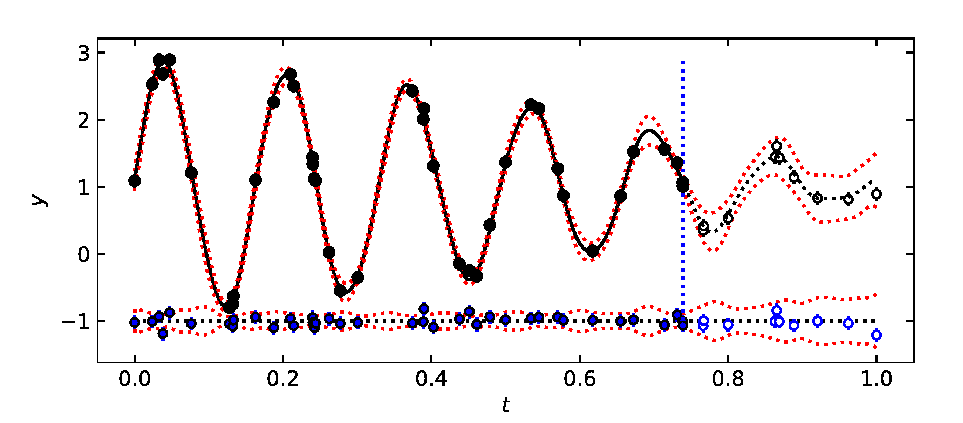}
        }
\vspace{-0.395\textwidth}
\centerline{\Large \bf 
  \hspace{0.85\textwidth}  \color{black}{(c)}
\hfill}
\vspace{0.33\textwidth}
\caption{DCM model forecast
  (Model 3 simulated {time series} for combination $n=50$ and $\SN=10$:
  first forty observations are forecasting data).
  (a) One signal model $g_{1,1,0}$ results.
  Forecasting data are
  $t_i$, $y_i$ and $\sigma_i$ ($n=40$ black dots)
  and 
  forecasted data are
  ten last observations $t_i'$, $y_i'$ and $\sigma_i'$ ($n=10$ open black dots).
  Continuous black line of
  forecasting data model $g(t,\bm{\beta}_{\mathrm{Fore}})$
  ends to vertical dotted blue line,
  where dotted black line of forecasted model
  $g'(t,\bm{\beta}_{\mathrm{Fore}})$ begins. 
  Dotted red line denotes  $\pm 3 \sigma_{g(t)}$ errors of both models
(Equation \ref{EqModelError}).
  Residuals of forecasting data model
  $\epsilon_i=y_i-g_i$ (blue dots),
  residuals of forecasted data model
  $\epsilon_i='y_i'-g_i'$ (open blue dots)
  and  $\pm 3 \sigma$ errors of both models (red dotted line)
  are offset to level -1 (blue dotted line).
  (b) Two signal model $g_{2,1,0}$ results. Otherwise as in ``a''.
  (c) Three signal model $g_{3,1,0}$ results. Otherwise as in ``a''.  
}
\label{FigEight}
\end{figure*}

\begin{table}
  \caption{DCM time series analysis
     between
    $P_{\mathrm{min}}=0.53$
    and
    $P_{\mathrm{max}}=4.80$
    for forecasting data
    (Model = 3 combination $n=50$ and $\SN=10$
    simulated data: first forty observations).
    (1) Model. (2) Forecasting data test statistic $z$ (Equation \ref{EqzChi}).
    (3) Forecasted data
    test statistic $z_{\mathrm{Fore}}$ (Equation \ref{EqForez}).
    (4) Data file.
    (5) Control file.
    (6) Figure where forecast is shown. }\label{TablezFore}
  \begin{center}
    \addtolength{\tabcolsep}{-0.12cm}
  \begin{tabular}{cccccc}
  \hline
  (1)                   & (2)   & (3)                         & (4)                                   & (5)             & (6)      \\
  Model              & $z$  &$z_{\mathrm{Fore}}$& Data file                           & Control file & Fig \\                          
  \hline
  $g_{1,1,0}$      & 2.27 &    8.61                   &\PR{Model3n40SN10.dat} & \PR{dcmModel3K110.dat} & 8a \\
  $g_{2,1,0}$      & 0.76 &   0.88                    &\PR{Model3n40SN10.dat} & \PR{dcmModel3K210.dat} & 8b \\
  $g_{3,1,0}$      & 0.70 &   0.91                   &\PR{Model3n40SN10.dat} & \PR{dcmModel3K310.dat} & 8c \\
\hline
\end{tabular}
    \addtolength{\tabcolsep}{+0.12cm}
\end{center}
\end{table}

\subsection{Forecast} \label{SectForecast}

There are numerous techniques for
forecasting a time series
\citep[e.g.,][]{Ham94,Has01,Kaz25}.
DCM model $g(t_i,\bm{\beta})$ can be used to forecast.
We divide 
all data into
 the forecasting data  and the forecasted data.
 The time points, the observations
and the errors of these samples are
\begin{itemize}

  \item[] $n$ forecasting data values $t_i$, $y_i$ and $\sigma_i$
  \item[] $n'$ forecasted data values $t_i'$, $y_i'$ and $\sigma_i'$
\end{itemize}
\noindent
DCM gives
the best forecasting data model
\begin{eqnarray}
  g_i=g(t_i,\bm{\beta}_{\mathrm{Fore}}),
  \label{EqForecastingDataModel}
\end{eqnarray}
where $\bm{\beta}_{\mathrm{Fore}}$ are the free parameter values.
The $t_{\mathrm{mid,Fore}}\!=\!(t_n\!+\!t_1)/2$
and
$\Delta T_{\mathrm{Fore}}\!=\!t_n\!-\!t_1$ values are
computed from the forecasting data time
points $t_i$.
The  forecasting model values at any arbitrary time $t$
  can be obtained from the
  $\bm{\beta}_{\mathrm{Fore}}$, $t_{\mathrm{mid,Fore}}$
and    $\Delta T_{\mathrm{Fore}}$  values.

The forecasted data model values are
\begin{eqnarray}
g'_i & = & g(t'_i,\bm{\beta}_{\mathrm{Fore}}) ,  
  \label{EqForecastingDataModel}
\end{eqnarray}
where
$t_{\mathrm{mid}}=t_{\mathrm{mid,Fore}}$
and
$\Delta T=\Delta T_{\mathrm{Fore}}$.
We do not compute ``new''
$t_{\mathrm{mid}}$ and $\Delta T$ values
from the forecasted data time points $t'_i$
because the correct $g'_i$ values are obtained
only from the $\bm{\beta}_{\mathrm{Fore}}$,
$t_{\mathrm{mid,Fore}}$ and $\Delta T_{\mathrm{Fore}}$ combination
of the forecasting model
(Equation \ref{EqForecastingDataModel}).
The $n'$ forecasted data model residuals
\begin{eqnarray}
  \epsilon'_i  =  y'_i-g'_i
\nonumber
\end{eqnarray}
give the {\it forecasted data test statistic}
\begin{eqnarray}
  z_{\mathrm{Fore}}  & = &
                           \textnormal{z test statistic for forecasted} \label{EqForez} \\
                              &    &
                                     \textnormal{data $t'_i$, $y'_i$ and $\sigma'_i$ 
                                     (Equation \ref{EqzR} or \ref{EqzChi}). } \nonumber
\end{eqnarray}
\noindent
This parameter $z_{\mathrm{Fore}}$
measures how well the forecast
(Equation \ref{EqForecastingDataModel}) obtained
from the forecasting data works for the forecasted data.
If DCM detects a new signal from the forecasting data,
there are two alternatives: 

\begin{itemize}
\item[] If this new signal is real,
  the $z_{\mathrm{Fore}}$ value of forecasted data decreases.  
\item[]  If this new signal is unreal,
  the $z_{\mathrm{Fore}}$ value of forecasted data increases.
 \end{itemize}
 This ``Forecast-test''
 technique revealed at least five real signals in
the sunspot record \citep{Jet25}

We compute the $z_{\mathrm{Fore}}$ parameter value
from
the {\it known} forecasted data
$t'_i$, $y'_i$ and $\sigma'_i$ values (Equation \ref{EqForez}).
Forecasts are possible even if
all $t'_i$, $y'_i$ or $\sigma'_i$ forecasted
data values are {\it unknown}.
In this case,
the $t_i$, $y_i$ and $\sigma_i$
values of all data can be used as forecasting data. 
The best DCM model $g(t_i,\bm{\beta}_{\mathrm{Fore}})$ for all data
determines the correct $\bm{\beta}_{\mathrm{Fore}}$,
$t_{\mathrm{mid,Fore}}=(t_n+t_1)/2$ and
$\Delta T_{\mathrm{Fore}}=t_n-t_1$ combination of the
forecasting model.
The $n'$ forecasted data time points $t_i'$ can be created for
any arbitrary chosen sample window $\Delta T'=t_n'-t_i'$.
The  $n'$ forecasted
$g_i'=g'_i (t_i',\bm{\beta}_{\mathrm{Fore}})$
values are obtained from
Equation 
  \ref{EqForecastingDataModel}.
These $g'_i$ values can be used, for example, to compute
the {\it forecasted mean level}
\begin{eqnarray}
  m_{\mathrm{Fore}}  = {{1}\over{n'}}\sum_{i=1}^{n'}g'_i
  \label{EqForeMean}
\end{eqnarray}
during the chosen  sample window $\Delta T'=t_n'-t_1'$.
\citet{Jet25}
used this $m_{\mathrm{Fore}}$ parameter to postdict the known
$\Delta T'$ time intervals of prolonged solar activity minima, like the
Maunder minimum between the years 1640 and 1720 \citep{Uso07}.
Since the known mean level of all sunspot data was $m$,
\citet{Jet25}  used 
these three criteria for
correct postdictions of past prolonged activity minima
$\Delta T'$ time intervals:
\begin{itemize}
\item[]  If DCM detects a real new signal in all data,
 the $m_{\mathrm{Fore}}$ value decreases.
\item[]  If  DCM detects many real signals in all data,
  the $m_{\mathrm{Fore}}$ value falls below  $m$.
\item[]  If DCM detects an unreal new signal in all data,
  the $m_{\mathrm{Fore}}$ value increases.
  \end{itemize}

We use Model 3 
combination $n=50$ and $\SN=10$
simulated data (Table \ref{TableModelThree}, column 2)
to illustrate DCM forecasting technique.
The black dots in Figures \ref{FigEight}a-c
are  the first $n=40$ forecasting data values 
$t_i$, $y_i$ and $\sigma_i$.
The open black dots denote the $n'=10$ forecasted data
values $t_i'$, $y_i'$ and $\sigma_i'$.
The continuous black line is the forecasting model
$g(t)$ and the dotted black line is the forecasted model $g'(t)$.
The red dotted line shows
the $\pm 3 \sigma$ errors of both models.
The blue dots are the forecasting data
$\epsilon_i=y_i-g_i$ residuals and
the open blue dots are
the forecasted data $\epsilon_i'=y_i'-g_i'$ residuals.

Model 3 is the sum of
two $P_1=0.16$ and $P_2=0.17$ pure sine signals
$(K_1=2, K_2=1)$ superimposed
on the constant mean level $M_0=1$ $(K_3=0)$.
We compute the $z$ and  $z_{\mathrm{Fore}}$ values for the
$g_{1,1,0}$,
$g_{2,1,0}$ and
$g_{3,1,0}$ models (Table \ref{TablezFore}).
These three models have the same $K_2=1$ and $K_3=0$
orders as Model 3,
but  their signal numbers $K_1=1, 2$ or 3 are different,
the $g_{2,1,0}$ model being the correct simulation Model 3.

The correct $g_{2,1,0}$ model gives
the smallest  $z_{\mathrm{Fore}}=0.88$
value (Table \ref{TablezFore}).
Therefore, it is a better forecasting model than the
$g_{1,1,0}$ and $g_{3,1,0}$ models.
The one signal $g_{1,1,0}$ model forecast fails
because the blue open circles denoting
the forecasted data residuals $\epsilon_i'=y_i'-g_i'$
show large deviations from 
to the blue dotted line offset level of $\epsilon_i'=-1$
(Figure \ref{FigEight}a). 
The two signal $g_{2,1,0}$ model forecast
succeeds because all open blue dots denoting
the forecasted data residuals stay close
to the blue dotted line offset level $\epsilon_i'=-1$, as well as
inside the red dotted $\pm 3\sigma$ model error limits
(Figure \ref{FigEight}b).

The three signal model $g_{3,1,0}$ periods are
$P_1=0.0643\pm0.0018$,
$P_2=0.1592\pm0.0012$ and
$P_3=0.1716\pm0.0012$.
The amplitudes of these pure sine signals are
$A_1=0.092\pm0.025$,
$A_2=1.879\pm0.16$ and
$A_3=1.82\pm0.15$.
The periods and the amplitudes of the
two strongest $P_2$ and $P_3$  signals are correct because they are
the same as in Model 3 simulation
(Table \ref{TableModelThree}, column 1).
Therefore,
the $g_{3,1,0}$ forecast appears
nearly as good as the $g_{2,1,0}$ forecast
because the residuals
$\epsilon_i'$ are close to the offset level
of $\epsilon_i'=-1$ (blue open dots),
and these residuals also
stay inside the red dotted $\pm 3 \sigma$ model error limits
(Figure \ref{FigEight}c).
The amplitude $A_1=0.092$
of the third $P_1=0.0643$ signal is very low.
Due this weak ``unreal'' $P_1$ signal,
the $g_{3,1,0}$ model has a larger
$z_{\mathrm{Fore}}=0.91$ value
than the correct  $g_{2,1,0}$ model
(Table \ref{TablezFore}).
Finally, we note that the three signal model
$g_{3,1,0}$ is not unstable (``$\UM$'') although
the simulated {time series} contains only two signals.

 \renewcommand{\Ki}{}
For the forecasting data,
  the extreme Fisher-test critical levels  $\QF < 10^{-13}$
  confirm that
  the two signal $g_{2,1,0}$ $(\chi^2=22.946)$
  and
  the three signal $g_{3,1,0}$  $(\chi^2=19.790)$ models
  are certainly better
  than the one signal $g_{1,1,0}$ $(\chi^2=206.3)$  model.
\Ki
The forecasting data parameters $n=40$,
$\eta_1=7$, $\eta_2=10$, $\chi_1=22.946$
and $\chi_2=19.790$
for the simple
$g_{2,1,0}$ model
and
the complex $g_{3,1,0}$ model 
give the Fisher-test critical level
$\QF=0.22>\gamma_{\mathrm{F}}=0.001$.
\Ki
The $g_{2,1,0}$ model beats the  $g_{3,1,0}$ model
because the $H_0$ hypothesis is not rejected.
\Ki
Hence, the Fisher-test confirms that the 
$g_{2,1,0}$ model is  the best model for the forecasting data.
\Ki
The best DCM model (Fisher-test, $g_{2,1,0}$ )
is also the correct DCM  model (Forecast-test, $g_{2,1,0}$).
\Ki
The double-check works for this particular Model 3
simulated time series!

DFT cannot detect the correct frequencies
for the forecasting data
because
the simulated frequencies are ``too close'' 
(Equation \ref{EqTooClose}).
\Ki
Therefore, DFT forecast cannot succeed.

The analyses of all seven complex {time series}
  indicate that the relative accuracy of
    amplitude and period estimates are lower than the relative accuracy
    of signal minimum and maximum epoch estimates
    (Tables \ref{TableModelOne} - \ref{TableModelSeven}).
    This statistical effect is the same when DCM is applied to
    the {time series} of any arbitrary phenomenon.
    This effect would, for example, explain why
    our solar cycle amplitude forecasts
    are less accurate than
    our solar cycle minimum and maximum
    epoch forecasts \citep{Jet25}.
    DCM detects the correct simulated period values,
    but DFT detects less accurate period values,
    which are not always correct.
      The leakage of DFT spectral power
      \citep{Kay81,Gha21}
      would explain why correct signal periods
    have not  been detected earlier from
    the sunspot data.

      \renewcommand{\Ki}{}
      DCM analysis results can be double-checked.
\Ki
      We use the Fisher-test to identify the best DCM model
from all tested DCM models (Equation \ref{EqFisher}).
\Ki
The correct model may not be among the tested models.
\Ki
Hence, the best model is not necessarily the correct one.
\Ki
This best model must be correct if
it passes the Forecast-test
(Equation \ref{EqForez}).
\Ki
This Fisher- and Forecast-test combination
double-checks DCM analysis results.
\Ki
The perfect result is that the best model is the correct model.
Therefore, the Forecast-test also prevents overfitting.

\section{Discussion of DCM performance} \label{SectDiscussion}  

\renewcommand{\Ki}{}

DCM proceeds through two stages.
\Ki
It computes the $R$ (Equation \ref{EqR})
    or $\chi^2$ (Equation  \ref{EqChi}) values for a
    massive number of LS fits (Equation \ref{EqMassive}).
    \Ki
    The Gau\ss-Markov  theorem ensures that
    DCM model having the lowest $R$ or $\chi^2$
    is inevitably {\it always}  found, and the result for
    the non-linear iteration is unique (Equation \ref{EqIteration}).
    \Ki
This first stage devours CPU.
\Ki
The second fast stage, the Fisher-test, then reveals
the best DCM model alternative (Equation \ref{EqFisher}).
\Ki
  Practical time series analysis applications,
like digital signal processing,
require fast computational algorithms
\citep{Kay81}.
\Ki
Long computation time is
the main \AL{} constraining DCM.
  \Ki
  For example, \citet{Jet25} computed
  about 4.5 million LS fits to
  search for four signals in
  the monthly sunspot data
  $(n=3287, K_1=4, K_2=1, K_3=0, 
     n_{\mathrm{L}}=100, n_{\mathrm{S}}=30, n_{\mathrm{B}}=20)$.
   \Ki
   This required several months of CPU.
   \Ki
   The parallel Python code computations took a few days.
\Ki
Nevertheless, the CPU used is beside the point
if  a well-posed computational solution can be found
to any challenging scientific ill-posed problem.  

\renewcommand{\Ki}{}
DCM computation time \AL{}{}
constraint is amply compensated
in real observations because
there is no need to wait for the repetition of the \mbox{signal(-s)}.
\Ki
The \WDE ~is the spearhead of DCM  because
the sample window $(\Delta T)$ can be infinitesimally short.
   \Ki
   Fast accurate observations are the best approach.
   \Ki
  If the noise $(\sigma \equiv \SN)$ cannot be eliminated
   it is always possible to increase the sample size $(n)$.

   It is time to summarise why the fifteen \AL{}s  (Section \ref{SectInt}) 
     of other frequency-domain parametric time series analysis methods
    do not constrain DCM.

\begin{enumerate}

\renewcommand{\Ki}{}
\item  Data errors (level of noise) are unknown. 
       
        \begin{itemize}
        \item[]   
            DCM can solve both alternatives:
            errors  $\sigma_i$ known or unknown
            (Equations \ref{EqzR}, \ref{EqzChi},
            \ref{EqFR} and  \ref{EqFChi}).
            \Ki
            All DCM model parameter 
            solutions in Tables \ref{TableModelOne}-\ref{TableModelSeven}
            converge to the correct values
            when the simulated data $n$ and/or $\SN$ increase
            ($\sigma_i$ decrease).
            \Ki
            Due to this \WDE,
            sufficient noise reduction always
            leads to the correct model detection. 
\end{itemize}

\renewcommand{\Ki}{}
            \item Data error information is not utilised.

       \begin{itemize}
       \item[]
         DCM utilises this information.
         \Ki
            If the errors $\sigma_i$ are known,
           DCM performs weighted LS fits which
           minimise the model $\chi^2$ for every tested frequency
           combination  (Equation \ref{EqChi}).
           \Ki
           This gives DCM test statistic $z$ (Equation \ref{EqzChi}).
           \Ki
           The Fisher-test utilises the $\chi^2$ values
           of all alternative tested DCM models
           to identify the best DCM model
           (Equation \ref{EqFChi}).

         \end{itemize}

\renewcommand{\Ki}{}
                 \item Data must be evenly spaced. 

       \begin{itemize}
       \item[]
         DCM performance is independent of data spacing.
         \Ki
         Even or uneven spacing
         is  irrelevant for all $n_{\mathrm{Lin}}$
         LS fits (Equation \ref{EqMassive}).
           \Ki
           However, we do admit that long gaps
           can mislead even these LS fits. 
         \end{itemize}

\renewcommand{\Ki}{}
       \item Model parameter errors are unknown. 
        
        \begin{itemize}
        \item[]
            \Ki
            DCM gives error estimates 
            for the model parameters of
            Equations \ref{EqP}-\ref{EqMk}.
         \Ki
          For non-linear models,
          such as DCM model,
          the analytical solution for the model parameter errors 
          is a highly complex effort
            \citep[][]{Fur20}.
           \Ki
          DCM solves these error estimates using
           the computational statistical bootstrap technique
            (Equation \ref{EqBoot}).
            \Ki
           \end{itemize}

\renewcommand{\Ki}{}
 \item Model and forecast errors are unknown.
   \begin{itemize}
   \item[]  The computational bootstrap technique
     (Equation \ref{EqBoot}) gives $n_B$ estimates
     for $\bm{\beta}_J$,
     $t_{\mathrm{mid,J}}$ and $\Delta T_{J}$,
     where $J=1, 2, ... n_B$.
     \Ki
     These estimates
     give $n_B$ values for  $g(t)$  at any time $t$
     inside and outside the sample window $\Delta T$.
     \Ki
     The standard deviation of these $n_B$ values gives the error limits
     $\sigma_{ g(t) }$ in Equation \ref{EqModelError}.
     \Ki
     The $n_B$ estimates
      for $\bm{\beta}_J$,
      $t_{\mathrm{mid,J}}$ and $\Delta T_{J}$
      also give the error limits for the
      functions $h(t), h_i(t)$ and $p(t)$ in DCM model $g(t)$.
    \end{itemize}

\renewcommand{\Ki}{}
\item
  Sample window is shorter than signal period(s). 
        
        \begin{itemize}
        \item[] DCM solves this problem.
          \Ki
           The sample window is shorter
           than the period(s) in Model 1, 2, 5 and 7 time series
           simulations.
           \Ki
           Regardless of this,
           all model parameter estimates converge
           to correct values when $n$ and/or $\SN$ increase
                     (Tables \ref{TableModelOne},
            \ref{TableModelTwo},
            \ref{TableModelFive} and \ref{TableModelSeven}). 
           \Ki
           Due to this \WDE,
            the sample window $\Delta T$ value is irrelevant for DCM.
           \Ki
            It can perceive the future and the past much
     ``earlier''   (from shorter $\Delta T$) than has been previously thought,
            like in Equations \ref{EqTooShort} and \ref{EqTooClose}.

          \end{itemize}

\renewcommand{\Ki}{}
\item Presence and shape of trend are unknown. 
        \begin{itemize}
        \item[] DCM can test any $K_1$, $K_2$ and $K_3$ model
          combination.
          \Ki
          The Fisher-test reveals the combination of the
          best DCM model.
           \Ki
           The Forecast-test can double-check this result.
         \Ki
           These tests give the correct $K_3$ trend order.
            \Ki
           Due to the \WDE,
           the polynomial trend coefficients $M_k$ converge
           to correct values in all simulated time series
          (Sections \ref{SectModelOne}-\ref{SectModelSeven}).
           \Ki
          The trend is absent (stationary time series) if
            $K_3=-1 \equiv$ $ p(t)=0$ or
            $K_3=0 \equiv$ $p(t)=M_0= \text{constant}$
            (Equations \ref{EqPolyOne}  and \ref{EqPolyTwo}).
            \Ki
         \end{itemize}

\renewcommand{\Ki}{}        
      \item Sample window causes leakage.
        \begin{itemize}
        \item[]
          Leakage does not constrain DCM because
          the sample window $\Delta T$ has no effect on
          the LS fits (Equation \ref{EqMassive}: $n_{\mathrm{Lin}}$).
          \Ki
          All frequencies  converge to exactly correct values when
          the $n$
          and/or $\SN$ values of simulated time series increase
            (Tables \ref{TableModelOne}-\ref{TableModelSeven}).
              \Ki
            This confirms the absence of leakage.
            \Ki
            Due to the \WDE,
            the performance of DCM does not depend
            on the sample window $\Delta T$.
             
        \end{itemize}

\renewcommand{\Ki}{}
          \item Leakage weakens frequency resolution. 
    \begin{itemize}
    \item[] 
      \Ki
      There is no  leakage because
      the sample window is irrelevant  (\AL{\ref{LimWindow}}).
      \Ki
      DCM frequency resolution is limited only by
       the sample size $n$ and the data accuracy $\sigma$.
      \Ki
      The \WDE ~ensures that
      all frequency detections for the simulated
      time series converge to exactly
      correct values 
      (Tables \ref{TableModelOne}-\ref{TableModelSeven}).
      
    \end{itemize}

\renewcommand{\Ki}{}
  \item Signal shapes are not pure sines.
    \begin{itemize}
    \item[] 
        DCM always finds the correct
        $K_1$, $K_2$ and $K_3$  values
        (\AL{\ref{LimTrend}}).
        \Ki
        The $K_2$ value determines the signal shape.
        \Ki
        The $K_2=1$ pure sine shape is the simplest.
        \Ki
        Higher $K_2$ values allow DCM modelling of
        more complex shapes.
        \Ki
        We demonstrate the $K_2=2$ double wave signal detections
        in Sections \ref{SectModelSix} and \ref{SectModelSeven}.      
    \end{itemize}

\renewcommand{\Ki}{}

  \item Number of signals is unknown. 
          \begin{itemize}
                    \item[] The correct
        $K_1$, $K_2$ and $K_3$  values
        can always be found
        (\AL{\ref{LimTrend}}).
        \Ki
        The $K_1$ value determines the number of signals.
        \Ki
        For all simulated time series,
        the detected signal frequencies and
        amplitudes converge to exactly correct 
        values 
        (Sections \ref{SectModelOne}-\ref{SectModelSeven}).
        \Ki
        Furthermore,
        \citet{Jet20} showed that
        DCM models having too few or too many signals
        are often unstable $("\UM")$.
      \end{itemize}

 \begin{figure*}[h]  
\vspace{0.02\textwidth}
\centerline{\hspace*{0.005\textwidth}
 \includegraphics[width=0.600\textwidth,clip=]{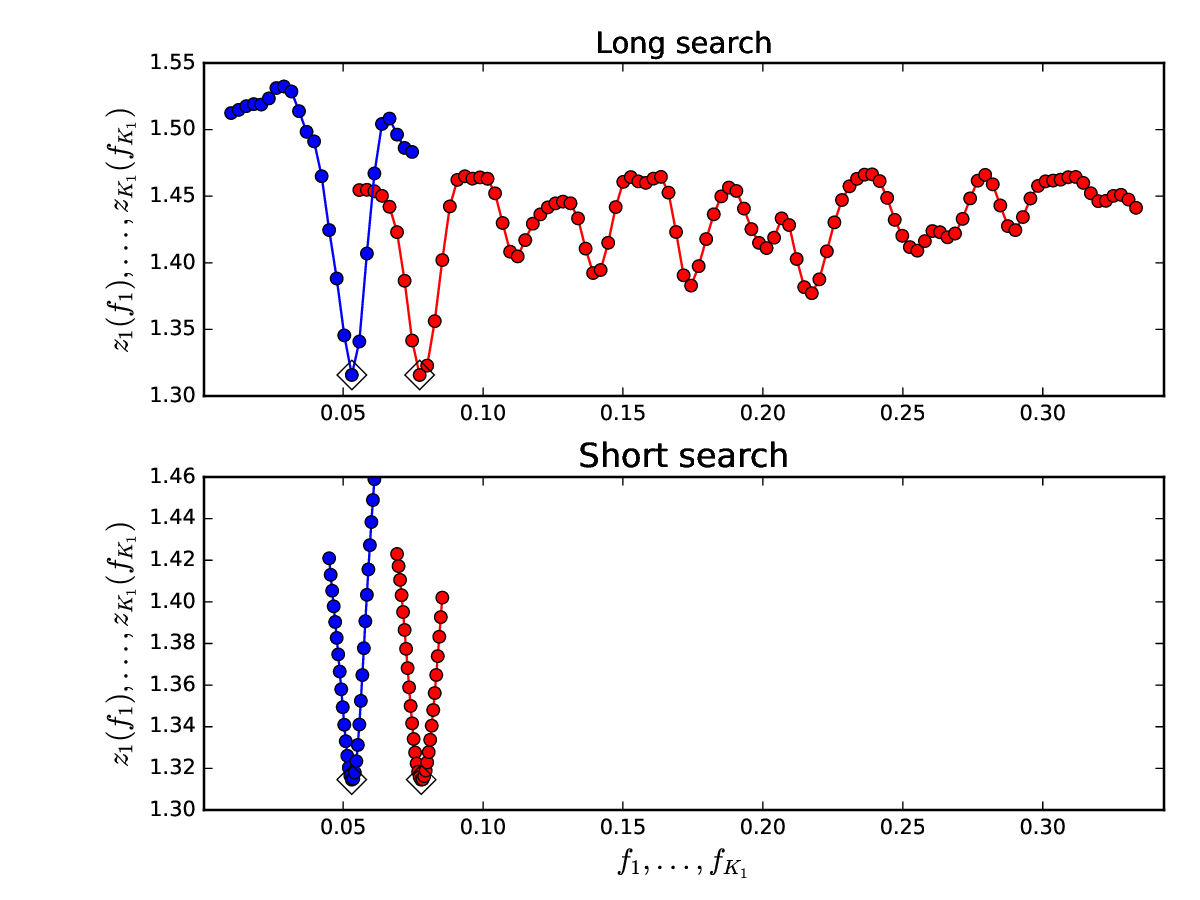}
 \hspace*{-0.01\textwidth}
 }
\vspace{-0.45\textwidth}
\centerline{\Large 
\hspace{0.75\textwidth}  \color{black}{(a)}
\hfill}
\vspace{0.45\textwidth}
\centerline{\hspace*{0.005\textwidth}
 \includegraphics[width=0.455\textwidth,clip=]{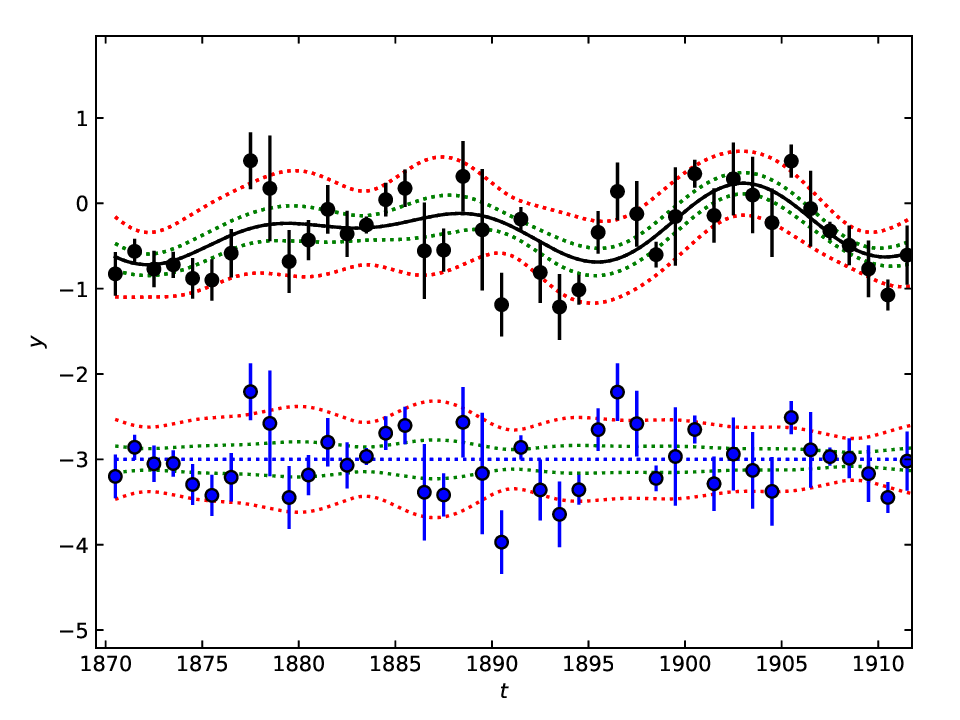}
 \hspace*{-0.01\textwidth}
 \includegraphics[width=0.455\textwidth,clip=]{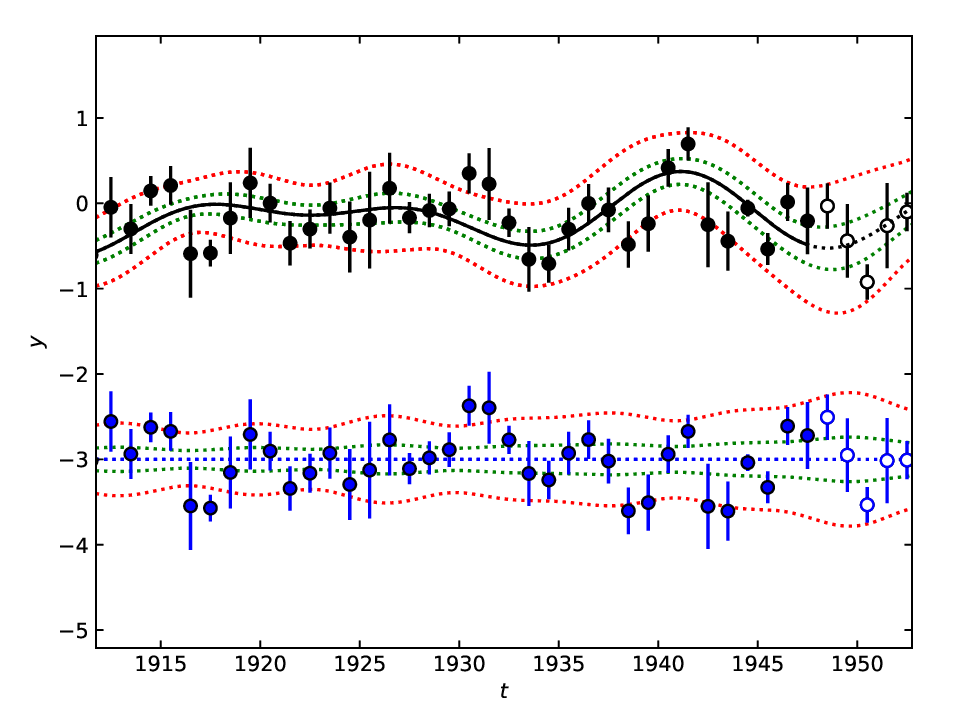}
}
\vspace{-0.36\textwidth}
\centerline{\Large 
\hspace{0.44\textwidth}  \color{black}{(b)}
\hspace{0.41\textwidth}  \color{black}{(c)}
\hfill}
\vspace{0.35\textwidth}
\centerline{\hspace*{0.005\textwidth}
 \includegraphics[width=0.455\textwidth,clip=]{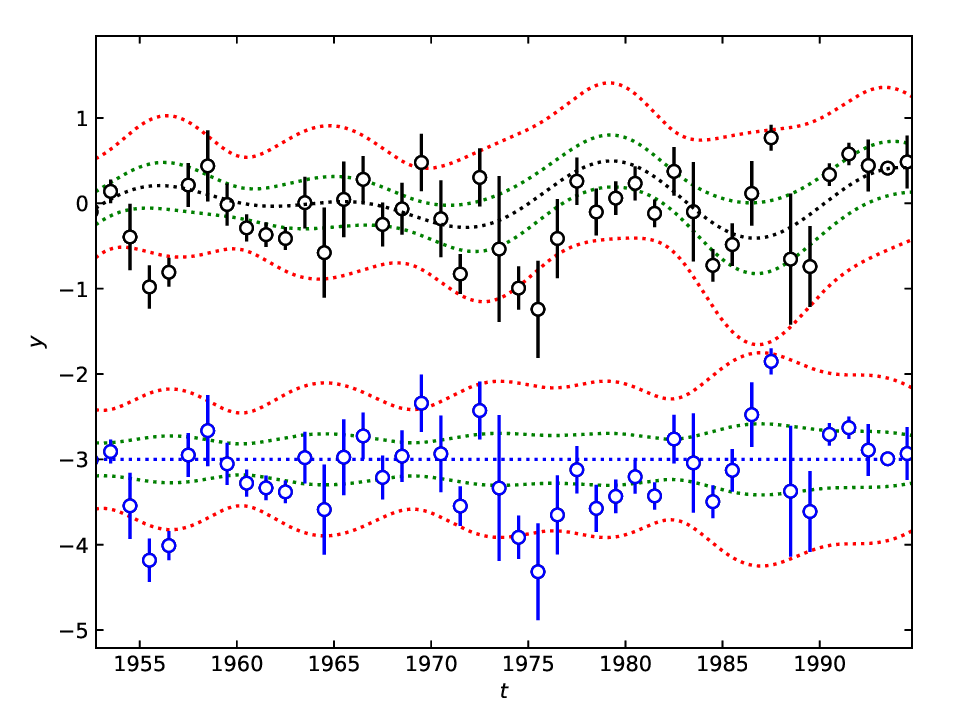}
 \hspace*{-0.01\textwidth}
 \includegraphics[width=0.455\textwidth,clip=]{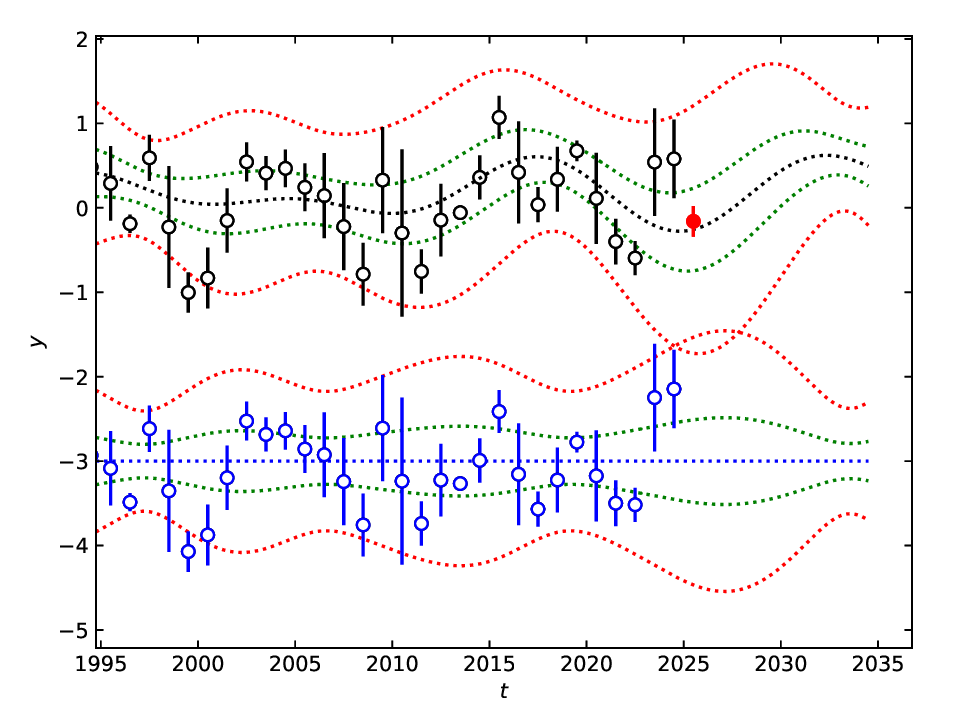}
}
\vspace{-0.36\textwidth}
\centerline{\Large 
\hspace{0.44\textwidth}  \color{black}{(d)}
\hspace{0.41\textwidth}  \color{black}{(e)}
\hfill}
\vspace{0.32\textwidth}
\caption{First half forecast (Table \ref{TableChalf}: \M=2):
  (a)  Periodogram (Equation  \ref{EqPeriodograms}).
  (b-e)
  Forecasting data (closed black circles) and forecasted data (open black circles).
  Red closed circle denotes 2025 yearly mean which is discussed
  later in Section  \ref{SectAllYears}.
  Green and red dottted lines denote one and three sigma model $g(t)$ errors.
  Residuals (closed and open blue circles) are offset to level of -3.
  Units are 
  x-axis: $[t_i]$=y, y-axis: $[y_i]=$ \Cc ~and error bars = $[1 \times \sigma_i]=$ \Cc.
}
\label{FighalfC14K211}
\end{figure*}

 \begin{figure*}[h]  
\vspace{0.02\textwidth}
\centerline{\hspace*{0.005\textwidth}
 \includegraphics[width=0.600\textwidth,clip=]{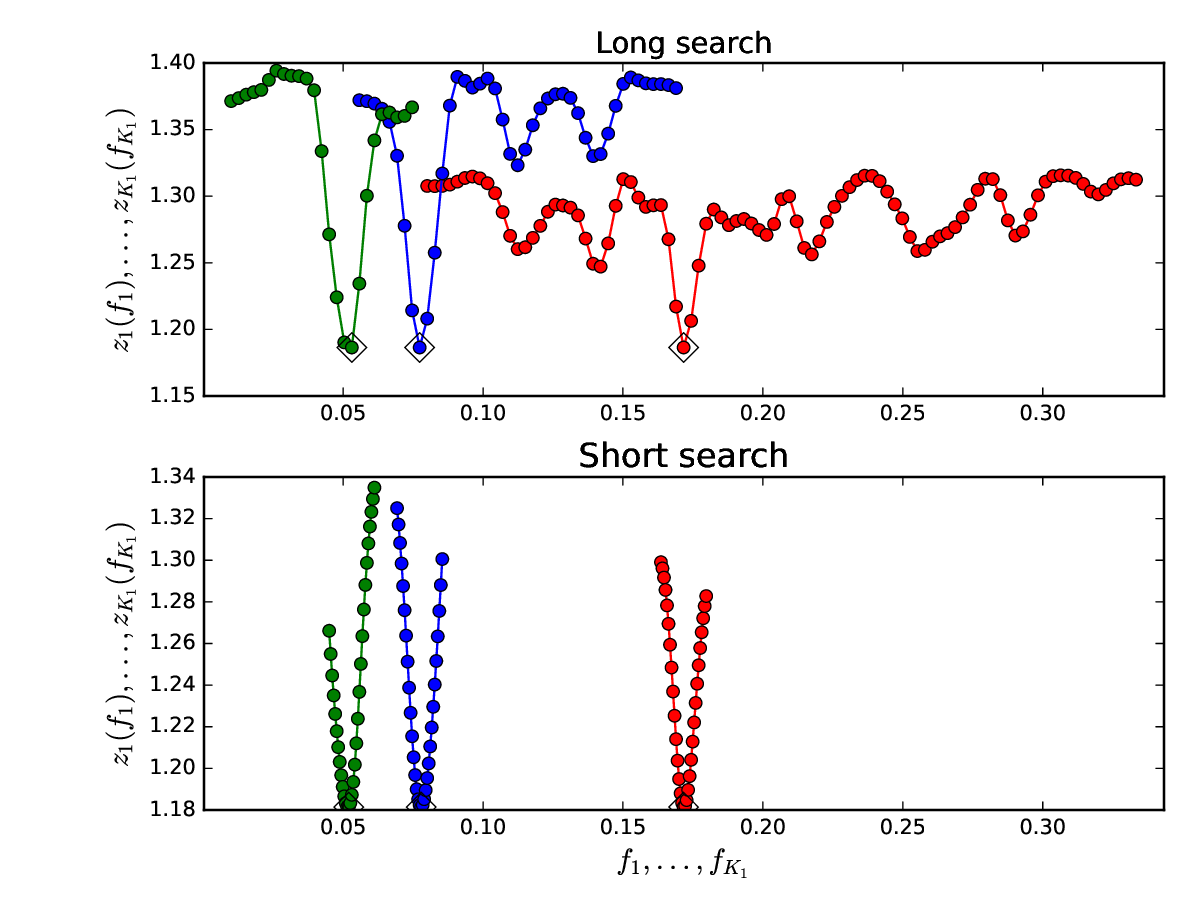}
 \hspace*{-0.01\textwidth}
 }
\vspace{-0.45\textwidth}
\centerline{\Large 
\hspace{0.75\textwidth}  \color{black}{(a)}
\hfill}
\vspace{0.45\textwidth}
\centerline{\hspace*{0.005\textwidth}
 \includegraphics[width=0.455\textwidth,clip=]{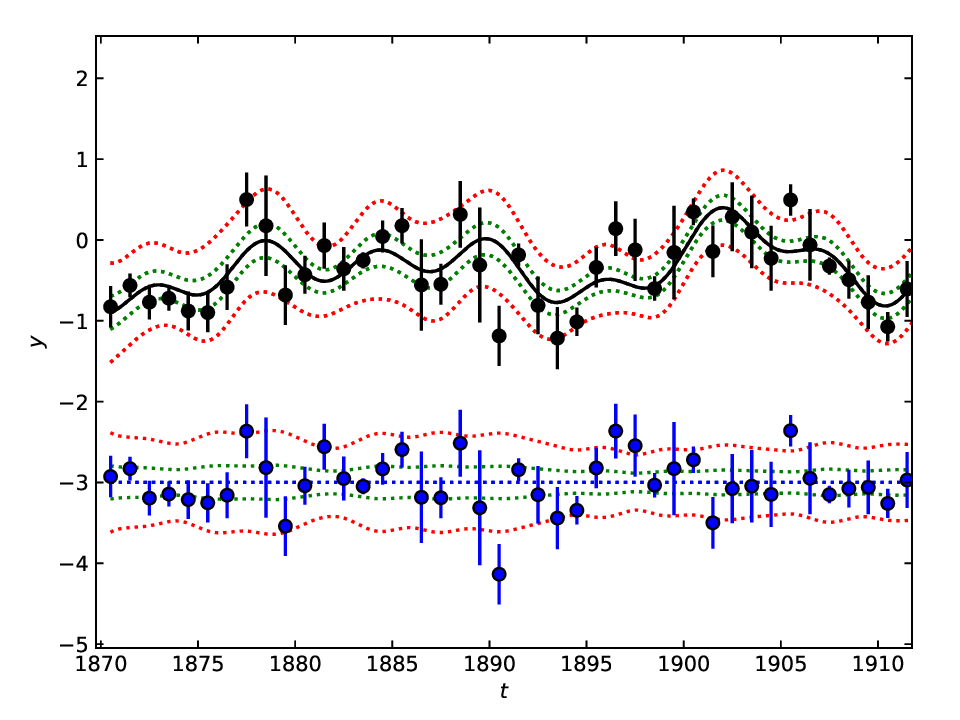}
 \hspace*{-0.01\textwidth}
 \includegraphics[width=0.455\textwidth,clip=]{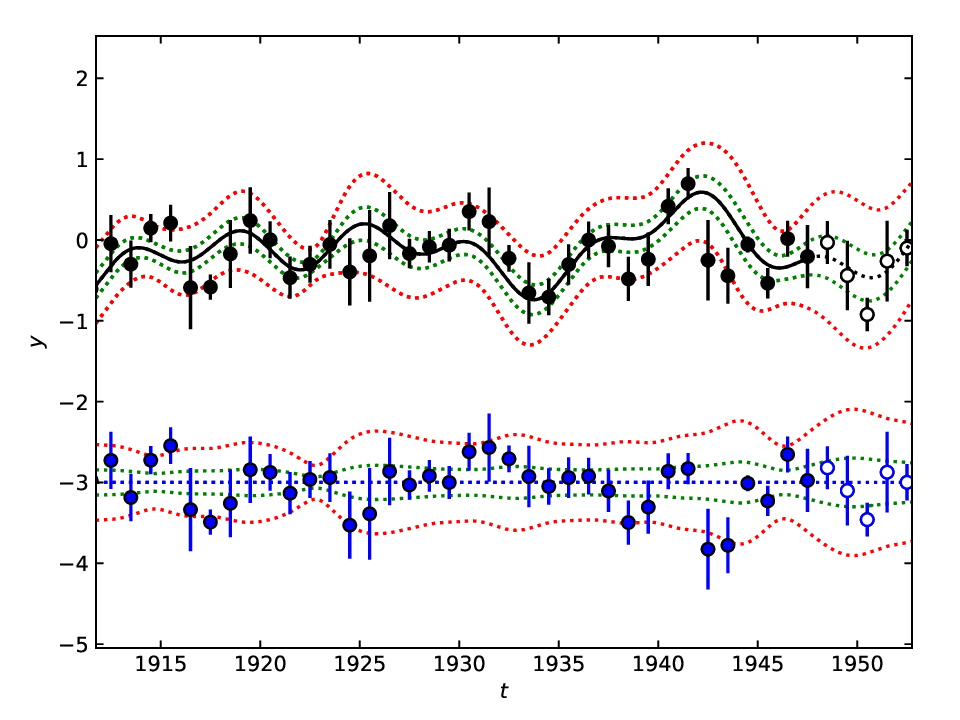}
}
\vspace{-0.36\textwidth}
\centerline{\Large 
\hspace{0.44\textwidth}  \color{black}{(b)}
\hspace{0.41\textwidth}  \color{black}{(c)}
\hfill}
\vspace{0.35\textwidth}
\centerline{\hspace*{0.005\textwidth}
 \includegraphics[width=0.455\textwidth,clip=]{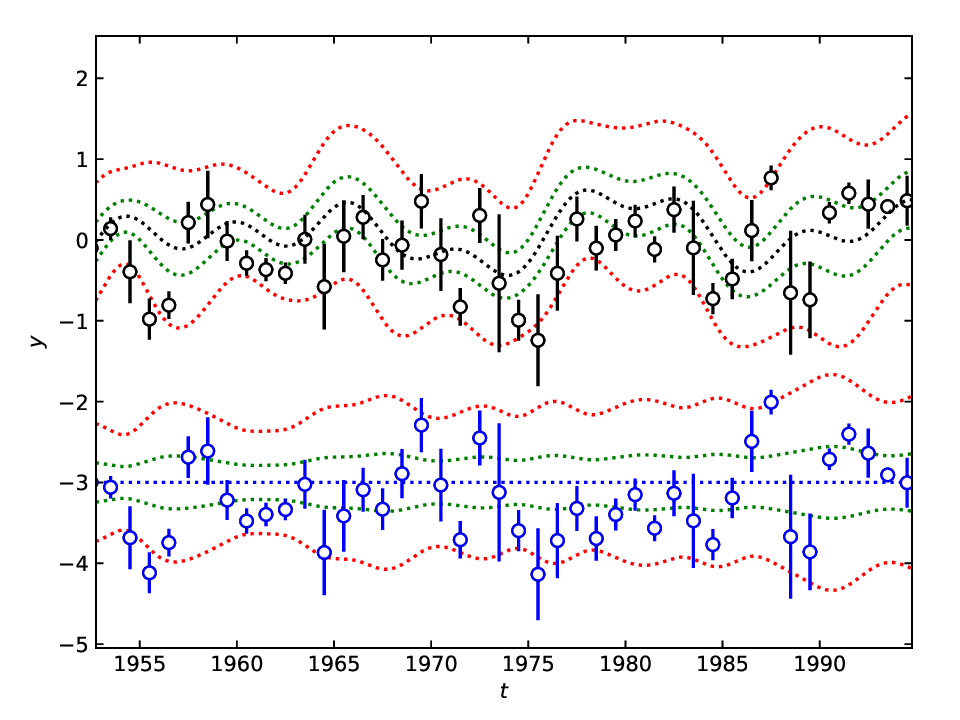}
 \hspace*{-0.01\textwidth}
 \includegraphics[width=0.455\textwidth,clip=]{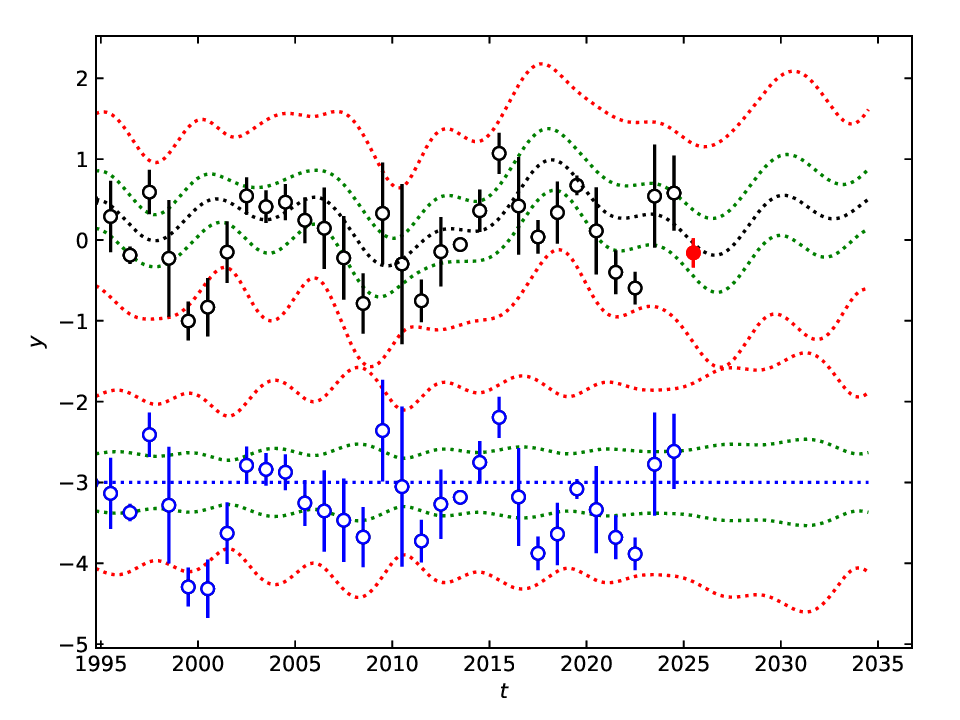}
}
\vspace{-0.36\textwidth}
\centerline{\Large 
\hspace{0.44\textwidth}  \color{black}{(d)}
\hspace{0.41\textwidth}  \color{black}{(e)}
\hfill}
\vspace{0.32\textwidth}
\caption{First half forecast (Table \ref{TableChalf}: \M=3).
  Otherwise as in Figure \ref {FighalfC14K211}.}
\label{FighalfC14K311}
\end{figure*}


\renewcommand{\Ki}{}
      \item Correct model alternative is unknown.
       \begin{itemize}
       \item[]The values of $K_1$, $K_2$ and $K_3$ 
         determine the correct model 
        (\AL{\ref{LimTrend}}).
        \Ki
         The
        $(n-\eta_2-1)/(\eta_2-\eta_1)$ penalty terms
        prevent overfitting
        when the Fisher-test is used to identify the best DCM model
        from all tested DCM models
(Equations \ref{EqFR} and \ref{EqFChi}).
           \Ki
           This best model is also the correct model  if it passes
           the Forecast-test  (Equation \ref{EqForez}).
           \Ki
           The \WDE ~ensures that better data can
           inevitably reveal the correct model.
       \end{itemize}

\renewcommand{\Ki}{}
\item Signal significances are unknown. 
        \begin{itemize}
        \item[]
          The Fisher-test gives the $Q_{\mathrm{F}}$ critical levels
          (signal significances) for all detected signals
          (Sections \ref{SectBestModel} and \ref{SectSignificance}).
          \Ki
          This $Q_{\mathrm{F}}$
          is the probability of falsely rejecting the $H_0$ hypothesis
          when it is in fact true.
          \Ki
          Thus, the $Q_{\mathrm{F}}$ value represents the probability of false
          signal detection.
          \Ki
          This detection is absolutely certain in the \REJ ~cases.
          \Ki
         Our pre-assigned  significance level for
         signal detection is $\gamma=0.001$
          (Equation \ref{EqFisher}).

       \end{itemize}

\renewcommand{\Ki}{}
\item Model solution is ill-posed. 
    \begin{itemize}
      \item[] The {\it analytical} solution for the non-linear DCM model
        is ill-posed (Equation \ref{Eqmodel}).
        \Ki
          We show that there exists a well-posed {\it computational}
solution (Section  \ref{SectIllPosed}).
\Ki
The \WDE ~ensures that
this solution can always be found by
increasing the sample size $(n)$ and/or
the signal to noise ratio $(\SN)$,
regardless of the {time series} complexity
($K_1$, $K_2$ and $K_3$ combination).
\Ki
The main DCM constraint is that
the solutions for
more complex models require more computation time
when the number of signals increases
(Equation \ref{EqMassive}).
\end{itemize}

\renewcommand{\Ki}{}
  \item Complex non-linear model forecasts fail. 
    \begin{itemize}
    \item[]  Our computational solution
      for the non-linear DCM model is well-posed
      (Section \ref{SectIllPosed}).
      \Ki
      The forecast of this well-posed solution
      can be double-checked
      (Section \ref{SectForecast}).
      \Ki
      First, the Fisher-test identifies the best DCM
      model from all tested DCM models (Section \ref{SectBestModel}).
      \Ki
      Then, the Forecast-test reveals if  this best DCM model is also the
      correct one (Equation \ref{EqForez}).
      \Ki
      Again, the \WDE ~ensures that the correct model and
      forecast can be found for any complex time series
      ($K_1$, $K_2$ and $K_3$ combination)
      when $n$ and/or $\SN$ increase.
      \Ki
      We conclude that DCM rises to meet the challenge
      of ``forecasting the evolution of complex systems''
      \citep{Che15}.

    \end{itemize}
   
  \end{enumerate}

 \renewcommand{\Ki}{}
    \Ki
     DCM turns things upside down.
    \Ki
There would be no need for time series analysis,
if the correct frequencies were already {\it known}.
\Ki
The tested frequencies are already {\it known}.
\Ki
DCM does not search for {\it unknown} frequencies,
it just tests {\it known} frequencies.
\Ki
DCM tests all possible frequency combinations. 
\Ki
The LS fit for every frequency
combination gives unique $R$ and $\chi^2$ values.
\Ki
The best frequency combination minimises $R$ or $\chi^2$.
\Ki
This gives the unique initial free parameter $\bm{\beta}$ values
for the non-linear iteration.
\Ki
The frequencies are never {\it unknown} in this process.
\Ki
The Gauss-Markov theorem, the well-posed computational model solution
  and the revolutionary \WDE ~make absolutely sure that DCM succeeds for any
    number of signals $(K_1)$,  all signal shapes $(K_2)$ 
    and every polynomial trend $(K_3)$.
  \Ki
  The Fisher-test  and the Forecast-test can double-check
  that DCM model solution is correct.
  \Ki
  DCM analysis of 
  intensive, large and accurate time series
  can see through time:
 a glimpse of the future and the past.

Our DCM is a remarkable method because it outperforms DFT.
  Gau{\ss} (1777–1855) developed an early form of
  the Fast Fourier Transform (FFT) algorithm for astronomical
  purposes, but did he not publish it.
  Perhaps our DCM should be re-named
  the Slow Gau{\ss} Transform (SGT).

 \begin{figure*}  
\vspace{0.02\textwidth}
\centerline{\hspace*{0.005\textwidth}
 \includegraphics[width=0.600\textwidth,clip=]{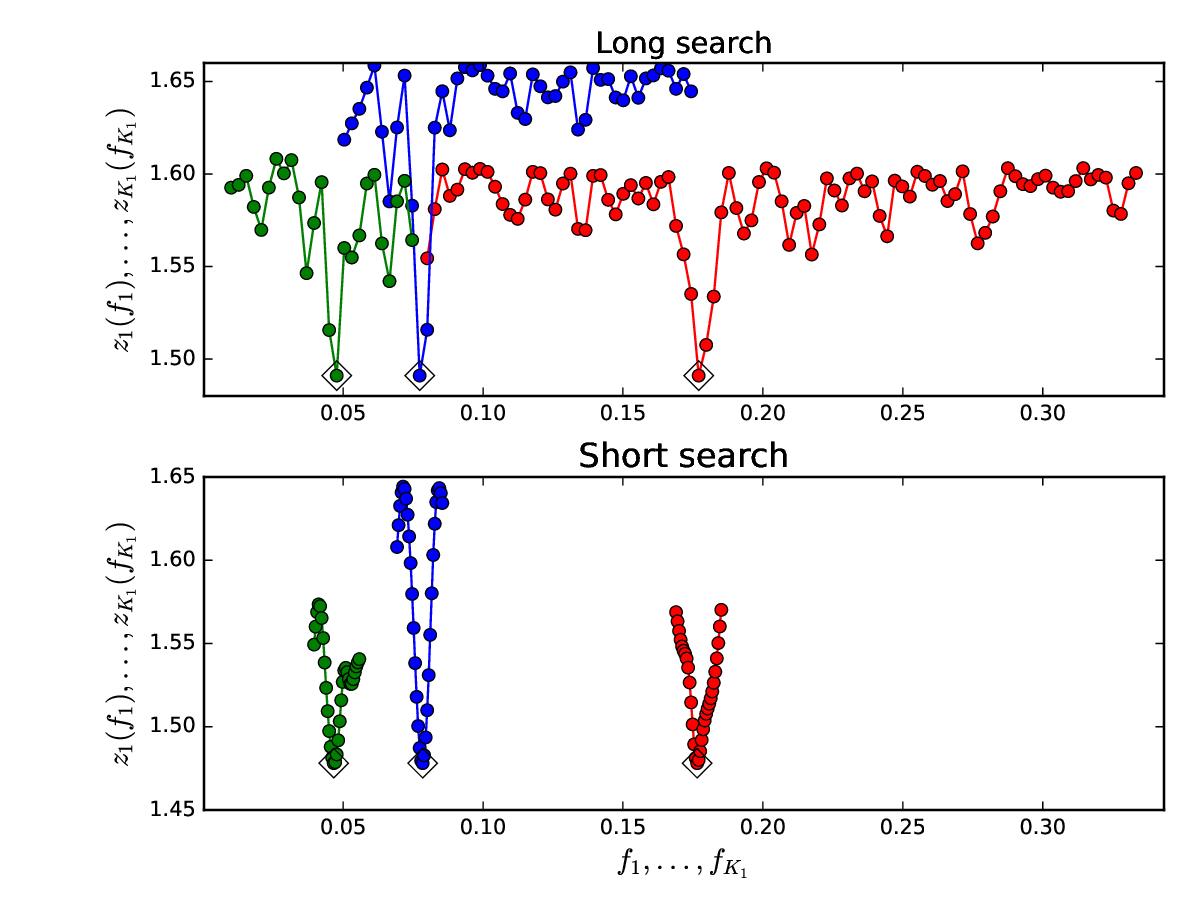}
 \hspace*{-0.01\textwidth}
 }
\vspace{-0.45\textwidth}
\centerline{\Large 
\hspace{0.75\textwidth}  \color{black}{(a)}
\hfill}
\vspace{0.45\textwidth}
\centerline{\hspace*{0.005\textwidth}
 \includegraphics[width=0.455\textwidth,clip=]{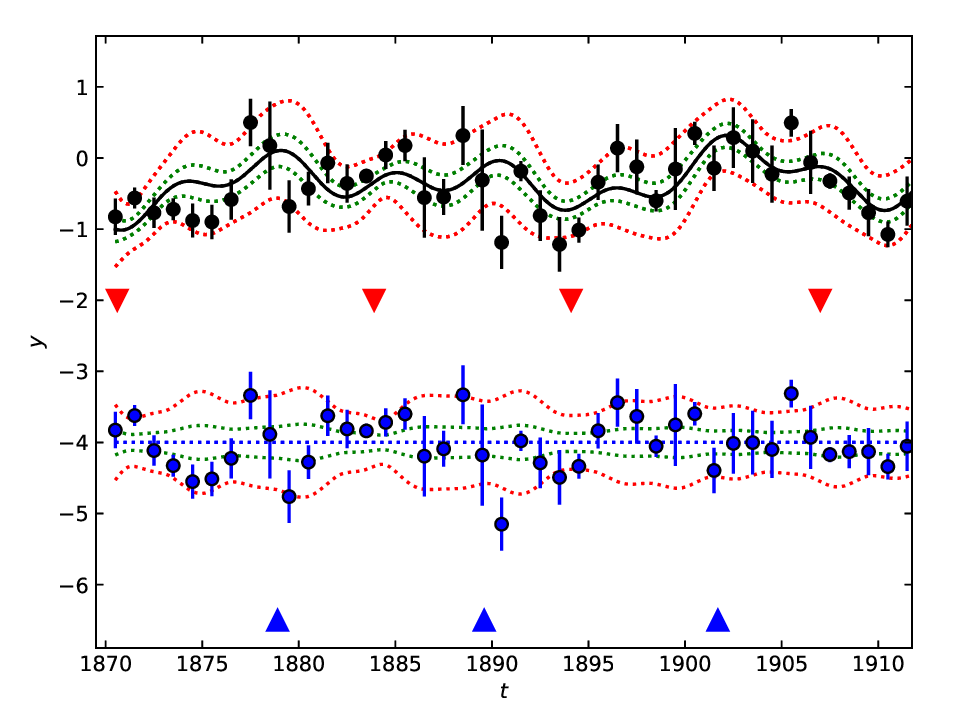}
 \hspace*{-0.01\textwidth}
 \includegraphics[width=0.455\textwidth,clip=]{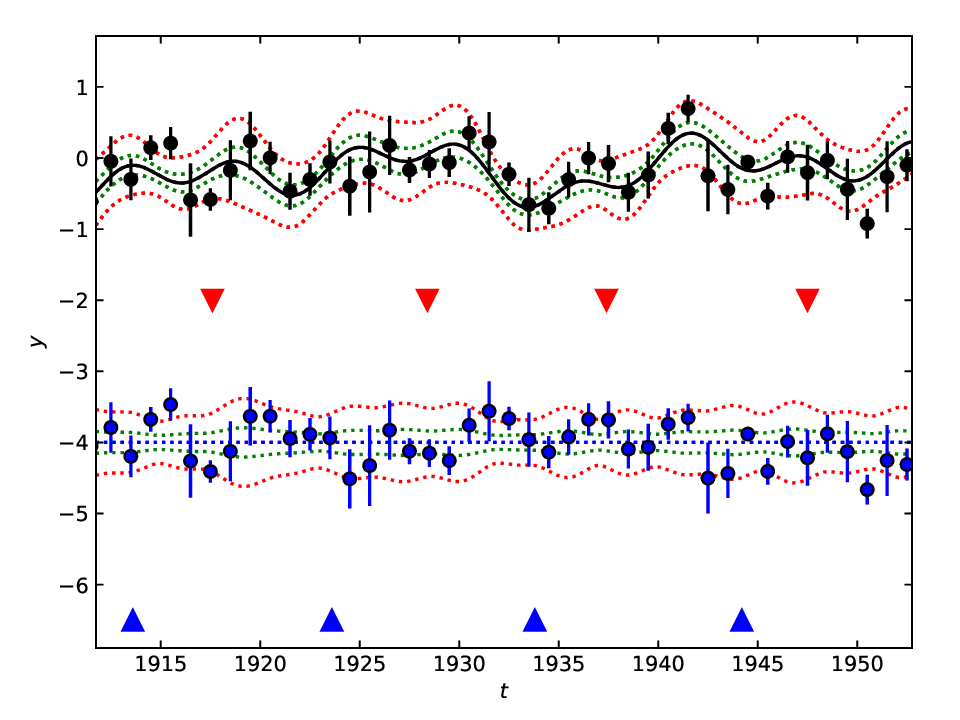}
}
\vspace{-0.36\textwidth}
\centerline{\Large 
\hspace{0.44\textwidth}  \color{black}{(b)}
\hspace{0.41\textwidth}  \color{black}{(c)}
\hfill}
\vspace{0.35\textwidth}
\centerline{\hspace*{0.005\textwidth}
 \includegraphics[width=0.455\textwidth,clip=]{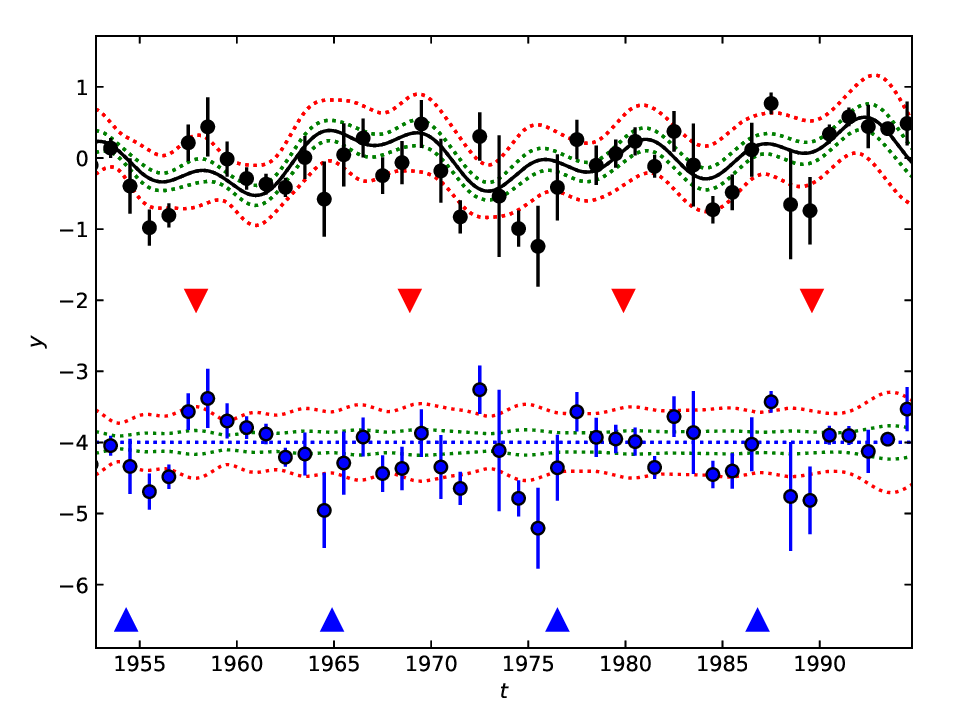}
 \hspace*{-0.01\textwidth}
 \includegraphics[width=0.455\textwidth,clip=]{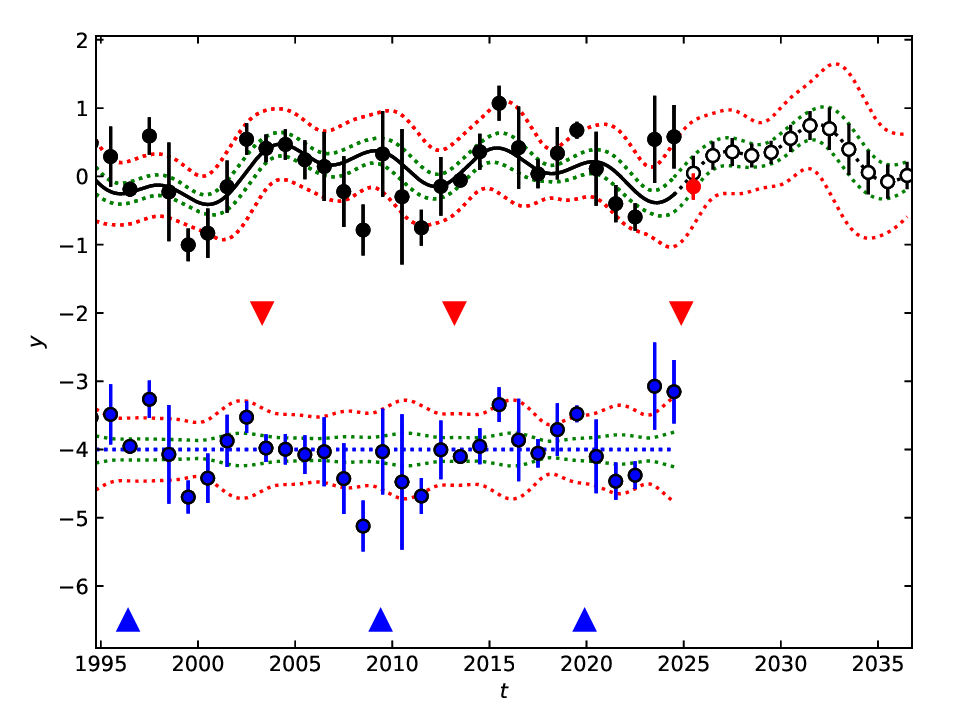}
}
\vspace{-0.36\textwidth}
\centerline{\Large 
\hspace{0.44\textwidth}  \color{black}{(d)}
\hspace{0.41\textwidth}  \color{black}{(e)}
\hfill}
\vspace{0.32\textwidth}
\caption{\BigWave ~for all weighted
  yearly mean data model \M=3 in Table \ref{TableClong}.
  Red circle is added
  {\it new observed} 2025 year value.
  Blue and red triangles
    denote sunspot minima and maxima \citep[][Table 3]{Had13}.
  Note how residuals more than $1 \sigma$ 
  below offset level of -4 concentrate close to sunspot minima
  (Section \ref{SectAllYears}: \Double).
  Otherwise as in Figure \ref{FighalfC14K211}.}
\label{FiglongC14K311}
\end{figure*}

 \begin{figure*}  
\vspace{0.02\textwidth}
\centerline{\hspace*{0.005\textwidth}
 \includegraphics[width=0.600\textwidth,clip=]{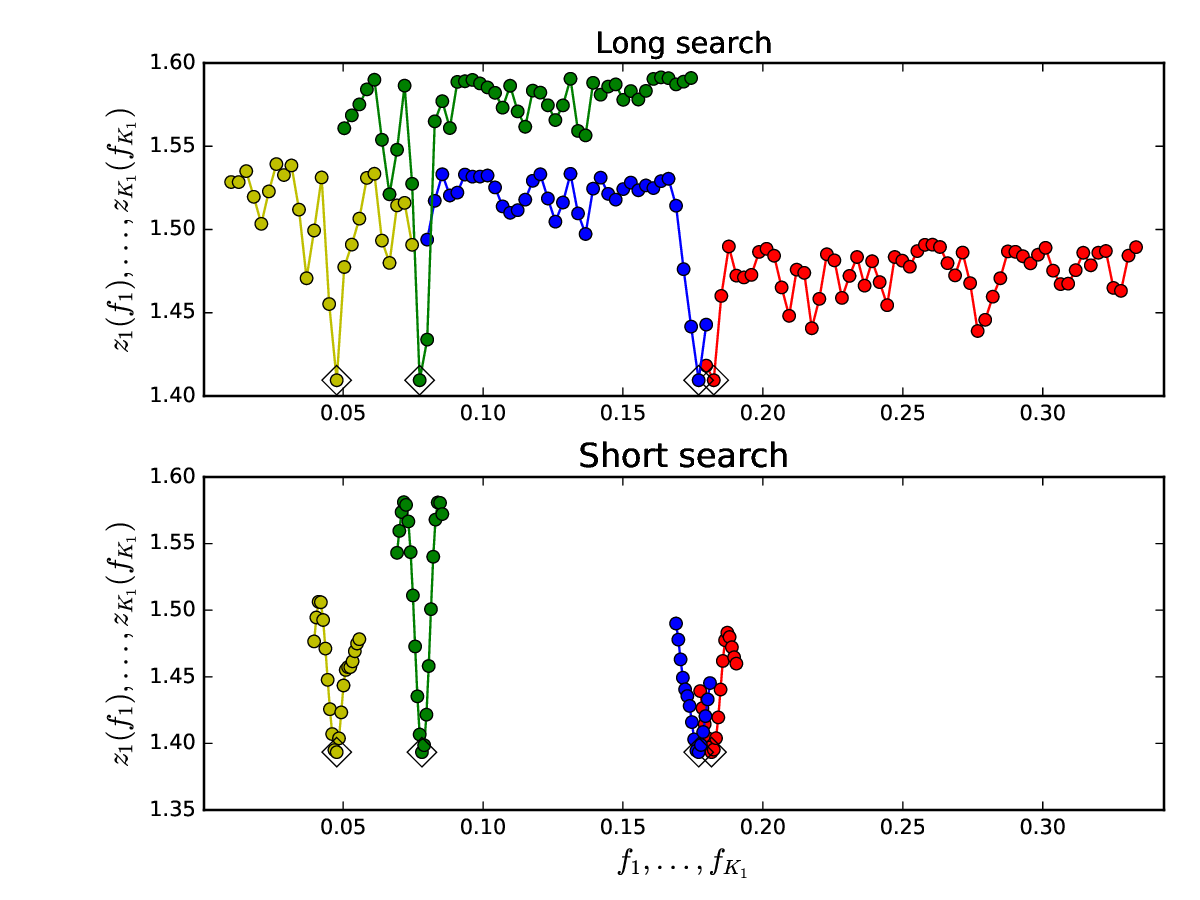}
 \hspace*{-0.01\textwidth}
 }
\vspace{-0.45\textwidth}
\centerline{\Large 
\hspace{0.75\textwidth}  \color{black}{(a)}
\hfill}
\vspace{0.45\textwidth}
\centerline{\hspace*{0.005\textwidth}
 \includegraphics[width=0.455\textwidth,clip=]{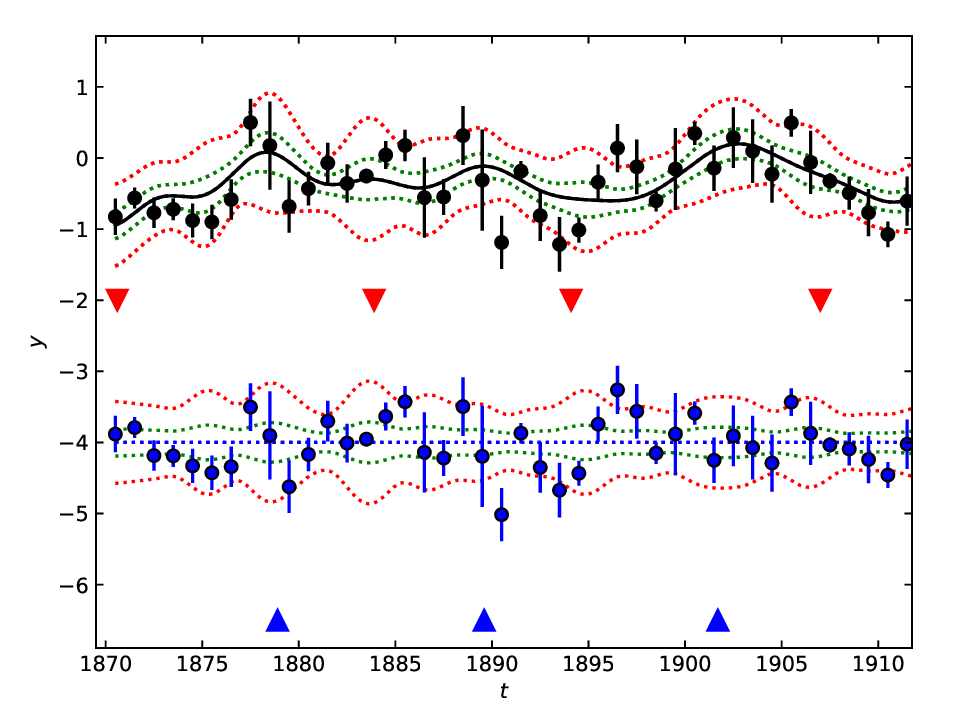}
 \hspace*{-0.01\textwidth}
 \includegraphics[width=0.455\textwidth,clip=]{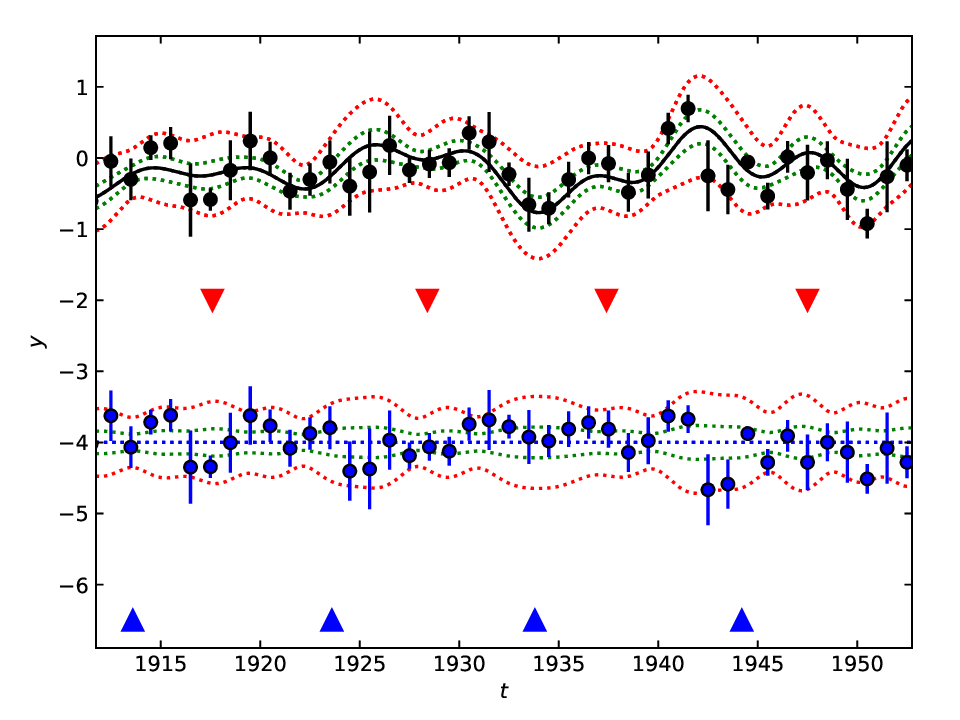}
}
\vspace{-0.36\textwidth}
\centerline{\Large 
\hspace{0.44\textwidth}  \color{black}{(b)}
\hspace{0.41\textwidth}  \color{black}{(c)}
\hfill}
\vspace{0.35\textwidth}
\centerline{\hspace*{0.005\textwidth}
 \includegraphics[width=0.455\textwidth,clip=]{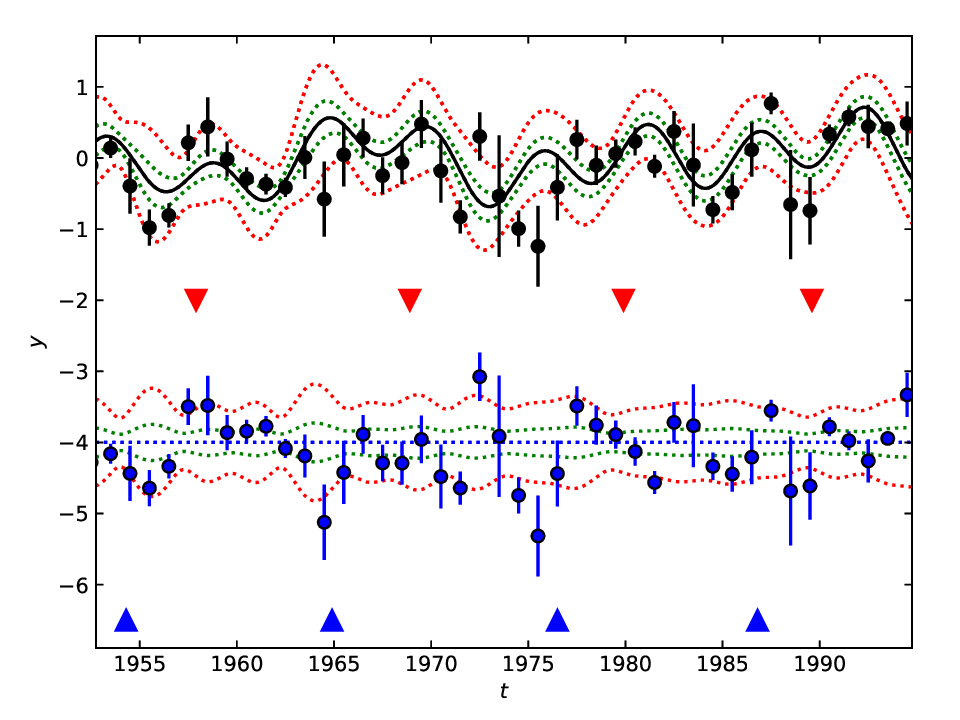}
 \hspace*{-0.01\textwidth}
 \includegraphics[width=0.455\textwidth,clip=]{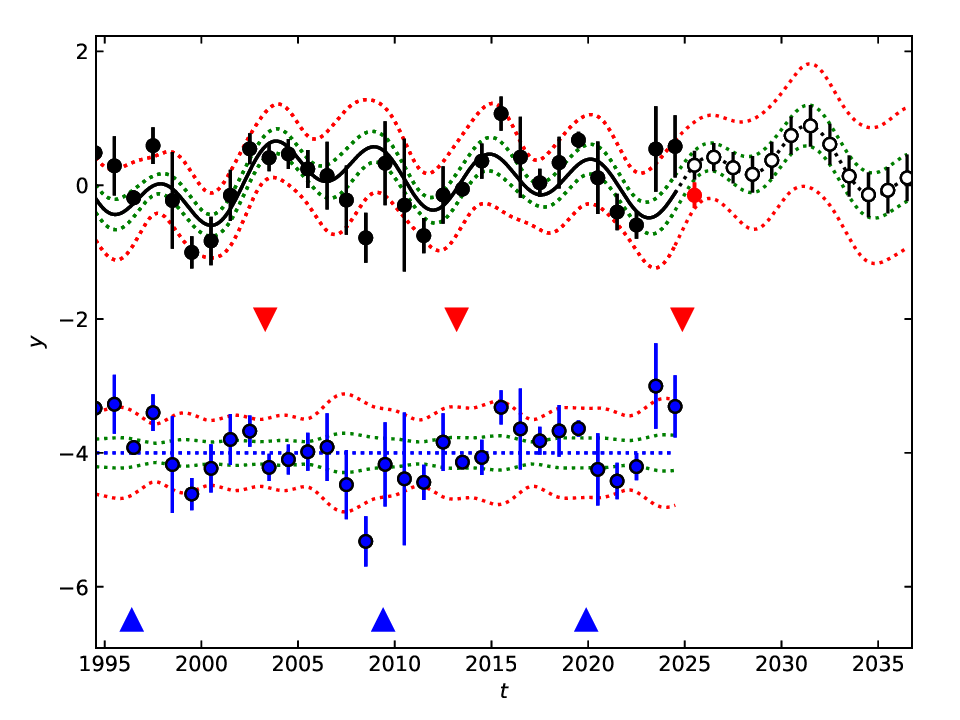}
}
\vspace{-0.36\textwidth}
\centerline{\Large 
\hspace{0.44\textwidth}  \color{black}{(d)}
\hspace{0.41\textwidth}  \color{black}{(e)}
\hfill}
\vspace{0.32\textwidth}
\caption{\BigWave ~for all weighted
  yearly mean data model \M=4 in Table \ref{TableClong}.
  Otherwise as in Figure \ref{FiglongC14K311}.
}
\label{FiglongC14K411}
\end{figure*}

\section{Results of DCM
    analysis for real data}\label{SectStress}

\subsection{Validation on complex
  terrestrial time series:
  El Ni\~no-southern oscillation}

\renewcommand{\Ki}{}
 We subject DCM to a rigorous stress test
   to evaluate its performance limits.
 \Ki
 DCM is applied to the \Data ~time series which
 measures the sea surface temperature (\sst) anomalies
  in the central-western equatorial Pacific Ocean.
  \Ki
  An \Axel ~event threshold is  
  that the \Data ~exceeds 0.5 \Cc
  ~\citep{Han03,Bun09,Ren11}.
   \Ki
 The mainstream climatological models treat
 \Axel ~as a non-linear, chaotic phenomenon.
 \Ki
 The forecasting of \Axel ~events is challenging
 and even the best forecasts fail after 18 months
 \citep{Lud13,Cai14,Tim18,Lia21,Liu23,Hu24,Thi24,Lu25}.
  
   \subsection{Data}
   
   We apply DCM to the
   \Data ~values from 
   \SampleText ~sample. 
  \Ki
  These \sst ~anomaly data from
\Link{https://psl.noaa.gov/data/timeseries/month/data/nino4.long.anom.data}
{NOAA}\footnote{https://psl.noaa.gov/data/timeseries/month/ \\ data/nino4.long.anom.dat}
were retrieved on January 29th, 2026.
For the last year 2025,
  four monthly mean values from September to December
  were missing.
Therefore, we analyse $n=155$ yearly mean values
and $n=1860$ monthly mean
values between January, 1870 and December, 2024.
All analysed data samples are summarised in Table \ref{TableSamples}.


\subsection{DCM analysis results}

\renewcommand{\Ki}{}
\subsubsection{Trend in all yearly mean data}
\label{SectTrend}

 For all yearly mean \Data ~data,
 we detect significant signals only
 from the weighted values
 (Table \ref{TableSamples}: \PRtext{Clong.dat}).
\Ki
Therefore,  the correct
$K_3$ order for the polynomial trend $p(t)$
(Equation \ref{EqPolyOne})                               
is solved from these weighted all yearly mean data.
\Ki
The weighted DCM analysis results for all yearly mean data
are given in Table \ref{TableTrend}.
\Ki
For one, two, three, and four signal models,
the best alternative is always the linear trend $K_3=1$.
\Ki
We use this $K_3=1$ trend  in all our subsequent
DCM analysis of \Data ~data. 
\Ki
For all yearly mean data,
the positive  linear temperature increase
trend for the best three signal
model \M=3 (Table \ref{TableTrend})
is  $2 M_1=0.56 \pm 0.20$ \Cc ~during $\Delta T=154$ years
  (Equation \ref{EqPolyTwo}).
  \Ki
  The value of this trend is approximately the same
  in all linear trend models.
  \Ki
  This positive linear trend
  pattern confirms 
  the climate change-induced global warming.
  
\subsubsection{Forecast of first half of all yearly mean data}
\label{SectHalfYears}

\renewcommand{\Ki}{}
We use the first half of all yearly mean
\Data ~data $(n=78)$ to forecast
the second half  $(n=77)$.
\Ki
The weighted DCM analysis results
for the first half
are
given in Table \ref{TableChalf}.
\Ki
DCM detects many signals.
\Ki
The sum of these interfering signals is hereafter called the \BigWave.
\Ki
The periods of the three strongest \BigWave ~signals are
$P_1=5.580\pm0.085$, $P_2=12.82\pm0.40$
and $P_3=19.30\pm0.83$
years (Table \ref{TableChalf}, \M=3). 
\Ki
For the pure white noise simple model
alternative, the complex model first signal significance
  is $Q_F=6.2 \times 10^{-5}$ (Table \ref{TableHalfPolynomials}, \M=1).
The two, three and four signal significances
  are $Q_F=0.0016$, 0.0025 and 0.013
  (Table \ref{TableChalf}, \M=1-3).
\Ki
For the pre-assigned significance level $\gamma=0.005$,
  the \M=2 model is the best.
\Ki
The \M=2 and \M=3 model periodograms show clearly defined,
very sharp minima
(Figures \ref{FighalfC14K211}a and \ref{FighalfC14K311}a).
\Ki
The simulations of Models 1-7
  (Sections \ref{SectModelOne}-\ref{SectModelSeven})
  have already confirmed
that only real signals cause such clear periodogram minima.
\Ki
The two and three signal model forecasts,
dotted black lines,
are shown in
Figures \ref{FighalfC14K211}b-e and \ref{FighalfC14K311}b-e.
\Ki
Both forecasts can reproduce the changes of forecasted yearly 
\Data ~means, the open black circles,  during the second half.
\Ki
For the second half data, 
the first half forecast $\chi^2$ values for
the one, two, three and four signal
models are
421, 336, 370 and 428, respectively. 
\Ki
Hence, the two signal \M=2 model gives the
best forecast for the second half of weighted yearly
mean data.
The red closed circles denoting
  the observed 2025 yearly mean 
  (Figures \ref{FighalfC14K211}e and \ref{FighalfC14K311}e)
  are discussed in greater detail in Section \ref{SectRealTime}.

\subsubsection{All weighted yearly mean data} \label{SectAllYears}

\renewcommand{\Ki}{}
DCM analysis results for all weighted yearly
mean \Data ~data are given in
Table \ref{TableClong}.
\Ki
We detect the same \BigWave ~periods that are already detected
from the first half of weighted yearly mean data.
\Ki
The three signal \M=3 model is the best.
\Ki
The \M=3 model DCM periodogram 
  shows three clearly defined, sharp minima
  (Figure \ref{FiglongC14K311}a).
\Ki
  The \BigWave ~signal periods are
$P_1=5.662\pm0.077$, $P_2=12.78\pm0.12$
and $P_3=21.3\pm1.5$ years.
  \Ki
  The peak to peak amplitudes are
  $A_1=0.40\pm0.13$ \Cc,
 $A_2=0.47\pm0.10$ \Cc ~and
 $A_3=0.37\pm0.12$ \Cc. 
 \Ki
   We show this \M=3 model 
   in Figures \ref{FiglongC14K311}b-e.
   \Ki
   The Fisher-test comparison between the
   simple pure white noise model
and
the complex one signal model
gives signal
significance  $\QF=3.9\TM10^{-13}$
(Table \ref{TableElNinoPolynomials}, \M=1).
\Ki
This $\QF$ value represents the probability that
the strongest signal, the first detected
$12.65$ years signal of model \M=1,
is not real.
\Ki
All other pure
polynomial model $g(t)=p(t)$ alternatives are also rejected
(Table \ref{TableElNinoPolynomials}, \M$\ge 2$).
\Ki
The second and third signal significances are
$\QF=2.7\TM10^{-7}$
and
$\QF=0.00038$
(Table \ref{TableClong}: \M=2-3). 
\Ki
We conclude that
the detections of all these three \BigWave ~signals are significant.

\renewcommand{\Ki}{}
The results for the unstable four signal model
(Table \ref{TableClong}, \M=4, $\UM$)
   are shown in Figures \ref{FiglongC14K411}a-e.
   \Ki
   The \M=3 and \M=4 model alternatives for all yearly
   mean weighted data give
   essentially the same \BigWave ~forecast values
   between 2025 and 2036 (Table \ref{TableClongOneNumbers}).

\begin{table}[h]
  \caption{\BigWave  yearly mean forecasts.
    (1) Year.
    (2-3) Figure \ref{FiglongC14K311}e: open circles.
    (4-5) Figure \ref{FiglongC14K411}e: open circles. 
  }
\label{TableClongOneNumbers}
\begin{scriptsize}
\begin{center}
  \begin{tabular}{crcrc}
    \hline
    & \multicolumn{2}{c}{Figure \ref{FiglongC14K311}e}
    & \multicolumn{2}{c}{Figure \ref{FiglongC14K411}e}      \\
    \hline
    (1)  &  (2)  & (3)       & (4)     & (5) \\
    $t$ & $y$ & $\pm 1 \sigma$ & $y$ & $\pm 1 \sigma$ \\
     (y) & (\Cc)  & (\Cc)  & (\Cc)  & (\Cc) \\
    \hline
  2025.5 &     0.05 &     0.25 &     0.30 &     0.22   \\
  2026.5 &     0.30 &     0.21 &     0.42 &     0.21   \\
  2027.5 &     0.36 &     0.20 &     0.26 &     0.23   \\
  2028.5 &     0.31 &     0.17 &     0.16 &     0.27   \\
  2029.5 &     0.35 &     0.17 &     0.37 &     0.28   \\
  2030.5 &     0.56 &     0.20 &     0.74 &     0.28   \\
  2031.5 &     0.74 &     0.21 &     0.89 &     0.31   \\
  2032.5 &     0.70 &     0.31 &     0.61 &     0.30   \\
  2033.5 &     0.40 &     0.37 &     0.14 &     0.31   \\
  2034.5 &     0.06 &     0.32 &    -0.14 &     0.33   \\
  2035.5 &    -0.08 &     0.25 &    -0.08 &     0.34   \\
  2036.5 &     0.01 &     0.20 &     0.11 &     0.35   \\
    \hline
 \end{tabular}
\end{center}
\end{scriptsize}
\end{table}
\renewcommand{\Ki}{}

\subsubsection{Real-time yearly mean forecast} \label{SectRealTime}
\renewcommand{\Ki}{}
The {\it old} \Data ~(\Ha) data
from
\Link{https://psl.noaa.gov/data/timeseries/month/data/nino4.long.anom.data}
{NOAA} database
were retrieved on  \Eilen.
\Ki
The 
four last monthly mean values were missing for the year 2025.
\Ki
Today, \Tanaan,  we have 
  retrieved those missing values.
\Ki
Our Table \ref{TableProof} gives
the {\it old} and {\it new} \Data ~(\Ha) data
after December 2024.

\renewcommand{\Ki}{}
   The data gap after December 2024 is a fortunate coincidence for us.
   \Ki
   The {\it new} monthly mean values in
   Table \ref{TableProof} (column 4)
   give the {\it new} 2025 yearly mean value 
   $y_i \pm 1 \sigma_i=-0.15 \pm 0.19$ \Cc.
   \Ki
   The red closed circle now
     highlights this {\it observed new} 2025 yearly value
     in Figures
     \ref{FighalfC14K211}e, \ref{FighalfC14K311}e, 
     \ref{FiglongC14K311}e and \ref{FiglongC14K411}e.
     \Ki
     All four figures confirm that the DCM forecast
       for this {\it new} 2025 yearly mean value succeeds!
       \Ki
       For  first half of data between 1870 and 1947,
       both DCM models \M=2 and \M=3 can,
       after a gap of 78 years,
       forecast the year 2025 {\it observed  new} value within $\pm 1 \sigma$ 
      (Figures   \ref{FighalfC14K211}e and \ref{FighalfC14K311}e).
     \Ki
      For the best \M=3 model of all yearly means, 
      the \BigWave ~forecast for the year 2025 (black open circle) and
      the {\it observed  new}
      2025 value (red closed circle) agree within $\pm 1\sigma$
      (Figure \ref{FiglongC14K311}e).
    \Ki
       Even the unstable \M=4 model for all yearly means
         gives an accurate 2025 forecast
           (Figure \ref{FiglongC14K411}e).
         \Ki
         In short, the {\it observed new} 2025 value not only
           confirms both all data forecasts, but it also confirms
          both first half data forecasts.
         \Ki
     All these forecasts can, of course, be dismissed
  just as lucky guesses.
  \Ki
  Who wants to take the risk that
  our severe \BigWave ~forecast for the year 2031
 is also ``lucky''   (Table \ref{TableClongOneNumbers})?

\begin{figure*}
  \centering
\includegraphics[width=1.0\textwidth,clip=]{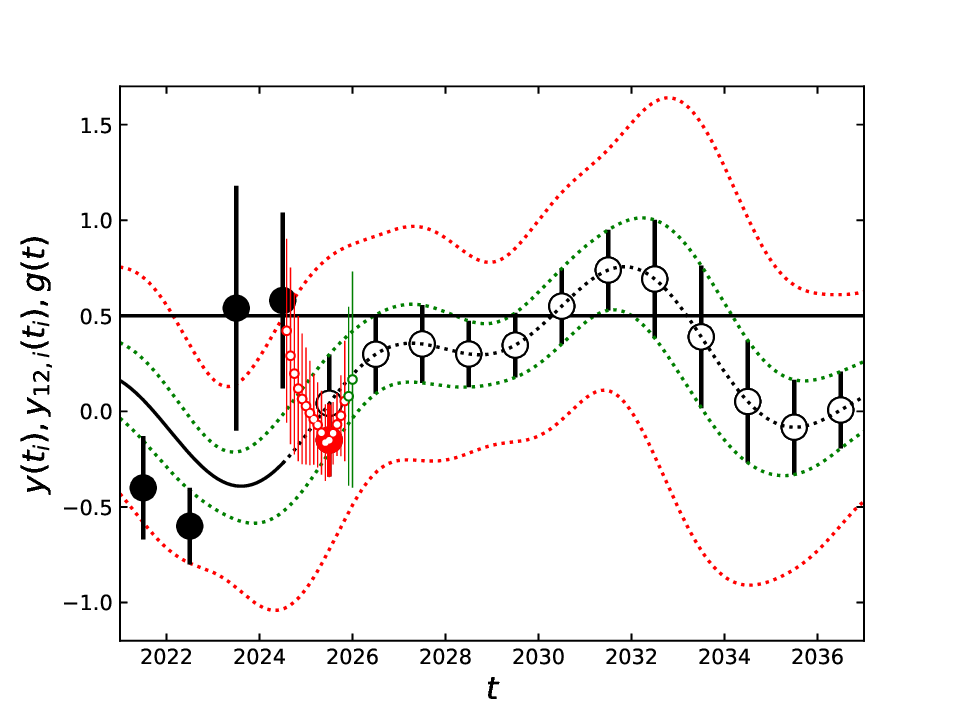}
\caption{Real-time forecast validation for  \M=3 model (Table \ref{TableClong}).
  Model curves, observed large closed black circles and
  forecasted large open
  white circles are as in
  Figure \ref{FiglongC14K311}e.
  {\it Observed new} 2025 yearly mean (red large closed circle) falls within
  $\pm 1 \sigma$ confidence interval of deterministic
 forecast for 2025 (large open black circle) generated
  from data before December 2024. 
  Yearly sliding means  $y_{i,12}$ from Table \ref{TableLastYears}
  are denoted with 
  small open red circles (\Ha) and small green open circles (\Er).
    Continuous horizontal black line shows  \Axel ~$+0.5$ \Cc ~limit. 
    Note that
    our real-time forecast for 2026 also
    successfully captures neutral conditions
      during winter, subsequent warming in spring,
      and reaches \Axel ~limit within $\pm 1 \sigma$ during summer.
    This demonstrates that our real-time forecast is currently
    synchronised with "Big Wave" periodicities
  although $y_{i,12}$ values for this year 2026 are still missing.
     All error bars are $\pm 1 \sigma$.
    Units are [y] (x-axis) and [\Cc] (y-axis) }
   \label{FigLastYears}
 \end{figure*}

 \renewcommand{\Ki}{}
  This best forecasting model \M=3 (Table \ref{TableClong})
  is computed for the $n=155$ yearly means
  (Figure \ref{FiglongC14K311}e,
  black closed circles).
\Ki
  We can check this real-time yearly mean forecast against only one
  new yearly mean value, the year 2025
  (Figure \ref{FiglongC14K311}e, red closed circle).
  \Ki
  However, every monthly mean $t_i$ and $y_i$ value 
after December 2024
can be used to compute
  the yearly sliding mean values
  $t_{12,i}$, $y_{12,1}$ and $\sigma_{12,i}$
  for the past 12 months.
  \Ki
  The observed yearly sliding means
  are given in Table \ref{TableLastYears}
  and displayed in Figure \ref{FigLastYears} (red small open circles).
  \Ki
   These yearly sliding mean values
   confirm
   that the \M=3 model (Table \ref{TableClong})
   real-time forecast after December 2024 has become true
   for the sixteen $y_{i,12}$ values between January 2025  
   and April 2026.

\renewcommand{\Ki}{}
\July{Today \Tanaan,}
we cannot compute  the yearly sliding mean values
  $t_{12,i}$, $y_{12,1}$ and $\sigma_{12,i}$ after April 2026
because  NOAA has not published
new monthly \Data ~\Ha ~values.
 \Ki
 However, there is another \Data ~which has been 
 updated during this year May - June 2026,
 the \Er ~index.\footnote{\Tanaan, \Er ~ values were retrieved  from \\
   \Link{https://psl.noaa.gov/data/correlation/nina4.anom.data}
{https://psl.noaa.gov/data/correlation/nina4.anom.data}.}
 \Ki
 The \Ha ~and \Er ~\Data ~values correlate
 because they measure the \sst ~anomalies of the
 same geographical area.
\Ki
 For the past two years 2024 and 2025,
 the maximum differences 
 for the monthly and yearly values of these two indices
 were 0.3 \Cc ~and 0.1~\Cc , respectively (Figure \ref{FigIndexCompare}).
 \Ki
 The \Er ~\Data ~monthly mean and sliding yearly mean values
 are given in Table \ref{TableLastYears}.
 \Ki
 The yearly sliding means for May and June 2026
 are shown in Figure  \ref{FigLastYears} (small green open circles).
 \Ki
 The  \Ha ~and \Er ~\Data ~yearly
 sliding means $y_{i,12}$ in Table \ref{TableLastYears}
 confirm that our
 real-time forecast after December 2024 has become
 true for 18 months in a row.

 \renewcommand{\Ki}{} 
 We can show that
   our sliding yearly mean $y_{i,12}$ forecasts
   will also come true for the next months after June 2026.
   \Ki
   The next \Ha ~\Data ~data values for May - August 2026
   will be published in October 2026.
   \Ki
   Therefore, we estimate the future sliding yearly means $y_{i,12}$
   by using the available \Er ~\Data ~values.
   \Ki
   Before doing this, we must establish
   how rapidly the observed historical \Er ~\Data ~values
   have changed between the years 1950 - 2025.
   \Ki
   The biggest observed
   \Er ~index rapid rises during one and two months are
   $0.68$ \Cc ~(April - May 1951)                
   and
   $0.89$ \Cc ~(April - June  1950).             
   \Ki
   The respective rapid falls during one and two months
   are
   $-0.72$ \Cc ~(September-October 1988) 
   and
   $-0.94$ \Cc ~(June - August 2010).         
   \Ki
   The highest value for the mean of two months is
   $1.52$ \Cc ~(December 2023 - January 2024).       
   \Ki
   The lowest two months mean is
   $-2.16$ \Cc ~(October - November 1975).              

 \renewcommand{\Ki}{} 
   Our yearly mean forecast for this year 2026 is
   $m \pm s= 0.30 \pm 0.21$ \Cc ~(Table \ref{TableClongOneNumbers}).
   \Ki
   The $\pm 1\sigma$ upper and lower limits are
   $m+s=0.51$ \Cc ~and $m-s=0.09$ \Cc.
   \Ki
  We use the relation
   \begin{eqnarray}
     T_{\pm}(J)={{12(m \pm s)- \sum_{j=i-(12-J)}^i y_j  }   \over {J}},
     \label{EqBreak}
   \end{eqnarray}
   where \July{$y_{j=i}=1.22$ \Cc} ~is the monthly mean
     value of June 2026.
   \Ki
   This relation gives the upper
   $T_+(J)$ and lower $T_{-}(J)$ limit for the average monthly
   \Er ~\Data ~values during the next $J=1,2, ..., 6$ months after June 2026.
   \Ki
   If the values given in Table \ref{TableBreak}
   are reached during the $J$:th month after June 2026,
   the yearly sliding mean $y_{i,12}$ value 
   exceeds the $\pm 1 \sigma$ limits of our yearly mean
   $m \pm s$ forecast for this year 2026.
\Ki
To accomplish that,
the July 2026 monthly mean should 
rise to \July{$y_i=4.15$ \Cc} ~or fall to \July{$-0.89$ \Cc
  ~(Table \ref{TableBreak}, $J=1$).}
  \Ki
  Both values are impossible 
  because the current \Er ~index
  value for June, 2026 is \July{$y_i=1.22$ \Cc.}
  \Ki
  It is also impossible that
  the average for July - August 2026 
  rises to \July{$T_{+}(2)=2.00$} \Cc ~because the
  highest historical two month mean
  is 1.32~\Cc ~(Table \ref{TableBreak}, $J=2$).
\Ki
The fall to \July{$T_{-}(2)=-0.52$} \Cc ~is also impossible
because largest historical fall during two months is $-0.94$~\Cc. 
  \Ki
  These numbers confirm that our real-time forecast will certainly
  hold for 20 months in a row, from January 2025 to August 2026.
  \Ki
  Our real-time $\pm 1 \sigma$ forecasts will succeed for two full years, 
  if the average for July - December 2026 monthly means does not
  rise to \July{$T_{+}(6)=0.41$} \Cc ~or fall to \July{$T_{-}(6)=-0.43$} \Cc
   ~(Table \ref{TableBreak}).
   \Ki
   Although the $y_{i,12}$ are still missing for this year 2026,
  the black dotted curve of
  our model forecasts correctly
  the neutral conditions during winter, the warming in spring and
  the \Axel ~level temperatures today \Tanaan  ~(Figure  \ref{FigLastYears}).

\begin{table}
  \caption{\July{Monthly \Er ~\Data ~limits for
    exceeding} $m \pm s = 0.30 \pm 0.21$ \Cc ~yearly
    mean forecast for 2026. 
    (1) Month.
    (2-4) $J$, $T_+$ and $T_{-}$ values of Equation \ref{EqBreak}.}
  \label{TableBreak}
  \begin{center}
\begin{tabular}{lrrr}
  \hline
  (1) &  (2) & (3) & (4)  \\
  Month & $J$ & $T_{+}(J)$ & $T_{-}(J)$ \\
  (-)   & (-)   & \Cc            &  \Cc          \\
  \hline
          July 2026        &     1 &   4.15 &  -0.89 \\
         August 2026    &     2 &   2.00 &  -0.52 \\
      September 2026 &     3 &   1.24 &  -0.44 \\
        October 2026   &     4 &   0.83 &  -0.43 \\
       November 2026 &     5 &   0.56 &  -0.45 \\
       December 2026 &     6 &   0.41 &  -0.43 \\
\hline
\end{tabular}
\end{center}
\end{table}

\renewcommand{\Ki}{}
The real-data forecast criterion in Equation \ref{EqBreak} gives
  the limits for the cases $|y_{i,12}-m| < s$, where
  $m=g(t_i)$ and $s=\sigma_{g(t)}$ (Table \ref{TableBreak}).
  \Ki
  It uses the forecasting model bootstrap error {\it estimate}.
  \Ki
  An alternative criterion for a successful real-time forecast would be
  $|y_{i,12}-m| <  s$,  where $m=g(t_i)$ and $s=\sigma_{i,12}$.
  \Ki
  It uses the  {\it observed} standard deviation for the yearly sliding mean.
  \Ki
  This alternative real-time forecast criterion will certainly hold
  long after June 2026 because the current standard deviations
  $\sigma_{i,12}$ are very large
  (Figure  \ref{FigLastYears}: green error bars).

\renewcommand{\Ki}{}
   While there is no fixed numerical  \Ha ~\Data ~limit for \Axel ~events,
the general academic standard is $+0.5$ {\Cc}~
\citep{Han03,Bun09,Ren11}.
\Ki
The \Data ~variance is much lower than the Ni\~no 3.4 index variance.
\Ki
Consequently, a lower \Data ~anomaly value
is statistically more significant.
\Ki
The forecasted $\pm 1 \sigma$
yearly mean values between the years
2026 and 2029 rise close to the \Axel ~limit
+0.5 \Cc ~(Figure \ref{FigLastYears}: continuous horizontal black line).
\Ki
We forecast a prolonged three year \Axel ~period, where
the yearly mean values may exceed the +0.5 \Cc ~limit 
(Table \ref{TableClongOneNumbers}, 2030-2032).
\Ki
However,  the yearly mean values of both
  \M=3 and \M=4 models rise
  $1\sigma$ above this limit only during the year 2031.
\Ki
A cooler four year period may begin in the year 2033.

 \renewcommand{\Ki}{}
 \Ki
  For the first
  half of yearly mean $n=78$ sample,
  both DCM models \M=2 and \M=3 give
  excellent forecasts for the second half
  (Figures \ref{FighalfC14K211}b-e and \ref{FighalfC14K311}b-e).
  \Ki
  During the first six years,
  the observed second half
  data $\pm 1 \sigma_i$ errors match the first half
  forecast $\pm 1 \sigma_{g(t)}$ errors.
  \Ki
  Therefore, our forecasts for 2025 - 2036 should be even more
  accurate because DCM can utilise
  a two times larger $n=155$ sample of all weighted yearly means
  (Table \ref{TableClongOneNumbers},
  Figures \ref{FiglongC14K311}e
  and
   \ref{FiglongC14K411}e).

\renewcommand{\Ki}{}
For the sake of consistency, we also give DCM results
for all non-weighted yearly \Data ~data (Table  \ref{TableRlong}).
\Ki
We detect the signatures of the same \BigWave ~signals,
but the signal significances $Q_{\mathrm{F}}$ are low.
\Ki
This confirms that the yearly data error information
is crucial for weighted DCM analysis.

\begin{figure*}  
\vspace{0.02\textwidth}
\centerline{\hspace*{0.005\textwidth}
 \includegraphics[width=0.600\textwidth,clip=]{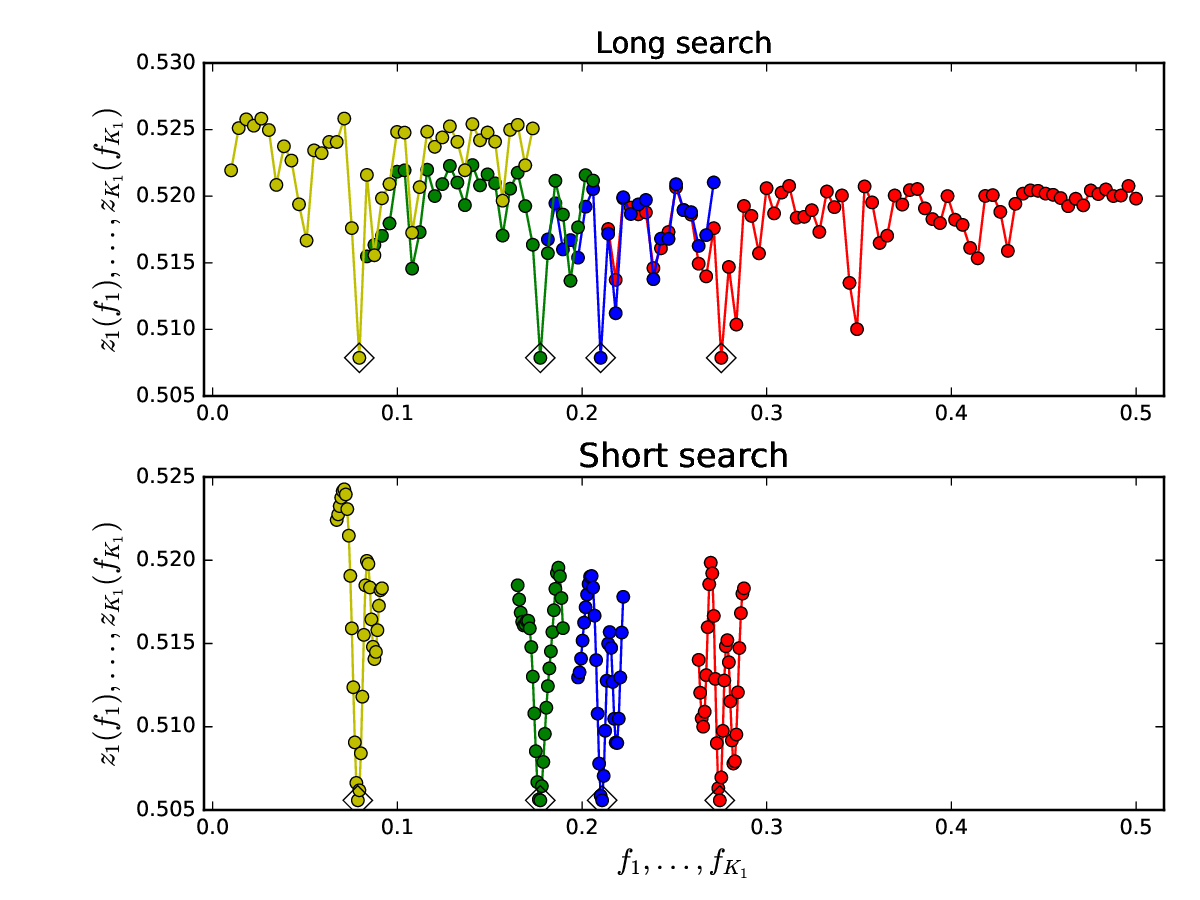}
 \hspace*{-0.01\textwidth}
 }
\vspace{-0.45\textwidth}
\centerline{\Large 
\hspace{0.75\textwidth}  \color{black}{(a)}
\hfill}
\vspace{0.45\textwidth}
\centerline{\hspace*{0.005\textwidth}
 \includegraphics[width=0.455\textwidth,clip=]{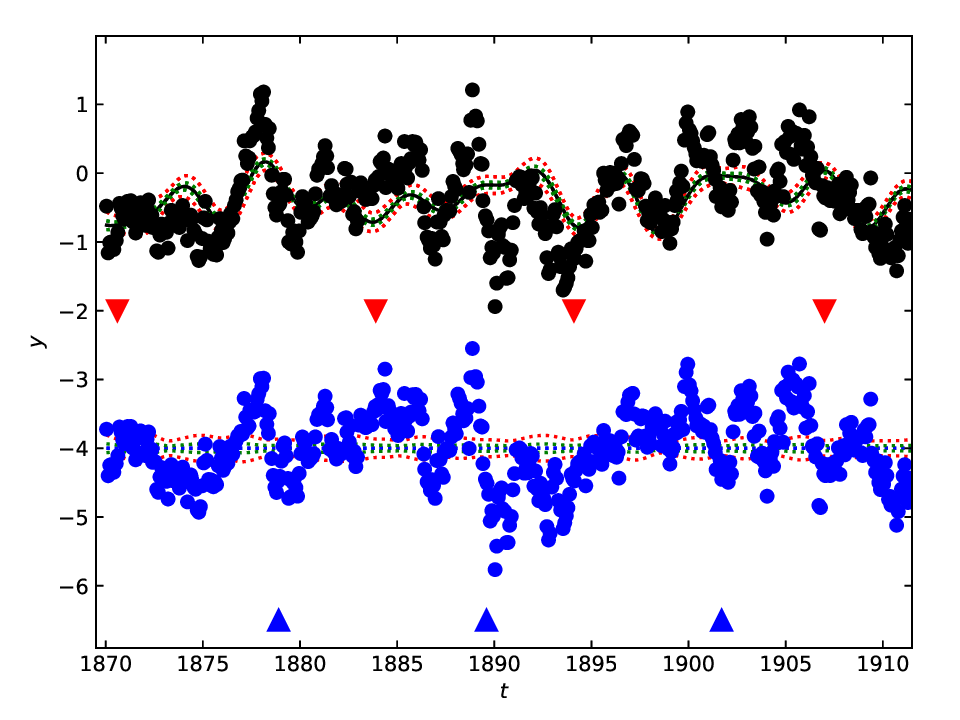}
 \hspace*{-0.01\textwidth}
 \includegraphics[width=0.455\textwidth,clip=]{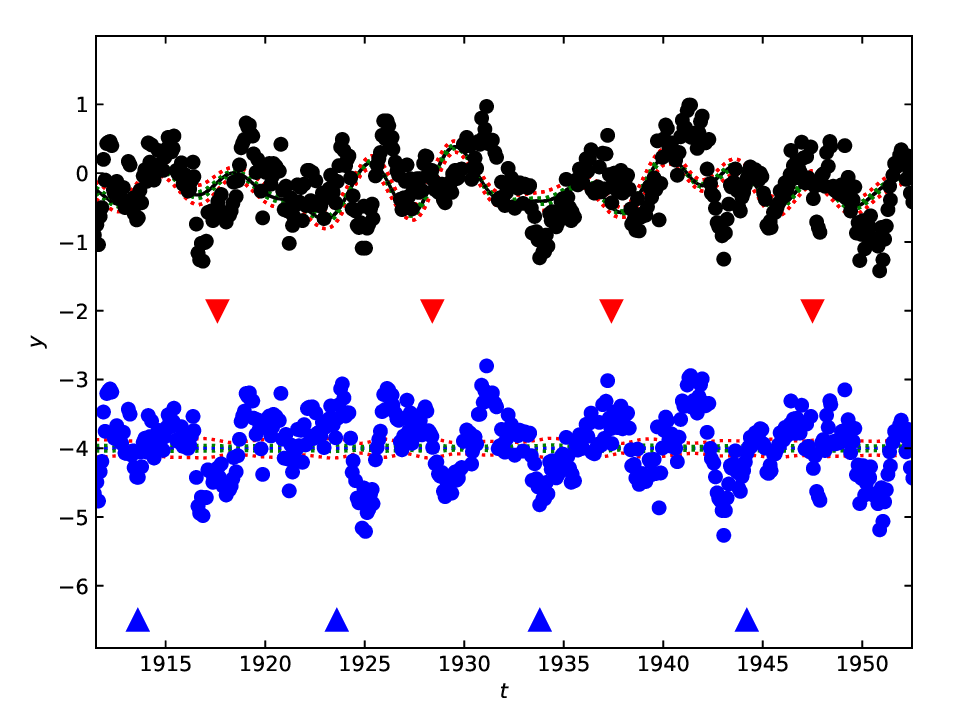}
}
\vspace{-0.36\textwidth}
\centerline{\Large 
\hspace{0.44\textwidth}  \color{black}{(b)}
\hspace{0.41\textwidth}  \color{black}{(c)}
\hfill}
\vspace{0.35\textwidth}
\centerline{\hspace*{0.005\textwidth}
 \includegraphics[width=0.455\textwidth,clip=]{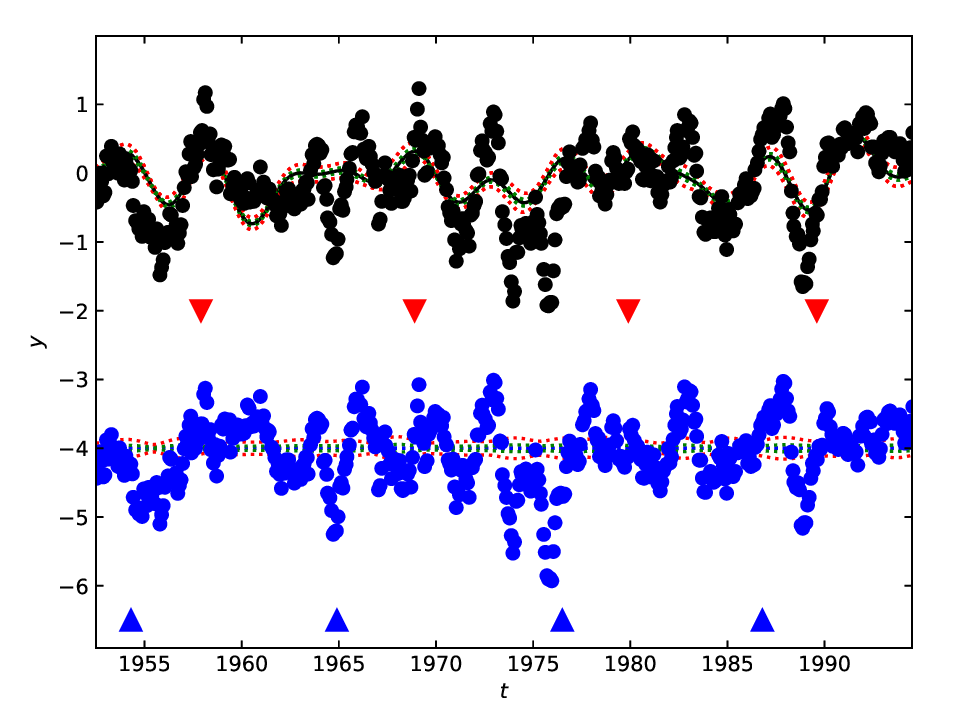}
 \hspace*{-0.01\textwidth}
 \includegraphics[width=0.455\textwidth,clip=]{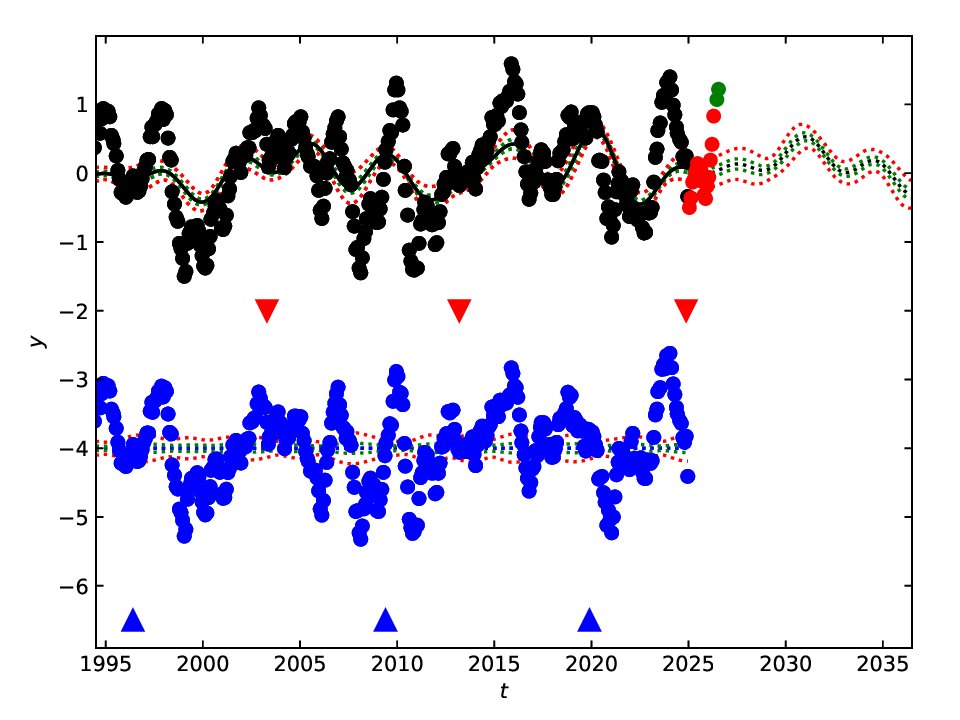}
}
\vspace{-0.36\textwidth}
\centerline{\Large 
\hspace{0.44\textwidth}  \color{black}{(d)}
\hspace{0.41\textwidth}  \color{black}{(e)}
\hfill}
\vspace{0.32\textwidth}
\caption{All monthly mean data model \M=4 in Table \ref{TablemAll}.
  Red and green dots are  new updated \Ha ~and \Er ~monthly mean
  \Data ~values from Table  \ref{TableProof}.
  Otherwise as in Figure  \ref{FiglongC14K311}.}
\label{FigmAll}
\end{figure*}

 \subsubsection{All monthly means}
\label{SectAllMonths}

\renewcommand{\Ki}{}
 Here, we  analyse
$n = 1860$ monthly mean \Data ~values between January, 1870 and
December, 2024.
\Ki
The computation of twelve month means may have
prevented the detection of shorter periods from yearly mean data.
\Ki
We search for shorter periods by lowering
$P_{\mathrm{min}}=3$ years to 2 years.
\Ki
The value $P_{\mathrm{max}}=100$ years remains the same.
\Ki
For the monthly means,
the four signal DCM model is the best one
(Table \ref{TablemAll}, \M=4).
\Ki
The signal periods are
$P_1=$ $3.6436\pm0.0046$,
$P_2=$ $4.7477\pm0.0080$,
$P_3=$ $5.647\pm0.011$,
and
$P_4=$ $12.704\pm0.063$ years. 
\Ki
The 12.7 years signal has the highest peak to peak amplitude
$A_4=$ $0.392\pm0.038$~\Cc.

 \renewcommand{\Ki}{}
Four prominent, sharp periodogram minima
(Figure \ref{FigmAll}a)  indicate the presence of true periodic signals.
\Ki
The first signal significance
is \REJ ~(Table \ref{TableElNinoPolynomials2}, \M=1-4).
\Ki
The second, third and fourth
signal significances are also \REJ ~(Table \ref{TablemAll}).
\Ki
All four signal 
detections are absolutely certain.
\Ki 
DCM search for the fifth signal
from the \M=4 model residuals 
gives $R=529$. 
\Ki
This confirms that these monthly
mean data contain {\it only four} signals
because the \M=4 model has $R=475$. 

%
\begin{figure}[h]
  \centering
\includegraphics[width=0.50\textwidth,clip=]{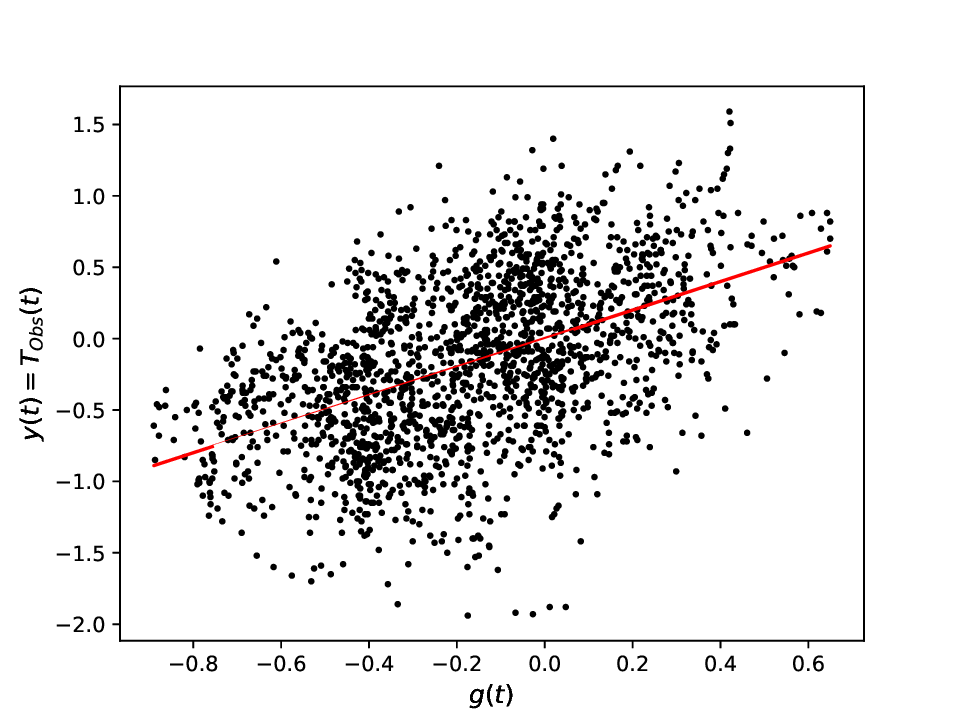}
\caption{Correlation between monthly mean
  $g(t_i)$ and $y(t_i)$ values.
  Linear correlation coefficient $r=0.501$ significance
  is $p=9.3 \times 10^{-119}$.
  Red line denotes
  $y(t)=a [g(t)] + b$, where $a=1.001\pm0.040$
  and $b=0.000\pm 0.013$.
  Units are [\Cc] (x- and y-axis).}
\label{FigCorrelationOne}
\end{figure}

%
\begin{figure}[h]
  \centering
\includegraphics[width=0.50\textwidth,clip=]{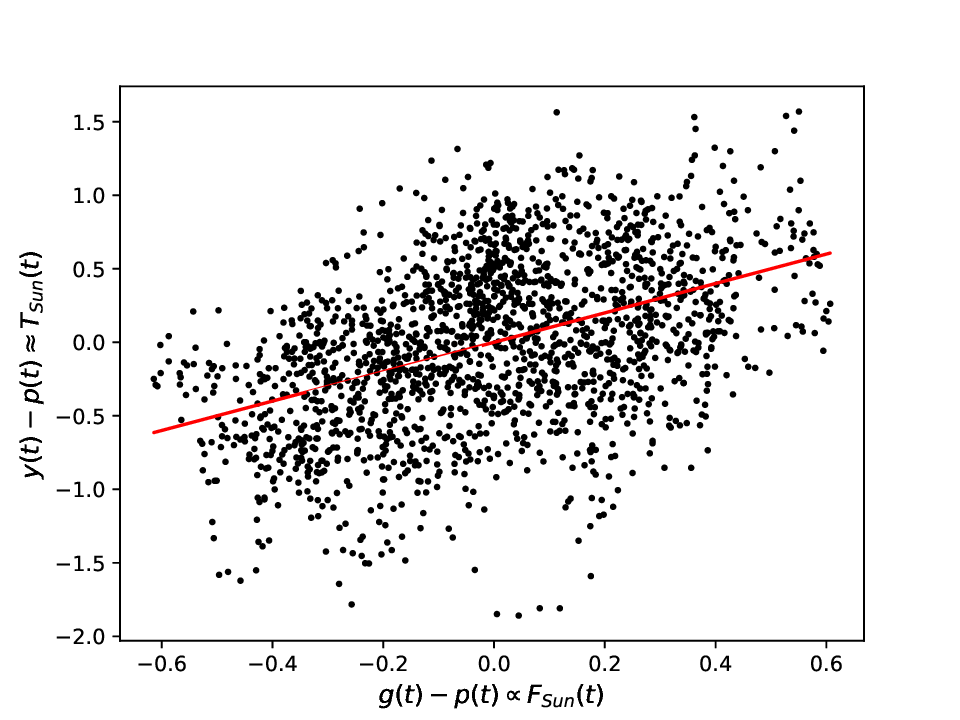}
\caption{Correlation between monthly
  $g(t_i)-p(t_i)$ and $y(t_i)-p(t_i)$ values.
  Linear correlation coefficient $r=0.447$ significance
  is $p=5.7 \times 10^{-92}$.
  Red line denotes
  $y(t)-p(t)=a [g(t)-p(t)] + b$, where $a=0.999\pm0.046$
  and $b=0.000\pm 0.012$.
  Units are [\Cc] (x- and y-axis).} 
\label{FigCorrelationTwo}
\end{figure}


\renewcommand{\Ki}{}
We show the best DCM model for all monthly mean data
in Figures \ref{FigmAll}b-e.
\Ki
The consecutive model residuals
(blue closed circles) show strong residual correlation (red noise)
which is typical for climatological time series
\citep{Zwi95,Gue14}.
\Ki
The DCM linear LS
model free parameter $\bm{\beta}$ estimates are unbiased
even when the data contains red noise 
\citep[][Chapter 12]{Was20}.
\Ki
However, due to this red noise,
the bootstrap procedure gives too small
model error $\sigma_{g(t)}$ estimates
(Equation  \ref{EqModelError}).
\Ki
For this reason, we give only the next decade
2025-2036 monthly mean forecast values,
but not their error estimates
(Table  \ref{TablemAllForecast}).
\Ki
The monthly mean values after December 2024  are not analysed,
but they are displayed in
Figure \ref{FigmAll}e (red and green closed circles).

 \renewcommand{\Ki}{}
 The correlating monthly mean model residuals
 (blue closed dots)
 consistently
show how rapidly and strongly the monthly $y(t_i)$
values react to the $g(t_i)$ peaks and valleys
(Figures \ref{FigmAll}b-e).
\Ki
The extrema of  $y(t_i)$
and $g(t_i)$ coincide.
\Ki
This is confirmed by
the extreme significance
$p=9.3 \TM 10^{-109}$
for their linear
correlation
(Figure \ref{FigCorrelationOne}).
\Ki
However,
the scatter of monthly  $y(t_i)$ values at any $g(t_i)$ value
is about $\pm 1$ \Cc.
\Ki
This means that DCM cannot forecast the monthly mean
\Data ~values.

\renewcommand{\Ki}{}
\subsubsection{Separate months}
\label{SectSeparate}

We also analyse separately all months 
from January to December (Tables \ref{TableJan}-\ref{TableDec}).
\Ki
The one year gap between consecutive $y_i$ values
can weaken the residual correlation problem.
\Ki
We highlight the observed and forecasted monthly
values after December 2024  (Figures \ref{FigJan}-\ref{FigDec}:
red and green closed circles versus black open circles).
\Ki
The forecasted separate monthly mean
values for the next decade are given
in Table \ref{TableCollect}.
\Ki
Note that this table
 also gives the forecast error estimates because
the residual correlation problem for separate months
is weak or totally absent.

\renewcommand{\Ki}{}
\subsection{Solar forcing hypothesis}

DCM analysis of \Data ~data reveals
  that some strictly periodic mysterious
  ``engine'' warms and cools the Pacific Ocean.
  \Ki
  It is highly improbable that some
  {\it internal} oceanic ``engine'' functions with such mechanical regularity.
  \Ki
  The primordial ``engine'' {\it outside} the ocean must be the Sun.
  \Ki
  We apply DCM to the observed anomalies of
  the sea surface temperatures
  $T_{\mathrm{Sea}}(t)$.
  \Ki
  If the ``engine'' is the Sun,
  the $T_{\mathrm{Sea}}(t)$ changes
\begin{eqnarray}
y(t) =  T_{\mathrm{Sea}}(t)=T_{\mathrm{Sol}}(t)+T_{\mathrm{Glo}}(t),
\end{eqnarray}
are the sum of solar forcing
$T_{\mathrm{Sol}}(t)$
and global warming
$T_{\mathrm{Glo}}(t)$,
where the units are \Cc.

\renewcommand{\Ki}{}
Our model $g(t)=h(t)+p(t)$ is the sum of a periodic $h(t)$ function
and an aperiodic $p(t)$ function.
\Ki
If the observed
linear polynomial trend $p(t)$ represents the global warming
\begin{eqnarray}
T_{\mathrm{Glo}}(t)=p(t),
\end{eqnarray}
we get
\begin{eqnarray}
y(t)-p(t)=T_{\mathrm{Sea}}(t) - p(t) = T_{\mathrm{Sol}}(t).
\end{eqnarray}
\noindent
\Ki
In this case,
\begin{eqnarray}
  T_{\mathrm{Sol}}(t) & =           & y(t)-p(t) \nonumber \\
                                & \approx & g(t)-p(t) = h(t) \propto F_{\mathrm{Sun}}(t),
\end{eqnarray}
where the  $F_{\mathrm{Sun}}(t)$ sum of solar forcing flux signals
  represents the \SolLC ~having the units ${\mathrm{W}} {\mathrm{m}^{-2}}$.
\Ki
The maximum $h(t)$ solar forcing effect is
$A_1+A_2+A_3=1.2 \pm 0.2$ \Cc ~for yearly
means (Table \ref{TableClong}) 
and 
$A_1+A_2+A_3+A_4=1.41\pm0.06$ \Cc ~for
monthly means (Table \ref{TablemAll}).
\Ki
The $\pm 1 \sigma$ limits of these two estimates overlap.

\renewcommand{\Ki}{}
We use the relation
\begin{eqnarray}
F_{\mathrm{Sun}}(t) \Trigger T_{\mathrm{Sol}}(t)
\label{EqTrigger}
\end{eqnarray}
to describe how $F_{\mathrm{Sun}}(t)$
triggers the $T_{\mathrm{Sol}}(t)$ changes.
\Ki
The symbol ``$\Trigger$'' contains all unknown climatological
processes that transform ${\mathrm{W}} {\mathrm{m}}^{-2}$ to \Cc.
\Ki
If the Pacific Ocean acts like a giant bolometer,
  the absorption (input) is the \SolLC ~and
  the resistance (output) is the \sst ~\BigWave.

\renewcommand{\Ki}{}
For the monthly mean values,
the linear correlation between
$g(t)-p(t) \propto F_{\mathrm{Sun}}(t)$
and $y(t)-p(t) \approx T_{\mathrm{Sol}}(t)$ 
is absolutely certain
(Figure \ref{FigCorrelationTwo}: $p=5.7 \times 10^{-92}$),
but
the $\pm 1$ \Cc ~scatter of monthly $y(t_i)-p(t_i)$ values
prevents accurate forecasting.
\Ki
For the yearly means, however, forecasts succeed
because the average of twelve monthly means
eliminates the chaotic
``$\Trigger$'' effect in Equation \ref{EqTrigger}.

\subsection{Deterministic Framework:
    El Niño, the Sun, and the Planets}

 \renewcommand{\Ki}{}
Here, we discuss the hypothesis that {\it external} solar forcing,
rather than {\it internal} climate dynamics,
governs the general pattern of Niño 4 index changes.
\Ki
Some studies
have found correlation between the solar cycle and 
the climate temperature
\citep{Fri91,Sch11A,Had13,Con21}.
\Ki
The mainstream scientific consensus is that
solar activity plays a minor role in global warming
\citep{Gra10,Kop11,Loc20}.
\Ki
For the period 1870 - 2024,
our DCM estimates for all yearly means give
$+~0.6$ \Cc ~for linear global warming trend
and $\pm ~0.6$ \Cc ~for solar forcing superimposed
  on this trend. 

Our arguments are  transparent.

\begin{itemize}

\item[]  \ArgOne: If the \BigWave ~is a real phenomenon,
  solar forcing is the only logical cause.

\item[] \ArgTwo: If the solar forcing causes the \BigWave,
  this forcing must be strictly periodic.

\item[] \ArgThree: If the solar forcing is strictly
  periodic, the solar dynamo must be strictly periodic.

\item[] \ArgFour: If the solar dynamo is strictly
  periodic, the planetary tides
  may be the cause.

\end{itemize}

The main counterarguments are

\begin{itemize}

\item[] \CArgOne:
  \Axel ~is an internal climate phenomenon,
  where the ocean and the atmosphere are dynamically coupled
  \citep{Bje69,Phi89}.
  
\item[] \CArgTwo:
  The solar dynamo is sto\-chas\-tic
  and non-stationary
  \citep{Uso17,Cha20}.

\item[] \CArgThree:
   Planetary gravitational fields are too weak
  to sustain a strictly periodic solar dynamo \citep{Kud22,Cha22}.
\end{itemize}

\subsubsection{\ArgOne}


 DCM detects the same strictly periodic signals from
 half of the yearly \Data ~means, all yearly \Data ~means
 and all monthly \Data ~means.
\Ki
 The signal significances are extreme, like \REJ ~for all
 four signals detected from monthly mean \Data ~values.
\Ki
DCM models give accurate forecasts.
\Ki
All these results confirm that
the \BigWave ~must be a real phenomenon.
\Ki
 If  \CArgOne ~were true, such a chaotic system would not cause the
 \BigWave.
 \Ki
If  the ocean or atmosphere cannot cause the \BigWave,
the only remaining alternative is the Sun.

\renewcommand{\Ki}{}
\Ki
The blue and read triangles highlight the 
epochs of sunspot cycle minima and maxima
in our Figures
\ref{FiglongC14K311}b-e,
\ref{FiglongC14K311}b-e and
\ref{FigmAll}b-e.
\Ki
The large 2008 yearly mean residual stands out
because it falls far below the expected minus
four degrees offset level
(Figures
\ref{FiglongC14K311}e and
\ref{FiglongC14K411}e).
\Ki
It coincides with the strongest sunspot
cycle minimum of modern times
which has been connected to simultaneous
climate cooling
\citep{Sch11A,Had13}.
\Ki
In fact, all pronounced DCM model negative residuals for the yearly means
are located close to the blue triangles denoting the sunspot cycle minima.
\Ki
If the Pacific Ocean cools down when the solar forcing is at its weakest,
the \sst ~changes are synchronised with the sunspot cycle.
\Ki
This causality moves our work from Statistics to Astrophysics.
\Ki
The positive residuals do not show similar
systematical deviations from the DCM model
at the sunspot cycle maxima
(Figures
\ref{FiglongC14K311}b-e and
\ref{FiglongC14K411}b-e: Red triangles).
\Ki
DCM models for \sst ~{\it anomalies}
reveal residual {\it anomalies}.
\Ki
These significant negative residuals aligning with sunspot cycle minima
are hereafter called the \Double.
\Ki
They confirm that the DCM models work,
but the models have significantly larger negative residuals
when the solar activity is at its weakest.

\subsubsection{\ArgTwo}

\renewcommand{\Ki}{}
\Ki
If the solar forcing were not strictly periodic,
 what other alternative phenomenon
 could cause the observed regular \BigWave ~in the Pacific Ocean?
\Ki
If the  solar forcing were irregular,
the global warming and climate response noise ``$\Trigger$"
(Equation \ref{EqTrigger})  could prevent
DCM detection of \BigWave.
For example,
  this forcing must be strictly periodic and stationary
  because the yearly means between the years 1870 and 1947
  can accurately forecast the 2025 yearly mean
(Figures \ref{FighalfC14K211}e and \ref{FighalfC14K311}e:  closed red circle).
The epochs of
  \Double ~confirm that the \SolLC ~is synchronised
  with the strictly multi-periodic
  \BigWave ~in the Pacific Ocean (Figures
\ref{FiglongC14K311}b-e and
\ref{FiglongC14K411}b-e: Red triangles).
The Pacific Ocean and the Sun can remain
synchronised for 154 years {\it if, and only if,}
the solar dynamo is also
strictly multi-periodic, as claimed  by \citet{Jet25}. 

\subsubsection{\ArgThree}

 \renewcommand{\Ki}{}
\Ki
If the solar dynamo is stochastic and non-stationary (\CArgTwo),
the solar forcing can not be stationary and strictly multi-periodic.
\Ki
Only a deterministic \SolLC~can cause the \BigWave ~measured
by the Earth's largest bolometer, the Pacific Ocean.
\Ki
There are dynamo models,
where the synchronised response
to planetary tides drives a stationary and multi-periodic solar dynamo
\citep{Kle23, Mou25,Ste25,Ste26}.
\Ki
The solar dynamo may be multi-periodic and stationary
because DCM has also detected the extremely significant, strictly periodic
\SignalOne, \SignalTwo ~and \SignalThree ~signals from the
sunspot record \citep{Jet25}.
\Ki
The above subscripts refer to the signal periods in years.

\subsubsection{\ArgFour}
\renewcommand{\Ki}{~ \\ \Bb{tilaa \\ ~\\}}
 \renewcommand{\Ki}{}

\CArgThree ~is not true 
if the synchronised  planetary tides can provide
the amplification mechanism for
the solar dynamo \citep{Abr12,Kle23, Mou25, Ste25,Ste26}.
\Ki
The unending interference of
planetary tides could explain
how the deterministic \SolLC ~causes
the regular \BigWave ~in the Pacific Ocean.
\Ki
Except for the planets, there is nothing else in the solar
system that can cause the regular signals in
the period range of the \BigWave. 
\Ki
The planetary tides could also explain the strictly periodic
\SignalOne, \SignalTwo ~and \SignalThree ~signals 
in the sunspot data \citep{Jet25}.
\Ki
It may not be a coincidence that
the \SignalThree ~signal period detected from the sunspot record
is equal to Jupiter's orbital period of 11.86 years.
\Ki
The tidally driven solar dynamo could explain both the  \BigWave ~in
the Pacific Ocean and the strictly periodic
signals in the sunspot record \citep{Jet25}.
\Ki
If the solar dynamo were indeed stochastic and non-stationary
(\CArgTwo),
our DCM would never detect extremely significant,
strictly periodic signals
in the Sun and the Pacific Ocean.

To be absolutely clear,
  we are not proposing a new
   theoretical physical amplification mechanism for the solar dynamo.
\Ki
  Instead, the purpose of this paper is
   to present an objective,
   empirical DCM detection of strictly periodic signals
   in the \sst ~anomalies of the Pacific Ocean.
  \Ki
  DCM discovers the deterministic signature of the amplification
  mechanism, whatever it may be.
  \Ki
   Regardless of future theoretical explanations,
  the time series analysis of 150 years of independent \Data ~data
  and real-time tracking of solar forcing by the Pacific Ocean provide
  an empirical proof of  the {\BigWave}'s existence.

\subsubsection{Damping and amplifying of solar forcing}

 \renewcommand{\Ki}{}
 DCM detected
 the \SignalTwo, \SignalOne ~and \SignalThree ~signals
 from the sunspot record \citep{Jet25}.
 \Ki
 These signals of \SolLC ~would
 amplify or dampen each other in the Pacific Ocean.
 \Ki
In this paper, we use the notation  \SignalNino ~to refer to
the strongest $12.78\pm0.12$ years \BigWave  ~signal
in the yearly mean \Data ~data
 (Table  \ref{TableClong}, \M=3).

 The beat frequency of two
arbitrary frequencies $f_1=1/p_1$ and $f_2=1/p_2$ is
\begin{eqnarray}
f_{1,2}=| f_1-f_2|.
\end{eqnarray}

One connection between 
  the sunspot signals \SignalTwo $(p_1=1/f_1)$, \SignalOne ~$(p_2=/f_2)$
  and \SignalThree ~$(p_3=1/f_3)$, and
  the \Axel ~signal \SignalNino ~$(p_4=1/f_4)$,
  is outlined in Table \ref{TableSynodic}.
\Ki
If we take the "beat" $f_{2,3}$ between $f_2$ and $f_3$,
and subtract this beat from the frequency $f_3$, we get
\begin{eqnarray}
  12.87                 & \equiv   & 0.0777 = f_3-f_{2,3} \\   
          \approx  f_4 & =         & 0.0782 \equiv 12.78  = p_4. \nonumber
\end{eqnarray}
This relation is well within the $\pm 1 \sigma = 0.12~ {\mathrm{y}}$ error limits.
\Ki
The $f_1$, $f_2$, $f_3$ and $f_4$
frequencies are nearly equidistant (Table \ref{TableSynodic}).
\Ki
The chain is
$10       \xrightarrow{\smash{-0.009}}$
$ 11        \xrightarrow{\smash{-0.007}}$
$ 11.86   \xrightarrow{\smash{-0.006}}$
$ 12.78$.
\Ki
The 12.78 period may represent the low frequency
caused by interference between the 11 and 11.86 year periods.
\Ki
For example,
\citet{Ste19} discuss
this \SignalNino ~signal 
in connection to their planetary tidal synchronisation
dynamo model.  

We show that the strongest \SignalNino ~signal in \Data ~data
  can be connected to the strongest \SignalOne ~signal
  and the third strongest \SignalThree ~signal which
DCM has detected in the sunspot data \citep{Jet25}.
\Ki
Therefore, it is unnecessary to begin a long discussion
about all possible connections between the signals
detected in the Pacific Ocean and the Sun.
\Ki
The main point is that DCM signal detections
in these two different physical systems can be connected.
\Ki
Furthermore, the \Double ~also confirm
  that the sunspot and \sst ~data are synchronised.

\subsubsection{General comments} \label{SectGeneral}

 \renewcommand{\Ki}{}
   Our {\it mathematical} model detects regularities that
   the {\it climatological} models fail to detect,
   let alone forecast.
   \Ki
   Mathematics detects signals that, according to Climatology, should not be there.
   \Ki
   The counterargument is that
   simple mathematics cannot explain the complex
   coupling between the ocean and atmosphere.
     \Ki
   Our simple physical model is that
   the \SolLC ~forcing causes the \BigWave. 
   \Ki
   The Pacific Ocean is the Earth's
     largest, and therefore the most robust,
     thermal reservoir for measuring this forcing in real-time.
   \Ki
     If the sum of many \SolLC ~signals moderates
     this forcing,
     it is no surprise that we
     have so far failed to discover this \BigWave ~interference
     and the \Double ~in the sea surface temperature anomalies.
     \Ki
     Furthermore, the \BigWave ~is buried under the global warming
     trend and the strong noise "$\Trigger$"
     caused by the climate response (Equation \ref{EqTrigger}).
     \Ki
     \Axel ~seems to work like a ``clock'' because
     our DCM can model and forecast the \BigWave.
     \Ki
     This result represents a potential
       paradigm shift in long-term climatological forecasting.
   \Ki
    Our \BigWave ~discovery  does not fully
    solve the \Axel ~riddle,
    like the $\pm 1$~\Cc ~chaotic \Data ~monthly means.
    \Ki
    However, 
    the \BigWave ~does give a good starting point for
    modelling the climatological ``$\Trigger$'' effects
    (Equation \ref{EqTrigger}).

\renewcommand{\Ki}{}
The mainstream dynamical models
simulate chaotic
ocean-atmosphere coupling using supercomputers.
  \Ki
  The ``spring predictability barrier'' prevents
  long-term forecasting.
  \Ki
  Even the best dynamical model \Axel ~forecasts fail
   in 1.5 years \citep{Zha24}.
  \Ki
  These dynamical models may be  
  over-complicated because
  our DCM analysis confirms that
  short-term weather does not affect
  long-term trends.
  \Ki
    Our mathematical DCM yearly mean forecast can 
    connect seven \BigWave ~decades to the next seven decades
    (Figures \ref{FighalfC14K211} and \ref{FighalfC14K311}).
    \Ki
    This should be impossible if
    the climatological chaos takes over within 1.5 years,
    not to mention the obstacle of
    breaking through over seventy ``spring predictability barriers''.
    \Ki
A reliable \Axel ~forecast
  even one year ahead would save trillions of
  dollars \citep{Hsi11,Cal23,Liu23A,Xu26}.
  \Ki
  Given these massive financial stakes,
  we urge the scientific community to verify our findings
  immediately.\footnote{See declaration ``Code and data availability''.} 
\Ki
An ordinary PC can detect the first two
\BigWave ~signals in about two minutes.\footnote{For  example,
  the detection of  first two signals in Table \ref{TableClong} requires
  these two commands. \\
  \PR{cp longC14K211.dat dcm.dat} \\
  \PR{python dcm.py} \\
  The bootstrap model parameter standard errors
    probably differ from those given in Table \ref{TableClong}.
    To stabilise these estimates, edit
    \PRtext{Rounds=50} to \PRtext{Rounds=1000}, re-run the code
    and wait
    (see Section \ref{SectDCM}, 4th last paragraph).
    If your PC cannot perform parallel computation,
      edit \PRtext{Parallel=1} to \PRtext{Parallel=0} in
  control file  \PRtext{longC14K211.dat}.
    \label{FootBoot}} 
\Ki
The third signal detection takes a few minutes.
\Ki
 William of Ockham (1287–1347) stated that
    if two competing hypotheses
    make the same predictions,
    the simplest hypothesis is usually the best one
    (Occam's razor).
    \Ki
    This comes true when an ordinary PC gives
    better \Axel ~forecasts than multi-million-dollar supercomputers.

 \renewcommand{\Ki}{}
   Since December 2024, DCM has achieved
   a critical milestone by forecasting in real-time the 2025 yearly mean
   within a $\pm 1 \sigma$ margin (Figure  \ref{FigLastYears}).
   \Ki
   Our real-time validation also accurately
   captured 18 months of sliding yearly mean $y_{i,12}$ changes,
     and will certainly continue to do so for the 19th and 20th month in a row.
     While there is a half year delay for obtaining the
     sliding yearly means $y_{i,12}$ for this year 2026,
     our real-time forecast also captured 
   the neutral conditions in early 2026,
   the warming trend at the end of spring 2026, and the current
   rising \sst ~trend today, \Tanaan.
   All these real-time  validity checks confirm DCM's
   predictive power of strictly periodic signals.
   \Ki
   These real-time results confirm that DCM can
   provide deterministic forecasts for complex climate transitions.
   \Ki
   The next months will show if the Pacific Ocean
   continues to dance the \BigWave ~tune.
   \Ki
   That would also finally
   confirm that the solar dynamo
   is multi-periodic and stationary \citep{Jet25}.
   \Ki
   It is fortunate for us that the \Data ~forecasts
   can be checked much faster than the sunspot cycle forecasts.

   \renewcommand{\Ki}{}
The ``notorious'' \Data ~time series is actually considerably
simpler than our simulated Model 7 time series discussed in
Section  \ref{SectModelSeven}.
\Ki
That time series is pathological
because both signal periods
are shorter than the data window.
\Ki
Both signals are not pure sines
and they are superimposed on a parabolic polynomial trend.
\Ki
This pathology does not hinder
the success of the Model 7 time series DCM analysis.
\Ki
Because of this robust performance, there is no rational reason
to doubt the results of our Niño 4 index DCM analysis.
\Ki
As shown in Figures  \ref{FighalfC14K211}a, \ref{FighalfC14K311}a,
\ref{FiglongC14K311}a,  and \ref{FigmAll}a,
the sharp periodogram minima unmistakably
indicate a successful DCM analysis.
\Ki
Official \Axel ~forecasts from NOAA and
similar agencies often cite probabilities in the 10\% – 90\% range.
  \Ki
  DCM probability that the four strongest monthly mean data
  ``Big wave''  signals are real, is larger than
  $1-10^{-16}=0.9~999~999~999~999~999$ (Table \ref{TablemAll}).
  \Ki
  If our DCM can decode the Earth's most complex non-linear
  phenomenon (\Axel) with this significance, it is powerful
  enough to handle complex astrophysical time series.
  \Ki
  Our \Axel ~stress test confirms that DCM is a robust,
  cross-disciplinary tool for detecting non-stationary signals in the
  presence of extreme variance.

   \section{Conclusions}

\renewcommand{\Ki}{}
The frequency-domain parametric time series analysis
methods suffer from many application limitations
(Section \ref{SectInt}:  \AL{\ref{LimNoise}} - \AL{\ref{LimForecast}}).
\Ki
We demonstrate thoroughly that none of those limitations
apply to DCM
(Sections  \ref{SectResults} - \ref{SectDiscussion}). 
\Ki
Due to the \WDE,
the time span of data window is
irrelevant
because the performance of DCM depends only on
the quantity and quality of data.
\Ki
By dividing the data into separate
forecasting and forecasted samples,
DCM is able to ``see through time''.
\Ki
The Fisher-test and the Forecast-test can
confirm the correct model for the data.
\Ki
DCM clearly outperforms DFT.

\renewcommand{\Ki}{}
The mainstream considers the \Axel ~phenomenon 
so chaotic and non-linear
that even multi-million-dollar supercomputer
long-term forecasts are doomed to fail.
\Ki
In our stress test,
we apply DCM to the \Data ~time series
(Section \ref{SectStress}).
\Ki
DCM can model and forecast \Axel.
\Ki
For the yearly means,
the DCM model for the first seven decades
can forecast the next seven decades.
\Ki
Despite the 78-year gap,
  both DCM models for the 1870 - 1947 yearly means
  can accurately predict the {\it observed new} 2025 yearly mean
  (Figures \ref{FighalfC14K211}e and \ref{FighalfC14K311}e: red closed circle).
\Ki
As for all {\it new} yearly sliding means $y_{i,12}$ after December 2024,
we perform a real-time DCM forecast validation.
\Ki
Our DCM real-time forecast for these $y_{i,12}$
data after December 2024
has now come true for 18 months in a row (Figure    \ref{FigLastYears}).
  \Ki
  This real-time forecast
  will certainly also come true for the 19th and 20th month.
\Ki
Although the $y_{i,12}$ values are still missing
  for this year 2026, our real-time forecast has already correctly
  captured the neutral winter conditions and the
  warming during spring.
\Ki
The rapid rise of sea surface temperatures
observed now, \Tanaan,
provides the latest striking real-time validation. 

\renewcommand{\Ki}{}
We detect the extremely significant and strictly periodic
\BigWave ~from
the yearly and monthly \Data ~data.
\Ki
No such precise clock can exist {\it inside} the Pacific Ocean.
\Ki
We present four arguments which show that
strictly periodic
solar forcing is the only possible {\it external} clock.
\Ki
The \Double ~of the \BigWave  ~represent the smoking gun
confirming
that  the Pacific Ocean is significantly colder close to
the sunspot minima
(Figures \ref{FiglongC14K311}b-e and \ref{FiglongC14K411}b-e).
\Ki
This {\it physical} causality confirms
that the Pacific Ocean \sst ~anomalies
and the solar forcing have been synchronised since 1870.
\Ki
The mainstream stochastic and non-stationary
solar dynamo  \citep{Cha20,Uso23}
cannot
cause the strictly periodic \SolLC ~forcing.
\Ki
Planetary tides may provide
such a precise multi-periodic solar dynamo clocking mechanism
\citep{Abr12,Kle23,Mou25,Ste19,Ste25,Jet25}.
\Ki
By definition, a system cannot be
  characterized as chaotic (e.g., \Axel)
  or stochastic (e.g., solar dynamo)
  if it produces strictly periodic,
  extremely significant signals
  that exhibit phase-stability for decades
  (e.g.,  red closed circles in Figures \ref{FighalfC14K211}e and
            \ref{FighalfC14K311}e).

\renewcommand{\Ki}{}
While our \Axel ~dynamics and solar dynamo hypotheses
  depart from the current scientific consensus,
  the proposed deterministic physical \SolLC ~forcing effect
  is currently empirically verifiable in real-time.
  \Ki
  The simple connection between the “Big Wave”, the Sun
    and the Planets may represent a foundational milestone in
    Climatology and Astrophysics.
    \Ki
    Yet, those three elements are merely physical phenomena.
    \Ki
    The mathematical DCM is the primary
    methodological advancement driving our discoveries.
\Ki
DCM has now discovered extremely
significant, strictly periodic signals in the
sunspot and sea surface temperature anomaly data.
\Ki
Both discoveries require that the solar dynamo is
stationary and strictly multi-periodic.
  \Ki
  However,
  all that scientific narrative is unimportant,
  for example, to commodity producers
  who deserve reliable real-time \Axel ~forecasts.
\Ki
Our scientific claims are now under
an objective, real-time ``review'' by the Pacific Ocean itself.

\section*{Acknowledgements}

This work has made use of NASA's 
Astrophysics Data System (ADS) services.
We thank Dr. Jouni R\"ais\"anen for his
comments of the manuscript.

  \section*{Code and data availability}

 All manuscript data files, control files and Python codes
are stored to \\
  \Link{https://zenodo.org/uploads/19854621}
  {https://zenodo.org/uploads/19854621} \\
  The manual for repeating all our analysis is also published there.


\clearpage 


\setcounter{section}{0}
\setcounter{figure}{0}
\setcounter{table}{0}
\setcounter{equation}{0}
\renewcommand{\thesection}{S\arabic{section}}
\renewcommand{\thefigure}{S\arabic{figure}}
\renewcommand{\thetable}{S\arabic{table}}
\renewcommand{\theequation}{S\arabic{equation}}

\begin{center}
{\bf \Large Supplementary material}
\end{center}

\begin{table}[h] 
  \caption{Samples.
    (1) Description. (2-3) Size and time span. (4) Errors available.
    (5) Electronic data file. } \label{TableSamples}
   \begin{center}
  \begin{scriptsize}
   \begin{tabular}{lrrcl}
     \hline
     (1)                          & (2)    & (3)           &  (4)             &   (5)                    \\
     Description             & $n$   & $\Delta T$& $\sigma_i$ & Data file            \\
     (-)                           & (-)     & (y)            & (y)             & (-)                     \\   
     \hline
     All yearly                 & 155  & 154           &  yes           & \PR{Clong.dat} \\
     First half all yearly   &  78   &  77            &  yes           & \PR{Chalf.dat} \\
     All yearly                 & 155  & 154           &  no            & \PR{Rlong.dat} \\
     All monthly             & 1860 & 154.9        &  no            & \PR{mAll.dat}   \\
     All January              &   155 & 154           & no             & \PR{Jan.dat}      \\
     All February            &   155 & 154           & no             & \PR{Feb.dat}      \\
     All March                &   155 & 154           & no             & \PR{Mar.dat}      \\
     All April                   &   155 & 154           & no            & \PR{Apr.dat}      \\
     All May                    &   155 & 154           & no            & \PR{May.dat}      \\
     All June                   &   155 & 154           & no             & \PR{Jun.dat}      \\
     All July                     &   155 & 154           & no            & \PR{Jul.dat}      \\
     All August               &   155 & 154           & no             & \PR{Aug.dat}      \\
     All September         &   155 & 154           & no             & \PR{Sep.dat}      \\
     All October              &   155 & 154           & no             & \PR{Oct.dat}      \\
     All November           &   155 & 154           & no            & \PR{Nov.dat}      \\
     All December           &   155 & 154           & no             & \PR{Dec.dat}      \\
     \hline
   \end{tabular}
 \end{scriptsize}
 \end{center}
   \end{table}

\begin{table*}
  \caption{Trend in all yearly mean weighted data. DCM analysis
    between $P_{\mathrm{min}}=3$ and  $P_{\mathrm{max}}=100$ years.
    We use $\gamma=0.001$ (Equation \ref{EqFisher}) for yearly data.
    Linear $K_3=1$ trend is the best alternative
      for one, two, three and four signal models.
    Otherwise as in Table  \ref{TableFisher} 
   }
  \label{TableTrend}
  \begin{tiny}
 \begin{center}
      \begin{tabular}{lccccccccc}
  \hline
 & & \multicolumn{4}{c}{Period analysis} & & \multicolumn{2}{c}{Fisher-test $(\gamma=0.001)$} &  \\
        \multicolumn{10}{c}{Data: Original weighted data ($n=155, \Delta T=154$: \PR{Clong.dat})} \\
(1)    & 
(2)    & 
(3)    & 
(4)    & 
(5)    & 
(6)    & 
       & 
(7)    & 
(8)    & 
(9)    \\ 
\M                    &
  &
$P_1$ (y)             &
$P_2$ (y)             &
$P_3$ (y)             &
$P_4$ (y)             &
                      &
\M=2                  &
\M=3                  &
                      \\ 
             &
$\eta$ (-)   &
$A_1$ (\Cc)    &
$A_2$ (\Cc)    &
$A_3$ (\Cc)    &
$A_4$ (\Cc)    &
             &
$F_{\chi}$ (-)    &
$F_{\chi}$ (-)    &
Control file \\ 
                       &
$\chi^2$ (-)                    &
$t_{\mathrm{min,1}}$ (y) &
$t_{\mathrm{min,1}}$ (y) &
$t_{\mathrm{min,1}}$ (y) &
$t_{\mathrm{min,1}}$ (y) &
                       &
$Q_F$ (-)              &
$Q_F$ (-)              &
       \\ 
\hline 
        \multicolumn{10}{c}{One signal} \\
\hline
        \M=1             &                  &$12.67\PM 0.16     $&     -                  & -             & -                     &&$\uparrow         $&$\uparrow        $&                                             \\
\CModel{1,1,0}     &       4         &$0.62\PM 0.15       $&     -                  & -             & -                     &&   26.9                & 14.0                  &               \PR{longC14K110.dat}  \\
                             &      516      &$1871.7\PM 1.0     $&     -                  & -             & -                     &&$6.7\TM10^{7}$&$2.7\TM10^{-6}$&                                             \\ 
\hline
\M=2                    &                   &$12.65\PM0.31       $& -                      & -            & -                     &&-                         &$\leftarrow        $&                                              \\
\CModel{1,1,1}    &        5         &$0.51 \PM0.13    $& -                      & -            & -                     &&-                         &  1.02                 &               \PR{longC14K111.dat}   \\
                            &    437         &$1871.8\PM1.5      $& -                      & -            & -                     &&-                          &  0.31                &                                               \\
\hline
\M=3                    &                  &$12.65\PM0.16       $&                         &              & -                    &&-                          & -                        &                                              \\
\CModel{1,1,2}    &       6         &$0.52\PM0.11         $&                         &              & -                    &&-                          & -                        &              \PR{longC14K112.dat}    \\
                            &     434       &$1871.92\PM0.92   $&                         &              & -                    &&-                          & -                        &                                               \\ 
\hline
        \multicolumn{10}{c}{Two signals} \\
                              &                  &                               &                        &                &                      && \M=5                 & \M=6                 &                                              \\
\hline 
\M=4                     &                  &$12.70\PM 0.18     $&$26.8\PM3.0   $& -             & -                     &&$\uparrow         $&$\uparrow        $&                                             \\
\CModel{2,1,0}     &       7         &$0.62\PM 0.14       $&$0.39\PM0.14 $& -             & -                     &&   27.4                & 13.6                 &               \PR{longC14K210.dat}  \\
                             &      456      &$1871.3\PM 1.1    $&$1870.8\PM4.4$& -            & -                     &&$5.6\TM10^{-7}$&$3.8\TM10^{-6}$&                                             \\ 
\hline
\M=5                    &                   &$5.677\PM0.066      $&$12.64\PM0.44$& -            & -                     &&-                         &$\leftarrow        $&                                              \\
\CModel{2,1,1}    &        8         &$0.39  \PM0.10       $&$0.47\PM0.12  $& -            & -                     &&-                         &  0                    &               \PR{longC14K211.dat}   \\
                            &     384        &$1876.15\PM.55      $&$1871.8\PM1.8$& -            & -                     &&-                          & 1                    &                                               \\
\hline
\M=6                    &                  &$5.677\PM0.097       $&$12.64\PM0.20$&              & -                    &&-                          & -                        &                                              \\
\CModel{2,1,2}    &       9         &$0.38\PM0.11           $&$0.47\PM0.11$&               & -                    &&-                          & -                        &              \PR{longC14K212.dat}    \\
                            &     384       &$1876.06\PM0.64    $&$1871.9\PM1.3$&              & -                    &&-                          & -                        &                                               \\ 
\hline
      \multicolumn{10}{c}{Three signals} \\
                              &                  &                               &                        &                &                      && \M=8                & \M=9                 &                                              \\
\hline 
\M=7                     &                  &$5.620\PM 0.091   $&$12.74\PM0.48$&$21.83\PM0.81$& -                     &&$\uparrow         $&$\uparrow        $&                                             \\
\CModel{3,1,0}     &      10         &$0.41\PM 0.13      $&$0.52\PM0.11 $& $0.40\PM0.12 $& -                     &&   29.2               & 15.0                   &     \PR{longC14K310.dat}  \\
                             &      407      &$1871.04\PM 0.77 $&$1870.7\PM1.7$&$1870.0\PM3.2$& -                     &&$2.6\TM10^{-7}$&$1.2\TM10^{-6}$&                                             \\ 
\hline
\M=8                    &                   &$5.662\PM 0.077     $&$12.78\PM 0.12   $&$21.3\PM 1.5  $& -                     &&-                        &$\leftarrow        $&                                              \\
\CModel{3,1,1}    &       11        &$0.40\PM 0.13        $&$0.47\PM 0.10     $&$0.37\PM 0.12    $& -                     &&-                         &  0.82                    &    \PR{longC14K311.dat}   \\
                            &     338       &$1870.77\PM 0.62    $&$1870.86\PM 0.88$&$1871.3\PM 3.3$& -                     &&-                         & 0.36                   &                                               \\
\hline
\M=9                    &                  &$5.661\PM 0.073  $&$12.76\PM 0.22   $&$21.2\PM 1.2   $& -                    &&-                          & -                        &                                              \\
\CModel{3,1,2}    &       12       &$0.40\PM 0.11      $&$0.49\PM 0.10     $&$0.39 \PM 0.12    $& -                    &&-                          & -                        &     \PR{longC1¤K312.dat}    \\
                            &     336       &$1870.8 \PM 0.74  $&$1871.1\PM 1.2   $&$1871.6\PM 2.8 $& -                    &&-                          & -                        &                                               \\ 
\hline
      \multicolumn{10}{c}{Four signals} \\
                              &                  &                               &                        &                &                      && \M=11                 & \M=12                 &                                              \\
\hline 
\M=10                   &                 &$5.597\PM 0.066   $&$12.77\PM0.19$&$19.4\PM1.6 ~\IF  $& $21.5\PM2.1 ~\IF$&&$\uparrow         $&$\uparrow        $&                                             \\
\CModel{4,1,0}     &      13         &$0.44\PM 0.10       $&$0.57\PM0.11 $& $0.4\PM1.3 ~\AD $& $0.4\PM1.4 ~AD$&&   28.1               & 14.2                   &        \PR{longC14K410.dat}  \\
$\UM$                   &      362      &$1872.06\PM 0.57 $&$1870.8\PM1.2$&$1873.3\PM3.7    $&$1872.3\PM3.3    $&&$4.5\TM10^{-7}$&$2.5\TM10^{-6}$&                                             \\ 
        \hline
\M=11                  &                     &$5.496\PM0.093~\IF$&$5.66\PM 0.13 ~\IF $&$12.76\PM 0.15   $&$21.1\PM 1.5 $&&-                        &$\leftarrow        $&                                              \\
\CModel{4,1,1}    &       14          &$0.3\PM 1.8 ~\AD    $&$0.4\PM 1.8 ~\AD    $&$0.49\PM 0.13    $&$0.39\PM 0.11 $&&-                         &  0.45                    &  \PR{longC14K411.dat}   \\
$\UM$                  &     300          &$1874.3\PM 1.2      $&$1870.7\PM 1.2        $&$1871.10\PM 0.97 $&$1871.8\PM 2.1$&&-                         & 0.50                    &                                               \\
\hline
\M=12                  &                  &$5.497\PM 0.087 ~\IF $&$0.566\PM0.062 ~\IF$&$12.75\PM 0.17   $&$21.0\PM1.0    $&&-                          & -                        &                                              \\
\CModel{4,1,2}    &       15       &$0.34\PM 0.94 ~\AD    $&$0.42\PM0.90 ~\AD $&$0.50 \PM 0.11    $&$0.398\PM0.084$&&-                          & -                        &         \PR{longC14K412.dat}    \\
$\UM$                   &     299       &$1874.3 \PM 1.0         $&$1870.68\PM 0.99   $&$1871.1\PM0.97 $&$1871.9\PM2.2$ &&-                          & -                        &                                               \\ 
        \hline
      \end{tabular}
    \end{center}
\end{tiny}
\addtolength{\tabcolsep}{+0.05cm}
\end{table*}

\begin{table}
  \caption{Significance of one signal model  \M=1
    in Table \ref{TableChalf}. Note that we use $\gamma=0.005$
      for this smallest sample.
    Otherwise as in Table \ref{TablePolynomials}.}
\label{TableHalfPolynomials}
\begin{center}
\begin{tabular}{lc}
  \hline
(1)                                                          & (2)                                    \\
                                                              & Fisher-test $(\gamma=0.005)$ \\
                                                             & $g_{1,1,1}, n=78$       \\
  Polynomial                                             & $\eta=5, \chi^2=168$   \\
  \hline
  \M=1                                                     &   $\uparrow$                 \\
  $g(t)=p(t,K_3=0)$                                &    $F=7.2$                 \\
  $\eta=1$, $\chi^2=235$                       &    $\QF=6.2\TM10^{-5}$                      \\
  \hline
  \M=2                                                     &   $\uparrow$                \\
  $g(t)=p(t,K_3=1)$                                 &   $5.14$           \\
  $\eta=2$, $\chi^2=204$                      &     $\QF=0.0028$                   \\
  \hline
  \M=3                                                     &    $\uparrow$                \\
  $g(t)=p(t,K_3=2)$                                  &     $F=7.3$               \\
  $\eta=3$, $\chi^2=202$                       &      $Q_F=0.0013$               \\
  \hline
  \M=4                                                     &     $\uparrow$               \\
  $g(t)=p(t,K_3=3)$                                   &   $F=12.86$                    \\ 
  $\eta=4$, $\chi^2=198$                         &  $\QF=0.00060$     \\
  \hline
  \M=5                                                     &    $\uparrow$              \\
  $g(t)=p(t,K_3=4)$                                   &    No test            \\
  $\eta=5$, $\chi^2=192$                   &     $\QF=1$                 \\
  \hline
  \M=6                                                     &     $\uparrow$             \\
  $g(t)=p(t,K_3=5)$                                   &     $F=-8.9$              \\
  $\eta=6$, $\chi^2=192$                   &     $\QF=1$          \\
  \hline
  \M=7                                                     &      $\uparrow$            \\
  $g(t)=p(t,K_3=6)$                                   &       $F=-4.2$            \\
  $\eta=7$, $\chi^2=191$                    &    $\QF=1$          \\
  \hline
  \M=8                                                     &     $\uparrow$              \\
  $g(t)=p(t,K_3=7)$                                   &      $F=-2.4$          \\
  $\eta=8$, $\chi^2=188$                      &   $\QF=1$                 \\
\hline
\end{tabular}
\end{center}
\end{table}

\begin{table*}
  \caption{Periods in first half yearly mean data. Otherwise as in
    Table \ref{TableTrend}. 
  }
  \label{TableChalf}
  \begin{tiny}
     \begin{center}
      \begin{tabular}{lcccccccccc}
  \hline
 & & \multicolumn{4}{c}{Period analysis} & & \multicolumn{3}{c}{Fisher-test $(\gamma=0.005)$} &  \\
        \multicolumn{11}{c}{Data: Original weighted data ($n=78, \Delta T=77$: \PR{Chalf.dat})} \\
(1)    & 
(2)    & 
(3)    & 
(4)    & 
(5)    & 
(6)    & 
       & 
(7)    & 
(8)    & 
(9)    &
         (10)        \\
        \M                    &
  &
$P_1$ (y)             &
$P_2$ (y)             &
$P_3$ (y)             &
$P_4$ (y)             &
                           &
 \M=2                  &
 \M=3                  &
 \M=4                  &
                      \\ 
             &
$\eta$ (-)   &
$A_1$ (\Cc)    &
$A_2$ (\Cc)    &
$A_3$ (\Cc)    &
$A_4$ (\Cc)    &
             &
$F_{\chi}$ (-)    &
$F_{\chi}$ (-)    &
$F_{\chi}$(-)    &
               Control file \\ 
                       &
$\chi^2$ (-)                    &
$t_{\mathrm{min,1}}$ (y) &
$t_{\mathrm{min,1}}$ (y) &
$t_{\mathrm{min,1}}$ (y) &
$t_{\mathrm{min,1}}$ (y) &
                       &
$Q_F$ (-)              &
$Q_F$ (-)              &
$Q_F$ (-)       \\ 
\hline 
        \multicolumn{10}{c}{One signal} \\
\hline
        \M=1             &                  &$19.0\PM 1.2         $&     -                  & -             & -                     &&$\uparrow         $&$\uparrow            $&$\uparrow$ &                          \\
\CModel{1,1,1}     &       5         &$0.44\PM 0.16       $&     -                  & -             & -                     &&   5.62                & 5.95                    &5.78             &               \PR{halfC14K111.dat}  \\
                             &    168        &$1874.2\PM 2.3     $&     -                  & -             & -                     &&  0.0016            &$5.0\TM10^{-5}$  &$7.5\TM10^{-6}$&                                        \\
\hline
       \multicolumn{10}{c}{Two signals} \\
        \hline
        \M=2                    &          &$12.81\PM0.56     $&$18.8\PM1.1         $& -            & -                     &&-                         &$\uparrow        $&$\uparrow$&                                              \\
\CModel{2,1,1}    &        8         &$0.44 \PM0.15      $&$0.53\PM0.12     $& -            & -                     &&-                         &  5.3                 &4.91 &               \PR{halfC14K211.dat}   \\
                            &     135        &$1870.9\PM1.7     $&$1874.8\PM1.9  $& -            & -                     &&-                         &0.0025             &0.00034&                                               \\ 
\hline
      \multicolumn{10}{c}{Three signals} \\
\hline 
\M=3                    &                   &$5.82\PM 0.14       $&$12.82\PM 0.43  $&$19.33\PM0.74     $& -                     &&-              & -                           &$\leftarrow$&                                              \\
\CModel{3,1,1}    &       11        &$0.42\PM 0.13        $&$0.49\PM 0.10   $&$0.50\PM 0.12$& -                     &&-              & -                         &3.88& \PR{halfC14K311.dat}   \\
                            &     109        &$1875.51\PM 0.94  $&$1874.0\PM 1.3  $&$1874.0\PM 1.8$& -                        &&-             & -                           &0.013&                                               \\
\hline
      \multicolumn{10}{c}{Four signals} \\
\hline 
\M=4                   &                    &$3.437\PM 0.052       $&$5.830\PM 0.047        $&$12.82\PM  0.45    $&$19.57\PM 0.59 $&&-                              & -        & -&                                              \\
\CModel{4,1,1}    &       14        &$0.30\PM0.12            $&$0.48\PM 0.13             $&$0.498\PM 0.097    $&$0.54\PM 0.12 $&&-                         & -                   &-&       \PR{halfC14K411.dat}   \\
                            &       92        &$1873.01\PM0.66      $&$1875.42\PM 0.39       $&$1870.7\PM 1.8     $&$1873.6\PM 2.2$&&-                            & -                   &-&                                               \\
\hline
\end{tabular}
\end{center}
\end{tiny}
\addtolength{\tabcolsep}{+0.05cm}
\end{table*}

\begin{table*}
  \caption{Periods in all yearly mean weighted data. Otherwise as in
    Table \ref{TableTrend}.}
  \label{TableClong}
  \begin{tiny}
     \begin{center}
      \begin{tabular}{lcccccccccc}
  \hline
 & & \multicolumn{4}{c}{Period analysis} & & \multicolumn{3}{c}{Fisher-test $(\gamma=0.001)$} &  \\
        \multicolumn{11}{c}{Data: Original weighted data ($n=155, \Delta T=154$: \PR{Clong.dat})} \\
(1)    & 
(2)    & 
(3)    & 
(4)    & 
(5)    & 
(6)    & 
        & 
(7)    & 
(8)    & 
(9)    &
(10)        \\ 
\M                    &
  &
$P_1$ (y)             &
$P_2$ (y)             &
$P_3$ (y)             &
$P_4$ (y)             &
                           &
 \M=2                  &
 \M=3                  &
 \M=4                  &
                      \\ 
             &
$\eta$ (-)   &
$A_1$ (\Cc)    &
$A_2$ (\Cc)    &
$A_3$ (\Cc)    &
$A_4$ (\Cc)    &
             &
$F_{\chi}$ (-)    &
$F_{\chi}$ (-)    &
$F_{\chi}$(-)    &
               Control file \\ 
                       &
$\chi^2$ (-)                    &
$t_{\mathrm{min,1}}$ (y) &
$t_{\mathrm{min,1}}$ (y) &
$t_{\mathrm{min,1}}$ (y) &
$t_{\mathrm{min,1}}$ (y) &
                       &
$Q_F$ (-)              &
$Q_F$ (-)              &
$Q_F$ (-)       \\ 
\hline 
        \multicolumn{10}{c}{One signal} \\
\hline
        \M=1             &                  &$12.65\PM 0.31     $&     -                  & -                        & -                     &&$\uparrow         $&$\uparrow        $&$\uparrow$&                                             \\
\CModel{1,1,1}     &       5         &$0.51\PM 0.13       $&     -                  & -                        & -                     &&   12.4                 & 6.98                  &7.10          &               \PR{longC14K111.dat}  \\
                             &      437      &$1871.8\PM 1.5     $&     -                  & -                        & -                     &&$2.7\TM10^{-7}$ &$1.6\TM10^{-6}$&$1.9\TM10^{-8}$&                                             \\
\hline
       \multicolumn{10}{c}{Two signals} \\
\hline 
\M=2                    &                   &$5.677\PM0.066   $&$12.64\PM0.44$& -                        & -                     &&-                         &$\uparrow        $&$\uparrow$&                                              \\
\CModel{2,1,1}    &        8         &$0.39  \PM0.10    $&$0.47\PM0.12  $& -                         & -                     &&-                         &  6.48                & 6,53         &               \PR{longC14K211.dat}   \\
                            &     384        &$1876.15\PM0.55 $&$1871.8\PM1.8$& -                         & -                     &&-                         & 0.00038           &$4.1\TM10^{-6}$&                                               \\
\hline
      \multicolumn{10}{c}{Three signals} \\
\hline 
\M=3                    &                   &$5.662\PM 0.077  $&$12.78\PM 0.12 $&$21.3\PM 1.5     $& -                     &&-              & -                           &$\uparrow$&                                              \\
\CModel{3,1,1}    &       11        &$0.40\PM 0.13       $&$0.47\PM 0.10   $&$0.37\PM 0.12   $& -                     &&-              & -                           & 5.91 & \PR{longC14K311.dat}   \\
                            &     338       &$1870.77\PM 0.62  $&$1870.86\PM0.88$&$1871.3\PM 3.3$& -                     &&-             & -                           & 0.00079 &                                               \\
\hline
      \multicolumn{10}{c}{Four signals} \\
\hline 
\M=4                    &                     &$5.496\PM 0.093 ~ \IF$&$5.66\PM 0.13~\IF$&$12.76\PM 0.15   $&$21.09\PM 1.5 $&&-                 & -             & - &                                              \\
\CModel{4,1,1}    &       14          &$0.3    \PM 1.8 ~\AD $&$0.4\PM 1.8 ~\AD   $&$0.49\PM 0.13  $&$0.39\PM 0.11 $&&-                   & -             &- &       \PR{longC14K411.dat}   \\
$\UM$                  &     300          &$1874.3\PM 1.2     $&$1870.7\PM1.1          $&$1871.1\PM 0.97 $&$1871.9\PM 22.$&&-                        & -             &- &                                               \\
\hline
\end{tabular}
\end{center}
\end{tiny}
\addtolength{\tabcolsep}{+0.05cm}
\end{table*}

\begin{table}
  \caption{Significance of one signal model  \M=1
    in Table \ref{TableClong}.
    Otherwise as in Table \ref{TablePolynomials}.}
\label{TableElNinoPolynomials}
\begin{center}
\begin{tabular}{lc}
  \hline
  (1)                                                          & (2)                                    \\
                                                                & Fisher-test $(\gamma=0.001)$ \\
                                                               & $g_{1,1,1}, n=155$       \\
  Polynomial                                             & $\eta=5, \chi^2=437$   \\
  \hline
  \M=1                                                     &   $\uparrow$                 \\
  $g(t)=p(t,K_3=0)$                                &    $F=19.9$                 \\
  $\eta=1$, $\chi^2=670$                       &    $\QF=3.9\TM10^{-13}$                         \\
  \hline
  \M=2                                                     &   $\uparrow$                \\
  $g(t)=p(t,K_3=1)$                                   &    $11.4$           \\
  $\eta=2$, $\chi^2=537$                      &      $\QF=9.2\TM10^{-7}$                   \\
  \hline
  \M=3                                                     &    $\uparrow$                \\
  $g(t)=p(t,K_3=2)$                                   &     $F=17.0$               \\
  $\eta=3$, $\chi^2=537$                       &      $\QF=2.1\TM10^{-7}$               \\
  \hline
  \M=4                                                     &     $\uparrow$               \\
  $g(t)=p(t,K_3=3)$                                   &   $F=33.8$                    \\
  $\eta=4$, $\chi^2=536$                         &  $\QF=3.6\TM10^{-8}$     \\
  \hline
  \M=5                                                     &    $\uparrow$              \\
  $g(t)=p(t,K_3=4)$                                   &    No test            \\
  $\eta=5$, $\chi^2=515$                   &     $\QF=1$                 \\
  \hline
  \M=6                                                     &     $\uparrow$             \\
  $g(t)=p(t,K_3=5)$                                   &     $F=-22$              \\
  $\eta=6$, $\chi^2=514$                   &     $\QF=1$          \\
  \hline
  \M=7                                                     &      $\uparrow$            \\
  $g(t)=p(t,K_3=6)$                                   &       $F=-8.2$            \\
  $\eta=7$, $\chi^2=492$                    &    $\QF=1$          \\
  \hline
  \M=8                                                     &     $\uparrow$              \\
  $g(t)=p(t,K_3=7)$                                   &      $F=-5.4$          \\
  $\eta=8$, $\chi^2=492$                      &   $\QF=1$                 \\
\hline
\end{tabular}
\end{center}
\end{table}

    \begin{table}[h]
      \caption{\Data ~\Ha ~and \Er ~values after December 2024.
        Sliding yearly means
        $t_{i,12}, y_{i,12}$ and $\sigma_{i,12}$
        are computed from each monthly mean
        $t_i$ and $y_i$, and eleven earlier monthly means.
        (1) Monthly $t_i$.
        (2) Yearly sliding mean $t_{i,12}$.
        (3-4) \Ha ~index $y_i$ values available on \Eilen ~and \Tanaan.
        (5-6) \Ha ~index sliding yearly mean $y_{i,12}$ and $\sigma_{i,12}$.
        (7) \Er ~index $y_i$ values available on \Tanaan.
        (8-9)  \Er ~index sliding yearly mean $y_{i,12}$ and $\sigma_{i,12}$.} \label{TableLastYears} \label{TableProof}
\begin{tiny}
  \begin{center}
  \begin{tabular}{ccccccccc}
    \hline
\multicolumn{2}{c}{Months}  &    \multicolumn{4}{c}{\Ha} & \multicolumn{2}{c}{\Er} \\
    (1)     & (2)            & (3)           & (4)            & (5)             & (6)                     & (7)            & (8)            & (9) \\
    $t_i$  & $t_{i,12}$  & $y_i$       & $y_i$        & $y_{i,12}$ & $\sigma_{1,12}$ & $y_i$         & $y_{i,12}$ & $\sigma_{1,12}$ \\
    (y)     &  (y)            & (\Cc)        & (\Cc)         & (\Cc)           & (\Cc)                 & (\Cc)          & (\Cc)       \\
   \hline

2025.042 &  2024.58 &    -0.50 &    -0.50 &     0.42 &     0.48 &    -0.55 &     0.43 &     0.52 \\
2025.125 &  2024.67 &    -0.36 &    -0.36 &     0.29 &     0.46 &    -0.52 &     0.28 &     0.51 \\
2025.208 &  2024.75 &    -0.13 &    -0.13 &     0.20 &     0.42 &    -0.34 &     0.17 &     0.49 \\
2025.292 &  2024.83 &    -0.09 &    -0.09 &     0.12 &     0.38 &    -0.17 &     0.08 &     0.44 \\
2025.375 &  2024.92 &     0.02 &     0.02 &     0.06 &     0.34 &     0.04 &     0.02 &     0.38 \\
2025.458 &  2025.00 &     0.14 &     0.14 &     0.03 &     0.31 &     0.17 &    -0.03 &     0.33 \\
2025.542 &  2025.08 &     0.05 &     0.05 &    -0.01 &     0.27 &     0.03 &    -0.07 &     0.29 \\
2025.625 &  2025.17 &     0.06 &     0.06 &    -0.04 &     0.24 &    -0.15 &    -0.12 &     0.25 \\
2025.708 &  2025.25 &          &    -0.12 &    -0.07 &     0.22 &    -0.27 &    -0.15 &     0.24 \\
2025.792 &  2025.33 &          &    -0.32 &    -0.11 &     0.22 &    -0.41 &    -0.19 &     0.23 \\
2025.875 &  2025.42 &          &    -0.37 &    -0.16 &     0.20 &    -0.52 &    -0.25 &     0.23 \\
2025.958 &  2025.50 &          &    -0.19 &    -0.15 &     0.19 &    -0.33 &    -0.25 &     0.23 \\
2026.042 &  2025.58 &          &    -0.06 &    -0.11 &     0.16 &    -0.08 &    -0.21 &     0.21 \\
2026.125 &  2025.67 &          &     0.19 &    -0.07 &     0.16 &     0.20 &    -0.15 &     0.22 \\
2026.208 &  2025.75 &          &     0.42 &    -0.02 &     0.21 &     0.43 &    -0.09 &     0.26 \\
2026.292 &  2025.83 &          &     0.83 &     0.05 &     0.31 &     0.81 &    -0.01 &     0.36 \\
2026.375 &  2025.92 &          &          &          &          &     1.07 &     0.08 &      0.47 \\
2026.458 &  2026.00 &          &          &          &          &     1.22 &     0.17 &      0.57 \\
    \hline
   \end{tabular}
\end{center}
\end{tiny}
\end{table}

\begin{table*}
  \caption{Periods in all non-weighted yearly mean data. Otherwise as in
    Table \ref{TableTrend}. 
  }
  \label{TableRlong}
  \begin{tiny}
     \begin{center}
      \begin{tabular}{lcccccccccc}
  \hline
 & & \multicolumn{4}{c}{Period analysis} & & \multicolumn{3}{c}{Fisher-test $(\gamma=0.001)$} &  \\
        \multicolumn{11}{c}{Data: Original non-weighted data ($n=155, \Delta T=154$: \PR{Rlong.dat})} \\
(1)    & 
(2)    & 
(3)    & 
(4)    & 
(5)    & 
(6)    & 
        & 
(7)    & 
(8)    & 
(9)    &
(10)        \\ 
\M                    &
  &
$P_1$ (y)             &
$P_2$ (y)             &
$P_3$ (y)             &
$P_4$ (y)             &
                           &
 \M=2                  &
 \M=3                  &
 \M=4                  &
                      \\ 
             &
$\eta$ (-)   &
$A_1$ (-)    &
$A_2$ (-)    &
$A_3$ (-)    &
$A_4$ (-)    &
             &
$F_{R}$ (-)    &
$F_{R}$ (-)    &
$F_{R}$(-)    &
               Control file \\ 
                       &
$R$ (-)                    &
$t_{\mathrm{min,1}}$ (y) &
$t_{\mathrm{min,1}}$ (y) &
$t_{\mathrm{min,1}}$ (y) &
$t_{\mathrm{min,1}}$ (y) &
                       &
$Q_F$ (-)              &
$Q_F$ (-)              &
$Q_F$ (-)       \\ 
\hline 
        \multicolumn{10}{c}{One signal} \\
\hline
        \M=1             &                  &$12.71\PM 0.20    $&     -                       & -                            & -                     &&$\leftarrow         $&$\leftarrow        $&$\leftarrow$&                                             \\
\RModel{1,1,1}     &       5         &$0.389\PM 0.084  $&     -                       & -                            & -                     && 3.67                 & 3.77                   &1.11&               \PR{longR14K111.dat}  \\
                             &     27.1      &$1871.7\PM 1.0    $&     -                       & -                            & -                     && 0.013                 & 0.0016             &0.36&                                             \\
\hline
       \multicolumn{10}{c}{Two signals} \\
\hline 
\M=2                    &                   &$5.651\PM0.078   $&$12.69\PM0.17     $& -                           & -                     &&-                         &$\leftarrow        $&$\leftarrow$&                                              \\
\RModel{2,1,1}    &        8         &$0.319 \PM0.084  $&$0.384\PM0.085   $& -                           & -                     &&-                         &  3.77               &-0.092&               \PR{longR14K211.dat}   \\
                            &      25.2      &$1870.76\PM0.61 $&$1871.83\PM0.97  $& -                           & -                     &&-                         & 0.0016              & 1&                                               \\
\hline
      \multicolumn{10}{c}{Three signals} \\
\hline 
\M=3                    &                   &$5.650\PM 0.032 $&$9.12\PM0.12       $&$12.69\PM 0.27    $& -                     &&-              & -                           &$\leftarrow$&                                              \\
\RModel{3,1,1}    &       11        &$0.327\PM 0.084  $&$0.299\PM 0.087  $&$0.381\PM 0.079 $& -                     &&-              & -                           &-3.50& \PR{longR14K311.dat}   \\
                            &    23.4        &$1870.78\PM0.46 $&$1873.0\PM1.2     $&$1871.9\PM 1.3   $& -                     &&-             & -                           & 1&                                               \\
\hline
      \multicolumn{10}{c}{Four signals} \\
\hline  
\M=4                    &                       &$5.496\PM  0.093~\IF$&$5.662\PM1.3~\IF   $&$12.76\PM 0.16               $&$21.1\PM1.5 $&&-                          &              &&                                              \\
\RModel{4,1,1}    &       14          &$0.3\PM 1.9~\AD  $&$0.4\PM 1.8  ~\AD $&$0.491\PM 0.49\PM0.13  $&$0.39\PM 0.11 $&&-                       &                &&       \PR{longR14K411.dat}   \\
      $\UM$            &     25.3         &$1874.3\PM 1.2        $&$1870.7\PM1.2         $&$1871.1\PM 0.97           $&$1871.8\PM 2.1$&&-                       &               &&                                               \\
\hline
\end{tabular}
\end{center}
\end{tiny}

\addtolength{\tabcolsep}{+0.05cm}
\end{table*}

\clearpage 

\begin{figure}[h]
  \centering
 \includegraphics[width=0.90\textwidth,clip=]{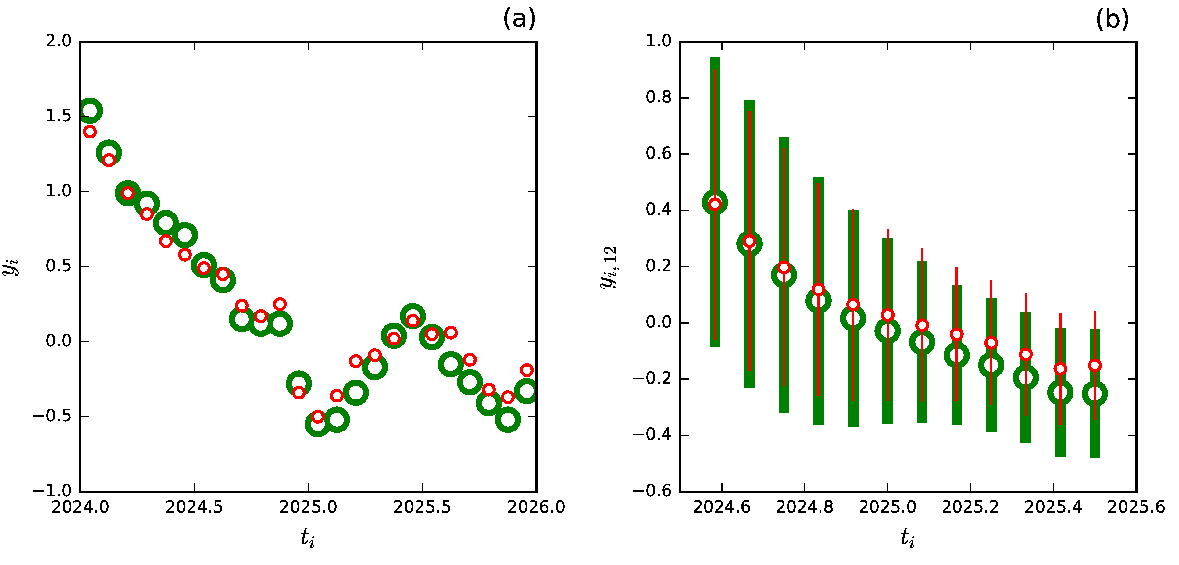}
  \caption{Comparison \Ha ~and \Er ~\Data ~values between January 2024
    and December 2025. (a) Monthly (\Ha ~= red, \Er~= green).
    (b) Yearly sliding means  (same symbol colours.).}
      \label{FigIndexCompare}
    \end{figure}

\begin{table*}
  \caption{Periods in non-weighted all months data. Otherwise as in
    Table \ref{TableTrend}.
  }
  \label{TablemAll}
  \begin{tiny}
       \addtolength{\tabcolsep}{-0.05cm}
       \begin{center}
       \begin{tabular}{lcccccccccc}
  \hline
 & & \multicolumn{4}{c}{Period analysis} & & \multicolumn{3}{c}{Fisher-test $(\gamma=0.001)$} &  \\
        \multicolumn{11}{c}{Data: Original non-weighted data ($n=1860, \Delta T=154.9$: \PR{mAll.dat})} \\
(1)    & 
(2)    & 
(3)    & 
(4)    & 
(5)    & 
(6)    & 
        & 
(7)    & 
(8)    & 
(9)    &
(10)        \\ 
\M                    &
  &
$P_1$ (y)             &
$P_2$ (y)             &
$P_3$ (y)             &
$P_4$ (y)             &
                           &
 \M=2                  &
 \M=3                  &
 \M=4                  &
                      \\ 
             &
$\eta$ (-)   &
$A_1$ (\Cc)    &
$A_2$ (\Cc)    &
$A_3$ (\Cc)    &
$A_4$ (\Cc)    &
             &
$F_{R}$ (-)    &
$F_{R}$ (-)    &
$F_{R}$(-)    &
               Control file \\ 
                       &
$R$ (-)                    &
$t_{\mathrm{min,1}}$ (y) &
$t_{\mathrm{min,1}}$ (y) &
$t_{\mathrm{min,1}}$ (y) &
$t_{\mathrm{min,1}}$ (y) &
                       &
$Q_F$ (-)              &
$Q_F$ (-)              &
$Q_F$ (-)       \\ 
\hline 
        \multicolumn{10}{c}{One signal} \\
\hline
        \M=1             &                  &$12.716\PM 0.052$&     -                  & -                        & -                     &&$\uparrow         $&$\uparrow        $&$\uparrow$&                                             \\
\RModel{1,1,1}     &       5         &$0.396\PM 0.031   $&     -                  & -                        & -                     &&    32.9                &    33.2          &   35.3          &               \PR{mAllR14K111.dat}  \\
                             &   556.8     &$1871.62\PM0.37  $&     -                  & -                        & -                     &&     \REJ                 &     \REJ             &    \REJ         &                                             \\
\hline
       \multicolumn{10}{c}{Two signals} \\
\hline 
\M=2                    &                   &$5.651\PM0.012$&$12.700\PM0.044$& -                       & -                     &&-                         &$\uparrow        $&$\uparrow$&                                              \\
\RModel{2,1,1}    &        8         &$0.348 \PM0.032   $&$0.389\PM0.039    $& -                       & -                     &&-                      &      31.8           &   34.7         &               \PR{mAllR14K211.dat}   \\
                            &   528.6       &$1870.75\PM0.19 $&$1871.76\PM0.34$& -                    & -                     &&-                         &     \REJ             &     \REJ     &                                               \\
\hline
      \multicolumn{10}{c}{Three signals} \\
\hline 
\M=3                    &                   &$4.7459\PM 0.0082$&$5.640\PM0.012 $&$12.694\PM0.043     $& -                     &&-              & -                           &$\uparrow$&                                              \\
\RModel{3,1,1}    &       11        &$0.335\PM 0.023      $&$0.340\PM0.034$&$0.388\PM 0.030      $& -                     &&-              & -                           &   35.7       & \PR{mAllR14K311.dat}            \\
                            &   502.6      &$1870.60\PM0.13  $&$1870.93\PM0.15 $&$1871.83\PM 0.29     $& -                     &&-             & -                            &  \REJ       &                                               \\
\hline
      \multicolumn{10}{c}{Four signals} \\
         \hline
         \M=4                    &                     &$3.6436\PM 0.0046  $&$4.7477\PM 0.0080 $&$5.647\PM 0.011   $&$12.704\PM 0.063 $&&-                 & -             & - &                                              \\
\RModel{4,1,1}    &       14          &$0.345 \PM 0.030     $&$0.332\PM 0.031     $&$0.342\PM 0.025  $&$0.392\PM 0.038 $&&-                   & -             &- &       \PR{mAllR14K411.dat}   \\
                            &     475.0       &$1872.68\PM 0.11    $&$1870.55\PM0.14    $&$1870.82\PM 0.19 $&$1871.74\PM0.39.$&&-                        & -             &- &                                               \\
\hline
\end{tabular}
\addtolength{\tabcolsep}{+0.05cm}
\end{center}
\end{tiny}
\addtolength{\tabcolsep}{+0.05cm}
\end{table*}

\begin{table} 
  \caption{Significance of one signal model  \M=1
    in Table \ref{TablemAll}.
    Otherwise as in Table \ref{TablePolynomials}.
  }\label{TableElNinoPolynomials2}
\begin{center}
  \begin{tabular}{lc}
  \hline
    (1)                                                          & (2)                                    \\
                                                                & Fisher-test $(\gamma=0.001)$ \\
                                                               & $g_{1,1,1}, n=1860$       \\
  Polynomial                                             & $\eta=5, R=557.1$   \\
  \hline
  \M=1                                                     &   $\uparrow$                 \\
  $g(t)=p(t,K_3=0)$                                &    $F=63.9$                 \\
  $\eta=1$, $R=634$                              &    \REJ                       \\
  \hline
  \M=2                                                     &   $\uparrow$                \\
  $g(t)=p(t,K_3=1)$                                   &    $F=39.8$           \\
  $\eta=2$, $R=593$                      &     \REJ                   \\
  \hline
  \M=3                                                     &    $\uparrow$                \\
  $g(t)=p(t,K_3=2)$                                   &     $F=59.7$               \\
  $\eta=3$, $R=593$                       &      \REJ              \\
  \hline
  \M=4                                                     &     $\uparrow$               \\
  $g(t)=p(t,K_3=3)$                                   &   $F=92.8$                    \\
  $\eta=4$, $R=585$                         &  \REJ     \\
  \hline
  \M=5                                                     &    $\uparrow$              \\
  $g(t)=p(t,K_3=4)$                                   &    No test           \\
  $\eta=5$, $R=584$                   &     $\QF=1$                 \\
  \hline
  \M=6                                                     &     $\uparrow$             \\
  $g(t)=p(t,K_3=5)$                                   &     $F=-34$              \\
  $\eta=6$, $R=584$                   &     $\QF=1$          \\
  \hline
  \M=7                                                     &      $\uparrow$            \\
  $g(t)=p(t,K_3=6)$                                   &       $F=-33.5$            \\
  $\eta=7$, $R=578$                    &    $\QF=1$          \\
  \hline
  \M=8                                                     &     $\uparrow$              \\
  $g(t)=p(t,K_3=7)$                                   &      $F=-22.3$          \\
  $\eta=8$, $R=578$                      &   $\QF=1$                 \\
\hline
\end{tabular}
\end{center}
\end{table}

\begin{table*}
  \caption{All months \BigWave  ~forecast (Figure \ref{FigmAll}).
(1) Year. (2-13) Month.
    Symbols \UU ~and \DD ~tell that preceding
    value is smaller or bigger.
  } \label{TablemAllForecast}
  \begin{tiny}
\addtolength{\tabcolsep}{-0.05cm}
\begin{center}
\begin{tabular}{rrrrrrrrrrrrrrrrrrrrrrrrrr}
  \hline
(1) & (2) & & (3) & & (4) & & (5) & & (6) & & (7) & & (8) & & (9) & & (10) & & (11) & & (12) & & (13) \\
  Year &
         \multicolumn{2}{c}{Jan} &
         \multicolumn{2}{c}{Feb} &
         \multicolumn{2}{c}{Mar} &
         \multicolumn{2}{c}{Apr} &
         \multicolumn{2}{c}{May} &
         \multicolumn{2}{c}{Jun} &
         \multicolumn{2}{c}{Jul} &
         \multicolumn{2}{c}{Aug} &
         \multicolumn{2}{c}{Sep} &
         \multicolumn{2}{c}{Oct} &
         \multicolumn{2}{c}{Nov} &
                                   \multicolumn{2}{c}{Dec} \\
  (y) & (\Cc) & & (\Cc) & & (\Cc) & & (\Cc) & & (\Cc) & & (\Cc) & & (\Cc) & & (\Cc) & & (\Cc) & & (\Cc) & & (\Cc) & & (\Cc) & \\ 
  \hline
  2025&  0.06& \UU&  0.05& \DD&  0.04& \DD&  0.03& \DD&
         0.02& \DD&  0.02& \DD&  0.01& \DD&  0.00& \DD&
        -0.00& \DD& -0.01& \DD& -0.01& \DD& -0.01& \UU\\
  2026& -0.00& \UU&  0.00& \UU&  0.01& \UU&  0.01& \UU&
         0.02& \UU&  0.03& \UU&  0.04& \UU&  0.05& \UU&
         0.06& \UU&  0.08& \UU&  0.09& \UU&  0.10& \UU\\
  2027&  0.11& \UU&  0.12& \UU&  0.12& \UU&  0.13& \UU&
         0.13& \UU&  0.13& \UU&  0.13& \UU&  0.13& \DD&
         0.13& \DD&  0.12& \DD&  0.12& \DD&  0.11& \DD\\
  2028&  0.10& \DD&  0.09& \DD&  0.09& \DD&  0.08& \DD&
         0.07& \DD&  0.06& \DD&  0.05& \DD&  0.05& \DD&
         0.04& \DD&  0.04& \DD&  0.04& \UU&  0.04& \UU\\
  2029&  0.05& \UU&  0.06& \UU&  0.07& \UU&  0.08& \UU&
         0.10& \UU&  0.12& \UU&  0.14& \UU&  0.17& \UU&
         0.19& \UU&  0.22& \UU&  0.25& \UU&  0.28& \UU\\
  2030&  0.31& \UU&  0.34& \UU&  0.37& \UU&  0.40& \UU&
         0.42& \UU&  0.45& \UU&  0.47& \UU&  0.49& \UU&
         0.50& \UU&  0.52& \UU&  0.52& \UU&  0.53& \UU\\
  2031&  0.53& \UU&  0.52& \DD&  0.52& \DD&  0.50& \DD&
         0.49& \DD&  0.47& \DD&  0.45& \DD&  0.42& \DD&
         0.39& \DD&  0.36& \DD&  0.33& \DD&  0.30& \DD\\
  2032&  0.27& \DD&  0.23& \DD&  0.20& \DD&  0.17& \DD&
         0.14& \DD&  0.12& \DD&  0.09& \DD&  0.07& \DD&
         0.05& \DD&  0.03& \DD&  0.02& \DD&  0.01& \DD\\
  2033&  0.01& \DD&  0.01& \DD&  0.01& \UU&  0.02& \UU&
         0.02& \UU&  0.03& \UU&  0.05& \UU&  0.06& \UU&
         0.08& \UU&  0.09& \UU&  0.11& \UU&  0.13& \UU\\
  2034&  0.14& \UU&  0.15& \UU&  0.17& \UU&  0.17& \UU&
         0.18& \UU&  0.19& \UU&  0.19& \UU&  0.18& \DD&
         0.18& \DD&  0.16& \DD&  0.15& \DD&  0.13& \DD\\
  2035&  0.11& \DD&  0.09& \DD&  0.06& \DD&  0.03& \DD&
         0.00& \DD& -0.03& \DD& -0.06& \DD& -0.10& \DD&
        -0.13& \DD& -0.16& \DD& -0.19& \DD& -0.22& \DD\\
\hline
\end{tabular}
\end{center}
\end{tiny}
\end{table*}

\clearpage

 \begin{figure*}  
\vspace{0.02\textwidth}
\centerline{\hspace*{0.005\textwidth}
 \includegraphics[width=0.455\textwidth,clip=]{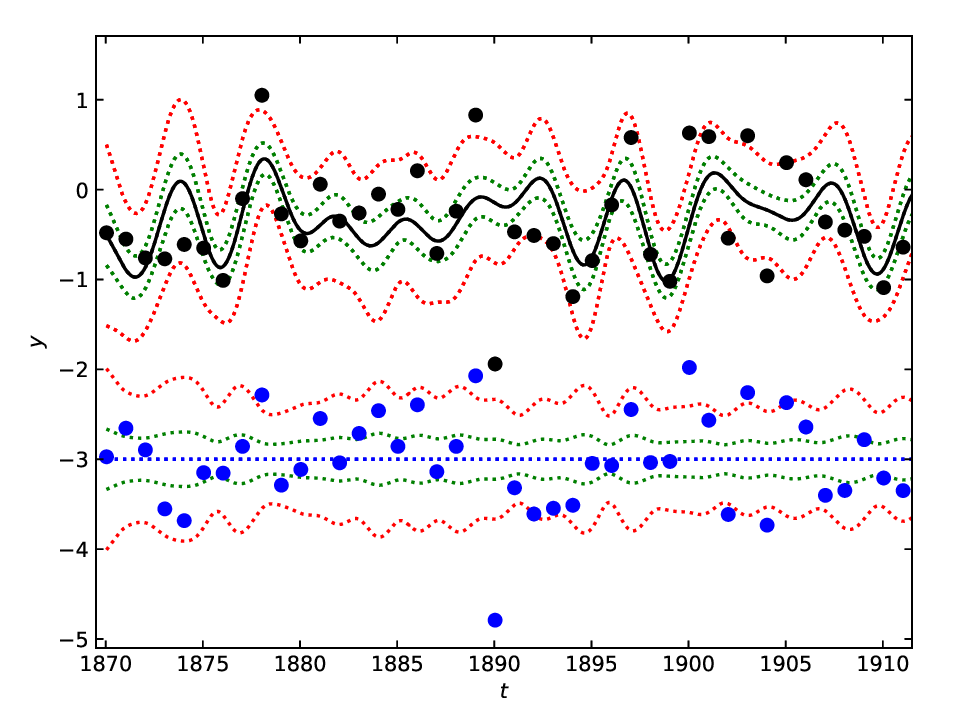}
 \hspace*{-0.01\textwidth}
 \includegraphics[width=0.455\textwidth,clip=]{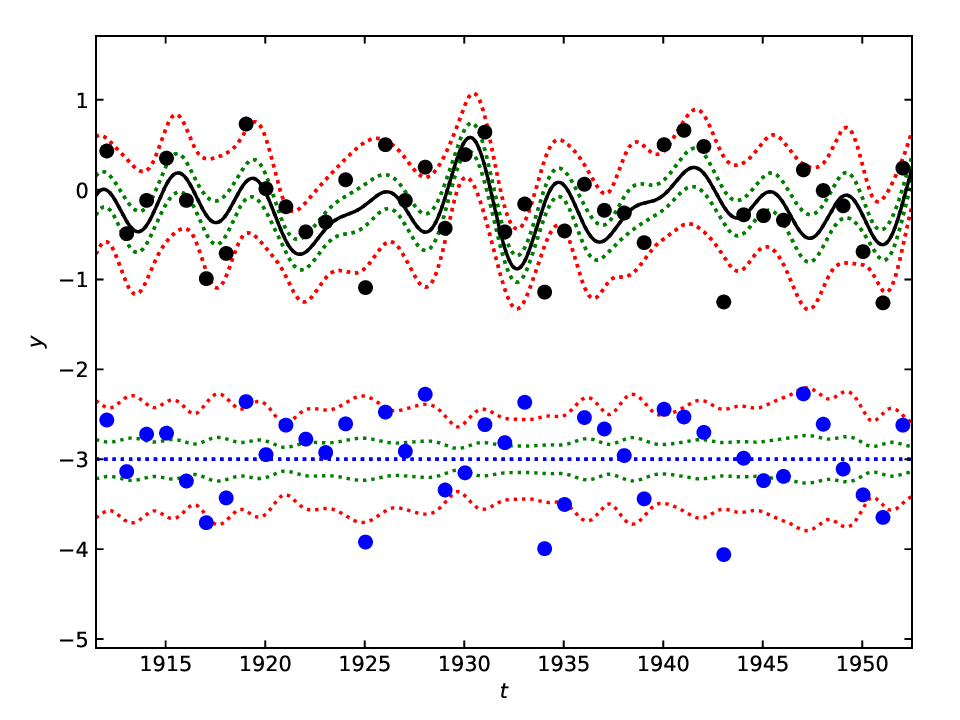}
 }
\vspace{-0.36\textwidth}
\centerline{\Large 
\hspace{0.44\textwidth}  \color{black}{(a)}
\hspace{0.41\textwidth}  \color{black}{(b)}
\hfill}
\vspace{0.35\textwidth}
\centerline{\hspace*{0.005\textwidth}
 \includegraphics[width=0.455\textwidth,clip=]{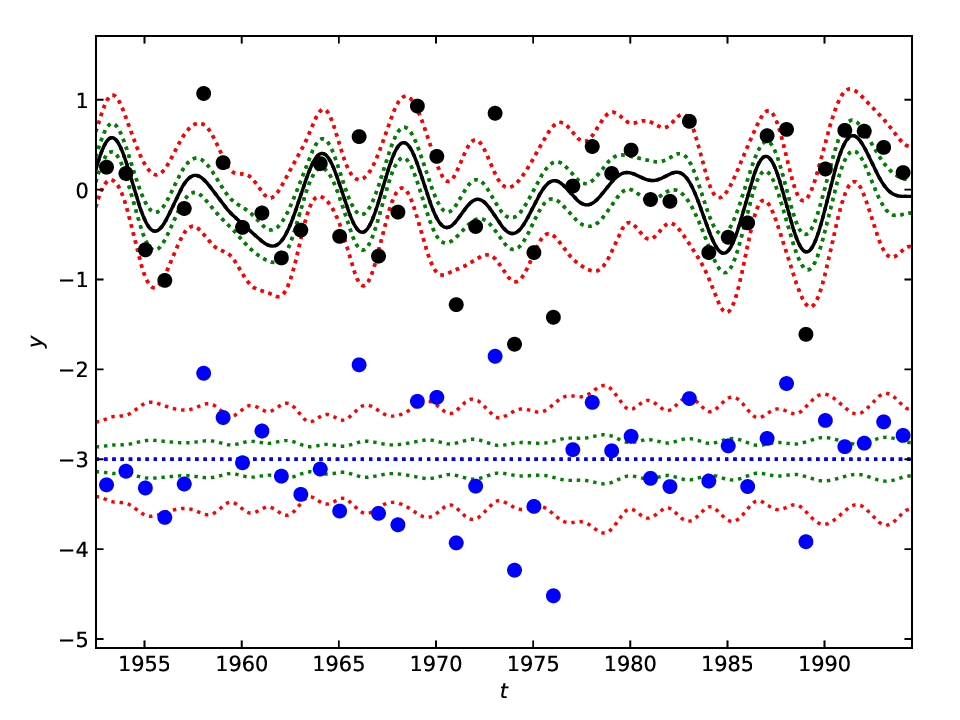}
 \hspace*{-0.01\textwidth}
 \includegraphics[width=0.455\textwidth,clip=]{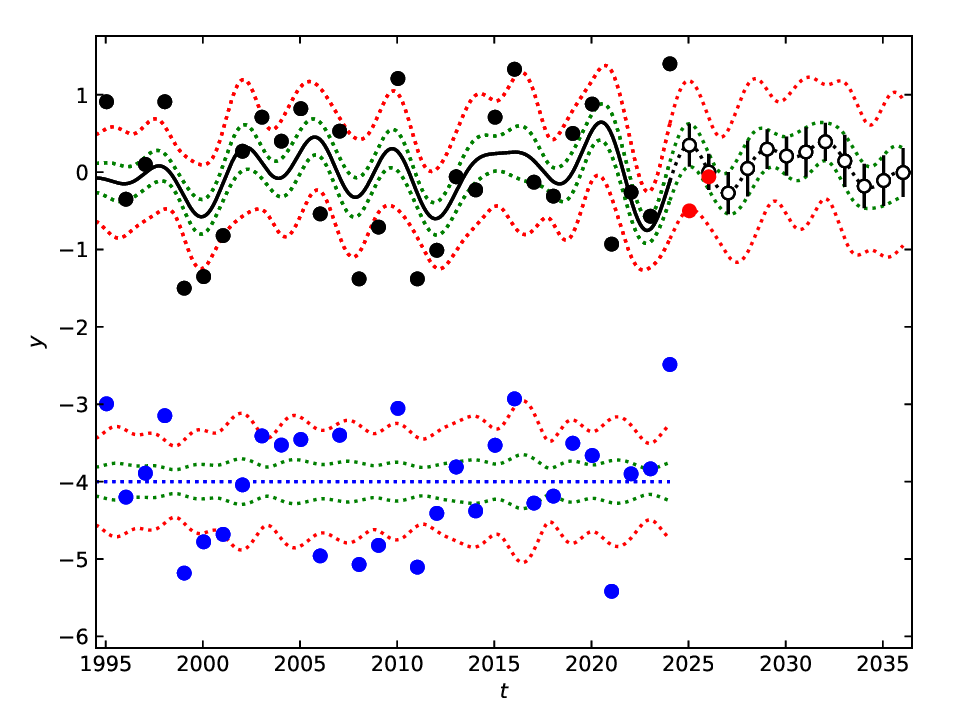}
}
\vspace{-0.36\textwidth}
\centerline{\Large 
\hspace{0.44\textwidth}  \color{black}{(c)}
\hspace{0.41\textwidth}  \color{black}{(d)}
\hfill}
\vspace{0.32\textwidth}
\caption{January data model \M=4 in Table \ref{TableJan}.
Red dots from Table \ref{TableProof}  (Column 4).
  Otherwise as in Figure \ref{FighalfC14K311}.
}
\label{FigJan}
\end{figure*}

\begin{table*}
  \caption{Periods in January data. Note that we use the pre-assigned significance
    level $\gamma=0.05$ (Equation \ref{EqFisher}) because data of
    separate months data are so noisy. Otherwise as in
    Table \ref{TableTrend}. }
  \label{TableJan}
  \begin{tiny}
     \begin{center}
      \begin{tabular}{lcccccccccc}
  \hline
 & & \multicolumn{4}{c}{Period analysis} & & \multicolumn{3}{c}{Fisher-test  $(\gamma=0.05)$} &  \\
        \multicolumn{11}{c}{Data: Original non-weighted data
        ($n=155, \Delta T=154$: \PR{Jan.dat})} \\
(1)    & 
(2)    & 
(3)    & 
(4)    & 
(5)    & 
(6)    & 
        & 
(7)    & 
(8)    & 
(9)    &
(10)        \\ 
\M                    &
  &
$P_1$ (y)             &
$P_2$ (y)             &
$P_3$ (y)             &
$P_4$ (y)             &
                           &
 \M=2                  &
 \M=3                  &
 \M=4                  &
                      \\ 
             &
$\eta$ (-)   &
$A_1$ (\Cc)    &
$A_2$ (\Cc)    &
$A_3$ (\Cc)    &
$A_4$ (\Cc)    &
             &
$F_{R}$ (-)    &
$F_{R}$ (-)    &
$F_{R}$(-)    &
               Control file \\ 
                       &
$R$ (-)                    &
$t_{\mathrm{min,1}}$ (y) &
$t_{\mathrm{min,1}}$ (y) &
$t_{\mathrm{min,1}}$ (y) &
$t_{\mathrm{min,1}}$ (y) &
                       &
$Q_F$ (-)              &
$Q_F$ (-)              &
$Q_F$ (-)       \\ 
\hline 
        \multicolumn{10}{c}{One signal} \\
\hline
        \M=1             &                  &$3.766\PM 0.027   $&     -                  & -                        & -                     &&$\uparrow         $&$\uparrow        $&$\uparrow$&                                             \\
\RModel{1,1,1}     &       5         &$0.46\PM 0.13       $&     -                  & -                        & -                     &&   3.33                &      3.40            &   3.26           &               \PR{JanR14K111.dat}  \\
                             &      64.0     &$1872.18\PM0.41  $&     -                  & -                        & -                     &&   0.021              &    0.0035          &    0.0012      &                                             \\
\hline
       \multicolumn{10}{c}{Two signals} \\
\hline 
\M=2                    &                   &$3.767\PM0.038   $&$4.734\PM0.050$& -                       & -                     &&-                         &$\uparrow        $&$\uparrow$&                                              \\
\RModel{2,1,1}    &        8         &$0.45 \PM0.14      $&$0.46\PM0.12    $& -                       & -                     &&-                         &    3.14               &   3.09    &               \PR{JanR14K211.dat}   \\
                            &     59.9        &$1872.18\PM0.46 $&$1871.02\PM0.64$& -                    & -                     &&-                         &   0.027              &  0.0071       &                                               \\
\hline
      \multicolumn{10}{c}{Three signals} \\
\hline 
\M=3                    &                   &$3.766\PM 0.045 $&$4.734\PM0.026$&$12.72\PM0.18     $& -                     &&-              & -                           &$\uparrow$&                                              \\
\RModel{3,1,1}    &       11        &$0.45\PM 0.10      $&$0.46\PM0.11   $&$0.44\PM 0.11      $& -                     &&-              & -                           &    2.91       & \PR{JanR14K311.dat}            \\
                            &     56.2      &$1872.19\PM0.44  $&$1871.02\PM0.57$&$1871.7\PM 1.4 $& -                     &&-             & -                            &   0.037      &                                               \\
\hline
      \multicolumn{10}{c}{Four signals} \\
\hline 
\M=4                    &                     &$3.766\PM 0.018 $&$4.737\PM 0.048 $&$5.643\PM 0.15   $&$12.68\PM 0.30 $&&-                 & -             & - &                                              \\
\RModel{4,1,1}    &       14          &$0.45  \PM 0.12     $&$0.44\PM 0.12    $&$0.41\PM 0.10  $&$0.43\PM 0.11 $&&-                   & -             &- &       \PR{JanR14K411.dat}   \\
                            &     52.9        &$1872.21\PM 0.44   $&$1870.94\PM0.62$&$1870.84\PM 0.73 $&$1871.9\PM1.4.$&&-                        & -             &- &                                               \\
\hline
\end{tabular}
\end{center}
\end{tiny}
\addtolength{\tabcolsep}{+0.05cm}
\end{table*}

\clearpage

 \begin{figure*}  
\vspace{0.02\textwidth}
\centerline{\hspace*{0.005\textwidth}
 \includegraphics[width=0.455\textwidth,clip=]{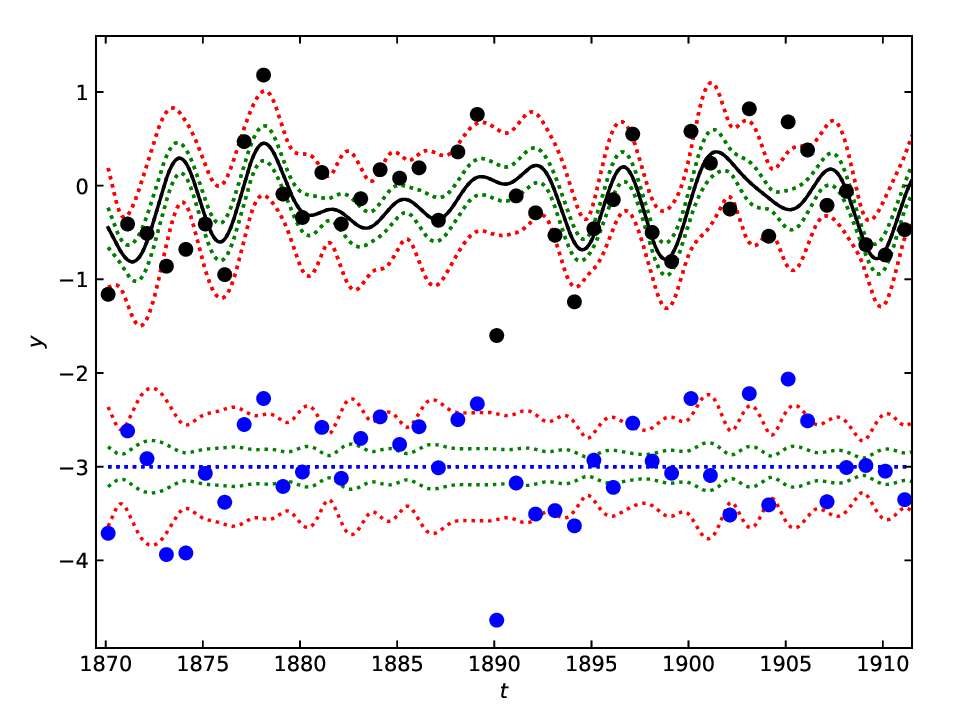}
 \hspace*{-0.01\textwidth}
 \includegraphics[width=0.455\textwidth,clip=]{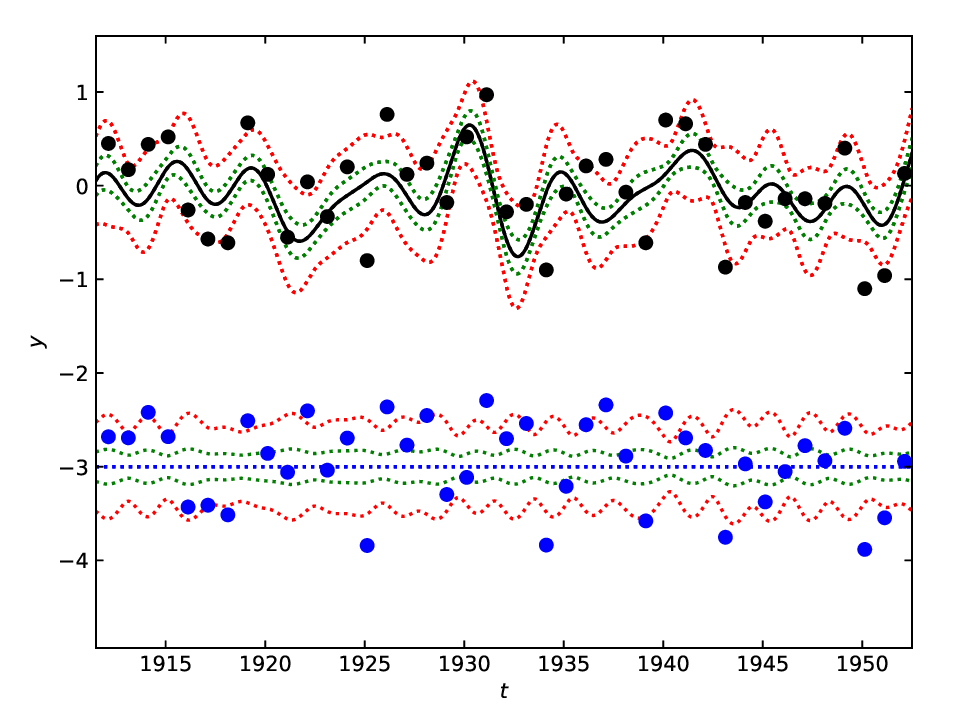}
 }
\vspace{-0.36\textwidth}
\centerline{\Large 
\hspace{0.44\textwidth}  \color{black}{(a)}
\hspace{0.41\textwidth}  \color{black}{(b)}
\hfill}
\vspace{0.35\textwidth}
\centerline{\hspace*{0.005\textwidth}
 \includegraphics[width=0.455\textwidth,clip=]{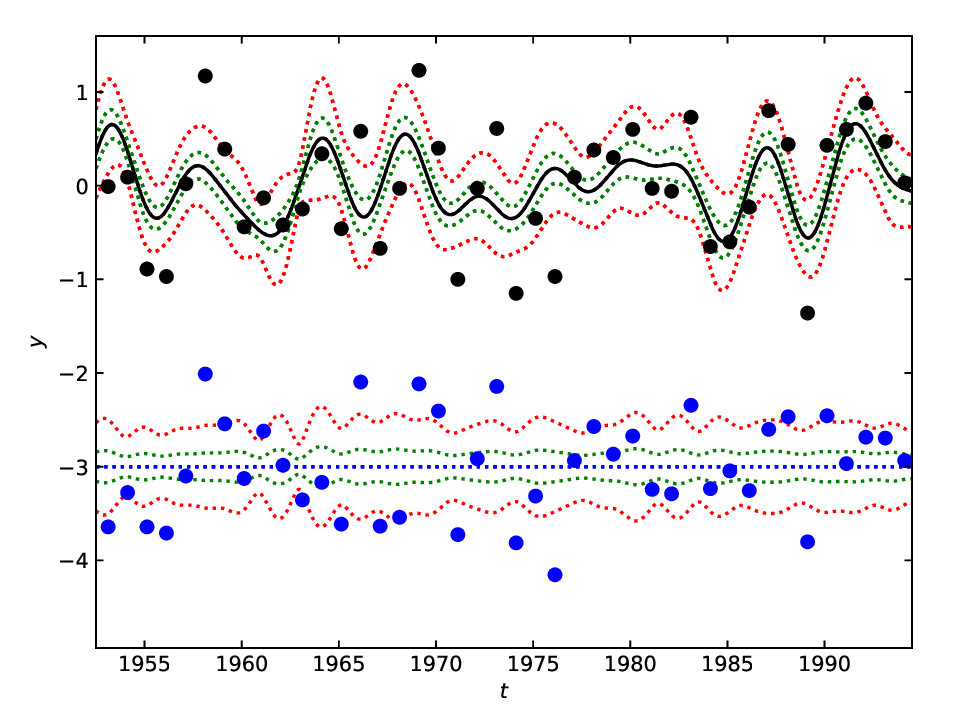}
 \hspace*{-0.01\textwidth}
 \includegraphics[width=0.455\textwidth,clip=]{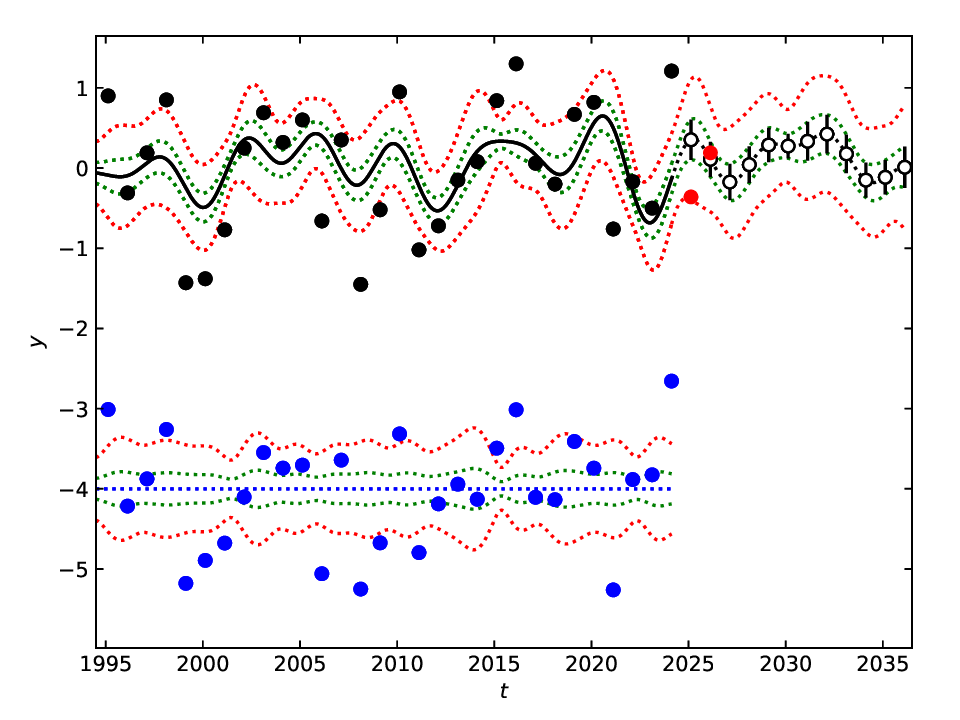}
}
\vspace{-0.36\textwidth}
\centerline{\Large 
\hspace{0.44\textwidth}  \color{black}{(c)}
\hspace{0.41\textwidth}  \color{black}{(d)}
\hfill}
\vspace{0.32\textwidth}
\caption{February data model \M=4 in Table \ref{TableFeb}.
Otherwise as in Figure \ref{FigJan}.
}
\label{FigFeb}
\end{figure*}

\begin{table*}
  \caption{Periods in February data. Otherwise as in
    Table \ref{TableJan}. 
  }
  \label{TableFeb}
  \begin{tiny}
     \begin{center}
      \begin{tabular}{lcccccccccc}
  \hline
 & & \multicolumn{4}{c}{Period analysis} & & \multicolumn{3}{c}{Fisher-test  $(\gamma=0.05)$} &  \\
        \multicolumn{11}{c}{Data: Original non-weighted data ($n=155, \Delta T=154$: \PR{Feb.dat})} \\
(1)    & 
(2)    & 
(3)    & 
(4)    & 
(5)    & 
(6)    & 
        & 
(7)    & 
(8)    & 
(9)    &
(10)        \\ 
        M                    &
  &
$P_1$ (y)             &
$P_2$ (y)             &
$P_3$ (y)             &
$P_4$ (y)             &
                           &
 \M=2                  &
 \M=3                  &
 \M=4                  &
                      \\ 
             &
$\eta$ (-)   &
$A_1$ (\Cc)    &
$A_2$ (\Cc)    &
$A_3$ (\Cc)    &
$A_4$ (\Cc)    &
             &
$F_{R}$ (-)    &
$F_{R}$ (-)    &
$F_{R}$(-)    &
               Control file \\ 
                       &
$R$ (-)                    &
$t_{\mathrm{min,1}}$ (y) &
$t_{\mathrm{min,1}}$ (y) &
$t_{\mathrm{min,1}}$ (y) &
$t_{\mathrm{min,1}}$ (y) &
                       &
$Q_F$ (-)              &
$Q_F$ (-)              &
$Q_F$ (-)       \\ 
\hline 
        \multicolumn{10}{c}{One signal} \\
\hline
        \M=1             &                  &$5.652\PM 0.040  $&     -                  & -                        & -                     &&$\uparrow         $&$\uparrow        $&$\uparrow$&                                             \\
\RModel{1,1,1}     &       5         &$0.477\PM 0.093   $&     -                  & -                        & -                     &&      3.55             &    3.36              &     3.23        &               \PR{FebR14K111.dat}  \\
                             &      55.8     &$1870.73\PM0.57  $&     -                  & -                        & -                     &&      0.016           &   0.0039           &     0.0013    &                                             \\
\hline
       \multicolumn{10}{c}{Two signals} \\
\hline 
\M=2                    &                   &$5.656\PM0.041   $&$12.74\PM0.34$& -                       & -                     &&-                         &$\uparrow        $&$\uparrow$&                                              \\
\RModel{2,1,1}    &        8         &$0.47 \PM0.12      $&$0.44\PM0.11  $& -                       & -                     &&-                   &     3.02              &     2.92      &               \PR{FebR14K211.dat}   \\
                            &     52.0        &$1870.68\PM0.65 $&$1871.3\PM1.6$& -                    & -                     &&-                           &     0.032            &     0.010      &                                               \\
\hline
      \multicolumn{10}{c}{Three signals} \\
\hline 
\M=3                    &                   &$4.741\PM 0.052 $&$5.644\PM0.089$&$12.73\PM0.69     $& -                     &&-              & -                           &$\uparrow$&                                              \\
\RModel{3,1,1}    &       11        &$0.40\PM 0.11      $&$0.45\PM0.11    $&$0.43\PM 0.101      $& -                     &&-              & -                           &    2.72     & \PR{FebR14K311.dat}            \\
                            &     48.9      &$1870.96\PM0.60  $&$1870.88\PM0.48$&$1871.37\PM 1.8 $& -                     &&-             & -                            &    0.046      &                                               \\
\hline
      \multicolumn{10}{c}{Four signals} \\
\hline 
\M=4                    &                     &$3.770\PM 0.034    $&$4.742\PM 0.046$&$5.647\PM 0.031   $&$12.73\PM 0.22 $&&-                            & -             & - &                                              \\
\RModel{4,1,1}    &       14          &$0.37    \PM 0.11    $&$0.398\PM 0.097  $&$0.45\PM 0.13       $&$0.436\PM 0.094 $&&-                          & -             &- &       \PR{FebR14K411.dat}   \\
                            &     46.2          &$1872.12\PM 0.44  $&$1870.93\PM0.67$&$1870.80\PM 0.47 $&$1871.31\PM 1.5$&&-                           & -             &- &                                               \\
\hline
\end{tabular}
\end{center}
\end{tiny}
\addtolength{\tabcolsep}{+0.05cm}
\end{table*}

\clearpage
 \begin{figure*}  
\vspace{0.02\textwidth}
\centerline{\hspace*{0.005\textwidth}
 \includegraphics[width=0.455\textwidth,clip=]{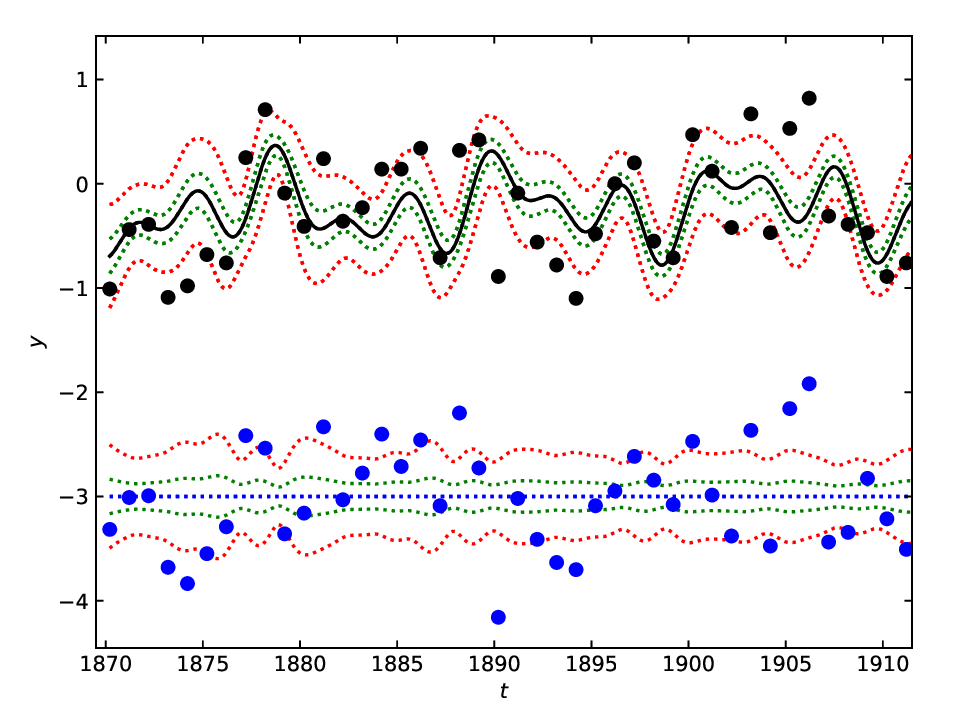}
 \hspace*{-0.01\textwidth}
 \includegraphics[width=0.455\textwidth,clip=]{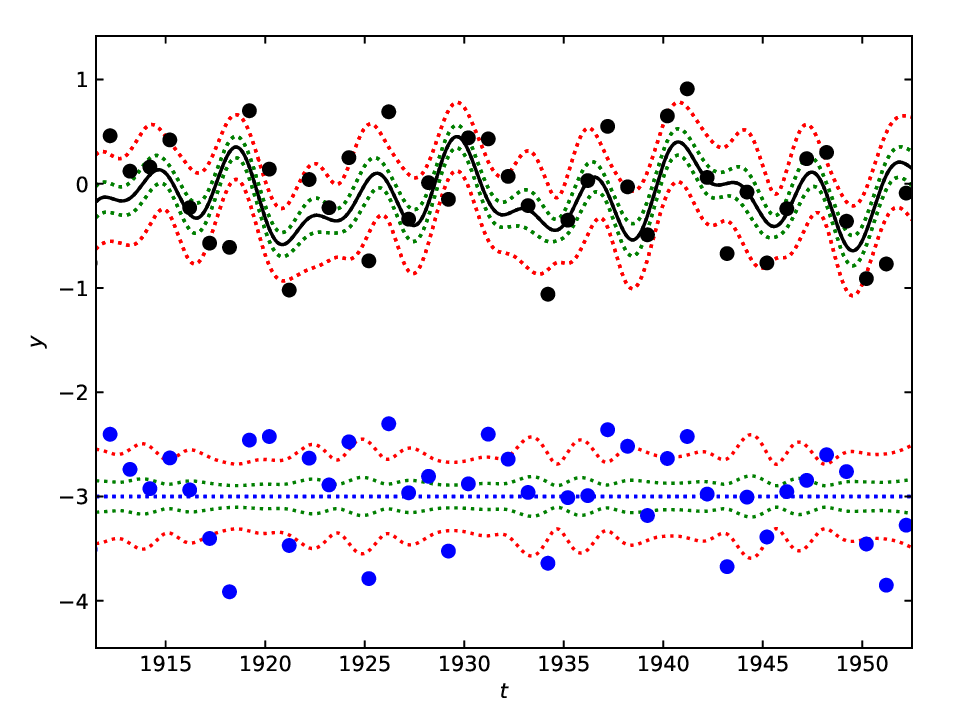}
 }
\vspace{-0.36\textwidth}
\centerline{\Large 
\hspace{0.44\textwidth}  \color{black}{(a)}
\hspace{0.41\textwidth}  \color{black}{(b)}
\hfill}
\vspace{0.35\textwidth}
\centerline{\hspace*{0.005\textwidth}
 \includegraphics[width=0.455\textwidth,clip=]{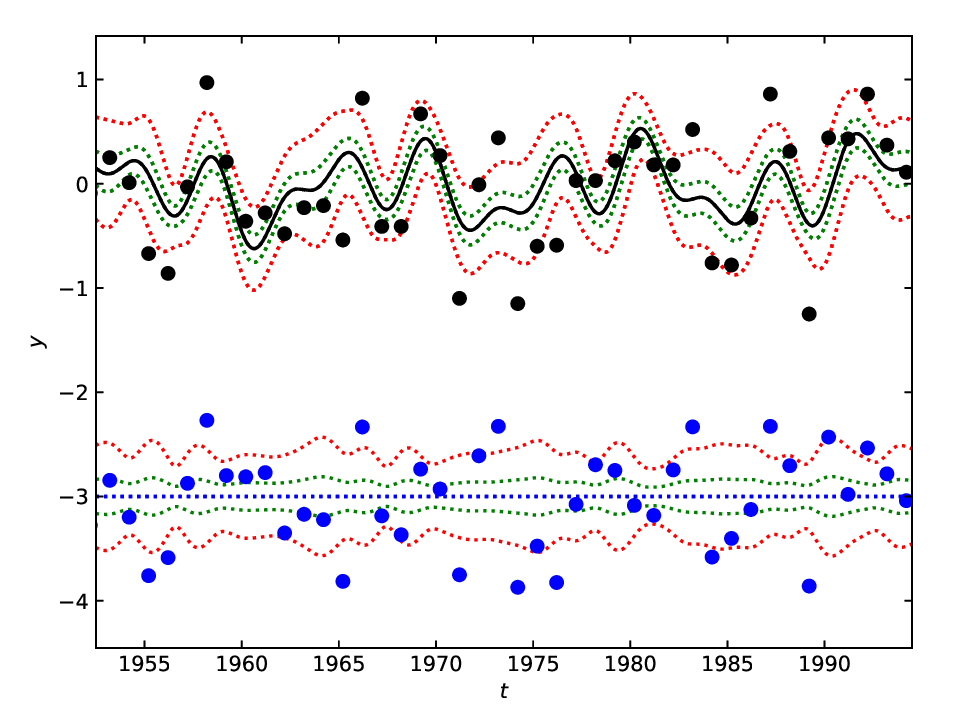}
 \hspace*{-0.01\textwidth}
 \includegraphics[width=0.455\textwidth,clip=]{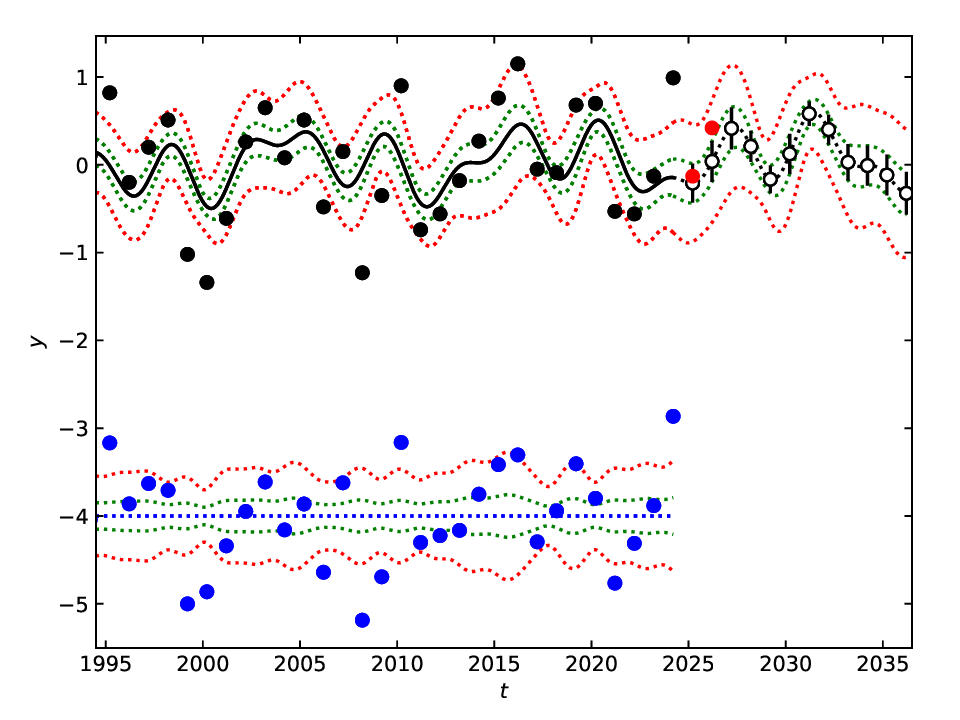}
}
\vspace{-0.36\textwidth}
\centerline{\Large 
\hspace{0.44\textwidth}  \color{black}{(c)}
\hspace{0.41\textwidth}  \color{black}{(d)}
\hfill}
\vspace{0.32\textwidth}
\caption{March data model \M=3 in Table \ref{TableMar}.
Otherwise as in Figure \ref{FigJan}.
}
\label{FigMar}
\end{figure*}

\begin{table*}
  \caption{Periods in March data. Otherwise as in
    Table \ref{TableJan}. 
  }
  \label{TableMar}
  \begin{tiny}
     \begin{center}
      \begin{tabular}{lcccccccccc}
  \hline
 & & \multicolumn{4}{c}{Period analysis} & & \multicolumn{3}{c}{Fisher-test  $(\gamma=0.05)$} &  \\
        \multicolumn{11}{c}{Data: Original non-weighted data ($n=155, \Delta T=154$: \PR{Mar.dat})} \\
(1)    & 
(2)    & 
(3)    & 
(4)    & 
(5)    & 
(6)    & 
        & 
(7)    & 
(8)    & 
(9)    &
(10)        \\ 
\M                    &
  &
$P_1$ (y)             &
$P_2$ (y)             &
$P_3$ (y)             &
$P_4$ (y)             &
                           &
 \M=2                  &
 \M=3                  &
 \M=4                  &
                      \\ 
             &
$\eta$ (-)   &
$A_1$ (\Cc)    &
$A_2$ (\Cc)    &
$A_3$ (\Cc)    &
$A_4$ (\Cc)    &
             &
$F_{R}$ (-)    &
$F_{R}$ (-)    &
$F_{R}$(-)    &
               Control file \\ 
                       &
$R$ (-)                    &
$t_{\mathrm{min,1}}$ (y) &
$t_{\mathrm{min,1}}$ (y) &
$t_{\mathrm{min,1}}$ (y) &
$t_{\mathrm{min,1}}$ (y) &
                       &
$Q_F$ (-)              &
$Q_F$ (-)              &
$Q_F$ (-)       \\ 
\hline 
        \multicolumn{10}{c}{One signal} \\
\hline
        \M=1             &                  &$5.665\PM 0.072 $&     -                  & -                        & -                     &&$\uparrow         $&$\uparrow        $&$\uparrow$&                                             \\
\RModel{1,1,1}     &       5         &$0.45\PM 0.12      $&     -                  & -                        & -                     &&  3.94                 &    4.27               &   3.71         &               \PR{MarR14K111.dat}  \\
                             &      44.1     &$1870.67\PM0.71  $&     -                  & -                        & -                     && 0.0097             &    0.00054          &   0.00033   &                                             \\
\hline
       \multicolumn{10}{c}{Two signals} \\
\hline 
\M=2                    &                   &$5.669\PM0.033   $&$12.68\PM0.31$& -                       & -                     &&-                         &$\uparrow        $&$\uparrow$&                                              \\
\RModel{2,1,1}    &        8         &$0.444 \PM0.098    $&$0.41\PM0.10$&   -                   & -                     &&-                           &    4.33              &    3.41        &               \PR{MarR14K211.dat}   \\
                            &     40.8        &$1870.61\PM0.54 $&$1871.9\PM1.3$& -                      & -                     &&-                           &  0.0059             &  0.0036       &                                               \\
\hline
      \multicolumn{10}{c}{Three signals} \\
\hline 
\M=3                    &                   &$3.633\PM 0.019 $&$5.671\PM0.0289$&$12.68\PM0.13     $& -                     &&-              & -                           &$\leftarrow$&                                              \\
\RModel{3,1,1}    &       11        &$0.42\PM 0.12      $&$0.45\PM0.10   $&$0.418\PM 0.090      $& -                     &&-              & -                      &    2.36         & \PR{MarR14K311.dat}            \\
                            &     37.4      &$1873.16\PM0.26  $&$1870.58\PM0.50$&$1871.85\PM0.77 $& -                     &&-             & -                            &   0.074       &                                               \\
\hline
      \multicolumn{10}{c}{Four signals} \\
\hline 
\M=4                    &                     &$3.634\PM 0.027 $&$4.585\PM 0.079  $&$5.671\PM 0.034   $&$12.68\PM 0.15 $&&-                 & -             & - &                                              \\
\RModel{4,1,1}    &       14          &$0.419  \PM 0.093$&$0.311\PM 0.084 $&$0.45\PM 0.10       $&$0.418\PM 0.080 $&&-                   & -             &- &       \PR{MarR14K411.dat}   \\
                            &     35.6          &$1873.14\PM0.17 $&$1870.51\PM0.48$&$1870.63\PM 0.65 $&$1871.9\PM 1.5$&&-                        & -             &- &                                               \\
\hline
\end{tabular}
\end{center}
\end{tiny}
\addtolength{\tabcolsep}{+0.05cm}
\end{table*}

\clearpage

 \begin{figure*}  
\vspace{0.02\textwidth}
\centerline{\hspace*{0.005\textwidth}
 \includegraphics[width=0.455\textwidth,clip=]{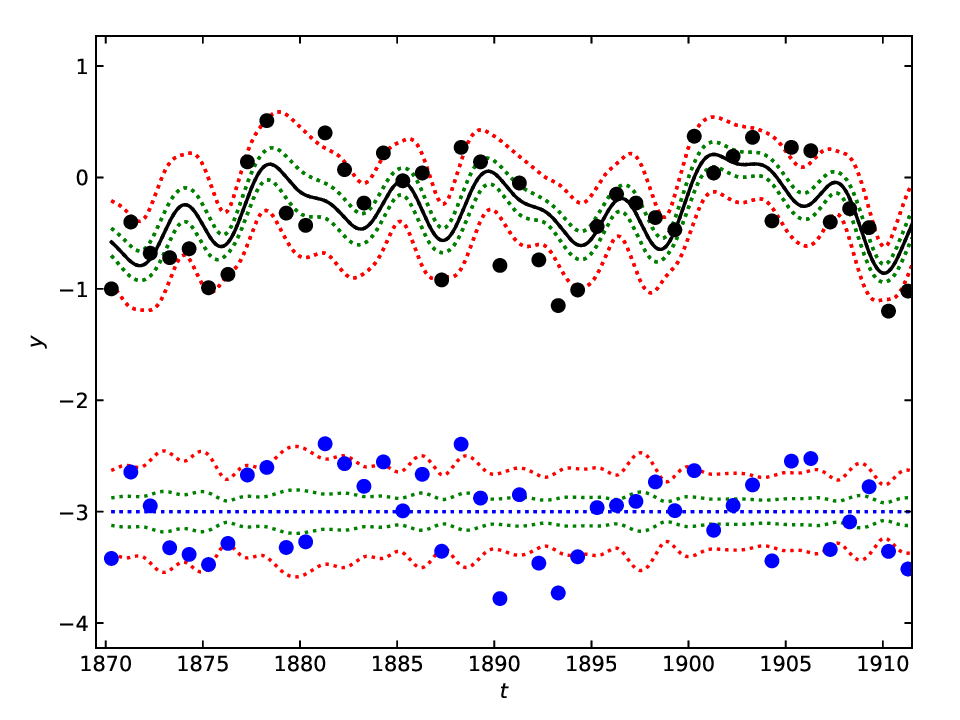}
 \hspace*{-0.01\textwidth}
 \includegraphics[width=0.455\textwidth,clip=]{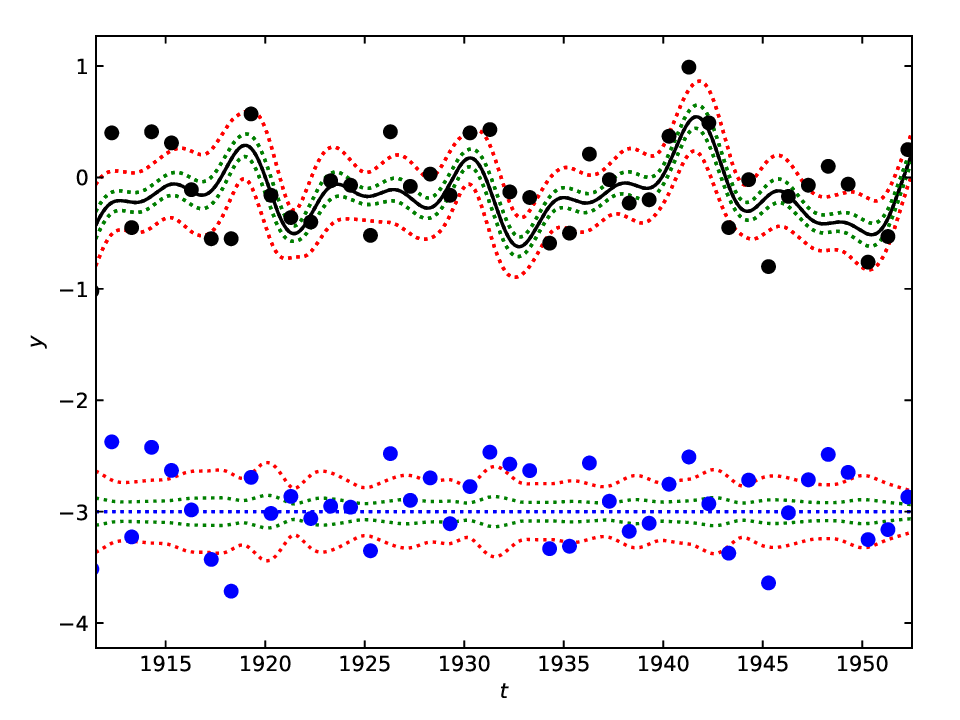}
 }
\vspace{-0.36\textwidth}
\centerline{\Large 
\hspace{0.44\textwidth}  \color{black}{(a)}
\hspace{0.41\textwidth}  \color{black}{(b)}
\hfill}
\vspace{0.35\textwidth}
\centerline{\hspace*{0.005\textwidth}
 \includegraphics[width=0.455\textwidth,clip=]{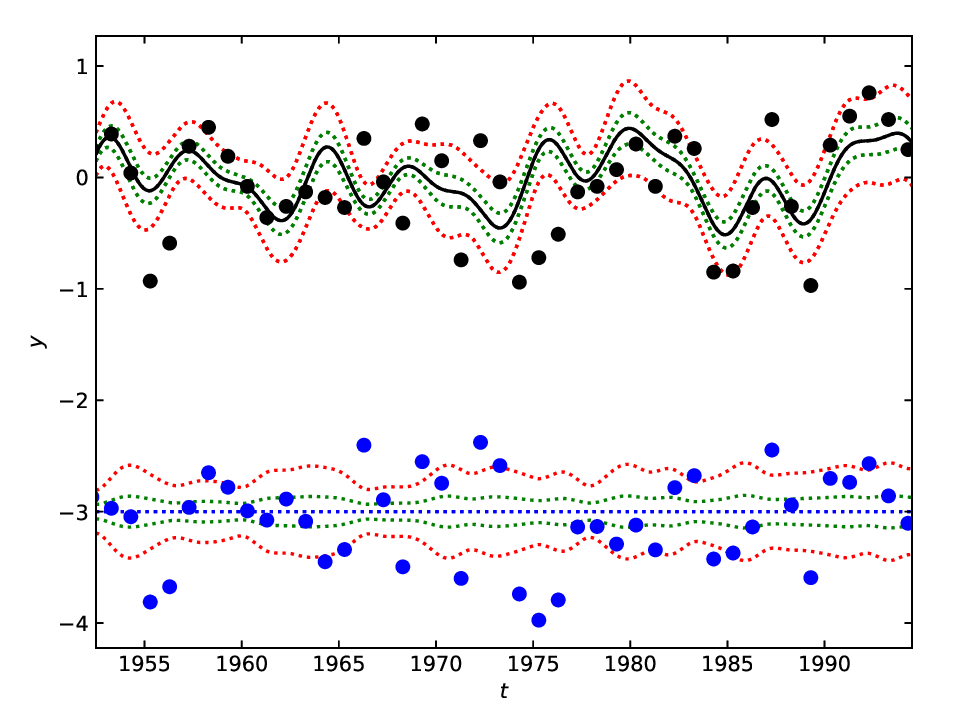}
 \hspace*{-0.01\textwidth}
 \includegraphics[width=0.455\textwidth,clip=]{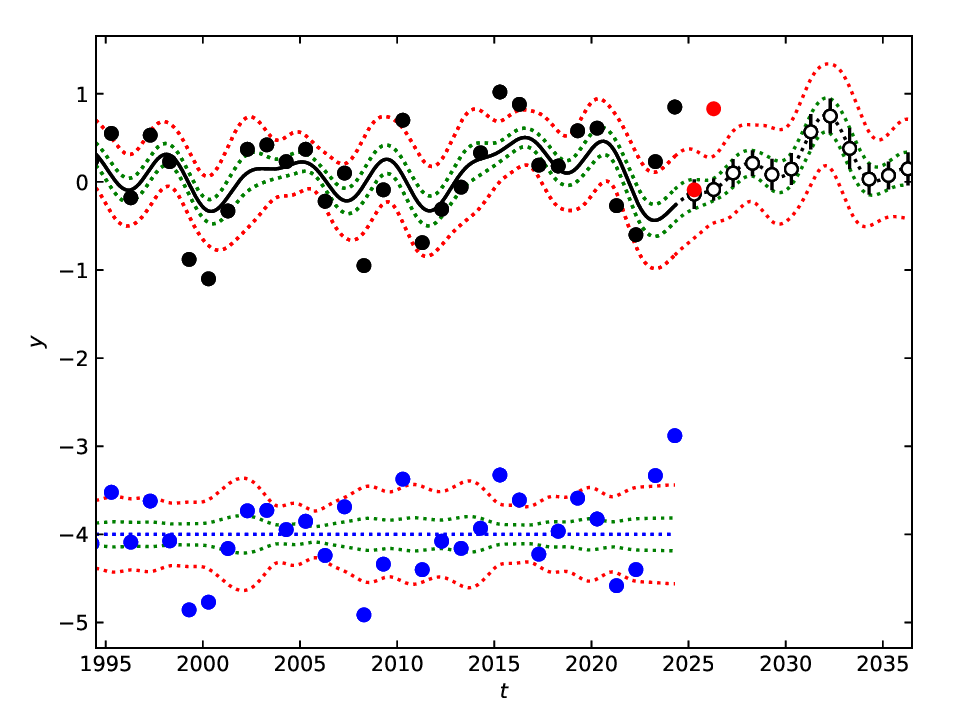}
}
\vspace{-0.36\textwidth}
\centerline{\Large 
\hspace{0.44\textwidth}  \color{black}{(c)}
\hspace{0.41\textwidth}  \color{black}{(d)}
\hfill}
\vspace{0.32\textwidth}
\caption{April data model \M=4 in Table \ref{TableApr}.
 Otherwise as in Figure \ref{FigJan}.
}
\label{FigApr}
\end{figure*}

\begin{table*}
  \caption{Periods in April data. Otherwise as in
    Table \ref{TableJan}.
  }
  \label{TableApr}
  \begin{tiny}
     \begin{center}
      \begin{tabular}{lcccccccccc}
  \hline
 & & \multicolumn{4}{c}{Period analysis} & & \multicolumn{3}{c}{Fisher-test  $(\gamma=0.05)$} &  \\
        \multicolumn{11}{c}{Data: Original non-weighted data ($n=155, \Delta T=154$: \PR{Apr.dat})} \\
(1)    & 
(2)    & 
(3)    & 
(4)    & 
(5)    & 
(6)    & 
        & 
(7)    & 
(8)    & 
(9)    &
(10)        \\ 
\M                    &
  &
$P_1$ (y)             &
$P_2$ (y)             &
$P_3$ (y)             &
$P_4$ (y)             &
                           &
 \M=2                  &
 \M=3                  &
 \M=4                  &
                      \\ 
             &
$\eta$ (-)   &
$A_1$ (\Cc)    &
$A_2$ (\Cc)    &
$A_3$ (\Cc)    &
$A_4$ (\Cc)    &
             &
$F_{R}$ (-)    &
$F_{R}$ (-)    &
$F_{R}$(-)    &
               Control file \\ 
                       &
$R$ (-)                    &
$t_{\mathrm{min,1}}$ (y) &
$t_{\mathrm{min,1}}$ (y) &
$t_{\mathrm{min,1}}$ (y) &
$t_{\mathrm{min,1}}$ (y) &
                       &
$Q_F$ (-)              &
$Q_F$ (-)              &
$Q_F$ (-)       \\ 
\hline 
        \multicolumn{10}{c}{One signal} \\
\hline
        \M=1             &                  &$12.69\PM 0.12     $&     -                  & -                        & -                     &&$\uparrow         $&$\uparrow        $&$\uparrow$&                                             \\
\RModel{1,1,1}     &       5         &$0.423\PM 0.078   $&     -                  & -                        & -                     &&          5.06           &     4.29           &    3.92       &               \PR{AprR14K111.dat}  \\
                             &      30.8     &$1871.90\PM0.87  $&     -                  & -                        & -                     &&       0.0023        &       0.00052     &   0.00018     &                                             \\
\hline
       \multicolumn{10}{c}{Two signals} \\
\hline 
\M=2                    &                   &$5.663\PM0.039   $&$12.66\PM0.21$& -                       & -                     &&-                         &$\uparrow        $&$\uparrow$&                                              \\
\RModel{2,1,1}    &        8         &$0.388 \PM0.088   $&$0.416\PM0.080$&   -                   & -                     &&-                           &    3.29            &       3.13    &               \PR{AprR14K211.dat}   \\
                            &     27.9        &$1870.73\PM0.48 $&$1872.06\PM0.88$& -                      & -                     &&-                           &     0.022          &    0.0065   &                                               \\
\hline
      \multicolumn{10}{c}{Three signals} \\
\hline 
\M=3                    &                   &$5.665\PM 0.017 $&$12.73\PM0.13$&$18.83\PM0.84     $& -                     &&-              & -                           &$\uparrow$&                                              \\
\RModel{3,1,1}    &       11        &$0.382\PM 0.097  $&$0.416\PM0.068 $&$0.31\PM 0.11      $& -                     &&-              & -                   & 2.82                  & \PR{AprR14K311.dat}            \\
                            &     26.1      &$1870.71\PM0.48  $&$1871.65\PM0.89$&$1874.0\PM 1.6 $& -                     &&-             & -                            &  0.040              &                                               \\
\hline
      \multicolumn{10}{c}{Four signals} \\
\hline 
\M=4                    &                     &$3.760\PM 0.039       $&$5.669\PM 0.069   $&$12.74\PM 0.12   $&$18.85\PM 0.37 $&&-                 & -             & - &                                              \\
\RModel{4,1,1}    &       14          &$0.276   \PM 0.074   $&$0.382\PM 0.096    $&$0.42\PM 0.13     $&$0.310\PM 0.090 $&&-                   & -             &- &       \PR{AprR14K411.dat}   \\
                            &     24.6         &$1872.27\PM 0.38     $&$1870.65\PM0.56  $&$1871.59\PM 0.89 $&$1873.9\PM 1.6$&&-                        & -             &- &                                               \\
\hline
\end{tabular}
\end{center}
\end{tiny}
\addtolength{\tabcolsep}{+0.05cm}
\end{table*}

\clearpage

 \begin{figure*}  
\vspace{0.02\textwidth}
\centerline{\hspace*{0.005\textwidth}
 \includegraphics[width=0.455\textwidth,clip=]{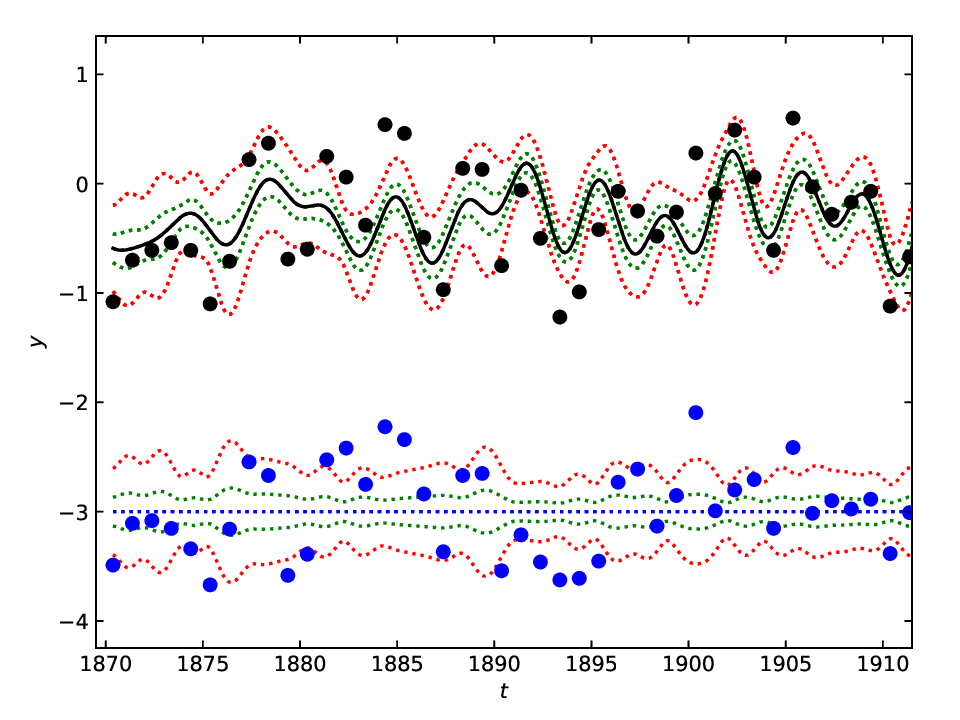}
 \hspace*{-0.01\textwidth}
 \includegraphics[width=0.455\textwidth,clip=]{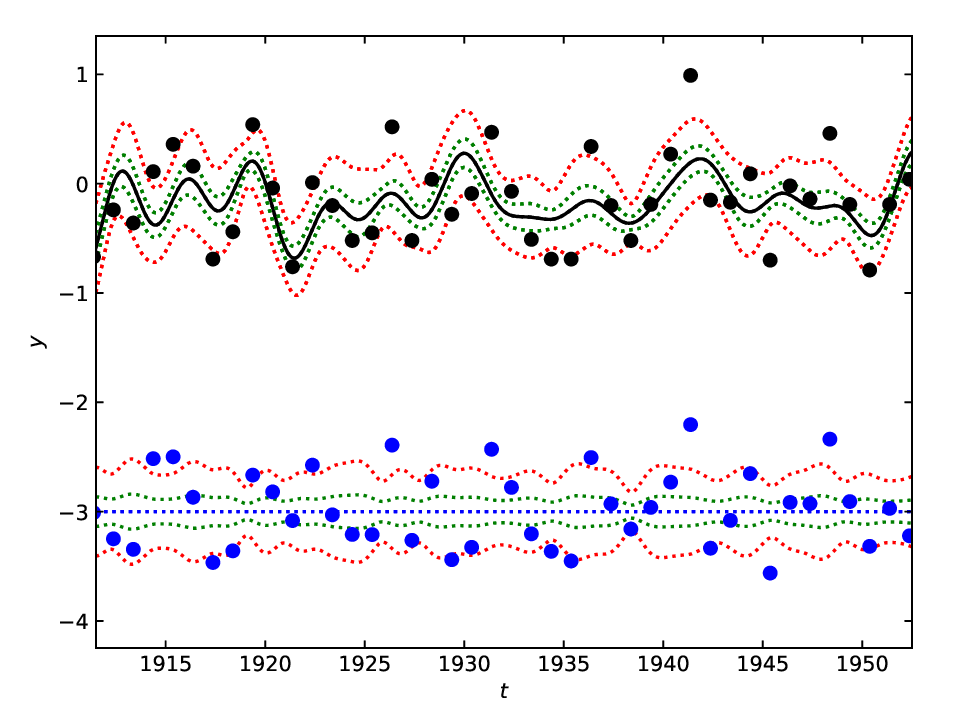}
 }
\vspace{-0.36\textwidth}
\centerline{\Large 
\hspace{0.44\textwidth}  \color{black}{(a)}
\hspace{0.41\textwidth}  \color{black}{(b)}
\hfill}
\vspace{0.35\textwidth}
\centerline{\hspace*{0.005\textwidth}
 \includegraphics[width=0.455\textwidth,clip=]{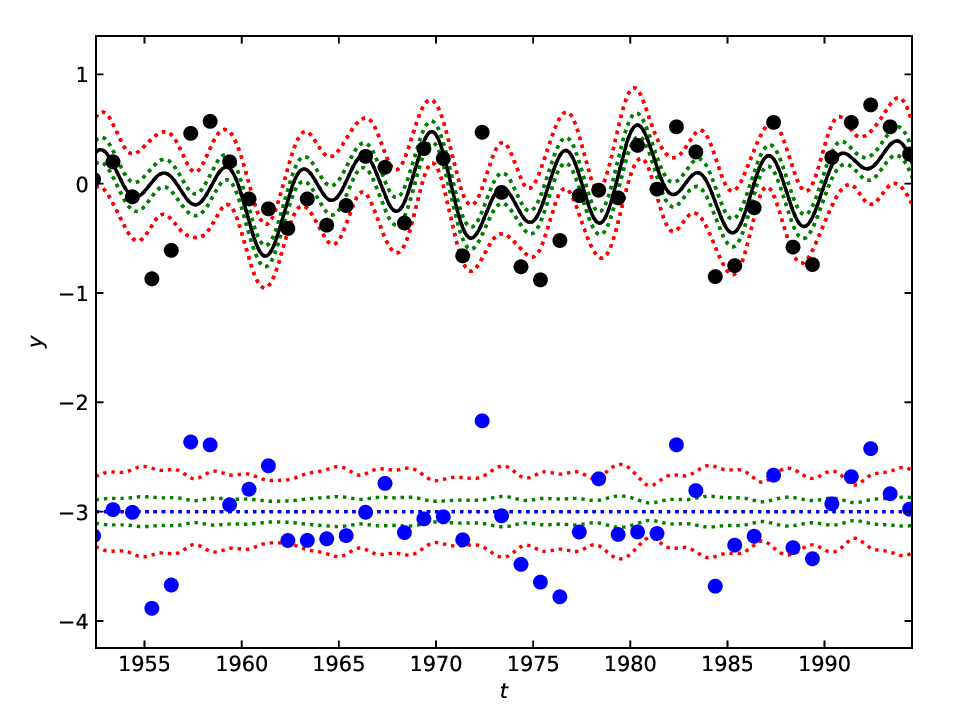}
 \hspace*{-0.01\textwidth}
 \includegraphics[width=0.455\textwidth,clip=]{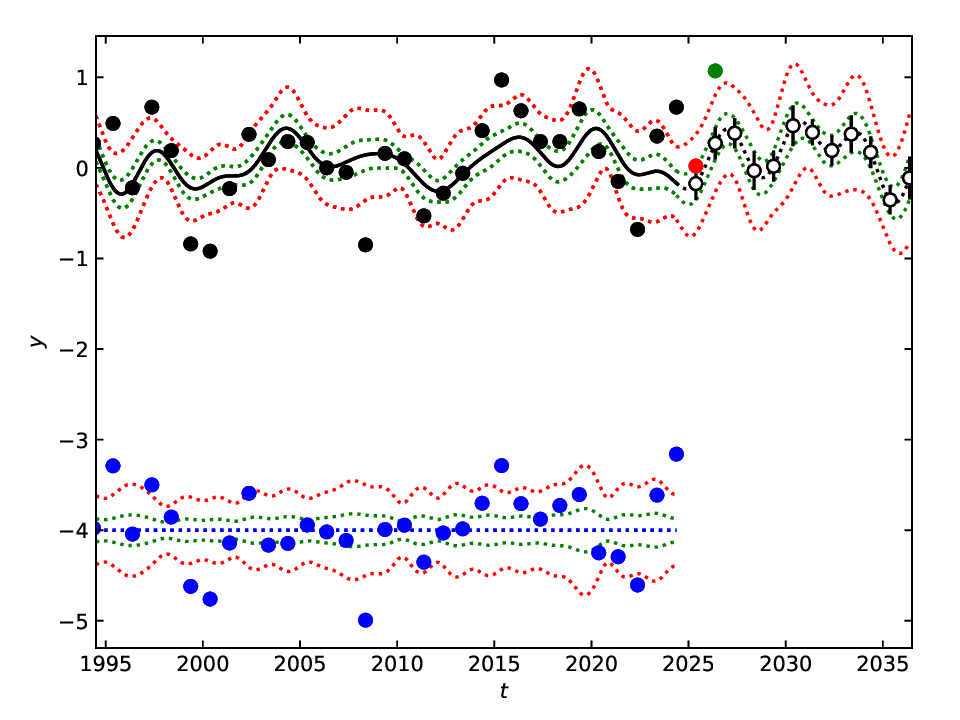}
}
\vspace{-0.36\textwidth}
\centerline{\Large 
\hspace{0.44\textwidth}  \color{black}{(c)}
\hspace{0.41\textwidth}  \color{black}{(d)}
\hfill}
\vspace{0.32\textwidth}
\caption{May data model \M=4 in Table \ref{TableMay}.
Green dot from Table \ref{TableProof}  (Column 7).
  Otherwise as in Figure \ref{FigJan}.
}
\label{FigMay}
\end{figure*}

\begin{table*}
  \caption{Periods in May data. Otherwise as in
    Table \ref{TableJan}.
  }
  \label{TableMay}
  \begin{tiny}
     \begin{center}
      \begin{tabular}{lcccccccccc}
  \hline
 & & \multicolumn{4}{c}{Period analysis} & & \multicolumn{3}{c}{Fisher-test  $(\gamma=0.05)$} &  \\
        \multicolumn{11}{c}{Data: Original non-weighted data ($n=155, \Delta T=154$: \PR{May.dat})} \\
(1)    & 
(2)    & 
(3)    & 
(4)    & 
(5)    & 
(6)    & 
        & 
(7)    & 
(8)    & 
(9)    &
(10)        \\ 
\M                    &
  &
$P_1$ (y)             &
$P_2$ (y)             &
$P_3$ (y)             &
$P_4$ (y)             &
                           &
 \M=2                  &
 \M=3                  &
 \M=4                  &
                      \\ 
             &
$\eta$ (-)   &
$A_1$ (\Cc)    &
$A_2$ (\Cc)    &
$A_3$ (\Cc)    &
$A_4$ (\Cc)    &
             &
$F_{R}$ (-)    &
$F_{R}$ (-)    &
$F_{R}$(-)    &
               Control file \\ 
                       &
$R$ (-)                    &
$t_{\mathrm{min,1}}$ (y) &
$t_{\mathrm{min,1}}$ (y) &
$t_{\mathrm{min,1}}$ (y) &
$t_{\mathrm{min,1}}$ (y) &
                       &
$Q_F$ (-)              &
$Q_F$ (-)              &
$Q_F$ (-)       \\ 
\hline 
        \multicolumn{10}{c}{One signal} \\
\hline
        \M=1             &                  &$12.67\PM 0.22     $&     -                  & -                        & -                     &&$\uparrow         $&$\uparrow        $&$\uparrow$&                                             \\
\RModel{1,1,1}     &       5         &$0.358\PM 0.075   $&     -                  & -                        & -                     &&      5.67              &     3.39             &      3.38       &               \PR{MayR14K111.dat}  \\
                             &     29.7     &$1872.4\PM1.3       $&     -                  & -                        & -                     &&    0.0010            &     0.0037          &   0.00088   &                                             \\
\hline
       \multicolumn{10}{c}{Two signals} \\
\hline 
\M=2                    &                   &$5.664\PM0.062   $&$12.65\PM0.24$& -                       & -                     &&-                         &$\uparrow        $&$\uparrow$&                                              \\
\RModel{2,1,1}    &        8         &$0.329 \PM0.073   $&$0.353\PM0.094$&   -                   & -                     &&-                           &    2.93            &     3.37        &               \PR{MayR14K211.dat}   \\
                            &     27.6        &$1870.76\PM0.56 $&$1872.5\PM1.3$& -                      & -                     &&-                           &    0.036            &    0.0060     &                                               \\
\hline
      \multicolumn{10}{c}{Three signals} \\
\hline 
\M=3                    &                   &$3.388\PM 0.025 $&$5.666\PM0.062$   &$12.65\PM0.30        $& -                     &&-              & -                           &$\uparrow$&                                              \\
\RModel{3,1,1}    &       11        &$0.288\PM 0.090  $&$0.330\PM0.057  $&$0.356\PM 0.080   $& -                     &&-              & -                           &     3.26      & \PR{MayR14K311.dat}            \\
                            &     26.0      &$1873.42\PM0.40  $&$1870.76\PM0.66$&$1872.55\PM 0.39 $& -                     &&-             & -                            &      0.023     &                                               \\
\hline
      \multicolumn{10}{c}{Four signals} \\
\hline 
        \M=4                    &             &$3.383\PM 0.024     $&$3.552\PM 0.042 $&$5.666\PM 0027   $&$12.65\PM 0.12 $&&-                      & -             & - &                                              \\
\RModel{4,1,1}    &       14          &$0.304  \PM 0.098   $&$0.30\PM 0.10     $&$0.332\PM 0.067  $&$0.356\PM 0.081 $&&-                   & -             &- &       \PR{MayR14K411.dat}   \\
                            &     24.3         &$1873.53\PM 0.33   $&$1872.03\PM0.37$&$1870.7\PM 0.47 $&$1872.5\PM 1.4$&&-                        & -             &- &                                               \\
\hline
\end{tabular}
\end{center}
\end{tiny}
\addtolength{\tabcolsep}{+0.05cm}
\end{table*}

\clearpage

 \begin{figure*}  
\vspace{0.02\textwidth}
\centerline{\hspace*{0.005\textwidth}
 \includegraphics[width=0.455\textwidth,clip=]{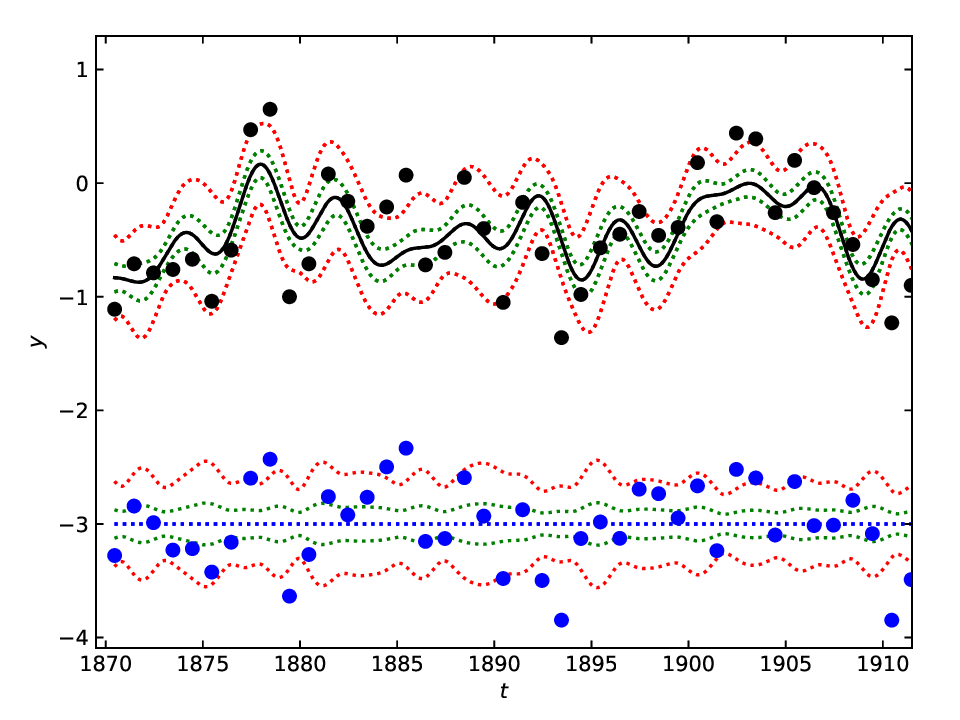}
 \hspace*{-0.01\textwidth}
 \includegraphics[width=0.455\textwidth,clip=]{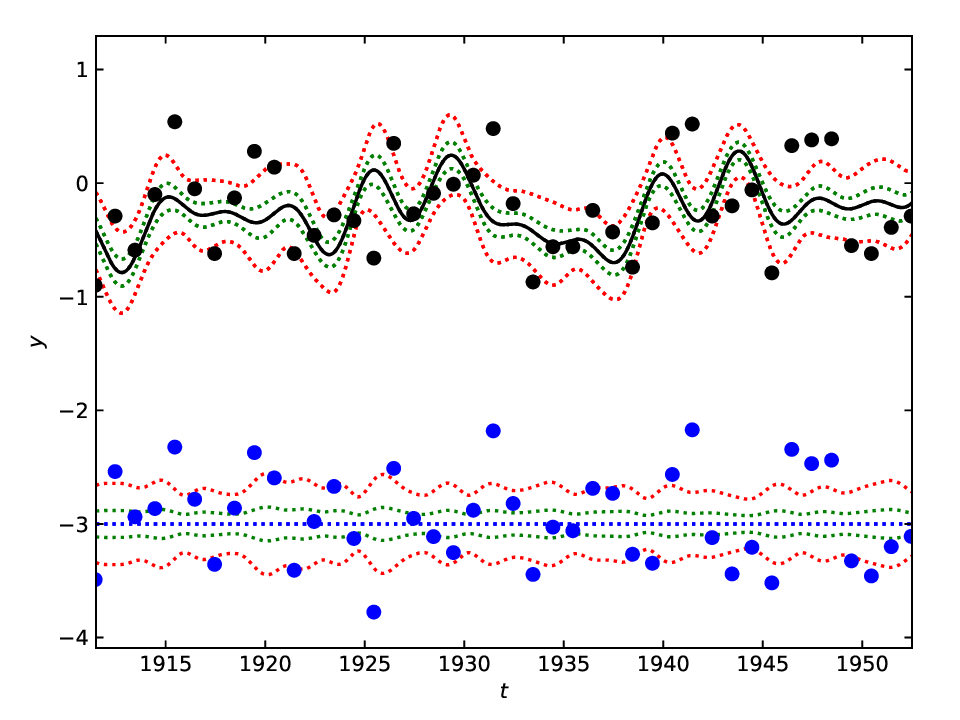}
 }
\vspace{-0.36\textwidth}
\centerline{\Large 
\hspace{0.44\textwidth}  \color{black}{(a)}
\hspace{0.41\textwidth}  \color{black}{(b)}
\hfill}
\vspace{0.35\textwidth}
\centerline{\hspace*{0.005\textwidth}
 \includegraphics[width=0.455\textwidth,clip=]{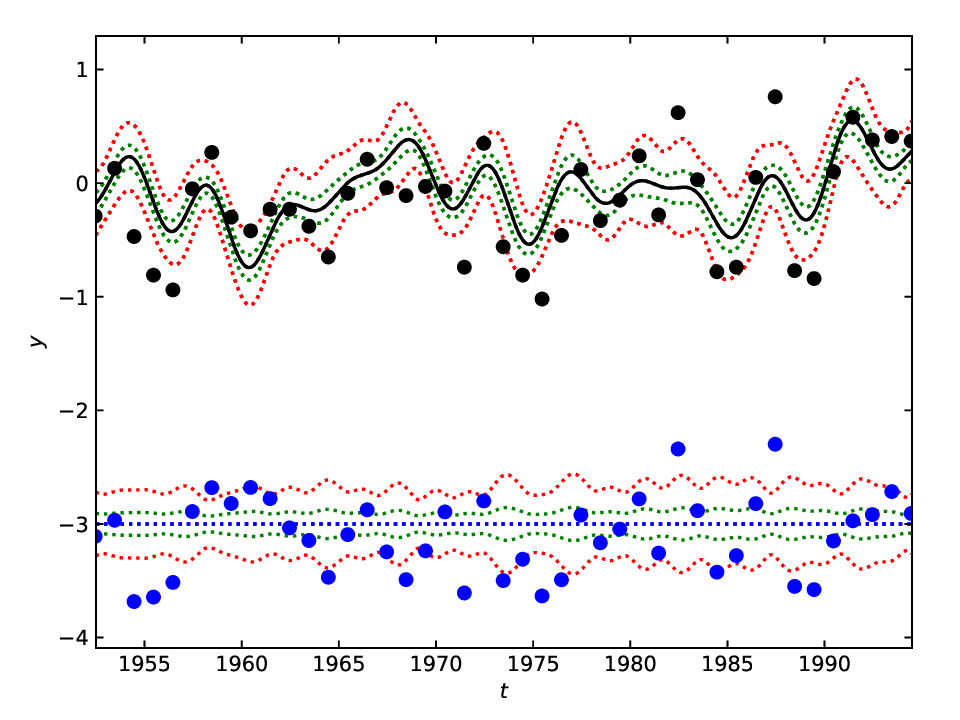}
 \hspace*{-0.01\textwidth}
 \includegraphics[width=0.455\textwidth,clip=]{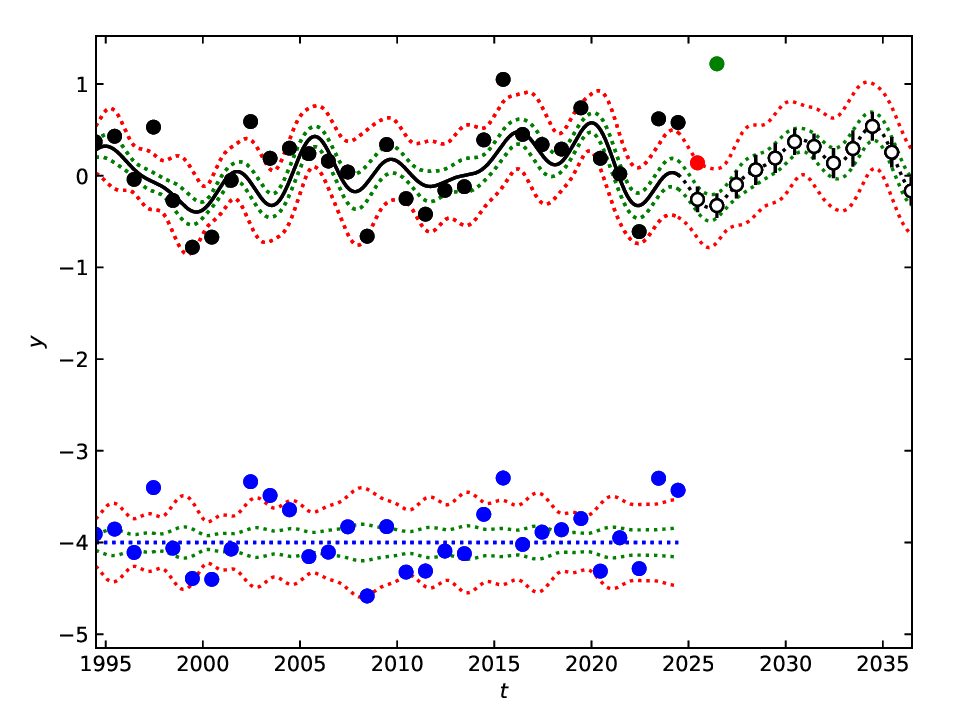}
}
\vspace{-0.36\textwidth}
\centerline{\Large 
\hspace{0.44\textwidth}  \color{black}{(c)}
\hspace{0.41\textwidth}  \color{black}{(d)}
\hfill}
\vspace{0.32\textwidth}
\caption{June data model \M=4 in Table \ref{TableJun}.
Green dot from Table \ref{TableProof}  (Column 7).
Otherwise as in Figure \ref{FigJan}.}
\label{FigJun}
\end{figure*}

\begin{table*}
  \caption{Periods in June data. Otherwise as in
    Table \ref{TableJan}.}
  \label{TableJun}
  \begin{tiny}
     \begin{center}
      \begin{tabular}{lcccccccccc}
  \hline
 & & \multicolumn{4}{c}{Period analysis} & & \multicolumn{3}{c}{Fisher-test  $(\gamma=0.05)$} &  \\
        \multicolumn{11}{c}{Data: Original non-weighted data ($n=155, \Delta T=154$: \PR{Jun.dat})} \\
(1)    & 
(2)    & 
(3)    & 
(4)    & 
(5)    & 
(6)    & 
        & 
(7)    & 
(8)    & 
(9)    &
(10)        \\ 
\M                    &
  &
$P_1$ (y)             &
$P_2$ (y)             &
$P_3$ (y)             &
$P_4$ (y)             &
                           &
 \M=2                  &
 \M=3                  &
 \M=4                  &
                      \\ 
             &
$\eta$ (-)   &
$A_1$ (\Cc)    &
$A_2$ (\Cc)    &
$A_3$ (\Cc)    &
$A_4$ (\Cc)    &
             &
$F_{R}$ (-)    &
$F_{R}$ (-)    &
$F_{R}$(-)    &
               Control file \\ 
                       &
$R$ (-)                    &
$t_{\mathrm{min,1}}$ (y) &
$t_{\mathrm{min,1}}$ (y) &
$t_{\mathrm{min,1}}$ (y) &
$t_{\mathrm{min,1}}$ (y) &
                       &
$Q_F$ (-)              &
$Q_F$ (-)              &
$Q_F$ (-)       \\ 
\hline 
        \multicolumn{10}{c}{One signal} \\
\hline
        \M=1             &                  &$12.72\PM 0.20     $&     -                  & -                        & -                     &&$\uparrow         $&$\uparrow        $&$\uparrow$&                                             \\
\RModel{1,1,1}     &       5         &$0.424\PM 0.091       $&     -                  & -                        & -                     &&     4.67                &    4.07              &    4.00     &               \PR{JunR14K111.dat}  \\
                             &     27.4     &$1872.0\PM1.2         $&     -                  & -                        & -                     &&     0.0038             &  0.00084          &  0.00014     &                                             \\
\hline
       \multicolumn{10}{c}{Two signals} \\
\hline 
\M=2                    &                   &$4.740\PM0.028  $&$12.72\PM0.30$& -                       & -                     &&-                         &$\uparrow        $&$\uparrow$&                                              \\
\RModel{2,1,1}    &        8         &$0.344 \PM0.081   $&$0.424\PM0.075  $&   -                   & -                     &&-                           &          3.26      &    3.42      &               \PR{JunR14K211.dat}   \\
                            &     25.0        &$1870.74\PM0.55 $&$1872.0\PM1.3$& -               & -                     &&-                           &       0.023    &  0.0035     &                                               \\
\hline
      \multicolumn{10}{c}{Three signals} \\
\hline 
\M=3                    &                   &$4.740\PM 0.018 $&$12.75\PM0.15 $   &$22.46\PM0.55        $& -                     &&-              & -                           &$\uparrow$&                                              \\
\RModel{3,1,1}    &       11        &$0.342\PM 0.072  $&$0.411\PM0.072  $&$0.297\PM 0.068   $& -                     &&-              & -                           &   3.42          & \PR{JunR14K311.dat}            \\
                            &     23.4      &$1870.73\PM0.47  $&$1871.9\PM1.0   $&$1891.5\PM 1.6 $& -                     &&-             & -                                 &  0.019        &                                               \\
\hline
      \multicolumn{10}{c}{Four signals} \\
\hline 
\M=4                    &                     &$3.647\PM 0.012     $&$4.741\PM 0.037 $&$12.76\PM 0.12   $&$22.44\PM 1.7 $&&-                 & -             & - &                                              \\
\RModel{4,1,1}    &       14          &$0.289  \PM 0.069     $&$0.338\PM 0.09$&$0.415\PM 0.077  $&$0.302\PM 0.086 $&&-                   & -             &- &       \PR{JunR14K411.dat}   \\
                            &     21.8         &$1872.53\PM 0.35   $&$1870.8\PM0.36$&$1871.81\PM 0.99 $&$1891.5\PM 1.8$&&-                        & -             &- &                                               \\
\hline
\end{tabular}
\end{center}
\end{tiny}
\addtolength{\tabcolsep}{+0.05cm}
\end{table*}

\clearpage

 \begin{figure*}  
\vspace{0.02\textwidth}
\centerline{\hspace*{0.005\textwidth}
 \includegraphics[width=0.455\textwidth,clip=]{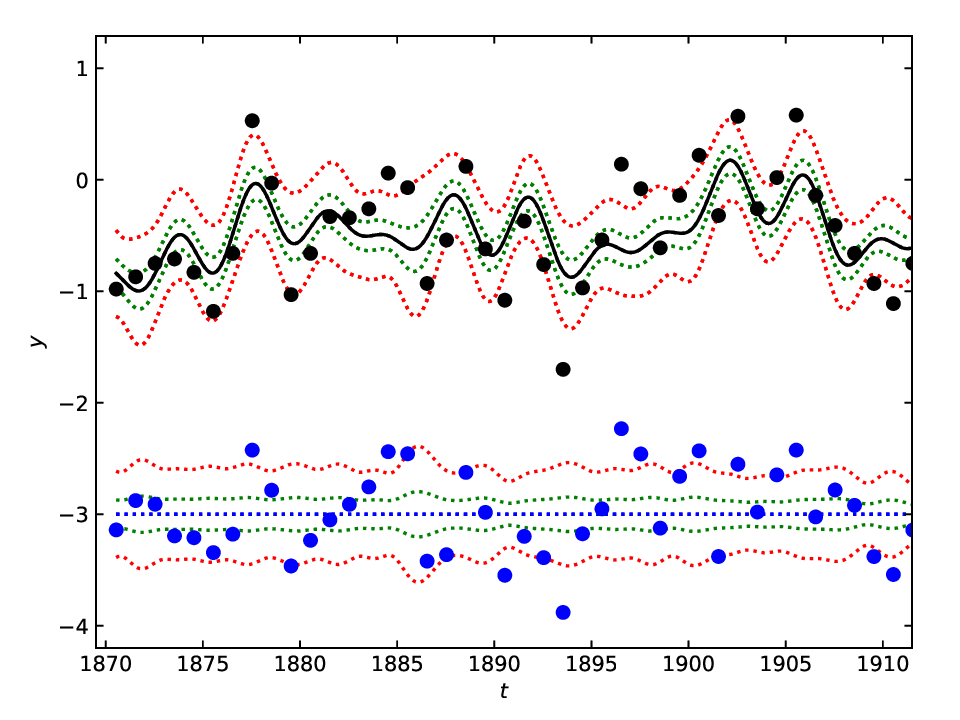}
 \hspace*{-0.01\textwidth}
 \includegraphics[width=0.455\textwidth,clip=]{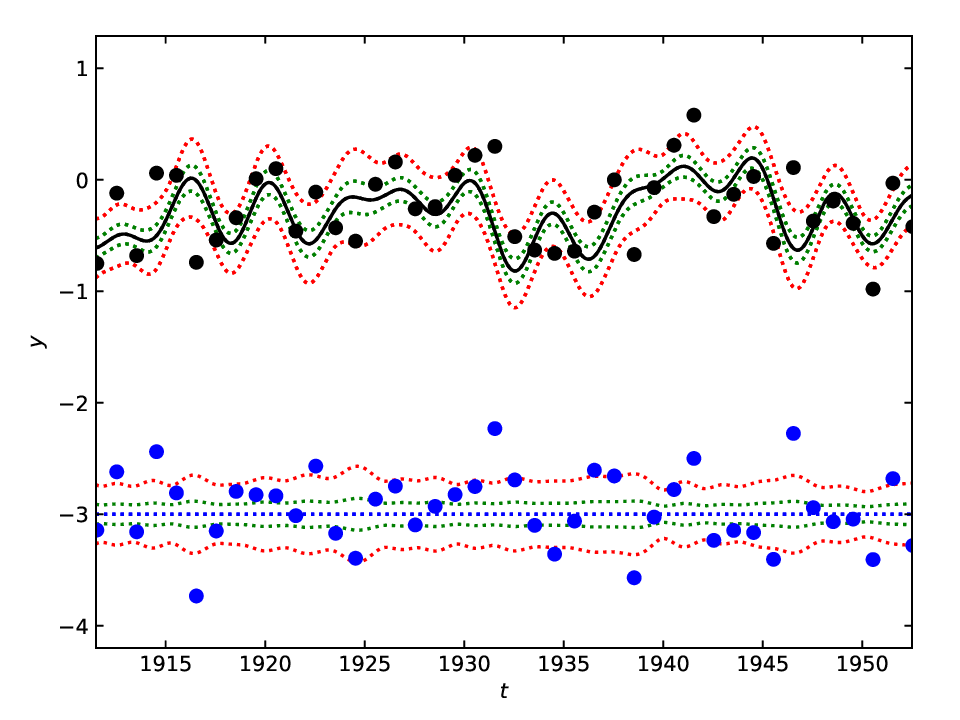}
 }
\vspace{-0.36\textwidth}
\centerline{\Large 
\hspace{0.44\textwidth}  \color{black}{(a)}
\hspace{0.41\textwidth}  \color{black}{(b)}
\hfill}
\vspace{0.35\textwidth}
\centerline{\hspace*{0.005\textwidth}
 \includegraphics[width=0.455\textwidth,clip=]{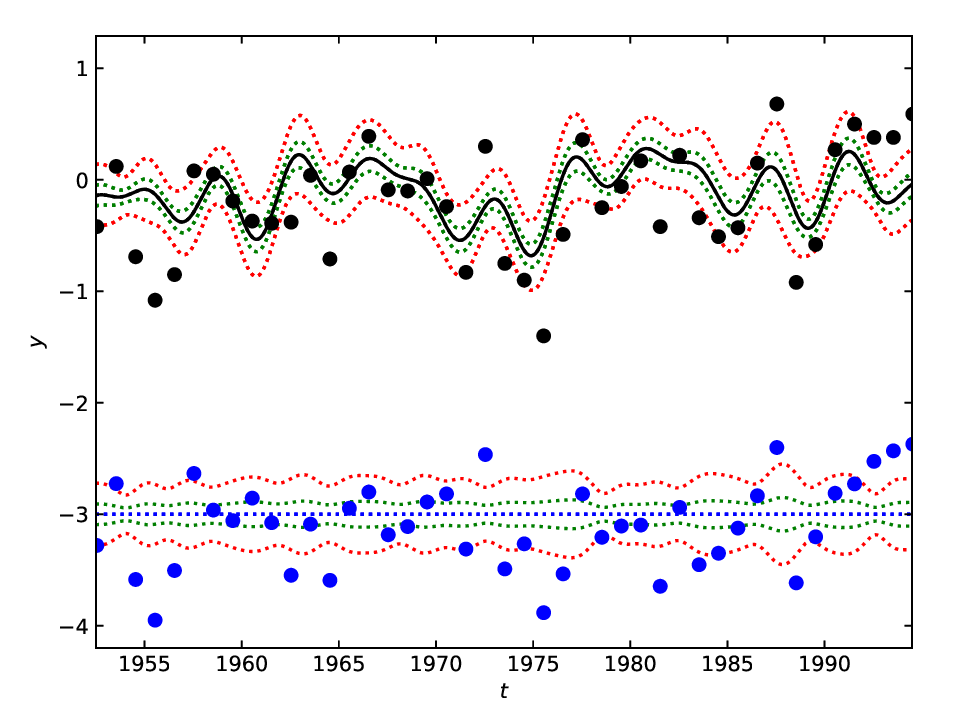}
 \hspace*{-0.01\textwidth}
 \includegraphics[width=0.455\textwidth,clip=]{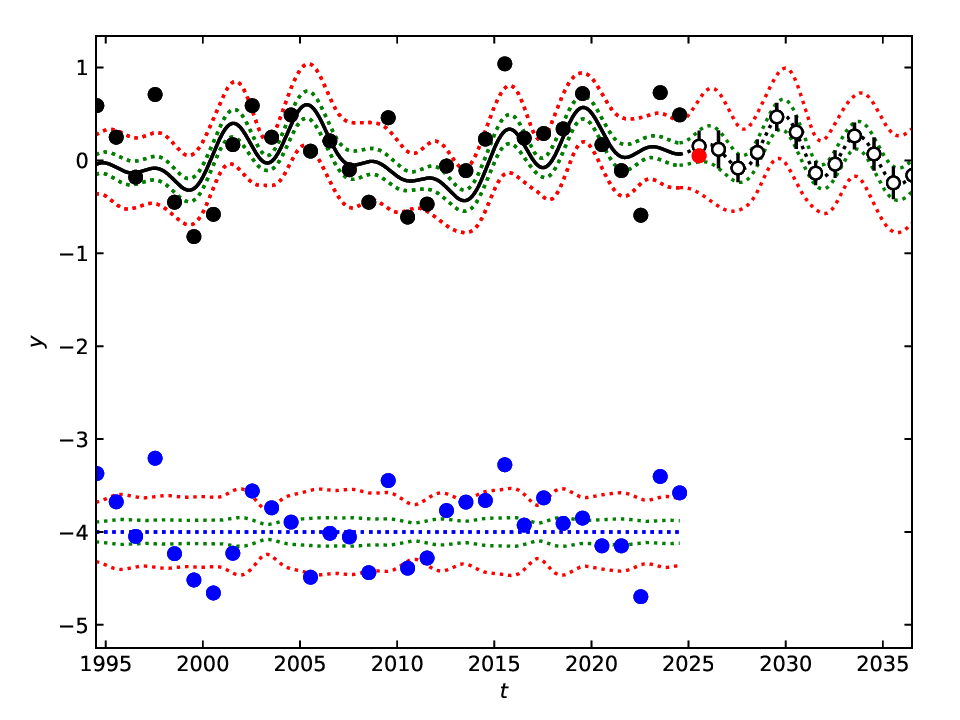}
}
\vspace{-0.36\textwidth}
\centerline{\Large 
\hspace{0.44\textwidth}  \color{black}{(c)}
\hspace{0.41\textwidth}  \color{black}{(d)}
\hfill}
\vspace{0.32\textwidth}
\caption{July data model \M=4 in Table \ref{TableJul}.
 Otherwise as in Figure \ref{FigJan}.}
\label{FigJul}
\end{figure*}

\begin{table*}
  \caption{Periods in July data. Otherwise as in
    Table \ref{TableJan}. }
  \label{TableJul}
  \begin{tiny}
     \begin{center}
      \begin{tabular}{lcccccccccc}
  \hline
 & & \multicolumn{4}{c}{Period analysis} & & \multicolumn{3}{c}{Fisher-test  $(\gamma=0.05)$} &  \\
        \multicolumn{11}{c}{Data: Original non-weighted data ($n=155, \Delta T=154$: \PR{Jul.dat})} \\
(1)    & 
(2)    & 
(3)    & 
(4)    & 
(5)    & 
(6)    & 
        & 
(7)    & 
(8)    & 
(9)    &
(10)        \\ 
\M                    &
  &
$P_1$ (y)             &
$P_2$ (y)             &
$P_3$ (y)             &
$P_4$ (y)             &
                           &
 \M=2                  &
 \M=3                  &
 \M=4                  &
                      \\ 
             &
$\eta$ (-)   &
$A_1$ (\Cc)    &
$A_2$ (\Cc)    &
$A_3$ (\Cc)    &
$A_4$ (\Cc)    &
             &
$F_{R}$ (-)    &
$F_{R}$ (-)    &
$F_{R}$(-)    &
               Control file \\ 
                       &
$R$ (-)                    &
$t_{\mathrm{min,1}}$ (y) &
$t_{\mathrm{min,1}}$ (y) &
$t_{\mathrm{min,1}}$ (y) &
$t_{\mathrm{min,1}}$ (y) &
                       &
$Q_F$ (-)              &
$Q_F$ (-)              &
$Q_F$ (-)       \\ 
\hline 
        \multicolumn{10}{c}{One signal} \\
\hline
        \M=1             &                  &$12.76\PM 0.16     $&     -                  & -                        & -                     &&$\uparrow         $&$\uparrow        $&$\uparrow$&                                             \\
\RModel{1,1,1}     &       5         &$0.372\PM 0.078       $&     -                  & -                        & -                 &&     4.07               &       4.23              &      4.24    &               \PR{JulR14K111.dat}  \\
                             &     28.5     &$1871.3\PM1.1      $&     -                  & -                        & -                     &&    0.0082             &   0.00059           & $7.1\TM10^{-5}$  &                                             \\
\hline
       \multicolumn{10}{c}{Two signals} \\
\hline 
\M=2                    &                   &$12.77\PM0.17      $&$19.99\PM0.68      $& -                       & -                     &&-                         &$\uparrow        $&$\uparrow$&                                              \\
\RModel{2,1,1}    &        8         &$0.372 \PM0.076   $&$0.330\PM0.092$&   -                   & -                     &&-                           &    4.14               &    4.06              &               \PR{JulR14K211.dat}   \\
                            &     26.3        &$1871.1\PM1.2     $&$1873.0\PM2.4  $& -                      & -                     &&-                           & 0.0076              & 0.00087        &                                               \\
\hline
      \multicolumn{10}{c}{Three signals} \\
\hline 
\M=3                    &                   &$4.738\PM 0.048 $&$12.77\PM0.16 $   &$19.90\PM0.71        $& -                     &&-              & -                           &$\uparrow$&                                              \\
\RModel{3,1,1}    &       11        &$0.331\PM 0.085  $&$0.369\PM0.078  $&$0.327\PM 0.069   $& -                     &&-              & -                           &     3.75             & \PR{JulR14K311.dat}            \\
                            &     24.2      &$1870.78\PM0.40  $&$1871.2\PM1.3   $&$1873.2\PM 1.6 $& -                     &&-             & -                            &        0.012  &                                               \\
\hline
      \multicolumn{10}{c}{Four signals} \\
\hline 
\M=4                    &                     &$3.544\PM 0.010       $&$4.740\PM 0.043 $&$12.77\PM 0.12   $&$19.97\PM 0.98 $&&-                 & -             & - &                                              \\
\RModel{4,1,1}    &       14          &$0.302   \PM 0.070     $&$0.333\PM 0.08 $&$0.370\PM 0.081  $&$0.329\PM 0.075$&&-                   & -             &- &       \PR{JulR14K411.dat}   \\
                            &     22.4          &$1872.25\PM 0.34   $&$1870.73\PM0.42$&$1871.2\PM 1.2 $&$1873.1\PM 2.1$&&-                        & -             &- &                                               \\
\hline
\end{tabular}
\end{center}
\end{tiny}
\addtolength{\tabcolsep}{+0.05cm}
\end{table*}

\clearpage

 \begin{figure*}  
\vspace{0.02\textwidth}
\centerline{\hspace*{0.005\textwidth}
 \includegraphics[width=0.455\textwidth,clip=]{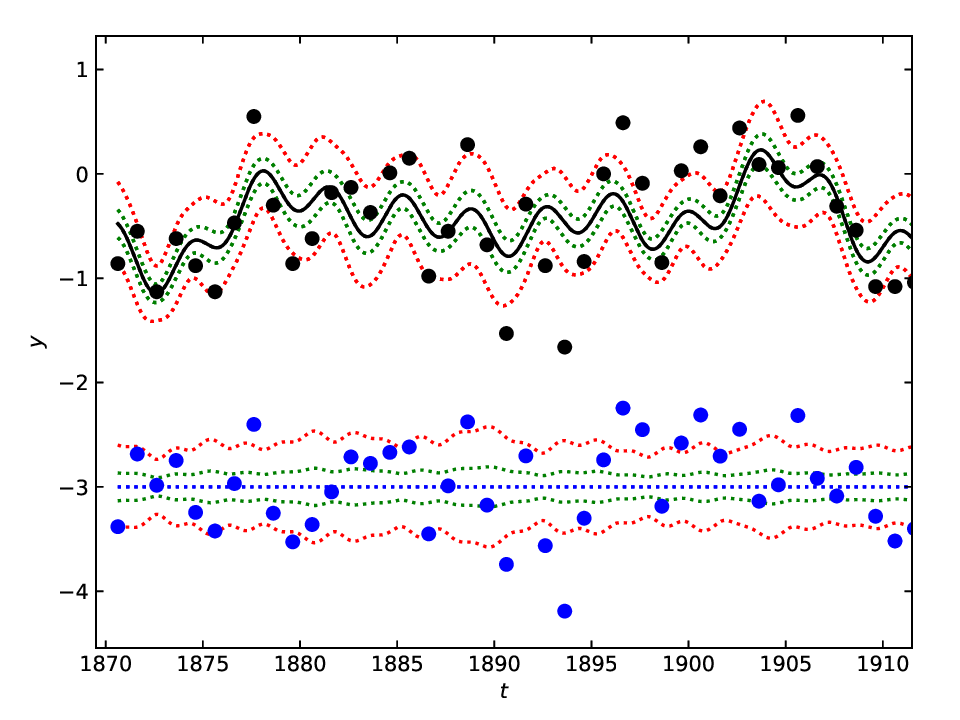}
 \hspace*{-0.01\textwidth}
 \includegraphics[width=0.455\textwidth,clip=]{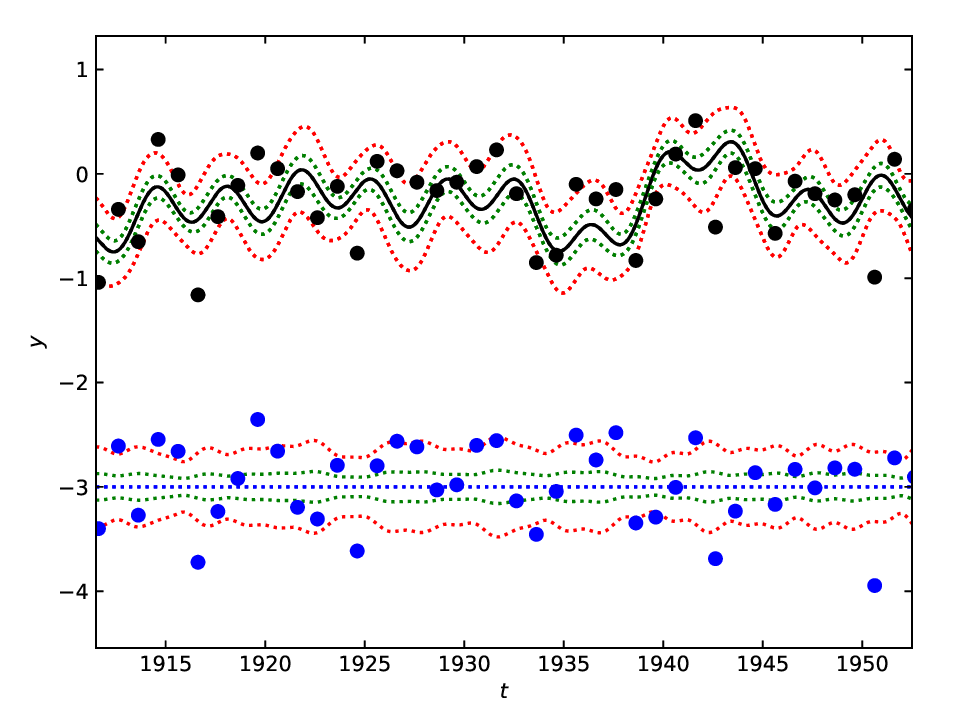}
 }
\vspace{-0.36\textwidth}
\centerline{\Large 
\hspace{0.44\textwidth}  \color{black}{(a)}
\hspace{0.41\textwidth}  \color{black}{(b)}
\hfill}
\vspace{0.35\textwidth}
\centerline{\hspace*{0.005\textwidth}
 \includegraphics[width=0.455\textwidth,clip=]{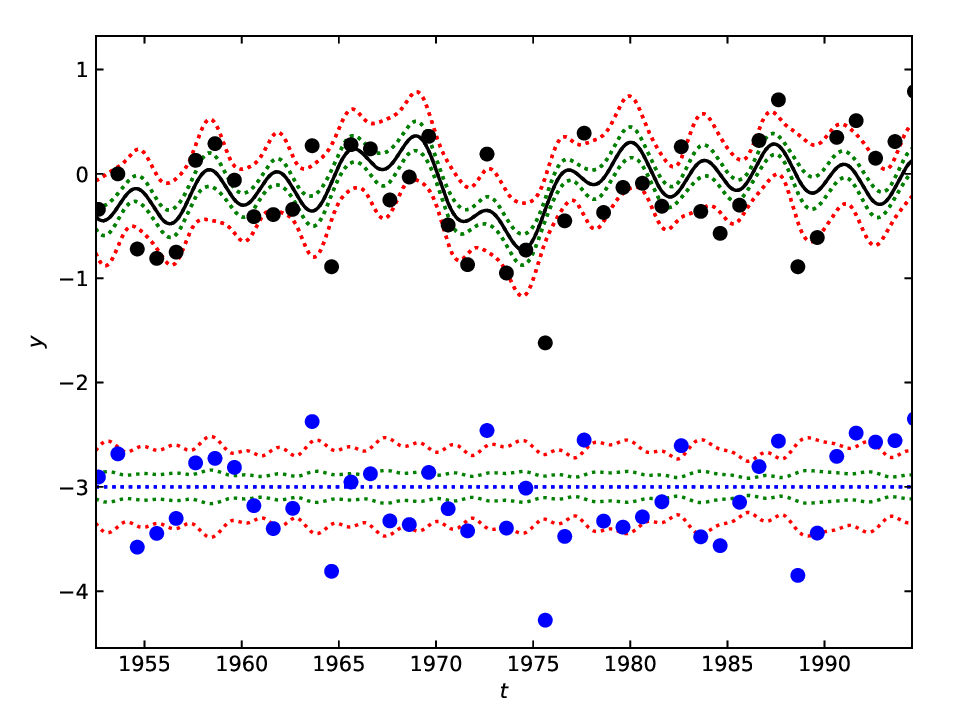}
 \hspace*{-0.01\textwidth}
 \includegraphics[width=0.455\textwidth,clip=]{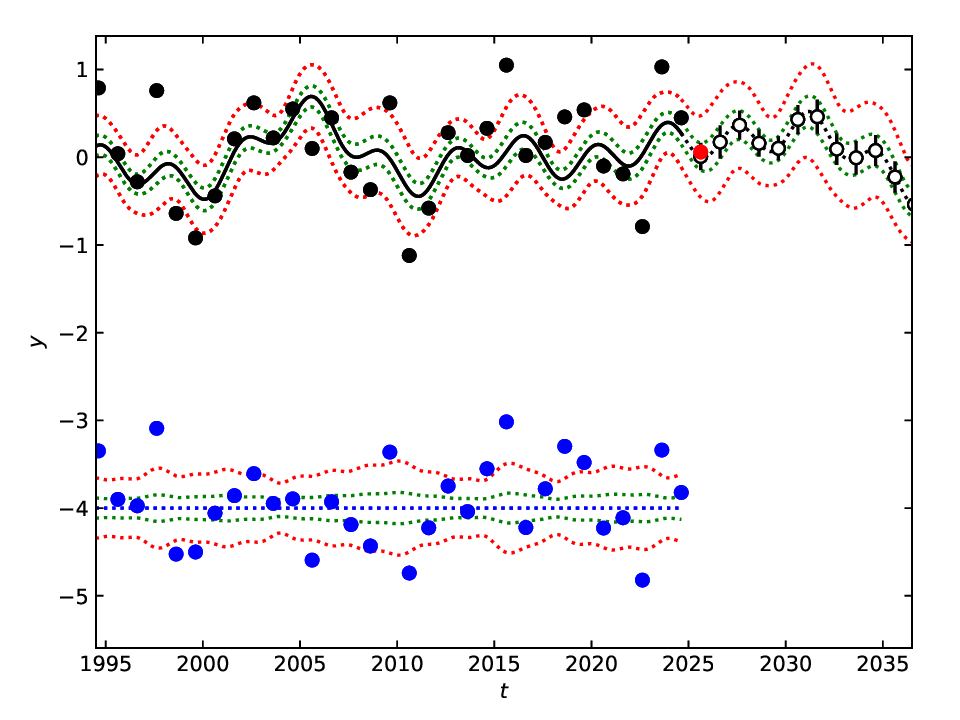}
}
\vspace{-0.36\textwidth}
\centerline{\Large 
\hspace{0.44\textwidth}  \color{black}{(c)}
\hspace{0.41\textwidth}  \color{black}{(d)}
\hfill}
\vspace{0.32\textwidth}
\caption{August data model \M=4 in Table \ref{TableAug}.
Otherwise as in Figure \ref{FiglongC14K311}
}
\label{FigAug}
\end{figure*}

\begin{table*}
  \caption{Periods in August data. Otherwise as in
    Table \ref{TableJan}.}
  \label{TableAug}
  \begin{tiny}
     \begin{center}
      \begin{tabular}{lcccccccccc}
  \hline
 & & \multicolumn{4}{c}{Period analysis} & & \multicolumn{3}{c}{Fisher-test  $(\gamma=0.05)$} &  \\
        \multicolumn{11}{c}{Data: Original non-weighted data ($n=155, \Delta T=154$: \PR{Aug.dat})} \\
(1)    & 
(2)    & 
(3)    & 
(4)    & 
(5)    & 
(6)    & 
        & 
(7)    & 
(8)    & 
(9)    &
(10)        \\ 
\M                    &
  &
$P_1$ (y)             &
$P_2$ (y)             &
$P_3$ (y)             &
$P_4$ (y)             &
                           &
 \M=2                  &
 \M=3                  &
 \M=4                  &
                      \\ 
             &
$\eta$ (-)   &
$A_1$ (\Cc)    &
$A_2$ (\Cc)    &
$A_3$ (\Cc)    &
$A_4$ (\Cc)    &
             &
$F_{R}$ (-)    &
$F_{R}$ (-)    &
$F_{R}$(-)    &
               Control file \\ 
                       &
$R$ (-)                    &
$t_{\mathrm{min,1}}$ (y) &
$t_{\mathrm{min,1}}$ (y) &
$t_{\mathrm{min,1}}$ (y) &
$t_{\mathrm{min,1}}$ (y) &
                       &
$Q_F$ (-)              &
$Q_F$ (-)              &
$Q_F$ (-)       \\ 
\hline 
        \multicolumn{10}{c}{One signal} \\
\hline
        \M=1             &                  &$3.648\PM 0.025$&     -                  & -                        & -                     &&$\uparrow         $&$\uparrow        $&$\uparrow$&                                             \\
\RModel{1,1,1}     &       5         &$0.384\PM 0.096       $&     -                  & -                        & -                     &&        3.86        &         37.4         &    3.56        &               \PR{AugR14K111.dat}  \\
                             &     35.4     &$1872.50\PM0.31   $&     -                  & -                        & -                     &&        0.011       & 0.0017           &    0.00051    &                                             \\
\hline
       \multicolumn{10}{c}{Two signals} \\
\hline 
\M=2                    &                   &$3.648\PM0.013      $&$20.58\PM0.92 $& -                       & -                     &&-                         &$\uparrow        $&$\uparrow$&                                              \\
\RModel{2,1,1}    &        8         &$0.384\PM0.092        $&$0.368\PM0.079$&   -                   & -                     &&-                           &  3.43               &  3.26                &               \PR{Aug14K211.dat}   \\
                            &     32.8        &$1872.51\PM0.29  $&$1871.8\PM2.6$& -                      & -                     &&-                           &   0.019             & 0.0050       &                                               \\
\hline
      \multicolumn{10}{c}{Three signals} \\
\hline 
\M=3                    &                   &$3.648\PM 0.017  $&$12.56\PM0.42 $   &$20.58\PM0.44        $& -                     &&-              & -                           &$\uparrow$&                                              \\
\RModel{3,1,1}    &       11        &$0.38\PM 0.10      $&$0.331\PM0.081  $&$0.367\PM 0.071   $& -                     &&-              & -                           &     2.92             & \PR{AugR14K311.dat}            \\
                            &     30.6      &$1872.51\PM0.32  $&$1873.0\PM1.5    $&$1871.9\PM 2.0 $& -                     &&-             & -                            &         0.036 &                                               \\
\hline
      \multicolumn{10}{c}{Four signals} \\
\hline 
\M=4                    &                     &$3.648\PM 0.010     $&$9.13\PM 0.19    $&$12.51\PM 0.31   $&$20.6\PM 1.2 $&&-                 & -             & - &                                              \\
\RModel{4,1,1}    &       14          &$0.382 \PM 0.084     $&$0.311\PM 0.064 $&$0.334\PM 0.084  $&$0.353\PM 0.092 $&&-                   & -             &- &       \PR{AugR14K411.dat}   \\
                            &     28.8         &$1872.51\PM 0.36   $&$1872.94\PM0.64$&$1873.6\PM 1.1     $&$1871.9\PM 1.9$&&-                        & -             &- &                                               \\
\hline
\end{tabular}
\end{center}
\end{tiny}
\addtolength{\tabcolsep}{+0.05cm}
\end{table*}

\clearpage

 \begin{figure*}  
\vspace{0.02\textwidth}
\centerline{\hspace*{0.005\textwidth}
 \includegraphics[width=0.455\textwidth,clip=]{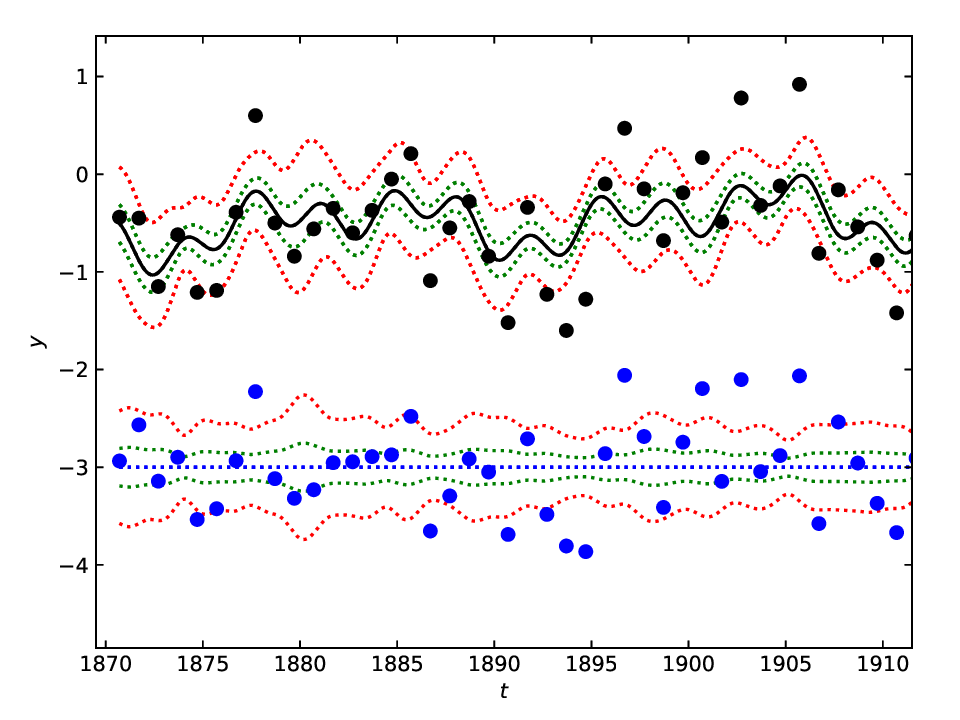}
 \hspace*{-0.01\textwidth}
 \includegraphics[width=0.455\textwidth,clip=]{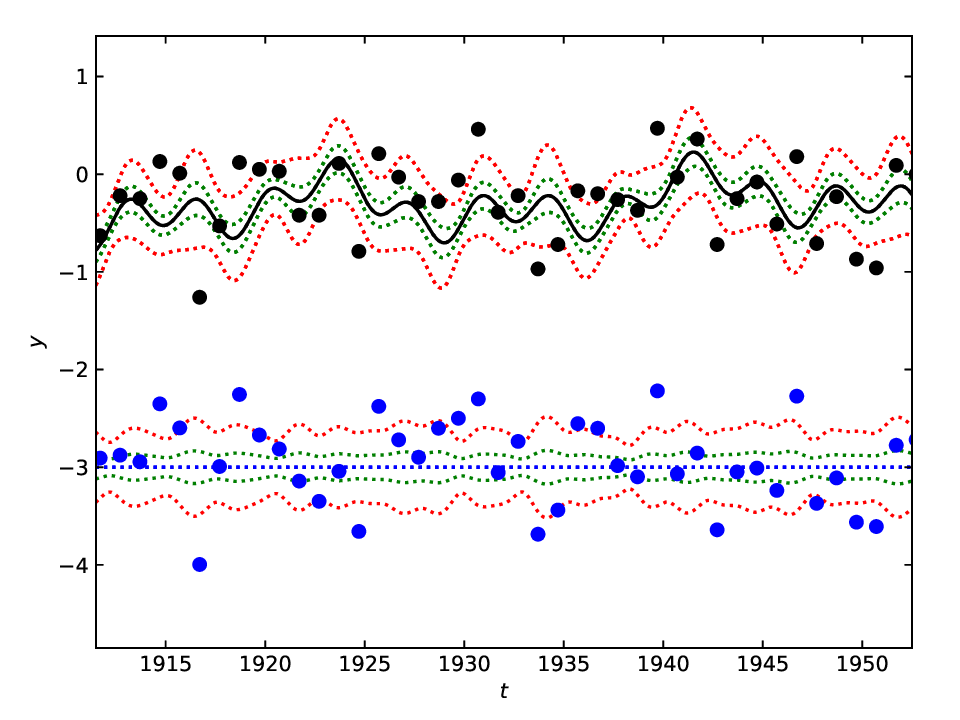}
 }
\vspace{-0.36\textwidth}
\centerline{\Large 
\hspace{0.44\textwidth}  \color{black}{(a)}
\hspace{0.41\textwidth}  \color{black}{(b)}
\hfill}
\vspace{0.35\textwidth}
\centerline{\hspace*{0.005\textwidth}
 \includegraphics[width=0.455\textwidth,clip=]{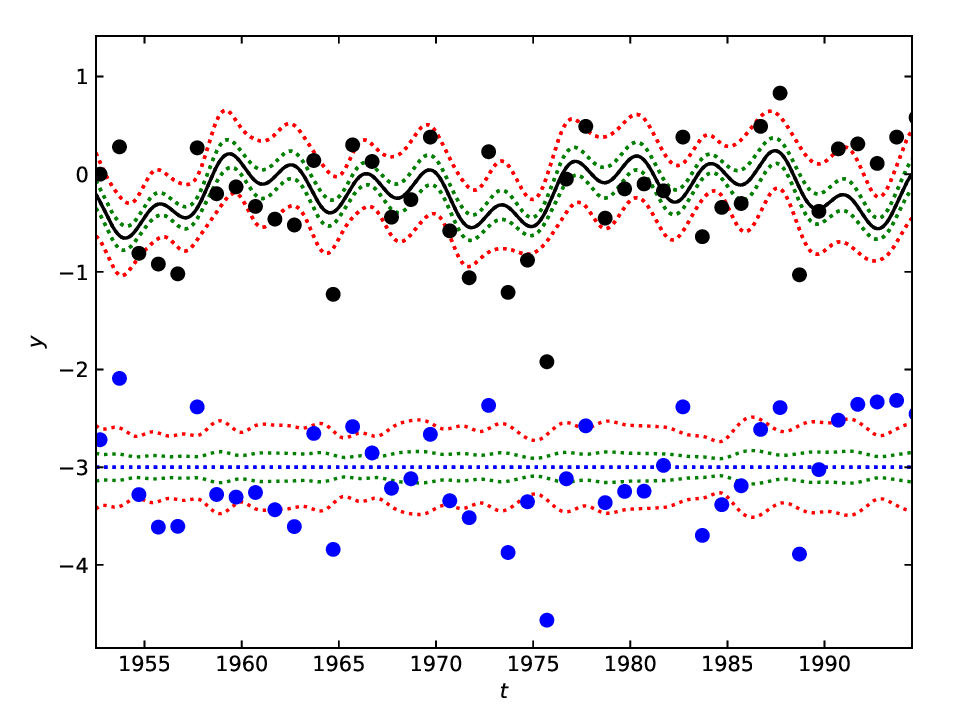}
 \hspace*{-0.01\textwidth}
 \includegraphics[width=0.455\textwidth,clip=]{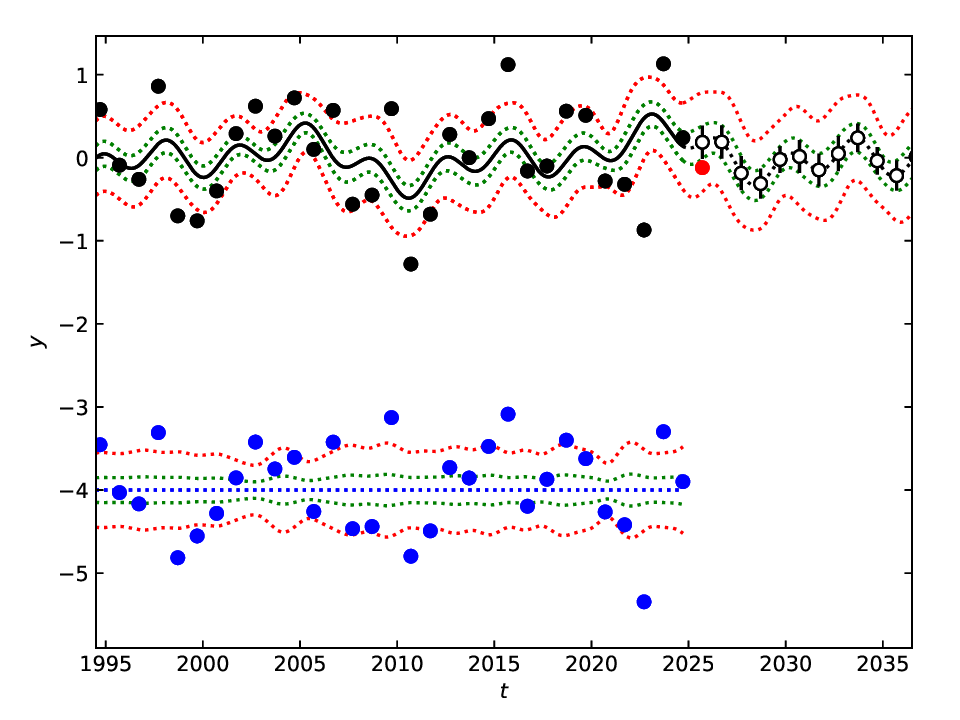}
}
\vspace{-0.36\textwidth}
\centerline{\Large 
\hspace{0.44\textwidth}  \color{black}{(c)}
\hspace{0.41\textwidth}  \color{black}{(d)}
\hfill}
\vspace{0.32\textwidth}
\caption{September data model \M=3 in Table \ref{TableSep}.
Otherwise as in Figure \ref{FigJan}.}
\label{FigSep}
\end{figure*}

\begin{table*}
  \caption{Periods in September data. Otherwise as in
    Table \ref{TableJan}.}
  \label{TableSep}
  \begin{tiny}
     \begin{center}
      \begin{tabular}{lcccccccccc}
  \hline
 & & \multicolumn{4}{c}{Period analysis} & & \multicolumn{3}{c}{Fisher-test  $(\gamma=0.05)$} &  \\
        \multicolumn{11}{c}{Data: Original non-weighted data ($n=155, \Delta T=154$: \PR{Sep.dat})} \\
(1)    & 
(2)    & 
(3)    & 
(4)    & 
(5)    & 
(6)    & 
        & 
(7)    & 
(8)    & 
(9)    &
(10)        \\ 
\M                    &
  &
$P_1$ (y)             &
$P_2$ (y)             &
$P_3$ (y)             &
$P_4$ (y)             &
                           &
 \M=2                  &
 \M=3                  &
 \M=4                  &
                      \\ 
             &
$\eta$ (-)   &
$A_1$ (\Cc)    &
$A_2$ (\Cc)    &
$A_3$ (\Cc)    &
$A_4$ (\Cc)    &
             &
$F_{R}$ (-)    &
$F_{R}$ (-)    &
$F_{R}$(-)    &
               Control file \\ 
                       &
$R$ (-)                    &
$t_{\mathrm{min,1}}$ (y) &
$t_{\mathrm{min,1}}$ (y) &
$t_{\mathrm{min,1}}$ (y) &
$t_{\mathrm{min,1}}$ (y) &
                       &
$Q_F$ (-)              &
$Q_F$ (-)              &
$Q_F$ (-)       \\ 
\hline 
        \multicolumn{10}{c}{One signal} \\
\hline
        \M=1             &                  &$9.12\PM 0.23       $&     -                  & -                        & -                     &&$\uparrow         $&$\uparrow        $&$\uparrow$&                                             \\
\RModel{1,1,1}     &       5         &$0.35 \PM 0.11     $&     -                  & -                        & -                     &&     2.84               &   2.81              &       2.78       &               \PR{SepR14K111.dat}  \\
                             &     43.5     &$1873.2\PM1.1      $&     -                  & -                        & -                     &&     0.040             &   0.013             &  0.0050        &                                             \\
\hline
       \multicolumn{10}{c}{Two signals} \\
\hline 
\M=2                    &                   &$3.546\PM0.032      $&$20.1\PM1.2        $& -                       & -                     &&-                         &$\uparrow        $&$\uparrow$&                                              \\
\RModel{2,1,1}    &        8         &$0.35 \PM0.11         $&$0.35\PM0.11$&   -                   & -                   &&-                         &  2.70                &         2.66    &               \PR{Sep14K211.dat}   \\
                            &     41.1        &$1872.32\PM0.35  $&$1872.3\PM2.4$& -                      & -                     &&-                           &  0.048               &    0.018  &                                               \\
\hline
      \multicolumn{10}{c}{Three signals} \\
\hline 
\M=3                    &                   &$3.546\PM 0.018 $&$9.13\PM0.21  $   &$19.99\PM0.89        $& -                     &&-              & -                           &$\leftarrow$&                                              \\
\RModel{3,1,1}    &       11        &$0.35\PM 0.11     $&$0.34\PM0.13      $&$0.341\PM 0.083   $& -                     &&-              & -                           &    2.53              & \PR{SepR14K311.dat}            \\
                            &     38.9      &$1872.31\PM0.47  $&$1873.18\PM0.81   $&$1872.6\PM 2.3 $& -                     &&-             & -                            &   0.060       &                                               \\
\hline
      \multicolumn{10}{c}{Four signals} \\
\hline 
\M=4                    &                     &$3.545\PM 0.025     $&$4.749\PM 0.071 $&$9.12\PM 0.11       $&$19.92\PM 0.93 $&&-                 & -             & - &                                              \\
\RModel{4,1,1}    &       14          &$0.35   \PM 0.12     $&$0.32\PM 0.11       $&$0.343\PM 0.079  $&$0.340\PM 0.082 $&&-                   & -             &- &       \PR{SepR14K411.dat}   \\
                            &     36.9          &$1872.35\PM 0.27   $&$1875.18\PM0.74$&$1873.2PM 1.4    $&$1872.7\PM 2.3$&&-                        & -             &- &                                               \\
\hline
\end{tabular}
\end{center}
\end{tiny}
\addtolength{\tabcolsep}{+0.05cm}
\end{table*}

\clearpage

 \begin{figure*}  
\vspace{0.02\textwidth}
\centerline{\hspace*{0.005\textwidth}
 \includegraphics[width=0.455\textwidth,clip=]{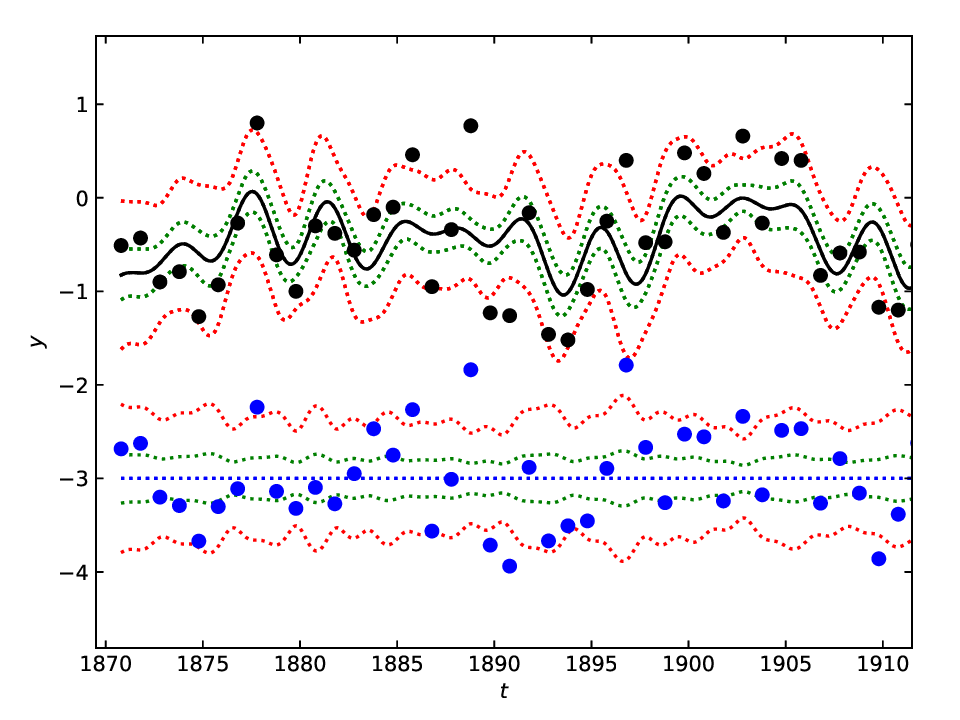}
 \hspace*{-0.01\textwidth}
 \includegraphics[width=0.455\textwidth,clip=]{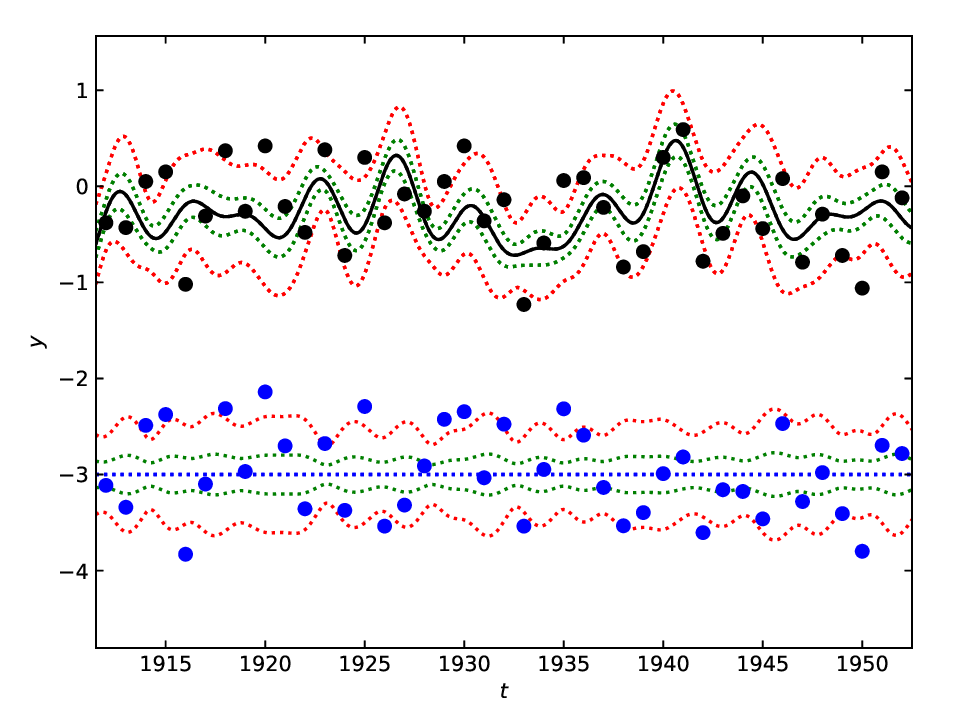}
 }
\vspace{-0.36\textwidth}
\centerline{\Large 
\hspace{0.44\textwidth}  \color{black}{(a)}
\hspace{0.41\textwidth}  \color{black}{(b)}
\hfill}
\vspace{0.35\textwidth}
\centerline{\hspace*{0.005\textwidth}
 \includegraphics[width=0.455\textwidth,clip=]{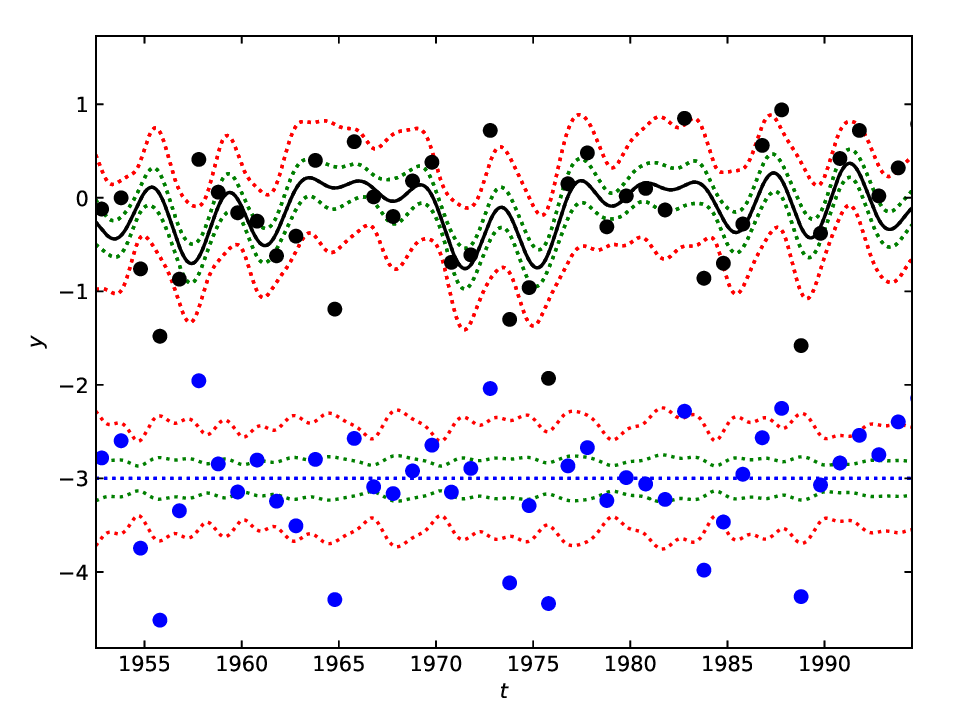}
 \hspace*{-0.01\textwidth}
 \includegraphics[width=0.455\textwidth,clip=]{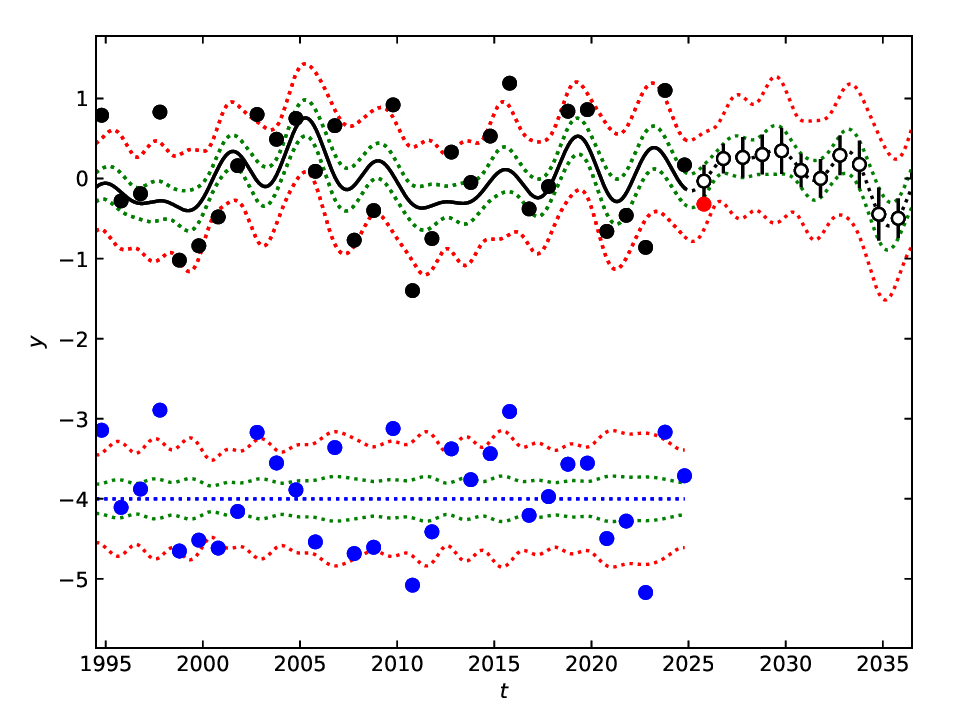}
}
\vspace{-0.36\textwidth}
\centerline{\Large 
\hspace{0.44\textwidth}  \color{black}{(c)}
\hspace{0.41\textwidth}  \color{black}{(d)}
\hfill}
\vspace{0.32\textwidth}
\caption{October data model \M=4 in Table \ref{TableOct}.
  Otherwise as in Figure \ref{FigJan}.}
\label{FigOct}
\end{figure*}

\begin{table*}
  \caption{Periods in October data. Otherwise as in
    Table \ref{TableJan}.}
  \label{TableOct}
  \begin{tiny}
     \begin{center}
      \begin{tabular}{lcccccccccc}
  \hline
 & & \multicolumn{4}{c}{Period analysis} & & \multicolumn{3}{c}{Fisher-test  $(\gamma=0.05)$} &  \\
        \multicolumn{11}{c}{Data: Original non-weighted data ($n=155, \Delta T=154$: \PR{Oct.dat})} \\
(1)    & 
(2)    & 
(3)    & 
(4)    & 
(5)    & 
(6)    & 
        & 
(7)    & 
(8)    & 
(9)    &
(10)        \\ 
\M                    &
  &
$P_1$ (y)             &
$P_2$ (y)             &
$P_3$ (y)             &
$P_4$ (y)             &
                           &
 \M=2                  &
 \M=3                  &
 \M=4                  &
                      \\ 
             &
$\eta$ (-)   &
$A_1$ (\Cc)    &
$A_2$ (\Cc)    &
$A_3$ (\Cc)    &
$A_4$ (\Cc)    &
             &
$F_{R}$ (-)    &
$F_{R}$ (-)    &
$F_{R}$(-)    &
               Control file \\ 
                       &
$R$ (-)                    &
$t_{\mathrm{min,1}}$ (y) &
$t_{\mathrm{min,1}}$ (y) &
$t_{\mathrm{min,1}}$ (y) &
$t_{\mathrm{min,1}}$ (y) &
                       &
$Q_F$ (-)              &
$Q_F$ (-)              &
$Q_F$ (-)       \\ 
\hline 
        \multicolumn{10}{c}{One signal} \\
\hline
        \M=1             &                  &$3.354\PM 0.028       $&     -                  & -                        & -                     &&$\uparrow         $&$\uparrow        $&$\uparrow$&                                             \\
\RModel{1,1,1}     &       5         &$0.38\PM 0.12          $&     -                  & -                        & -                     &&    2.51                 &      2.70            &   2.78              &               \PR{OctR14K111.dat}  \\
                             &     56.0     &$1872.38\PM0.46      $&     -                  & -                        & -                     &&     0.039               &     0.016             & 0.0049       &                                             \\
\hline
       \multicolumn{10}{c}{Two signals} \\
\hline 
\M=2                    &                   &$3.542\PM0.037      $&$4.576\PM0.046 $& -                       & -                     &&-                         &$\leftarrow        $&$\uparrow$&                                              \\
\RModel{2,1,1}    &        8         &$0.38  \PM0.12        $&$0.40\PM0.11$&   -                   & -                          &&-                            &   2.46               &    2.65       &               \PR{Oct14K211.dat}   \\
                            &     52.9        &$1872.41\PM0.56  $&$1874.92\PM0.61$& -                      & -                     &&-                           &   0.065              &  0.018       &                                               \\
\hline
      \multicolumn{10}{c}{Three signals} \\
\hline 
\M=3                    &                   &$3.542\PM 0.022 $&$4.578\PM0.047 $   &$12.82\PM0.47        $& -                     &&-              & -                           &$\uparrow$&                                              \\
\RModel{3,1,1}    &       11        &$0.381\PM 0.093  $&$0.41\PM0.11      $&$0.367\PM 0.093   $& -                     &&-              & -                           &   2.75               & \PR{OctR14K311.dat}            \\
                            &     50.3      &$1872.42\PM0.40  $&$1874.88\PM0.48   $&$1883.4\PM 2.0 $& -                     &&-             & -                            &       0.045    &                                               \\
\hline
      \multicolumn{10}{c}{Four signals} \\
\hline 
\M=4                    &                     &$3.542\PM 0.035     $&$5.579\PM 0.058 $&$12.84\PM 0.39   $&$20.4\PM 1.0 $&&-                 & -             & - &                                              \\
\RModel{4,1,1}    &       14          &$0.38   \PM 0.10     $&$0.40\PM 0.15      $&$0.38\PM 0.13  $&$0.38\PM 0.11 $&&-                   & -             &- &       \PR{OctR14K411.dat}   \\
                            &     47.5          &$1872.44\PM 0.46   $&$1874.86\PM0.47$&$1883.0\PM 1.4 $&$1872.0\PM 2.7$&&-                        & -             &- &                                               \\
\hline
\end{tabular}
\end{center}
\end{tiny}
\addtolength{\tabcolsep}{+0.05cm}
\end{table*}

\clearpage

 \begin{figure*}  
\vspace{0.02\textwidth}
\centerline{\hspace*{0.005\textwidth}
 \includegraphics[width=0.455\textwidth,clip=]{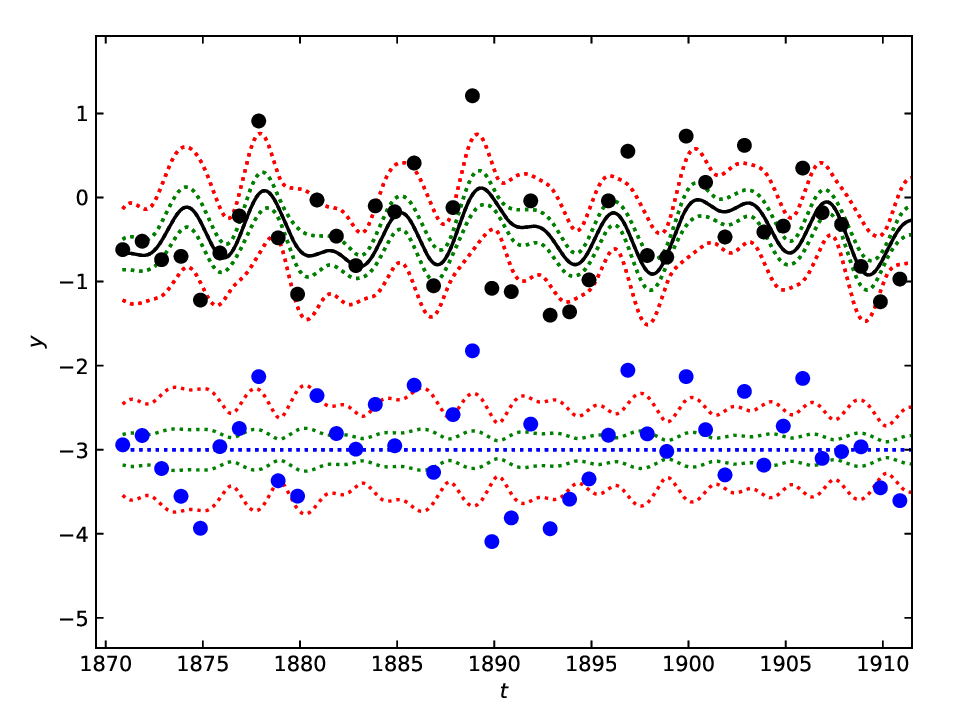}
 \hspace*{-0.01\textwidth}
 \includegraphics[width=0.455\textwidth,clip=]{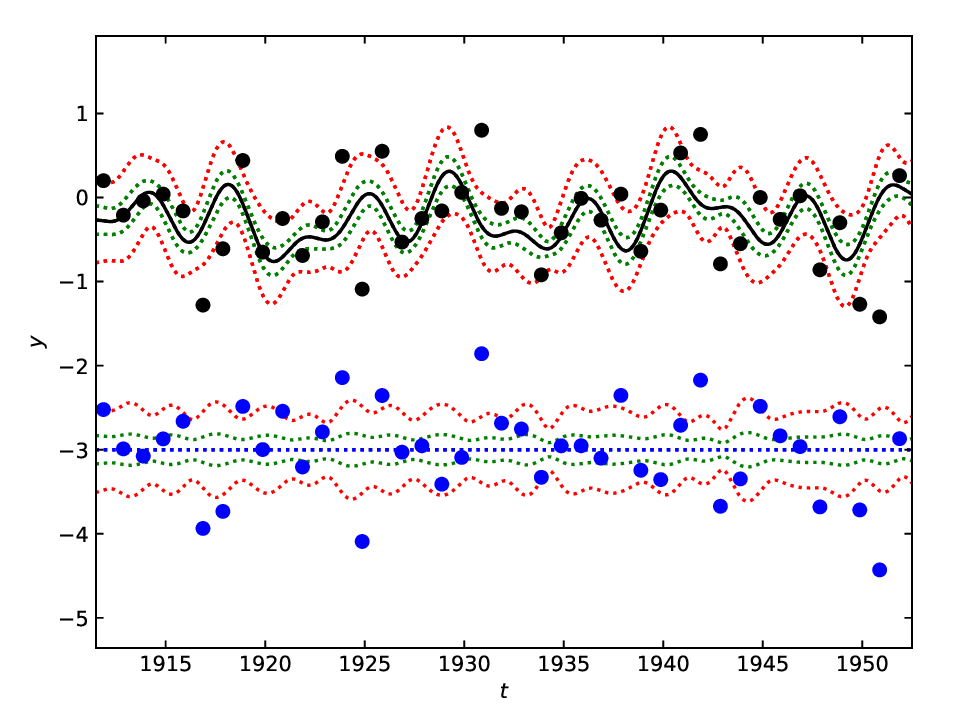}
 }
\vspace{-0.36\textwidth}
\centerline{\Large 
\hspace{0.44\textwidth}  \color{black}{(a)}
\hspace{0.41\textwidth}  \color{black}{(b)}
\hfill}
\vspace{0.35\textwidth}
\centerline{\hspace*{0.005\textwidth}
 \includegraphics[width=0.455\textwidth,clip=]{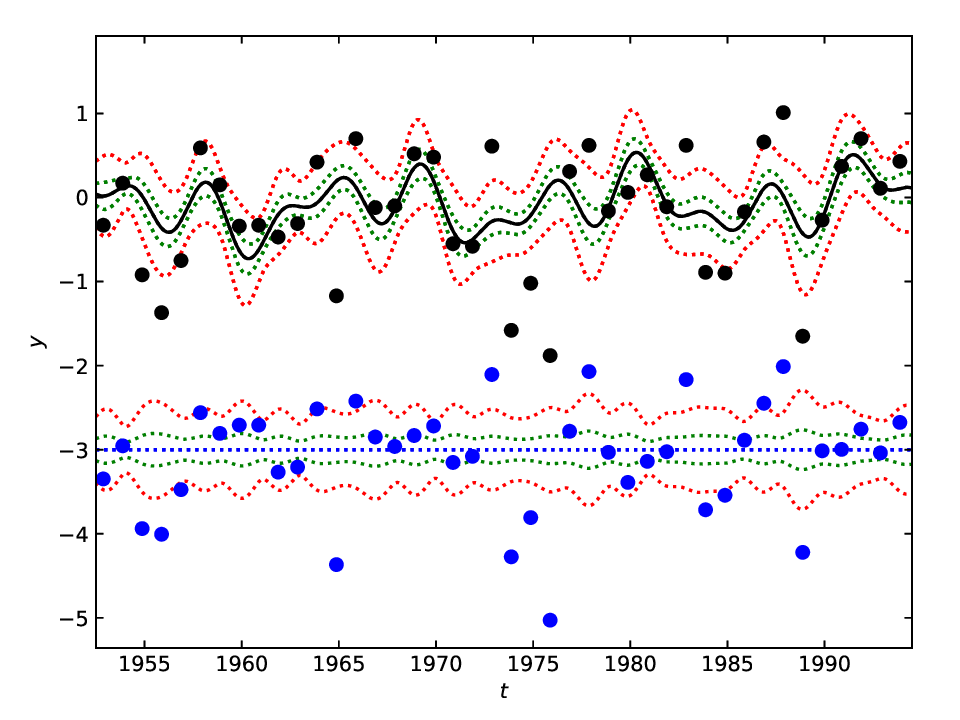}
 \hspace*{-0.01\textwidth}
 \includegraphics[width=0.455\textwidth,clip=]{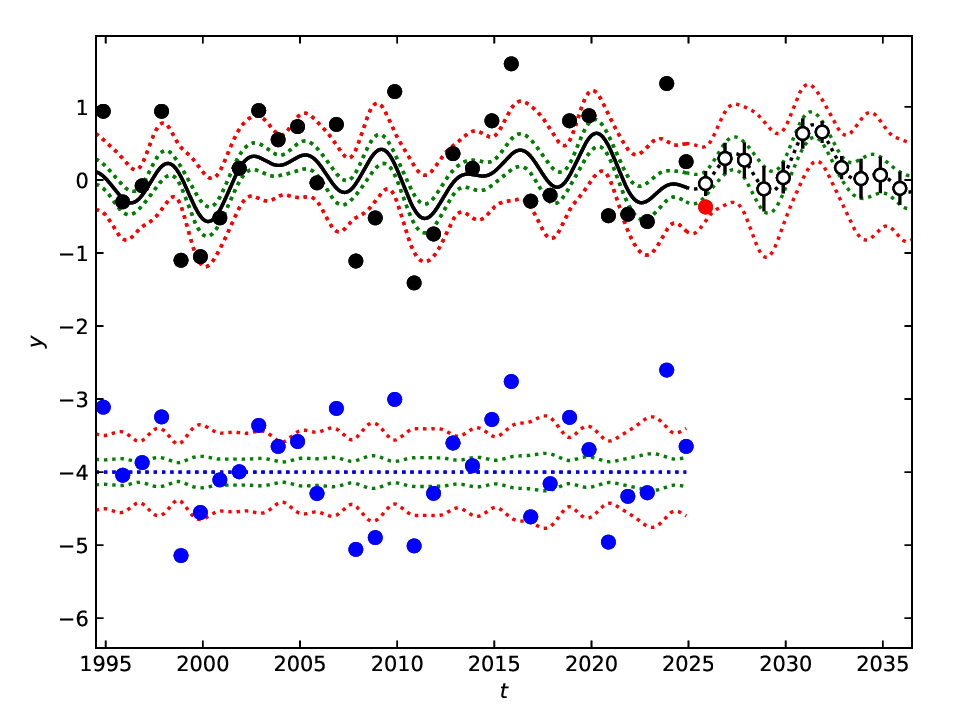}
}
\vspace{-0.36\textwidth}
\centerline{\Large 
\hspace{0.44\textwidth}  \color{black}{(c)}
\hspace{0.41\textwidth}  \color{black}{(d)}
\hfill}
\vspace{0.32\textwidth}
\caption{November data model \M=3 in Table \ref{TableNov}.
Otherwise as in Figure \ref{FigJan}.
}
\label{FigNov}
\end{figure*}

\begin{table*}
  \caption{Periods in November data. Otherwise as in
    Table \ref{TableJan}.}
  \label{TableNov}
  \begin{tiny}
     \begin{center}
      \begin{tabular}{lcccccccccc}
  \hline
 & & \multicolumn{4}{c}{Period analysis} & & \multicolumn{3}{c}{Fisher-test  $(\gamma=0.05)$} &  \\
        \multicolumn{11}{c}{Data: Original non-weighted data ($n=155, \Delta T=154$: \PR{Nov.dat})} \\
(1)    & 
(2)    & 
(3)    & 
(4)    & 
(5)    & 
(6)    & 
        & 
(7)    & 
(8)    & 
(9)    &
(10)        \\ 
\M                    &
  &
$P_1$ (y)             &
$P_2$ (y)             &
$P_3$ (y)             &
$P_4$ (y)             &
                           &
 \M=2                  &
 \M=3                  &
 \M=4                  &
                      \\ 
             &
$\eta$ (-)   &
$A_1$ (\Cc)    &
$A_2$ (\Cc)    &
$A_3$ (\Cc)    &
$A_4$ (\Cc)    &
             &
$F_{R}$ (-)    &
$F_{R}$ (-)    &
$F_{R}$(-)    &
               Control file \\ 
                       &
$R$ (-)                    &
$t_{\mathrm{min,1}}$ (y) &
$t_{\mathrm{min,1}}$ (y) &
$t_{\mathrm{min,1}}$ (y) &
$t_{\mathrm{min,1}}$ (y) &
                       &
$Q_F$ (-)              &
$Q_F$ (-)              &
$Q_F$ (-)       \\ 
\hline 
        \multicolumn{10}{c}{One signal} \\
\hline
        \M=1             &                  &$5.665\PM 0.054       $&     -                  & -                        & -                     &&$\uparrow         $&$\uparrow        $&$\uparrow$&                                             \\
\RModel{1,1,1}     &       5         &$0.47\PM 0.12           $&     -                  & -                        & -                     &&  2.86                  &    3.02             &    3.08        &               \PR{NovR14K111.dat}  \\
                             &     64.8      &$1876.05\PM0.66      $&     -                  & -                        & -                     &&   0.039                &   0.0081           & 0.0021       &                                             \\
\hline
       \multicolumn{10}{c}{Two signals} \\
\hline 
\M=2                    &                   &$3.647\PM0.032      $&$5.668\PM0.046 $& -                       & -                     &&-                         &$\uparrow        $&$\uparrow$&                                              \\
\RModel{2,1,1}    &        8         &$0.43  \PM0.14        $&$0.47\PM0.11$&   -                   & -                     &&-                           &         3.07       &           3.06    &               \PR{Nov14K211.dat}   \\
                            &     61.2 &$1872.54\PM0.48  $&$1876.01\PM0.68$& -                      & -                     &&-                           &          0.039       &        0.0076  &                                               \\
\hline
      \multicolumn{10}{c}{Three signals} \\
\hline 
\M=3                    &                   &$3.647\PM 0.041 $&$5.670\PM0.049 $   &$12.86\PM0.63    $& -                     &&-              & -                           &$\leftarrow$&                                              \\
\RModel{3,1,1}    &       11        &$0.44\PM 0.14      $&$0.46\PM0.12        $&$0.43\PM 0.13    $& -                     &&-              & -                           & 2.67          & \PR{NovR14K311.dat}            \\
                            &     57.5      &$1872.56\PM0.44  $&$1875.96\PM0.59  $&$1883.1\PM1.2 $& -                     &&-             & -                            &  0.050       &                                               \\
\hline
      \multicolumn{10}{c}{Four signals} \\
\hline 
\M=4                    &                     &$3.647\PM 0.029     $&$5.667\PM 0.072  $&$9.095\PM 0.090$&$12.90\PM0.31 $&&-                 & -             & - &                                              \\
\RModel{4,1,1}    &       14          &$0.43   \PM 0.13     $&$0.47\PM 0.15       $&$0.42\PM 0.12  $&$0.432\PM 0.096 $&&-                   & -             &- &       \PR{NovR14K411.dat}   \\
                            &     54.1          &$1872.55\PM 0.41 $&$1876.01\PM0.67$&$1873.54\PM 0.78 $&$1882.8\PM 1.7$&&-                        & -             &- &                                               \\
\hline
\end{tabular}
\end{center}
\end{tiny}
\addtolength{\tabcolsep}{+0.05cm}
\end{table*}

\clearpage

 \begin{figure*}  
\vspace{0.02\textwidth}
\centerline{\hspace*{0.005\textwidth}
 \includegraphics[width=0.455\textwidth,clip=]{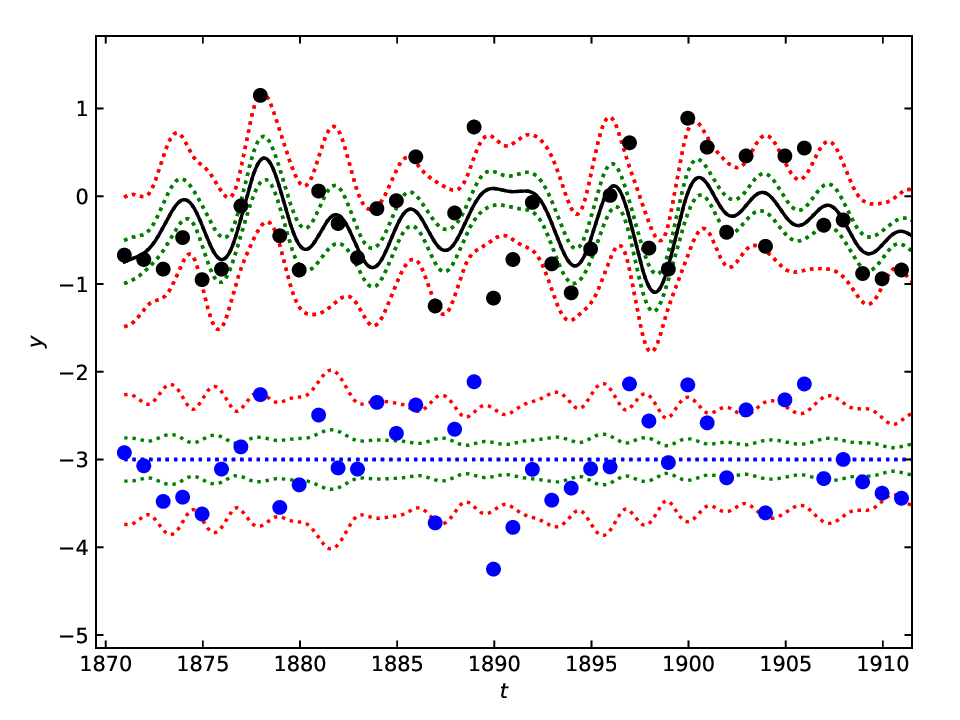}
 \hspace*{-0.01\textwidth}
 \includegraphics[width=0.455\textwidth,clip=]{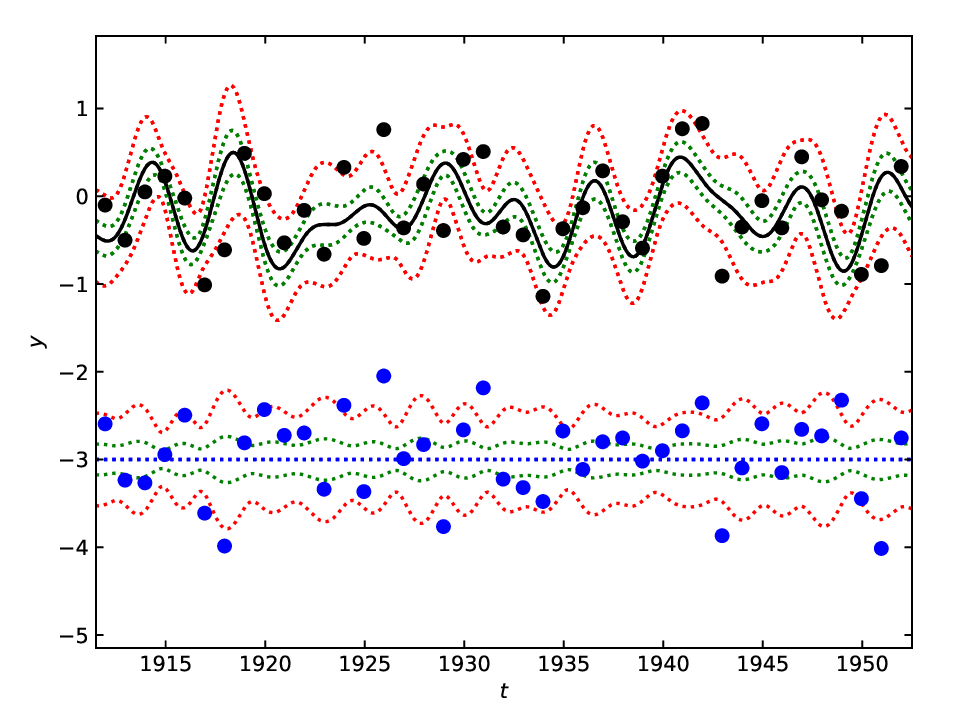}
 }
\vspace{-0.36\textwidth}
\centerline{\Large 
\hspace{0.44\textwidth}  \color{black}{(a)}
\hspace{0.41\textwidth}  \color{black}{(b)}
\hfill}
\vspace{0.35\textwidth}
\centerline{\hspace*{0.005\textwidth}
 \includegraphics[width=0.455\textwidth,clip=]{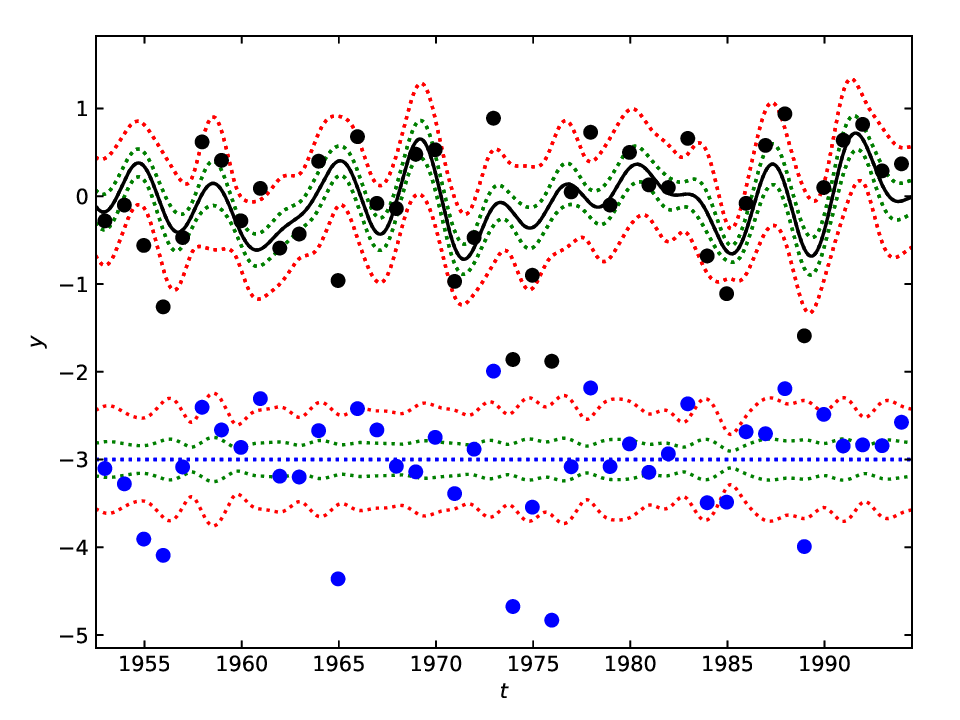}
 \hspace*{-0.01\textwidth}
 \includegraphics[width=0.455\textwidth,clip=]{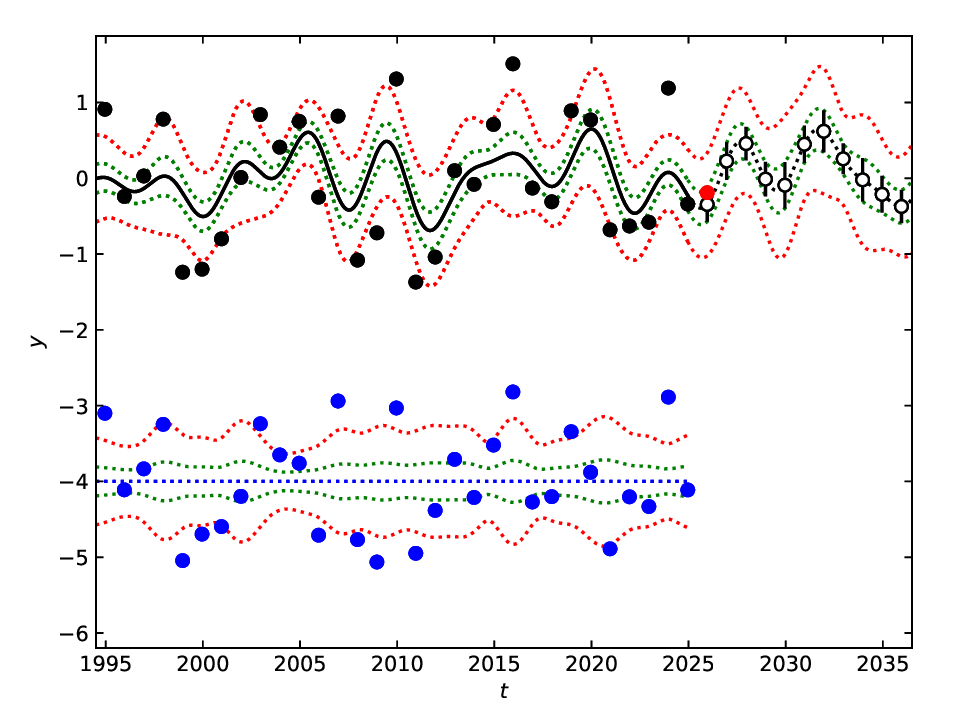}
}
\vspace{-0.36\textwidth}
\centerline{\Large 
\hspace{0.44\textwidth}  \color{black}{(c)}
\hspace{0.41\textwidth}  \color{black}{(d)}
\hfill}
\vspace{0.32\textwidth}
\caption{December data model \M=4 in Table \ref{TableDec}.
  Otherwise as in Figure \ref{FigJan}.
}
\label{FigDec}
\end{figure*}

\begin{table*}
  \caption{Periods in December data. Otherwise as in
    Table \ref{TableJan}.   }
  \label{TableDec}
  \begin{tiny}
     \begin{center}
      \begin{tabular}{lcccccccccc}
  \hline
 & & \multicolumn{4}{c}{Period analysis} & & \multicolumn{3}{c}{Fisher-test  $(\gamma=0.05)$} &  \\
        \multicolumn{11}{c}{Data: Original non-weighted data ($n=155, \Delta T=154$: \PR{Dec.dat})} \\
 (1)    & 
(2)    & 
(3)    & 
(4)    & 
(5)    & 
(6)    & 
        & 
(7)    & 
(8)    & 
(9)    &
(10)        \\ 
M                    &
  &
$P_1$ (y)             &
$P_2$ (y)             &
$P_3$ (y)             &
$P_4$ (y)             &
                           &
 \M=2                  &
 \M=3                  &
 \M=4                  &
                      \\ 
             &
$\eta$ (-)   &
$A_1$ (\Cc)    &
$A_2$ (\Cc)    &
$A_3$ (\Cc)    &
$A_4$ (\Cc)    &
             &
$F_{R}$ (-)    &
$F_{R}$ (-)    &
$F_{R}$(-)    &
               Control file \\ 
                       &
$R$ (-)                    &
$t_{\mathrm{min,1}}$ (y) &
$t_{\mathrm{min,1}}$ (y) &
$t_{\mathrm{min,1}}$ (y) &
$t_{\mathrm{min,1}}$ (y) &
                       &
$Q_F$ (-)              &
$Q_F$ (-)              &
$Q_F$ (-)       \\ 
\hline 
        \multicolumn{10}{c}{One signal} \\
\hline
        \M=1             &                  &$3.645\PM 0.031       $&     -                  & -                        & -                     &&$\uparrow         $&$\uparrow        $&$\uparrow$&                                             \\
\RModel{1,1,1}     &       5         &$0.48\PM 0.13           $&     -                  & -                        & -                     &&      4.06              &   3.71               &  3.61               &               \PR{DecR14K111.dat}  \\
                             &     63.7     &$1872.66\PM0.46      $&     -                  & -                        & -                     &&     0.0084            &   0.0018          &  0.00045        &                                             \\
\hline
       \multicolumn{10}{c}{Two signals} \\
\hline 
\M=2                    &                   &$3.645\PM0.028      $&$12.75\PM0.42 $& -                       & -                     &&-                         &$\uparrow        $&$\uparrow$&                                              \\
\RModel{2,1,1}    &        8         &$0.49  \PM0.15        $&$0.50\PM0.11   $&   -                   & -                     &&-                           &     3.20                   &     3.20             &               \PR{Dec14K211.dat}   \\
                            &     58.8        &$1872.67\PM0.43  $&$1871.3\PM1.4$& -                      & -                     &&-                           &      0.025               &    0.0056   &                                               \\
\hline
      \multicolumn{10}{c}{Three signals} \\
\hline 
\M=3                    &                   &$3.643\PM 0.029 $&$5.664\PM0.090 $   &$12.74\PM0.18        $& -                     &&-              & -                           &$\uparrow$&                                              \\
\RModel{3,1,1}    &       11        &$0.49\PM 0.10     $&$0.43\PM0.10       $&$0.50\PM 0.11   $& -                     &&-              & -                           &        3.06          & \PR{DecR14K311.dat}            \\
                            &     55.1      &$1872.69\PM0.38  $&$1876.19\PM0.61   $&$1871.35\PM 0.98 $& -                     &&-             & -                            &      0.030      &                                               \\
\hline
      \multicolumn{10}{c}{Four signals} \\
\hline 
\M=4                    &                     &$3.644\PM 0.040     $&$4.566\PM 0.020 $&$5.664\PM 0.058$&$12.74\PM 0.32 $&&-                 & -             & - &                                              \\
\RModel{4,1,1}    &       14          &$0.49   \PM 0.10      $&$0.42\PM 0.10     $&$0.428\PM 0.092  $&$0.50\PM 0.12 $&&-                   & -             &- &       \PR{DecR14K411.dat}   \\
                            &     51.7          &$1872.71\PM 0.38   $&$1875.35\PM0.59$&$1876.22\PM 0.80 $&$1871.4\PM 1.2$&&-                        & -             &- &                                               \\
\hline
\end{tabular}
\end{center}
\end{tiny}
\addtolength{\tabcolsep}{+0.10cm}
\end{table*}

\begin{table*}
  \caption{Separate months. \BigWave ~forecast
    (Figures \ref{FigJan}-\ref{FigDec}: open circles).
 Otherwise as in Table  \ref{TablemAllForecast}. } \label{TableCollect}
  \begin{tiny}
    \addtolength{\tabcolsep}{-0.15cm}
    \begin{center}
   \begin{tabular}{rrrrrrrrrrrrrrrrrrrrrrrrr}
     \hline
     (1) & (2) & & (3) & & (4) & & (5) & & (6) & & (6) & & (8) & & (9) & & (10) & & (11) & & (12) & & (13) & \\
      Year & Jan & & Feb & & Mar & & Apr & & May & & Jun & & Jul & & Aug & & Sep & & Oct & & Nov & & Dec \\
      (y)        &(\Cc)&&(\Cc)&  &(\Cc)& &(\Cc)&  &(\Cc)&  & (\Cc)& &(\Cc)& &(\Cc)& &(\Cc)& &(\Cc)& &(\Cc)& &(\Cc) \\
     \hline
   2025 &$    0.3 \PM  0.3 $&  \UU &$    0.4 \PM  0.3 $&  \UU &
       $   -0.2 \PM  0.2 $&  \DD &$   -0.1 \PM  0.2 $&  \UU &
       $   -0.2 \PM  0.2 $&  \DD &$   -0.3 \PM  0.1 $&  \DD &
       $    0.2 \PM  0.2 $&  \UU &$    0.0 \PM  0.2 $&  \DD &
       $    0.2 \PM  0.2 $&  \UU &$   -0.0 \PM  0.2 $&  \DD &
       $   -0.0 \PM  0.2 $&  \DD &$   -0.3 \PM  0.2 $&  \DD \\
  2026 &$    0.0 \PM  0.2 $&  \UU &$    0.1 \PM  0.2 $&  \UU &
       $    0.0 \PM  0.2 $&  \DD &$   -0.1 \PM  0.1 $&  \DD &
       $    0.3 \PM  0.2 $&  \UU &$   -0.3 \PM  0.1 $&  \DD &
       $    0.1 \PM  0.2 $&  \UU &$    0.2 \PM  0.2 $&  \UU &
       $    0.2 \PM  0.2 $&  \UU &$    0.2 \PM  0.2 $&  \UU &
       $    0.3 \PM  0.2 $&  \UU &$    0.2 \PM  0.2 $&  \DD \\
  2027 &$   -0.3 \PM  0.3 $&  \DD &$   -0.2 \PM  0.2 $&  \UU &
       $    0.4 \PM  0.2 $&  \UU &$    0.1 \PM  0.2 $&  \DD &
       $    0.4 \PM  0.2 $&  \UU &$   -0.1 \PM  0.1 $&  \DD &
       $   -0.1 \PM  0.2 $&  \UU &$    0.4 \PM  0.2 $&  \UU &
       $   -0.2 \PM  0.2 $&  \DD &$    0.3 \PM  0.3 $&  \UU &
       $    0.3 \PM  0.2 $&  \UU &$    0.5 \PM  0.2 $&  \UU \\
  2028 &$    0.0 \PM  0.4 $&  \DD &$    0.0 \PM  0.2 $&  \DD &
       $    0.2 \PM  0.2 $&  \UU &$    0.2 \PM  0.1 $&  \UU &
       $   -0.0 \PM  0.2 $&  \DD &$    0.1 \PM  0.2 $&  \UU &
       $    0.1 \PM  0.1 $&  \UU &$    0.2 \PM  0.2 $&  \UU &
       $   -0.3 \PM  0.2 $&  \DD &$    0.3 \PM  0.2 $&  \UU &
       $   -0.1 \PM  0.3 $&  \DD &$   -0.0 \PM  0.2 $&  \UU \\
  2029 &$    0.3 \PM  0.3 $&  \UU &$    0.3 \PM  0.2 $&  \DD &
       $   -0.2 \PM  0.2 $&  \DD &$    0.1 \PM  0.2 $&  \UU &
       $    0.0 \PM  0.2 $&  \DD &$    0.2 \PM  0.2 $&  \UU &
       $    0.5 \PM  0.1 $&  \UU &$    0.1 \PM  0.1 $&  \DD &
       $   -0.0 \PM  0.2 $&  \DD &$    0.3 \PM  0.3 $&  \UU &
       $    0.0 \PM  0.2 $&  \DD &$   -0.1 \PM  0.3 $&  \DD \\
  2030 &$    0.2 \PM  0.3 $&  \UU &$    0.3 \PM  0.2 $&  \UU &
       $    0.1 \PM  0.2 $&  \DD &$    0.1 \PM  0.2 $&  \UU &
       $    0.5 \PM  0.2 $&  \UU &$    0.4 \PM  0.1 $&  \DD &
       $    0.3 \PM  0.2 $&  \DD &$    0.4 \PM  0.2 $&  \UU &
       $    0.0 \PM  0.2 $&  \DD &$    0.1 \PM  0.2 $&  \UU &
       $    0.6 \PM  0.2 $&  \UU &$    0.4 \PM  0.2 $&  \DD \\
  2031 &$    0.3 \PM  0.3 $&  \DD &$    0.3 \PM  0.2 $&  \UU &
       $    0.6 \PM  0.1 $&  \UU &$    0.6 \PM  0.2 $&  \DD &
       $    0.4 \PM  0.1 $&  \DD &$    0.3 \PM  0.1 $&  \DD &
       $   -0.1 \PM  0.1 $&  \DD &$    0.5 \PM  0.2 $&  \UU &
       $   -0.1 \PM  0.2 $&  \DD &$    0.0 \PM  0.2 $&  \UU &
       $    0.7 \PM  0.1 $&  \UU &$    0.6 \PM  0.3 $&  \DD \\
  2032 &$    0.4 \PM  0.2 $&  \DD &$    0.4 \PM  0.2 $&  \UU &
       $    0.4 \PM  0.2 $&  \DD &$    0.7 \PM  0.2 $&  \UU &
       $    0.2 \PM  0.2 $&  \DD &$    0.1 \PM  0.2 $&  \DD &
       $   -0.0 \PM  0.1 $&  \DD &$    0.1 \PM  0.2 $&  \UU &
       $    0.0 \PM  0.2 $&  \DD &$    0.3 \PM  0.2 $&  \UU &
       $    0.2 \PM  0.2 $&  \DD &$    0.3 \PM  0.2 $&  \UU \\
  2033 &$    0.1 \PM  0.3 $&  \DD &$    0.2 \PM  0.2 $&  \UU &
       $    0.0 \PM  0.2 $&  \DD &$    0.4 \PM  0.2 $&  \UU &
       $    0.4 \PM  0.2 $&  \DD &$    0.3 \PM  0.2 $&  \DD &
       $    0.3 \PM  0.1 $&  \DD &$   -0.0 \PM  0.2 $&  \DD &
       $    0.2 \PM  0.2 $&  \UU &$    0.2 \PM  0.3 $&  \DD &
       $    0.0 \PM  0.3 $&  \DD &$   -0.0 \PM  0.3 $&  \DD \\
  2034 &$   -0.2 \PM  0.3 $&  \DD &$   -0.2 \PM  0.2 $&  \UU &
       $   -0.0 \PM  0.2 $&  \UU &$    0.0 \PM  0.2 $&  \UU &
       $    0.2 \PM  0.2 $&  \UU &$    0.5 \PM  0.2 $&  \UU &
       $    0.1 \PM  0.2 $&  \DD &$    0.1 \PM  0.2 $&  \UU &
       $   -0.0 \PM  0.2 $&  \DD &$   -0.4 \PM  0.3 $&  \DD &
       $    0.1 \PM  0.2 $&  \UU &$   -0.2 \PM  0.2 $&  \DD \\
  2035 &$   -0.1 \PM  0.3 $&  \UU &$   -0.1 \PM  0.2 $&  \DD &
       $   -0.1 \PM  0.2 $&  \DD &$    0.1 \PM  0.2 $&  \UU &
       $   -0.4 \PM  0.2 $&  \DD &$    0.3 \PM  0.2 $&  \UU &
       $   -0.2 \PM  0.2 $&  \DD &$   -0.2 \PM  0.2 $&  \UU &
       $   -0.2 \PM  0.2 $&  \UU &$   -0.5 \PM  0.2 $&  \DD &
       $   -0.1 \PM  0.2 $&  \UU &$   -0.4 \PM  0.2 $&  \DD \\
  2036 &$   -0.0 \PM  0.3 $&  \UU &$    0.0 \PM  0.3 $&  \UU &
       $   -0.3 \PM  0.2 $&  \DD &$    0.1 \PM  0.2 $&  \UU &
       $   -0.1 \PM  0.2 $&  \DD &$   -0.2 \PM  0.2 $&  \DD &
       $   -0.2 \PM  0.2 $&  \UU &$   -0.5 \PM  0.1 $&  \DD &
       $    0.0 \PM  0.2 $&  \UU &$    0.0 \PM  0.3 $&  \UU &
       $   -0.1 \PM  0.2 $&  \DD &$   -0.1 \PM  0.2 $&  \UU \\
    \hline
    \end{tabular}
\end{center}
  \end{tiny}
\end{table*}

\begin{table}
  \caption{Signal \SignalOne, \SignalTwo, \SignalThree ~and
  \SignalNino ~connections.}
  \label{TableSynodic}
\begin{center}
  \begin{tabular}{cccccc}
    \hline
    \SignalTwo           & \SignalOne               & \SignalThree          &   \SignalNino    \\
    $p_1$                  &  $p_2$                      & $p_3$                   & $p_4$               \\
    $10 \Yy$              & $11\Yy$                   & $11.86\Yy$           & $12.78\Yy$      \\
    \hline
    $f_1$                    & $f_2$                       & $f_3$                    & $f_4$               \\
    $0.1000\Yy^{-1}$& $0.0909\Yy^{-1}$   & $0.0843\Yy^{-1}$ & $0.0782\Yy^{-1}$\\
    \hline
                              & $f_1-f_2$                &$f_2-f_3$                   & $f_3-f_4$     \\
                              & $0.0091\Yy^{-1}$  & $0.0066\Yy^{-1}$   &  $0.0061\Yy^{-1}$   \\
    \hline
  \end{tabular}
\end{center}
\end{table}

\end{document}